\documentclass{aa}

\usepackage{color}
\usepackage{graphicx}
\usepackage[caption=false]{subfig} 

\usepackage{amssymb,amsmath,wasysym}
\usepackage[utf8]{inputenc}
\usepackage{txfonts}

\usepackage{comment}

\newcommand*\kms{\ensuremath{\,{\rm km}\,{\rm s}^{-1}}}
\newcommand*\uvd{$uv$-data}
\def\Mspy {\ifmmode {M_{\odot} {\rm yr}^{-1}} \else $M_{\odot}$~yr$^{-1}$\fi}
\def\Mdot {\ifmmode {\dot M} \else $\dot M$\fi}

\newcommand*\ratran{\mbox{\textsc{ratran}}}
\newcommand*\codex{\mbox{\textsc{codex}}}
\newcommand*\codexplt{\mbox{\texttt{CODEX}}}

\definecolor{myblue}{RGB}{0,0,192}

\newlength{\singlemapwidth}
\newlength{\channelmapwidth}
\newlength{\firstchannelmapwidth}
\newlength{\miriadmapwidth}
\newlength{\modelwidth}
\newlength{\multispecwidth}
\newlength{\codexmodelheight}
\newlength{\codexspecheight}
\newlength{\spectraheight}
\begin{document}

\title{Resolving the extended atmosphere and the inner wind of \\ Mira ($o$ Ceti) with long ALMA baselines}
\author{K. T. Wong\inst{1}\thanks{Member of the International Max Planck Research School (IMPRS) for Astronomy and Astrophysics at the Universities of Bonn and Cologne.}
\and T. Kami\'{n}ski\inst{1,2}
\and K. M. Menten\inst{1}
\and F. Wyrowski\inst{1}}

\institute{Max-Planck-Institut f\"{u}r Radioastronomie, Auf dem H\"{u}gel 69, 53121, Bonn, Germany \\
\email{[ktwong; kmenten; wyrowski]@mpifr-bonn.mpg.de}
\and ESO, Alonso de C\'{o}rdova 3107, Vitacura, Casilla 19001, Santiago, Chile \\
\email{tkaminsk@eso.org}}

\titlerunning{Resolving the extended atmosphere and the inner wind of Mira ($o$ Ceti) with long ALMA baselines}
\authorrunning{Wong et al.}

\date{Received ; accepted}



\abstract
{High angular resolution (sub)millimetre observations of Asymptotic Giant Branch (AGB) stars, now possible with the Atacama Large Millimeter/submillimeter Array (ALMA), allow direct imaging of these object's photospheres. The physical properties of the molecular material around these regions, which so far had only been studied by imaging of maser emission and (spatially unresolved) absorption spectroscopy, can be probed with radiative transfer modelling and compared to hydrodynamical model predictions. The prototypical Mira variable, $o$~Cet (Mira), has been observed as a Science Verification target in the 2014 ALMA Long Baseline Campaign, offering the first opportunity to study these physical conditions in detail.}
{With the longest baseline of 15\,km, ALMA produces clearly resolved images of the continuum and molecular line emission/absorption at an angular resolution of ${\sim}30\,{\rm mas}$ at 220\,GHz. Models are constructed for Mira's extended atmosphere to investigate the physics and molecular abundances therein.}
{We image the data of $^{28}$SiO~${\varv}=0,~2$~$J=5-4$ and H$_2$O~$v_2=1$~$J_{K_a,K_c}=5_{5,0}-6_{4,3}$ transitions and extract spectra from various lines-of-sight towards Mira's extended atmosphere. In the course of imaging the emission/absorption, we encountered ambiguities in the resulting images and spectra that appear to be related to the performance of the CLEAN algorithm when applied to a combination of extended emission, and compact emission and absorption. We address these issues by a series of tests and simulations. We derive the gas density, kinetic temperature, molecular abundance and outflow/infall velocities in Mira's extended atmosphere by modelling the SiO and H$_2$O lines.}
{We resolve Mira’s millimetre continuum emission and our data are consistent with a radio photosphere with a brightness temperature of $2611 \pm 51\,{\rm K}$. In agreement with recent results obtained with the Very Large Array, we do not confirm the existence of a compact region ($<5\,{\rm mas}$) of enhanced brightness. Our modelling shows that SiO gas starts to deplete beyond $4\,R_{\star}$ and at a kinetic temperature of ${\lesssim}600\,{\rm K}$. The inner dust shells are probably composed of grain types other than pure silicates. During this ALMA observation, Mira's atmosphere generally exhibited infall motion, with a shock front of velocity ${\lesssim}12\,\kms$ outside the radio photosphere. Despite the chaotic nature of Mira's atmosphere, the structures predicted by the hydrodynamical model, {\codex}, can reproduce the observed spectra in astonishing detail; while some other models fail when confronted with the new data.}
{For the first time, millimetre-wavelength molecular absorption against the stellar continuum has been clearly imaged. Combined with radiative transfer modelling, the ALMA data successfully demonstrates the ability to reveal the physical conditions of the extended atmospheres and inner winds of AGB stars in unprecedented detail. Long-term monitoring of oxygen-rich evolved stars will be the key to understand the unsolved problem of dust condensation and the wind-driving mechanism.}

\maketitle


\section{Introduction}
\label{sec:intro}

Mira A ($o$ Ceti; Mira) is an oxygen-rich, long-period variable star on the asymptotic giant branch (AGB). Together with Mira B (VZ Ceti), possibly a white dwarf \citep{mirab_wd}, they form the symbiotic binary system Mira AB. Mira A is the prototype of Mira variables. Its period of visual brightness variation is about 332 days and the visual $V$-band magnitude of the star varies by up to about 8.1 mag (a factor of $>1700$) in each cycle (based on the data in the American Association of Variable Star Observers, AAVSO, International Database). The large variation in the visual magnitude is caused by a combined effect of stellar pulsation and variable opacity of metal oxides whose abundance changes with the effective temperature of the star \citep{reid2002}. The distance of the Mira AB system was estimated to be $110{\pm}9$\,pc \citep{haniff1995}, which is based on the period-luminosity relation derived by \citet{feast1989}, the infrared $K$-band magnitude from \citet{robertson1981} and the period of the visual $V$ variation from the GCVS \citep{kholopov1987}. We will adopt this, which is roughly consistent with the revised Hipparcos value of $92 \pm 10\,{\rm pc}$ \citep{hipparcos}, throughout this article.

Traditionally, AGB star atmospheres have been probed by molecular absorption spectroscopy, which delivers spatially unresolved line-of-sight information. Examples include the detection of the near-infrared H$_2$O absorption band from the warm molecular forming layer (known as the MOLsphere) around M giant stars and Mira variables with the Infrared Space Observatory (ISO) \citep[e.g.][]{tsuji1997,woitke1999,tsuji2000}. In addition, mid-infrared interferometry with the Very Large Telescope Interferometer (VLTI) can also probe the molecular layers and dust shells around these stars \citep[e.g.][]{ohnaka2005,karovicova2011}. SiO and/or H$_2$O maser emission in the extended atmospheres of Mira variables has been imaged, see, for examples, \citet{cotton2004} and \citet{perrin2015} with the Very Long Baseline Array (VLBA), and \citet{rm2007} with the Very Large Array (VLA).

In order to test the predictions by existing hydrodynamical models for the extended atmospheres of Mira variables, which typically has a radius of only a few $R_{\star}$ (a few tens of milli-arcseconds for Mira A), high angular and spectral resolution observations of the molecular emission and absorption from these regions are mandatory. The Atacama Large Millimeter/submillimeter Array (ALMA) with long baselines thus allows us to reach the required angular resolution at high sensitivity and to study the detailed kinematics of the innermost envelope of Mira A. Observations of radio and (sub)millimetre wavelength molecular line emission/absorption, in particular the rotational transitions not exhibiting strong masers, may be used to compare and test the predicted structures of the extended atmospheres by hydrodynamical models. Through modelling the radiative transfer of the transition lines with the predicted atmospheric structures as the inputs, synthesised spectra can be produced and compared to the observed ones. 

In this article, we present the new ALMA observations of the Mira AB system, which was selected as one of the Science Verification (SV) targets in the 2014 ALMA Long Baseline Campaign to demonstrate the high angular resolution capability of ALMA \citep{lbc2014}. Based on the visual magnitude data reported by the AAVSO, the stellar phase of Mira A is ${\sim}0.45$ at the time of this observation, and we will adopt this phase throughout the article. In Section 2, we describe the SV observation of Mira AB and the data processing. In Section 3, we present the results including the radio continuum data of Mira A and B in the SV dataset, and the images and spectra of the SiO and H$_2$O lines from Mira A as covered in the observations. In Section 4, we present our radiative transfer modelling results of the SiO and H$_2$O spectra of Mira A. In Section 5, we discuss the implications of our modelling results for our understanding of Mira A's extended atmosphere, including the structures, dust condensation process, shock dissipation and the kinematics and compare with predictions from hydrodynamical models.


\section{Observations and data processing}
\label{sec:obs}

The Mira AB system was observed with ALMA on 2014 October 17 and 25 (ALMA Band 3) and on 2014 October 29 and November 1 (ALMA Band 6) as part of the 2014 ALMA Long Baseline Campaign Science Verification with the longest baseline of $15.24\,{\rm km}$. \citep{lbc2014}. By referring to the AAVSO visual data for Mira, we find that the ALMA observations took place between the visual phases 0.42 (2014 Oct 17) and 0.47 (2014 Nov 01)\footnote{The period of Mira changes by a few days on the time span of decades \citep{templeton2009} and the dates of maximum need to be determined observationally for each cycle to provide a reliable phase scaling. In recent cycles, however, the phases preceding and during the maxima cannot be observed in the optical because Mira is too close to the Sun. To obtain the phases of the ALMA observations in 2014, we first phased the AAVSO data for the cycles in 2000--2013 with a period of 333 days and the date of maximum on JD\,2\,452\,161.0 which had a bright and very well-defined maximum (Kami\'{n}ski et al. in prep.). Then, the data covering the cycle of the 2014 ALMA observations were phased with the same period but their date of maximum (or phase 0) was determined by a phase shift of about $+0.05$ with respect to JD\,2\,452\,161.0 (as adopted for the 2000--2013 cycles). In this way we are able to match the data in the 2014 cycle with the phased data from 2000--2013 and therefore we obtained slightly later phases for the ALMA observations than those calculated from a periodicity scale given by AAVSO and used in \citet{mrm2015}.}.

The shortest baselines (and the maximum number of antennae) in the observations of Bands 3 and 6 are $29.07\,{\rm m}$ (38) and $15.23\,{\rm m}$ (39), respectively. The maximum recoverable scales\footnote{\label{footnote:mrs}defined to be $0.6 \times ({\text{wavelength}}/{\text{shortest baseline}})$ \citep{almaprimer}.} of the SiO lines in Bands 3 and 6 are therefore ${\sim}14{\farcs}8$ and ${\sim}11{\farcs}3$, respectively, and that of the H$_2$O $v_2=1$ line in Band 6 is $10{\farcs}5$. In Band 3, three continuum windows at 88.2, 98.2 and 100.2\,GHz were observed, in addition to four spectral line windows of 58.6\,MHz bandwidth around the transitions of $^{28}$SiO ${\varv} = 0,1,2$ $J=2-1$ and $^{29}$SiO ${\varv} = 0$ $J=2-1$. The channel width of the spectral windows is 61.0\,kHz (${\sim}0.21\,\kms$). In Band 6, a continuum window at 229.6\,GHz together with six spectral line windows of 117.2\,MHz bandwidth around $^{28}$SiO ${\varv} = 0,1,2$ $J=5-4$, $^{29}$SiO ${\varv} = 0$ $J=5-4$, H$_2$O $v_2=1$ $J_{K_a,K_c}=5_{5,0}-6_{4,3}$, and the H30$\alpha$ recombination line were observed. The channel width of the four SiO windows is 122.1\,kHz (${\sim}0.17\,\kms$) and that of the H$_2$O and H recombination line windows is 61.0\,kHz (${\sim}0.08\,\kms$). Table \ref{tab:lines} summarises the observed spectral lines and their parameters.

\begin{table*}[!htbp]
\caption{Observed spectral lines in ALMA Band 6.}
\label{tab:lines}
\centering
\begin{tabular}{lclrcc}
\hline\hline
Species & Spectral Line & Rest Frequency   & $E_{\rm up}/k$   & $\Delta V$ \\
        &               & \hspace{4ex} GHz & K \hspace{1ex}   & $\kms$     \\ \hline
SiO     & ${\varv} = 0, J=5-4$ & 217.104919  & 31.26    & 0.169 \\ 
SiO     & ${\varv} = 1, J=5-4$ & 215.596018  & 1800.17  & 0.170 \\ 
SiO     & ${\varv} = 2, J=5-4$ & 214.088575  & 3551.97  & 0.169 \\ 
$^{29}$SiO & ${\varv} = 0, J=5-4$ & 214.385752 & 30.87 & 0.171 \\ 
H$_2$O  & $\nu_2=1, J_{K_a,K_c}=5_{5,0}-6_{4,3}$ & 232.68670 & 3461.88  & 0.079 \\ 
\hline
\end{tabular}
\tablefoot{
Columns are, from left to right, molecule, quantum numbers of observed transition, rest frequency, energy level of the upper state, and channel width in velocity of the raw data. The SiO and $^{29}$SiO transition frequencies were taken form the Cologne Database for Molecular Spectroscopy (CDMS). They are based on laboratory data presented by \citet{mueller2013cdms} and have an accuracy of about 1\,kHz, except for the SiO ${\varv} = 0, J=5-4$ transition which has an accuracy of 2\,kHz. The H$_2$O transition frequency is taken from \citet{belov1987} and has an accuracy of 50\,kHz.}
\end{table*}

The SV data had been calibrated by staff members of the Joint ALMA Observatory (JAO) and the ALMA Regional Centres (ARCs), with the Common Astronomy Software Applications \citep[CASA;][]{casa} package\footnote{\url{http://casa.nrao.edu/}} version 4.2.2. Detailed calibration scripts, preliminarily calibrated data products (i.e., without self-calibration), and self-calibration solutions for both continuum and spectral line data are available at the ALMA Science Portal\footnote{\url{http://www.almascience.org/almadata/sciver/MiraBand6/}}. Self-calibration solutions were derived from the continuum data for the continuum data itself; and from the strongest spectral channels of the $^{28}$SiO ${\varv} = 1$ data, which exhibits strong maser emission, for the spectral line data. We downloaded the self-calibration solutions from the ALMA Science Portal and applied them to the preliminarily calibrated data, and then imaged the continuum and spectral line data. We use CASA version 4.2.2 for the self-calibration and imaging (except for the image binning task as mentioned below), and the Miriad package\footnote{\url{http://www.atnf.csiro.au/computing/software/miriad/}} for our continuum analysis (Sect. \ref{sec:result_cont} and Appendix \ref{sec:appendix_cont} \citep{miriad}).

%
%
We have determined the centre of Mira A's continuum emission to be at $(\text{RA}, \text{Dec}) = (02^{\rm h}19^{\rm m}20{\fs}795, -02{\degr}58{\arcmin}43{\farcs}05) = (34{\fdg}836\,646, -2{\fdg}978\,625)$ by fitting its image, produced from the visibility data before self-calibration, in the 229.6\,GHz continuum windows (i.e., spectral windows $\texttt{spw}=0, 7$ in the SV dataset). We adopt these coordinates as the absolute position of Mira A. The position and proper motion of Mira A in the Hipparcos Catalogue are $(34{\fdg}836\,611, -2{\fdg}977\,061)$ at the Julian epoch 1991.25 and $(9.33 \pm 1.99, -237.36 \pm 1.58)\,{\rm mas}\,{\rm yr}^{-1}$, respectively \citep{hipparcos}. At the epoch of the ALMA SV observation, ${\sim}2014.83$ (JD 2\,456\,959.6 and JD 2\,456\,962.7), the expected coordinates of Mira A due to proper motion should be $(34{\fdg}836\,667, -2{\fdg}978\,615) \pm (0{\fdg}000\,013, 0{\fdg}000\,010)$. So the observed absolute position of Mira A is within $2\sigma$ from the predicted position of the Hipparcos Catalogue.

We then produced two sets of spectral line images, with and without subtraction of the continuum, respectively. The continuum was subtracted, with the \texttt{uvcontsub} task in CASA, by fitting a linear polynomial to the real and imaginary parts of the visibility data of the line-free (i.e., continuum-only) channels in each spectral window. Our selection of the line-free channels was slightly different from that in the example imaging script provided along with the SV data\footnote{available from the CASA Guides: \url{http://casaguides.nrao.edu/index.php?title=ALMA2014_LBC_SVDATA&oldid=18077}}.

The spectral line image data cubes of the SiO and H$_2$O lines in ALMA Band 6 were created by the image deconvolution task, \texttt{clean}, in CASA. The task performs an inverse Fourier transform to the visibility data (``\uvd'') and creates a raw image data cube (the ``DIRTY'' image), then deconvolves the ALMA point-spread function from each frequency plane of the image with the Clark image deconvolution algorithm \citep{hogbomclean,clarkclean} (the ``CLEAN'' process). The product of image deconvolution for each frequency is a set of point sources (the CLEAN component models) which, in aggregate, reproduce the same input ``DIRTY'' image when convolving with the array's point-spread function. The task finally restores the CLEAN component models with a restoring beam (the CLEAN beam) of parameters either determined from fitting the point-spread function (taking its FWHM) or specified by the user. The local standard of rest (LSR) velocities covered by the image cubes range from $26.7\,\kms$ to $66.7\,\kms$, centred at the systemic (centre-of-mass) LSR velocity of $46.7\,\kms$, which corresponds to $57.0\,\kms$ in the heliocentric rest frame. We have determined the systemic velocity from the mid-point of the total velocity ranges of the entire line-emitting/absorbing region, assuming the global infall or expansion motions at the extreme velocities are symmetric about the stellar systemic velocity. We weighted the visibilities with a robust (Briggs) parameter of $\mathcal{R}_{\rm Briggs}=0.5$ and CLEANed the images down to a threshold of ${\sim}2$--$3\,{\rm mJy}\,{\rm beam}^{-1}$, which is about 1.5 times the rms noise level. We restored the images with a circular beam of FWHM $0{\farcs}032$ for the $^{28}$SiO and $^{29}$SiO lines, and of FWHM $0{\farcs}030$ for the H$_2$O $v_2=1$ line. The FWHM of the beams are the geometric means of the major- and minor-axes of the elliptical point-spread functions fitted by the \texttt{clean} task. 

At 220\,GHz, the primary beam FWHM of the 12-m array is about $28\arcsec$, which is much larger than the size of the line emission/absorption, and therefore no primary beam correction is needed. Figure \ref{fig:siov0_full} is the only primary beam-corrected image in this article, which shows remote emission in the vibrational ground state $^{28}$SiO up to a distance of ${\sim}3{\arcsec}$. There is no significant difference in the flux of detectable emission from the image without primary beam correction.

The spectral line channel maps presented in Sect. \ref{sec:result_maps} and Appendix \ref{sec:appendix_maps} are further binned, with the image binning task \texttt{imrebin} (new from version 4.3.0) in CASA version 4.4.0. We use Python to generate plots, with the aid of the \texttt{matplotlib} plotting library (version 1.4.3) \citep{matplotlib}, the PyFITS module (version 3.3), and the Kapteyn Package\footnote{\url{http://www.astro.rug.nl/software/kapteyn/}} (version 2.3) \citep{kapteyn}. 

The cell size and total size of the images are $0{\farcs}005 \times 0{\farcs}005$ and $15{\arcsec} \times 15{\arcsec}$, respectively. Figure \ref{fig:siov0_full} shows the map of the SiO ${\varv} = 0$ $J=5-4$ line at the systemic velocity channel over a $7{\farcs}5 \times 7{\farcs}5$ region centred at Mira A. As shown in this figure, there is remote, arc-like emission extending up to about 3 arcsec from the star between the LSR velocities of $43.7\,\kms$ and $49.7\,\kms$. 

Moreover, we use the images without continuum subtraction for our spectral line modelling (Sects. \ref{sec:result_spec} and later), instead of the continuum-subtracted images as in the reference images in the ALMA Science Portal. As a result, our images only contain emission from spectral line and/or the radio continuum from Mira A and B throughout, without any real negative signals. As we will explain in further detail in Appendix \ref{sec:appendix_contsub}, we find spurious ``bumps'' in the absorption profiles of continuum-subtracted spectra. We believe that the image deconvolution of the strong line emission surrounding Mira's radio photosphere may have impaired the deconvolution of the region showing line absorption against the background continuum. The images (and hence the spectra) deconvolved without continuum subtracted should better represent the real emission and absorption of the SiO and H$_2$O lines.



\section{Results}
\label{sec:results}

\subsection{Continuum}
\label{sec:result_cont}

\citet{mrm2015} and \citet{vro2015} have independently analysed the continuum data of the Mira AB system in both ALMA Bands 3 (96\,GHz) and 6 (229\,GHz) of this SV dataset. Additionally, \citet{mrm2015} include the continuum data from their Q-band (46\,GHz) observation with the Karl G. Jansky Very Large Array (VLA) in 2014; \citet{vro2015} also include the continuum data from ALMA Band 7 (338\,GHz), of which some results have been reported by \citet{ramstedt2014}. From this SV dataset, both \citet{mrm2015} and \citet{vro2015} found that the visibilities of Mira A in the continuum of Band 6 can be better fitted with a two-component model, consisting of an elliptical uniform disk plus an additional Gaussian component, than a single-component model. Moreover, \citet{vro2015} found that the additional Gaussian component to be a compact, bright hotspot with a FWHM of ${\sim}4.7\,{\rm mas}$ and a brightness temperature of ${\sim}10\,000\,{\rm K}$.

We have conducted a similar analysis of the continuum data of Mira A and B as these authors for the Band 6 data. From the continuum map, the total flux of Mira A is $149.70 \pm 0.04\,{\rm mJy}$ and that of Mira B is $11.19 \pm 0.04\,{\rm mJy}$. In our model fitting, the elliptical uniform disk component for Mira A has a size of about $(51.2 \pm 0.1)\,{\rm mas} \times (41.0 \pm 0.1)\,{\rm mas}$, ${\rm PA} = -45.0\degr \pm 0.5\degr$ and a flux of $S_{229.6\,{\rm GHz}} = 102 \pm 9\,{\rm mJy}$. This corresponds to a brightness temperature of $1630 \pm 175\,{\rm K}$. For the additional Gaussian component of Mira A, the fitted flux is about $47 \pm 9\,{\rm mJy}$ and its FWHM is about $(26.4 \pm 0.2)\,{\rm mas} \times (22.4 \pm 0.2)\,{\rm mas}$, ${\rm PA} = 34.0\degr \pm 1.7\degr$, which is much larger than the size of the purported $4.7\,{\rm mas}$-hotspot. The brightness temperature of this Gaussian component corresponds to $1856 \pm 419\,{\rm K}$, which is much smaller than $10\,000\,{\rm K}$. Our elliptical Gaussian model for Mira B has a FWHM of about $(25.5 \pm 0.3)\,{\rm mas} \times (22.5 \pm 0.3)\,{\rm mas}$, ${\rm PA} = 72.7\degr \pm 3.6\degr$ and a flux of about $(11.3 \pm 0.5) \,{\rm mJy}$. Our results are in general consistent with those reported by \citet{mrm2015}. However, we did not find any evidence of the compact hotspot or reproduce the similar results of the visibility fitting as reported by \citet{vro2015}. We present our detailed continuum analysis of the visibility data in Appendix \ref{sec:appendix_cont}.

In this section, we only present our model fitting results using a single model component for Mira, i.e., an elliptical uniform disk or an elliptical Gaussian, but not both. Table \ref{tab:singlecontinuum} shows the results of our single-component fitting in the continuum window centred at 229.6\,GHz. The brightness temperature of the uniform disk model of Mira A is found to be $2611 \pm 51\,{\rm K}$.


\begin{table*}[!htbp]
\caption{Photospheric parameters from fitting the continuum visibility data, with the \texttt{uvfit} task in the Miriad software, of Mira A and B in the continuum window at 229.6\,GHz of ALMA Band 6.}
\label{tab:singlecontinuum}
\centering
\begin{tabular}{lccccc}
\hline\hline
Object & $S_{229.6\,{\rm GHz}}$ & $\theta_{\rm maj}$ & $\theta_{\rm min}$ & P.A. & $T_b$ \\
 & (mJy) & (mas) & (mas) & ($\degr$) & (K) \\
\hline
\multicolumn{6}{c}{Mira A Uniform Disk Model + Mira B Gaussian Model} \\
\hline
Mira A (Disk)     & $148.0 \pm 0.5$ & $45.96 \pm 0.03$ & $41.25 \pm 0.03$ & $-36.4 \pm 0.3$ & $2611 \pm 51$ \\
Mira B (Gaussian) & $ 11.4 \pm 0.5$ & $25.70 \pm 0.26$ & $21.38 \pm 0.30$ & $65.7 \pm 2.7$ & --- \\
\hline
\multicolumn{6}{c}{Mira A Gaussian Model + Mira B Gaussian Model} \\
\hline
Mira A (Gaussian) & $151.7 \pm 0.5$ & $30.50 \pm 0.02$ & $26.93 \pm 0.02$ & $-45.6 \pm 0.3$ & --- \\
Mira B (Gaussian) & $ 11.1 \pm 0.5$ & $25.73 \pm 0.28$ & $22.14 \pm 0.34$ & $78.2 \pm 2.9$ & --- \\
\hline
\end{tabular}
\tablefoot{The columns are (from left to right): the object being fitted and the type of model (uniform disk or Gaussian), total flux at 229.6\,GHz ($S_{229.6\,{\rm GHz}}$, mJy), major axis ($\theta_{\rm maj}$, mas), minor axis ($\theta_{\rm min}$, mas), the position angle (P.A., degrees) of each elliptical component, and the brightness temperature ($T_b$, K) of the uniform disk model for Mira A. The quoted uncertainties are all the formal quantities resulting from model fitting. The real uncertainty of the flux is dominated by the absolute flux calibration, which is estimated to be of order 20\%. The uncertainty of the brightness temperature including the contribution from the absolute flux calibration is about $565\,{\rm K}$. The total integrated fluxes of Mira A and B in the 229.6-GHz continuum map are $149.70 \pm 0.04\,{\rm mJy}$ and $11.19 \pm 0.04\,{\rm mJy}$, respectively.}
\end{table*}


In addition, we have also created continuum images by integrating all the line-free channels in each of the four SiO and one H$_2$O spectral line windows. By calculating the total flux, $S_{\nu}$, from Mira A within a $0{\farcs}25$-radius circle (which safely includes all possible continuum emission from Mira A, but does not contain any emission from Mira B) at respective frequencies, $\nu$, we derive an independent spectral index (using the spectral line windows in Band 6 only) of $1.82 \pm 0.33$, which is consistent with the value ($1.86$) derived by \citet{rm1997}.


\subsection{Images}
\label{sec:result_maps}

\begin{figure*}[!htbp]
\centering
\includegraphics[trim=0.0cm 0.2cm 0.0cm 0.2cm, clip, width=\singlemapwidth]{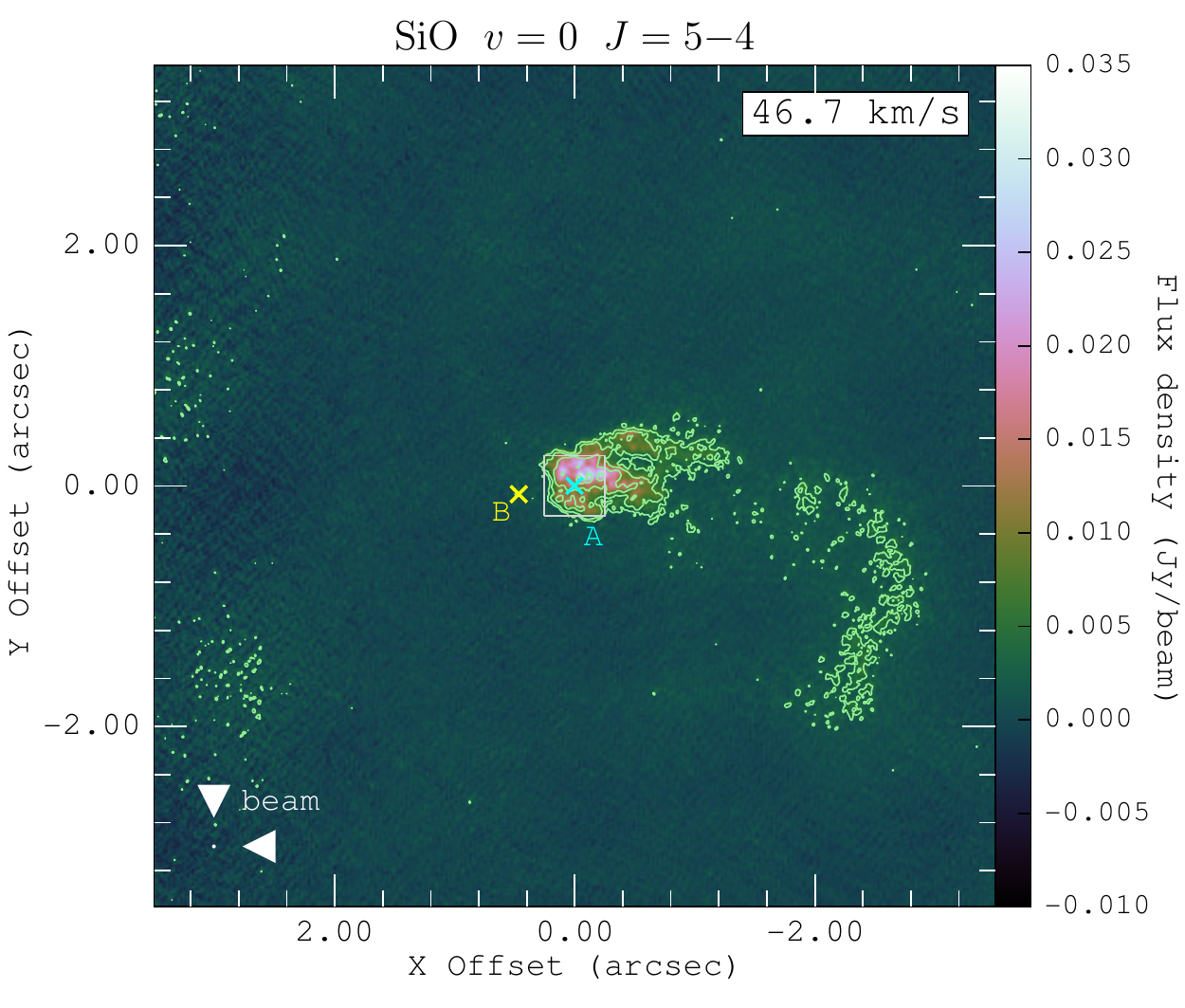}
\caption[]{The map of SiO ${\varv} = 0$ $J=5-4$ (with the continuum) at the channel of the systemic velocity ($46.7\,\kms$) with a channel width of $1.0\,\kms$. The positions of Mira A ($o$ Ceti; cyan cross) and Mira B (VZ Ceti; yellow cross) are indicated in the image. The horizontal and vertical axes are the relative offsets (arcsec) in the directions of right ascension ($X$) and declination ($Y$), respectively, with respect to the continuum centre of Mira A. 
The white box centred at the fitted position of Mira A indicates the $0{\farcs}50 \times 0{\farcs}50$ region as shown in Fig. \ref{fig:array}, within which we extract the SiO and H$_2$O line spectra from an array of positions. 
The horizontal and vertical axes are the relative offsets (arcsec) with respect to the Mira A in right ascension and declination, respectively. 
The light green contours represent 4, 8, 16, and $32\sigma$ of the SiO emission from the gas near Mira A, where the map rms noise is $\sigma = 0.80\,{\rm mJy}\,{\rm beam}^{-1}$. 
The circular restoring beam of $0{\farcs}032$ FWHM for the SiO image is indicated in white at the bottom-left.}
\label{fig:siov0_full}
\end{figure*}

Figure \ref{fig:siov0_full} shows the map of the SiO ${\varv} = 0$ $J=5-4$ transition in the LSR velocity of $46.7\,\kms$, which is the systemic (centre-of-mass) velocity of Mira A, over a $7{\farcs}5 \times 7{\farcs}5$ box centred at Mira A. The position of Mira B, $02^{\rm h}19^{\rm m}20{\fs}826, -02{\degr}58{\arcmin}43{\farcs}12$, as determined by fitting its image produced from the {\uvd} before self-calibration, is also indicated on the map. To the west of Mira A, the SiO vibrational ground state emission extends to a larger projected radial distance than other directions. This emission feature emerges from the west and north-west of Mira A and appears as an arc-like feature, which turns south at around $2{\arcsec}$ west of the star and reaches a maximum projected distance of ${\sim}3{\arcsec}$.

As we will explain in Appendix \ref{sec:appendix_contsub}, there are spurious ``bumps'' in the spectra extracted from the line-of-sight towards the continuum of Mira in the maps produced from the data continuum-subtracted before imaging. Since we are more confident in the quality of the image deconvolution without the subtraction of the continuum, we extract the spectra from the maps retaining the continuum (``full data maps'') for our radiative transfer modelling in Sects. \ref{sec:result_spec} and later. These full data maps are presented in Appendix \ref{sec:appendix_maps}. In this section, we only show the maps that are first imaged with the continuum, and \emph{then} continuum-subtracted with the CASA task \texttt{imcontsub}. Such post-imaging continuum subtraction can avoid the spurious features seen in pre-imaging continuum-subtracted images (and also the spectra).

\begin{figure*}[!htbp]
\centering
\includegraphics[trim=0.0cm 1.1cm 0.0cm 0.9cm, clip, width=\firstchannelmapwidth]{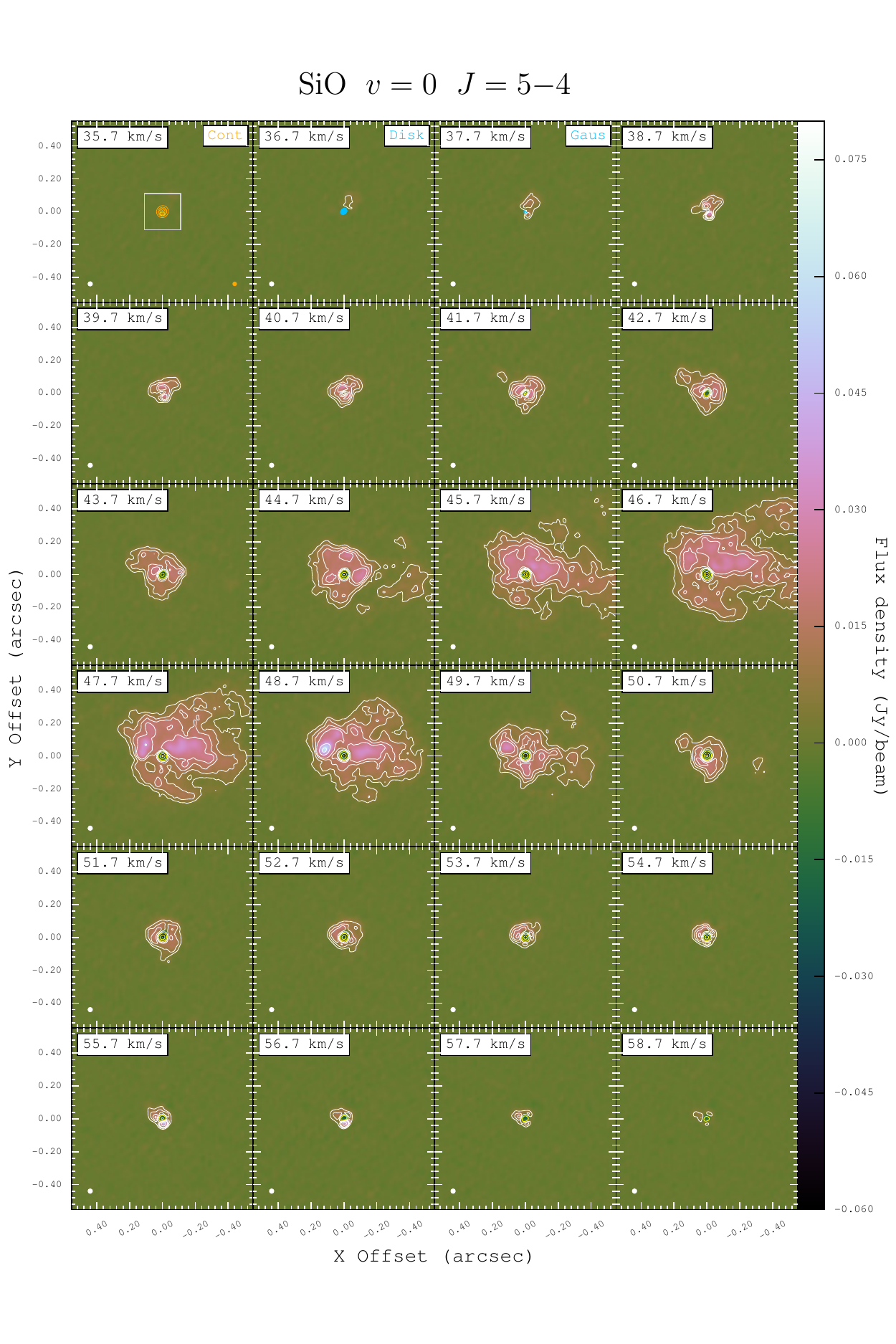}
\caption[]{Channel maps of post-imaging continuum-subtracted SiO ${\varv} = 0$ $J=5-4$ from LSR velocity $35.7\,\kms$ to $58.7\,\kms$, with a channel width of $1.0\,\kms$. The systemic velocity is $46.7\,\kms$. The horizontal and vertical axes indicate the relative offsets (arcsec) in the directions of right ascension ($X$) and declination ($Y$), respectively, with respect to the fitted absolute position of Mira A. 
The white contours represent 6, 12, 18, 24, 48, and $72\sigma$ and yellow contours represent $-60$, $-36$ and $-6\sigma$, where $\sigma = 0.80\,{\rm mJy}\,{\rm beam}^{-1}$ is the map rms noise. 
The circular restoring beam of $0{\farcs}032$ FWHM for the SiO image is indicated in white at the bottom-left in each panel. 
In the first panel of the top row, orange contours at 0.1, 0.3, 0.5, 0.7, and 0.9 times the peak flux density ($73.4\,{\rm mJy}\,{\rm beam}^{-1}$) of the 229-GHz continuum emission are also drawn and the corresponding restoring beam of $0{\farcs}028$ FWHM is indicated in orange at the bottom-right. 
The white box centred at Mira A indicates the $0{\farcs}22 \times 0{\farcs}22$ region of the zoomed maps of SiO ${\varv} = 0$ (Fig. \ref{fig:siov0chanzoomed_csub}), ${\varv} = 2$ (Fig. \ref{fig:siov2chan_csub}) and H$_2$O $v_2=1$ (Fig. \ref{fig:h2ov1chan_csub}). 
In the second and third panels of the top row, the sizes of the uniform disk and Gaussian models, respectively, in our continuum analysis are drawn in blue.}
\label{fig:siov0chan_csub}
\end{figure*}

Figure \ref{fig:siov0chan_csub} shows the continuum-subtracted channel maps of the SiO ${\varv} = 0$ $J=5-4$ transition in the LSR velocity range of $35.7$--$58.7\,\kms$ over a $1{\farcs}1 \times 1{\farcs}1$ box centred at Mira A (Mira hereafter). Contour lines at the $-36$, $-6$, $6$, $12$, $24$, $48$, and $72\sigma$ levels, where $\sigma = 0.80\,{\rm mJy}\,{\rm beam}^{-1}$, are drawn to indicate the region with significant line absorption (yellow contours; negative signals) or emission (white contours; positive signals). 

\begin{figure*}[!htbp]
\centering
\includegraphics[trim=0.0cm 1.1cm 0.0cm 0.9cm, clip, width=\channelmapwidth]{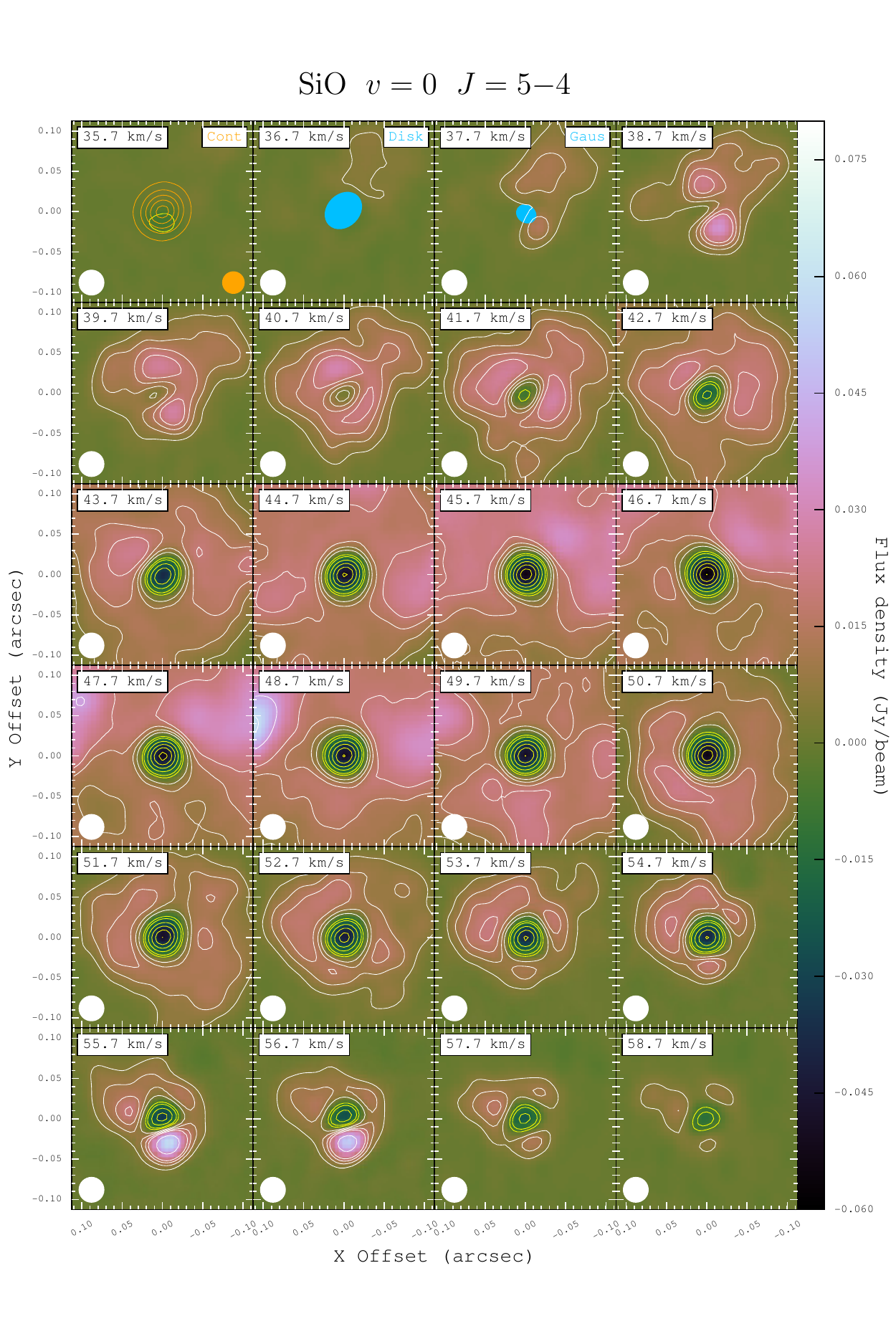}
\caption[]{Same as Fig. \ref{fig:siov0chan_csub} for the zoomed ($0{\farcs}22 \times 0{\farcs}22$) channel maps of post-imaging continuum-subtracted SiO ${\varv} = 0$ $J=5-4$. 
The white contours represent 6, 12, 18, 24, 48, and $72\sigma$ and yellow contours represent $-72$, $-60$, $-48$, $-36$, $-24$, $-12$, and $-6\sigma$, where $\sigma = 0.80\,{\rm mJy}\,{\rm beam}^{-1}$ is the map rms noise.}
\label{fig:siov0chanzoomed_csub}
\end{figure*}

Figure \ref{fig:siov0chanzoomed_csub} shows the same channel maps as Fig. \ref{fig:siov0chan_csub}, but zoomed in to show the inner $0{\farcs}22 \times 0{\farcs}22$ region around Mira. While, globally speaking, the emission of the vibrational ground state SiO line in the inner winds of Mira (${\lesssim}0{\farcs}2$) appears to be spherically symmetric, we find significant inhomogeneities with stronger emission from clumps that are localised in relatively small regions and which stretch over limited velocity ranges. 

\begin{figure*}[!htbp]
\centering
\includegraphics[trim=0.0cm 1.1cm 0.0cm 0.9cm, clip, width=\channelmapwidth]{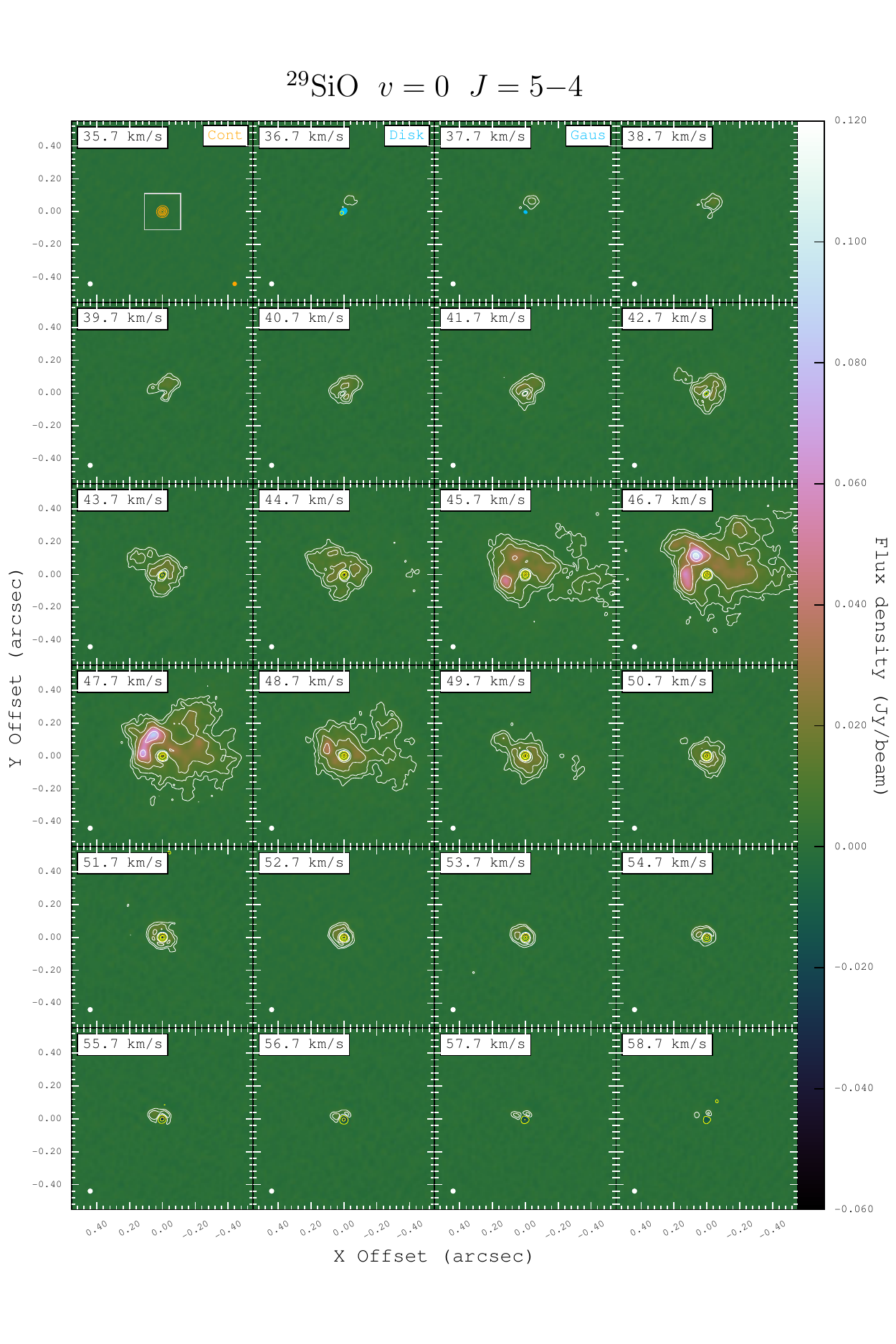}
\caption[]{Same as Fig. \ref{fig:siov0chan_csub} for the channel maps of post-imaging continuum-subtracted $^{29}$SiO ${\varv} = 0$ $J=5-4$. 
The white contours represent 6, 12, 24, 48, 96, and $144\sigma$ and yellow contours represent $-72$, $-54$, $-36$, and $-6\sigma$, where $\sigma = 0.65\,{\rm mJy}\,{\rm beam}^{-1}$ is the map rms noise.}
\label{fig:29siochan_csub}
\end{figure*}

As shown in Fig. \ref{fig:29siochan_csub}, the absorption and emission in the $J=5-4$ transition of the vibrational ground state of the $^{29}$SiO isotopologue appears to have a very similar extent as that observed in the analogous line to the main isotope of SiO. On larger scales, the $^{29}$SiO emission also appears to extend to the west of Mira, while its intensity falls off much more rapidly with increasing radius and no significant emission is seen beyond ${\sim}0.5\arcsec$. This is expected because the isotopic ratio of $^{28}$Si/$^{29}$Si in oxygen-rich giants is ${\gtrsim}13$ \citep[e.g.][]{tsuji1994,decin2010,ohnaka2014}. The $^{29}$SiO emission within ${\sim}0{\farcs}2$ also exhibits (1) general spherical symmetry and (2) localised, clumpy structures with more intense emission. While the maps in both isotopologues have the similar overall morphology, the peaks in the $^{29}$SiO emission do not all coincide with the $^{28}$SiO peaks.

\begin{figure*}[!htbp]
\centering
\includegraphics[trim=0.0cm 1.1cm 0.0cm 0.9cm, clip, width=\channelmapwidth]{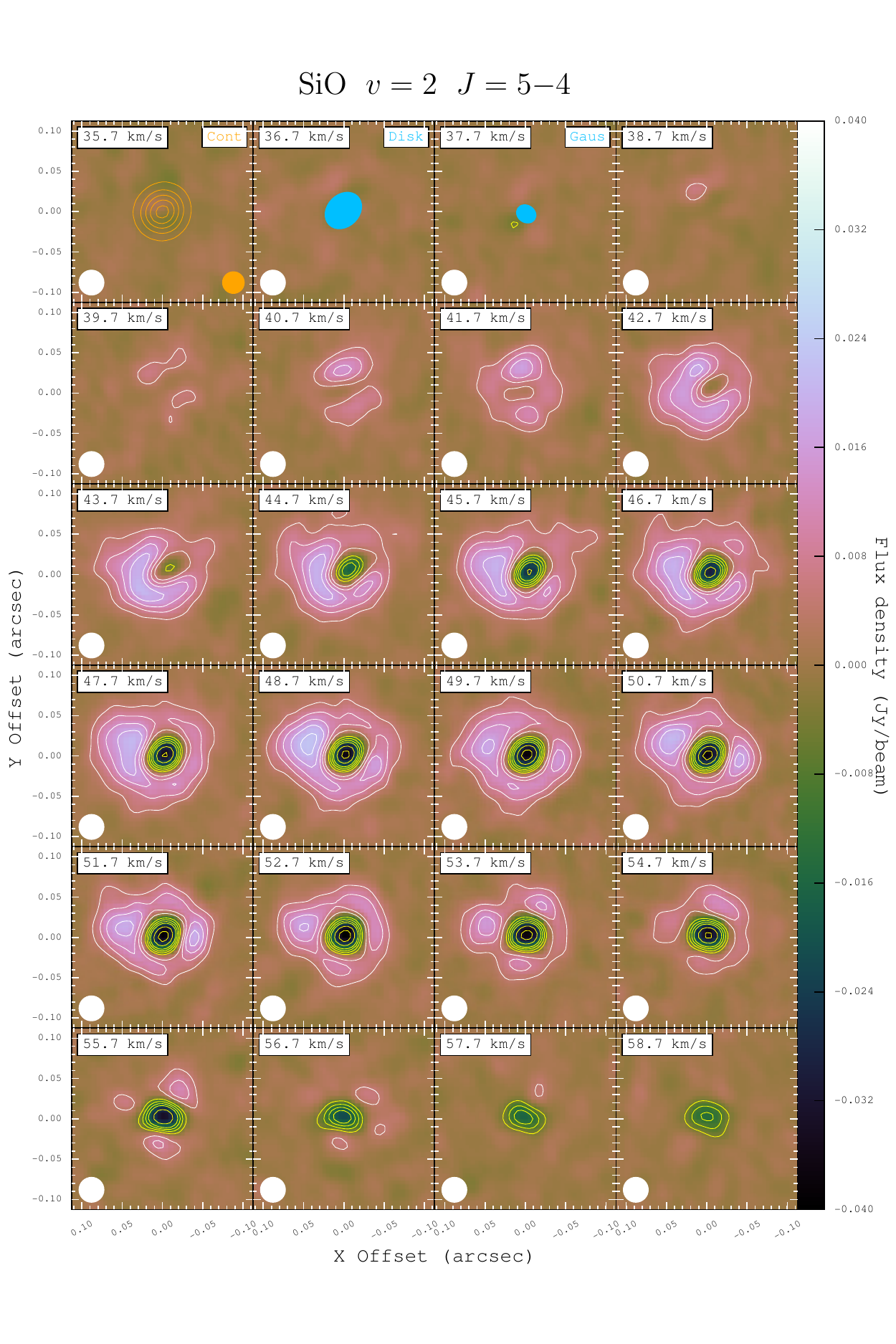}
\caption[]{Same as Fig. \ref{fig:siov0chan_csub} for the zoomed ($0{\farcs}22 \times 0{\farcs}22$) channel maps of post-imaging continuum-subtracted SiO ${\varv} = 2$ $J=5-4$. 
The white contours represent 6, 12, 18, 24, and $30\sigma$ and yellow contours represent $-48$, $-36$, $-24$, $-18$, $-12$, and $-6\sigma$, where $\sigma = 0.72\,{\rm mJy}\,{\rm beam}^{-1}$ is the map rms noise.}
\label{fig:siov2chan_csub}
\end{figure*}

\begin{figure*}[!htbp]
\centering
\includegraphics[trim=0.0cm 1.1cm 0.0cm 0.9cm, clip, width=\channelmapwidth]{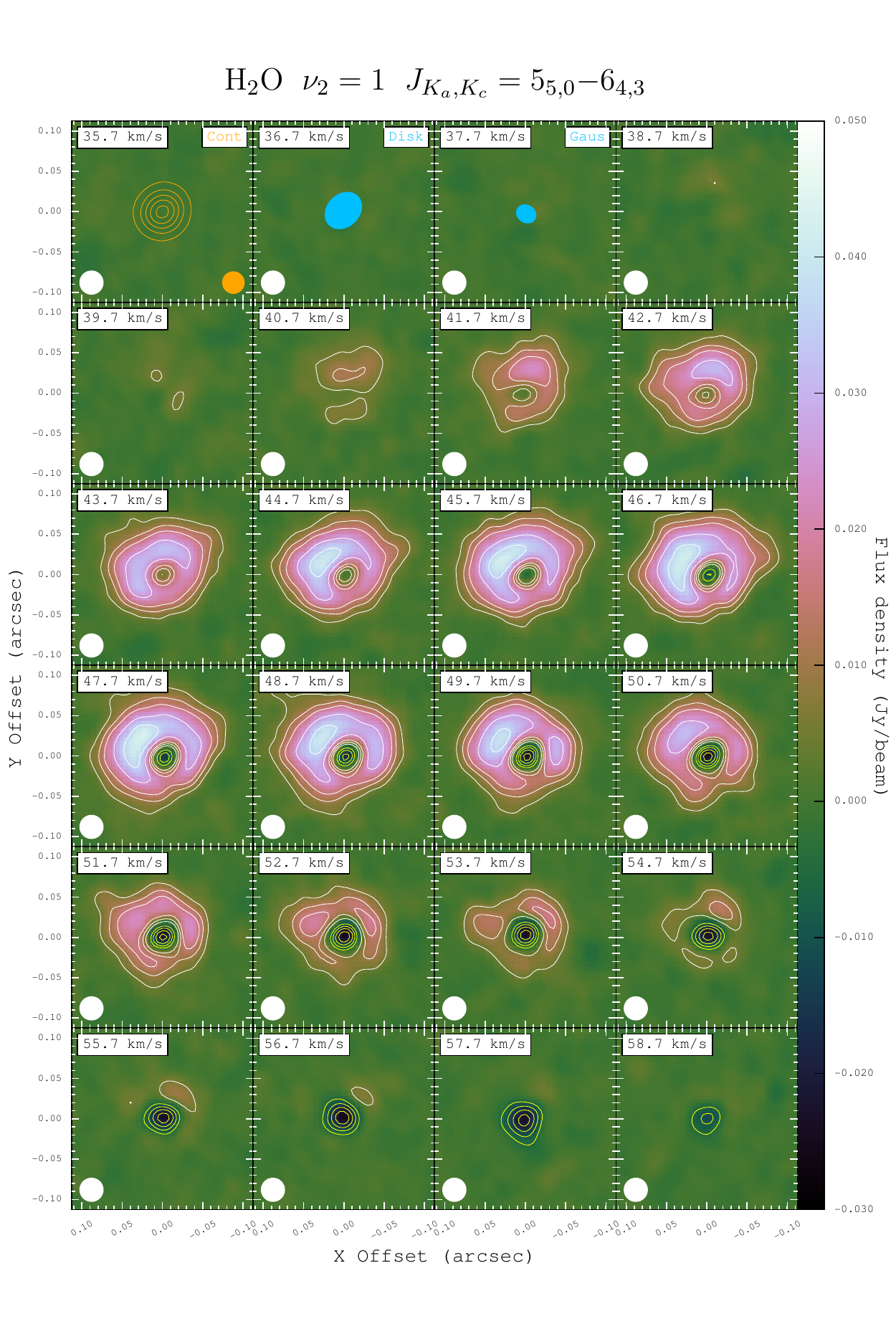}
\caption[]{Same as Fig. \ref{fig:siov0chan_csub} for the zoomed ($0{\farcs}22 \times 0{\farcs}22$) channel maps of post-imaging continuum-subtracted H$_2$O $v_2=1$ $J_{K_a,K_c}=5_{5,0}-6_{4,3}$. 
The white contours represent 6, 12, 18, 30, and $42\sigma$ and yellow contours represent $-24$, $-18$, $-12$, and $-6\sigma$, where $\sigma = 0.85\,{\rm mJy}\,{\rm beam}^{-1}$ is the map rms noise. 
The circular restoring beam of $0{\farcs}030$ FWHM for the H$_2$O images is indicated in white at the bottom-left in each panel. }
\label{fig:h2ov1chan_csub}
\end{figure*}

Figures \ref{fig:siov2chan_csub} and \ref{fig:h2ov1chan_csub} show the continuum-subtracted maps of the SiO ${\varv} = 2$ $J=5-4$ and H$_2$O $v_2=1$ $J_{K_a,K_c}=5_{5,0}-6_{4,3}$ lines, respectively. Since the emission of these two lines is more smoothly distributed than that in the vibrational ground state SiO and $^{29}$SiO lines, we can clearly see ring-like emission structures around the line absorption against Mira's continuum in the velocity channels around the systemic velocity ($46.7\,\kms$). In most velocity channels, the emission from both lines are confined well within $0{\farcs}1$ from the centre of the continuum, and there is no remote emission beyond ${\sim}0{\farcs}1$ as in the ground state SiO lines.

Close to the systemic velocity, there is a clump at about 0$\farcs$05 to the east of Mira which strongly emits in both the SiO ${\varv} = 2$ and H$_2$O $v_2=1$ lines. The brightness temperatures of the SiO ${\varv} = 2$ and H$_2$O $v_2=1$ emission are ${\sim}600$\,K and ${\sim}1000$\,K, respectively. This eastern clump is not prominent, however, in the vibrational ground state SiO and $^{29}$SiO lines which have very low excitation energies (i.e., the upper-state energy, $E_{\rm up}$). This clump therefore probably contains shock-heated gas at a high kinetic temperature ($T_{\rm kin} \gtrsim 1000$\,K). On the other hand, the intensely-emitting clumps in the ground state of SiO or $^{29}$SiO lines do not appear in the highly excited SiO ${\varv} = 2$ and H$_2$O $v_2=1$ lines, which have excitation energies of $E_{\rm up}/k \gtrsim 3500\,{\rm K}$.



\subsection{Spectra}
\label{sec:result_spec}

\begin{figure*}[!htbp]
\centering
\includegraphics[trim=0.0cm 0.2cm 0.0cm 0.2cm, clip, width=\singlemapwidth]{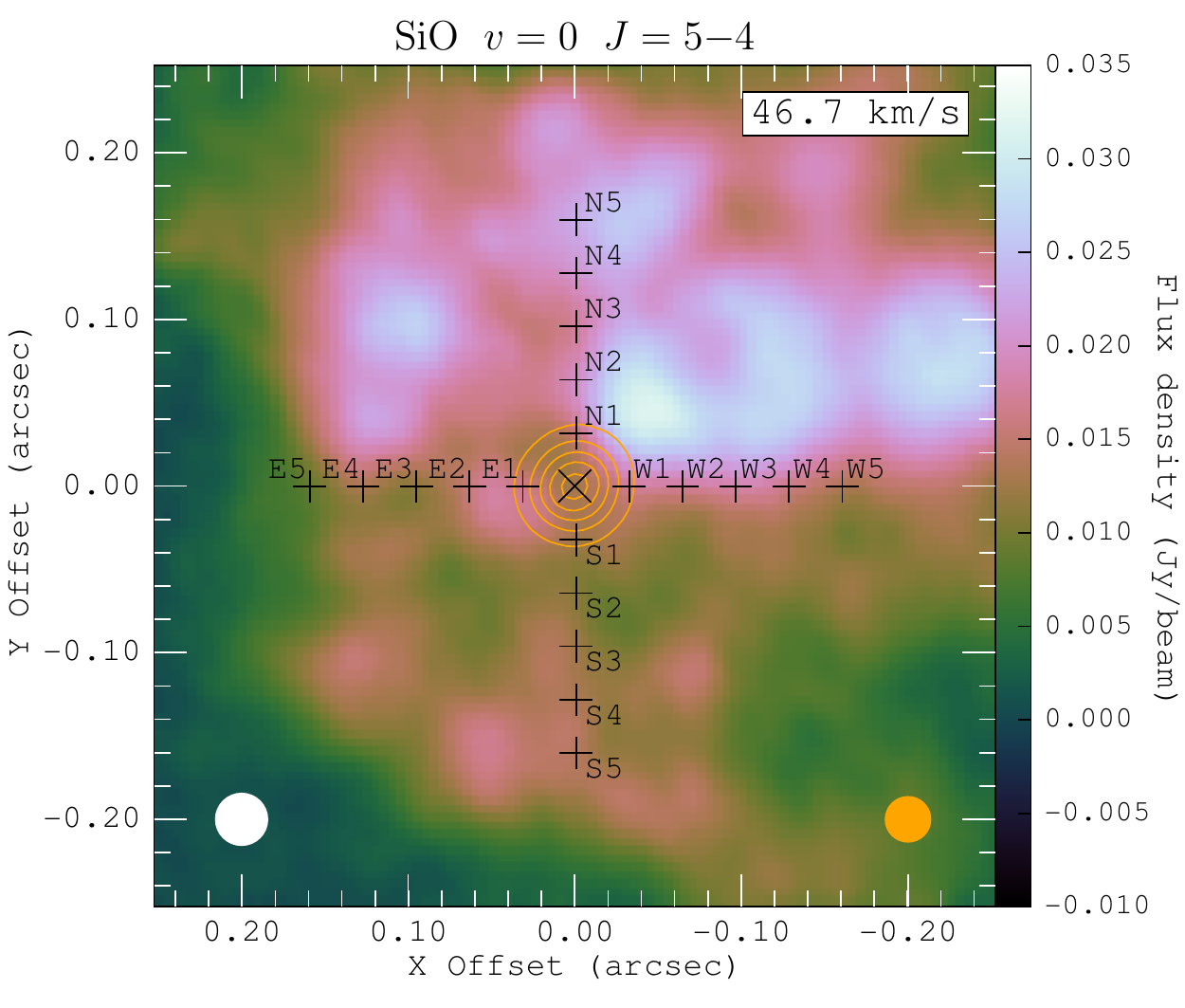}
\caption[]{The map of SiO ${\varv} = 0$ $J=5-4$ (with the continuum) at the channel of the systemic velocity ($46.7 \kms$) with a channel width of $1\,\kms$. The centre of Mira's continuum is marked as a black cross. Orange contours are drawn, representing 10\%, 30\%, 50\%, 70\%, and 90\% of the peak continuum flux ($73.4\,{\rm mJy}\,{\rm beam}^{-1}$). The black plus signs ($+$) indicate the positions at which SiO and H$_2$O spectra are sampled and modelled in Sect. \ref{sec:modelling}. Along each arm of this array of points, the sampling positions are separated by 32\,mas. The circular restoring beam of $0{\farcs}032$ FWHM for the SiO image is indicated in white at the bottom-left and that of $0{\farcs}028$ FWHM for the 229-GHz continuum contours is indicated in orange at the bottom-right.}
\label{fig:array}
\end{figure*}

We have extracted the SiO and H$_2$O spectra from the centre of Mira's continuum, and from an array of positions at radii 0{\farcs}032, 0{\farcs}064, 0{\farcs}096, 0{\farcs}128 and 0{\farcs}160 from the centre, along the legs at PA = {0\degr}, {90\degr}, {180\degr}, and {270\degr}. The positions are shown in Fig. \ref{fig:array}, which is the map of SiO ${\varv} = 0$ $J=5-4$, without subtraction of the continuum, in the channel of the stellar systemic velocity ($v_{\rm LSR}=46.7\,\kms$). The full set of the spectra will be presented along with the modelling results in Sect. \ref{sec:modelling}. In view of the fact that the inner envelope around Mira is partially filled with intense clumpy emission, we did not compute the azimuthally-averaged spectra in order to avoid the averaged spectra from being contaminated by isolated intense emission and to obtain a more representative view of the general physical conditions of the envelope.

\begin{figure*}[!htbp]
\centering
\includegraphics[height=\spectraheight]{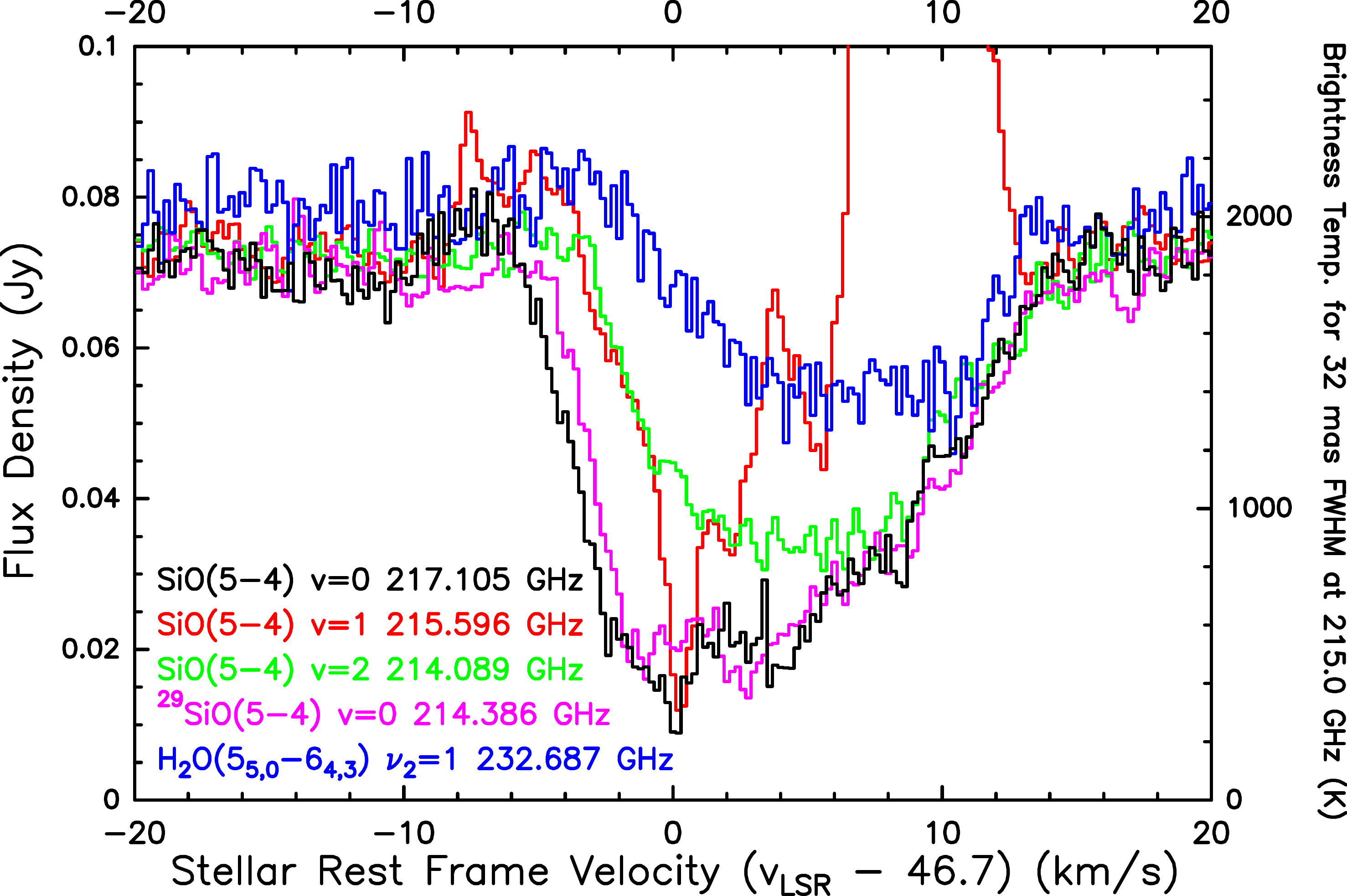}
\caption{Spectral lines in ALMA Band 6 extracted from the line-of-sight towards the centre of Mira's continuum. The SiO ${\varv} = 1$ $J=5-4$ transition (red colour) shows intense maser emission around $+10\,\kms$, with the peak flux density of $1.73\,{\rm Jy}$ at $+8.8\,\kms$. The maser spectrum above $0.10\,{\rm Jy}$ is not shown in this figure.}
\label{fig:band6lines}
\end{figure*}

Figure \ref{fig:band6lines} shows the spectra of various lines in ALMA Band 6 extracted from the centre of the continuum. As we did not subtract the continuum from the data, the flat emission towards the low- and high-velocity ends of the spectra represents the flux from the radio continuum of Mira near the frequencies of the respective spectral lines. In Appendix \ref{sec:appendix_contsub}, we will show the spectra with the continuum subtracted  (in the visibility data) before imaging.

The SiO ${\varv} = 1$ $J=5-4$ transition shows strong maser emission across a large range of LSR velocities, which introduces sharp spikes in its spectrum. For other lines which do not show strong maser emission (i.e., all except SiO ${\varv} = 1$), absorption against the continuum ranges between the offset velocity (relative to the stellar LSR velocity) of approximately $-4\,\kms$ and $+14\,\kms$. The absorption is in general redshifted relative to the systemic velocity. This indicates that the bulk of the material in the inner envelope is infalling towards Mira during the ALMA SV observation (near stellar phase 0.45). Infall motion at phase 0.45 is expected for another oxygen-rich Mira variable, W Hya, based on the detailed modelling of the CO $\Delta {\varv} = 3$ line profiles, as observed by \citet{lebzelter2005}, presented in the paper of \citet{nowotny2010}. The CO $\Delta {\varv} = 3$ lines probe the pulsation-dominated layers of the atmospheres of Mira variables, and therefore the radial velocity variation of these lines would indicate the infall or expansion velocities of the global motion of the extended atmospheres below the dust formation (and circumstellar wind acceleration) regions \citep[e.g.][]{hinkle1982,nowotny2005a}.

The spectra of $^{28}$SiO ${\varv} = 0$ $J=5-4$ and $^{29}$SiO ${\varv} = 0$ $J=5-4$ appear to be virtually identical. From the similarity of the line profiles and considering the high expected isotopic ratio of $^{28}$Si/$^{29}$Si ($\gtrsim 13$), the vibrational ground state $^{28}$SiO and $^{29}$SiO lines we see in Fig. \ref{fig:band6lines} are likely to be both very optically thick (saturated) and thermalised.

In Fig. \ref{fig:band6lines}, we can also see trends in the width and depth of the absorption profiles with excitation. The vibrationally excited SiO ${\varv} = 2$ and H$_2$O $v_2=1$ lines show narrower and shallower absorption than the two ground state SiO lines. This suggests that the vibrationally excited energy levels are less readily populated than the ground state levels, and hence the kinetic temperature of the bulk of the infalling material should be much lower than $3500$\,K, which corresponds to the excitation energies of SiO ${\varv} = 2$ and H$_2$O $v_2=1$ lines. This also explains the small radial extent of these two lines as shown in the channel maps because the kinetic temperature (and hence the excitation) in general falls off with the radial distance from the star. Because the SiO ${\varv} = 2$ and H$_2$O $v_2=1$ lines have very similar excitation energy ($E_{\rm up}/k \sim 3500$\,K), the difference in their line profiles is probably due to differences in the molecular abundance and molecular parameters such as the (de-)excitation rate coefficients.

There are two features in the spectra that strongly constrain our modelling in Sect. \ref{sec:modelling}. The first one is the small blueshifted emission feature at the offset velocities between $-10$ and $-3 \kms$. The size of the synthesised beam under robust weighting ($\mathcal{R}_{\rm Briggs}=0.5$) is about $0{\farcs}03$, which is comparable to that of the disk of the continuum emission (with minor axis about $0{\farcs}04$). Hence, some emission from the hottest inner layers of the envelope just outside the edge of the continuum disk is expected to ``leak'' into the beam. Since the innermost envelope shows global infall kinematics, the flux ``leakage'' should appear as excess blueshifted emission, i.e., an inverse P Cygni profile. We have also checked the spectra at different offset positions (some of which are modelled in Sect. \ref{sec:modelling}) and found that over the same blueshifted velocity range, the excess emission becomes more prominent as the continuum level decreases towards outer radial distances. For the H$_2$O transition, we also find a (much weaker) emission component near the offset velocity of $-3 \kms$, and a similar check at different offset positions also indicates that the component is likely to be real.

The other feature is presented by the redshifted wings in the offset velocity range between $+10$ and $+14 \kms$ of the $^{28}$SiO ${\varv} = 0$ and $2$, and the $^{29}$SiO ${\varv} = 0$ lines, which do not show strong maser emission. As shown in Fig. \ref{fig:band6lines}, the redshifted part of the absorption profiles of all these lines appears to be nearly identical. The lines could be in the optically thin regime only if the isotopic ratio of $^{28}$SiO/$^{29}$SiO is close to unity, which is not expected. So we believe that all the lines are in the optically thick regime in this velocity range. The brightness temperatures of the redshifted wings thus give an indication of the kinetic temperature of the coolest gas around the corresponding (infall) velocities.


\section{Radiative transfer modelling}
\label{sec:modelling}

We have modelled the H$_2$O and SiO spectra with the radiative transfer code {\ratran}\footnote{\url{http://www.sron.rug.nl/~vdtak/ratran/}} \citep{ratran}. The public version of the code accepts one-dimensional input models only. Despite the clumpy structures of the inner envelope, we find that the line spectra exhibit general spherical symmetry within ${\sim}0{\farcs}16$ and therefore 1-D modelling is applicable. Since the ALMA SV observations only provide a snapshot of Mira's extended atmosphere in its highly variable pulsation cycles, and the hydrodynamical models that we will compare and discuss in Sect. \ref{sec:discuss-hydrodyn} are also one-dimensional, using multi-dimensional radiative transfer code probably does not lead to better understanding of the general physical conditions of Mira's extended atmosphere. {\ratran} solves the coupled level population and radiative transfer equations with the Monte Carlo method and generates an output image cube for each of the modelled lines. We then convolved the image cubes with the same restoring beam as in our image processing and extracted the modelled spectra from the same set of positions as the observed spectra (Figure \ref{fig:array}). In the following subsections, we describe the details of our modelling, including the molecular data of H$_2$O and SiO, and the input physical models for the inner envelope with the continuum.


\subsection{H$_2$O molecular data}

The molecular data include information about all the energy levels considered in our radiative transfer model, and all possible transitions among these levels. The energies and statistical weights of the energy levels, and the Einstein $A$ coefficients (the rates of spontaneous emission), frequencies, upper level energies and collisional rate coefficients at various kinetic temperatures of the transitions are stored in a molecular datafile. The molecular datafile of H$_2$O is retrieved from the Leiden Atomic and Molecular DAtabase\footnote{\url{http://home.strw.leidenuniv.nl/~moldata/}} \citep[LAMDA;][]{lamda}. The LAMDA H$_2$O datafile includes rovibrational levels up to about $E_{\rm up}/k = 7190\,{\rm K}$ \citep{h2olinelvs}. In our modelling, we only include 189 energy levels up to 5130\,K in order to speed up the calculation. The selection includes 1804 radiative transitions and 17\,766 downward collisional transitions. The numbers of energy levels and transitions were reduced by more than half and three quarters, respectively, as compared to the original LAMDA file. Experiments have shown that such truncation of the datafile only has minute effects on the modelled spectra. The Einstein $A$ coefficients were provided by the BT2 water line list\footnote{\url{http://www.exomol.com/data/molecules/H2O/1H2-16O}} \citep{h2olinelist} and the collisional rate coefficients of H$_2$O with ortho-H$_2$ and para-H$_2$ were calculated by \citet{h2ocollrates}. The rates for ortho-H$_2$ and para-H$_2$ were weighted in the method described in \citet{lamda}.


\subsection{SiO molecular data}

Our radiative transfer modelling of SiO lines considers the molecule's vibrational ground and the first two excited states (${\varv} = 0, 1$ and $2$) up to an upper-state energy, $E_{\rm up}/k$, of about 5120\,K, similar to that for our H$_2$O modelling. There are totally 167 rotational energy levels in these vibrational states, where $J({\varv} = 0) \le 69$, $J({\varv} = 1) \le 56$, and $J({\varv} = 2) \le 39$. Among these energy levels are 435 radiative transitions (subject to the dipole selection rule $\Delta J = \pm 1$) and 13\,861 downward collisional transitions. The energies and statistical weights of the energy levels, and the line frequencies and Einstein $A$ coefficients of the radiative transitions are obtained from the EBJT SiO line list\footnote{\url{http://www.exomol.com/data/molecules/SiO/28Si-16O}} \citep{exomol2013}. These values are similar to those in the Cologne Database for Molecular Spectroscopy (CDMS)\footnote{\url{http://www.astro.uni-koeln.de/cdms/}} (version Jan 2014) \citep{cdms2001,cdms2005,mueller2013cdms}.

The rate coefficients for collisions between SiO and H$_2$ molecules in the vibrational ground state (${\varv} = 0 \rightarrow 0$) are extrapolated from the scaled (by 1.38) rate coefficients between SiO--He collisions as derived by \citet{dayou2006}. The SiO--He rate coefficients only include rotational levels up to $J({\varv} = 0) = 26$ and H$_2$ gas temperature up to 300\,K \citep{dayou2006}. We extrapolate the ${\varv} = 0 \rightarrow 0$ rate coefficients to higher $J$ and $T$ with the methods presented in Appendix \ref{sec:appendix_purerotrates}. Our temperature-extrapolated rate coefficients are consistent, within the same order of magnitude, with the corresponding values in the LAMDA SiO datafile \citep{lamda}. Rate coefficients of the rotational transitions involving vibrationally excited states (i.e., ${\varv} = 1,2$, where $\Delta {\varv} = 0,1$) can be computed with the infinite-order sudden (IOS) approximation \citep[e.g.][]{goldflam1977}, of which the parameters are given by \citet{bg1983a,bg1983b} for $J({\varv}) \le 39$ and $1000\,\rm{K} \le T \le 3000\,\rm{K}$. We extrapolate the parameters of \citet{bg1983a,bg1983b} to higher $J$ (see Appendix \ref{sec:appendix_vibrotrates}) and assume the temperature dependence of the parameters for $T < 1000\,\rm{K}$ and $T > 3000\,\rm{K}$ to be the same as that for $1000\,\rm{K} \le T \le 3000\,\rm{K}$. For ${\varv} = 2 \rightarrow 0$ transitions, we simply assume the rate coefficients from ${\varv} = 2 \rightarrow 0$ to be 10\% of those from ${\varv} = 2 \rightarrow 1$ transitions. We note that these coefficients in general do not affect the radiative transfer significantly \citep[e.g.][]{lw1984,le1992}.

Our extrapolation scheme of the SiO--H$_2$ collisional rate coefficients (Appendix \ref{sec:appendix_siorates}) is different from that described by \citet{doel1990}, on which the rate coefficients adopted by \citet{doel1995} and \citet{h96sio} are based. In particular, their extrapolation of the rate coefficients (including those for ${\varv} = 0 \rightarrow 0$ transitions) was based entirely on the set of parameters given by \citet{bg1983a,bg1983b}, which was the most complete and accurate one available at that time; they also refrained from further extrapolating the parameters beyond $J({\varv}) = 39$ for ${\varv} = 0, 1, \ldots, 4$ and beyond the temperature range considered by \citet{bg1983a,bg1983b} \citep[for detailed discussion, see Sect. 7.2 of][]{doel1990}.

We use the Python libraries, NumPy\footnote{\url{http://www.numpy.org}} (version 1.9.2) \citep{numpy} and SciPy\footnote{\url{http://www.scipy.org}} (version 0.15.1) \citep{scipy} in our extrapolation of the SiO collisional rate coefficients and compilation of the molecular datafile. Line overlapping between SiO and H$_2$O transitions, which may significantly affect the pumping of SiO masers \citep[e.g.][and references therein]{desmurs2014}, is neglected. 


\subsection{Continuum emission}

We include the continuum emission in the modelling. In {\ratran}, however, the ray-tracing code (\mbox{\textsc{sky}}) assumes that the size of the continuum is much smaller than the pixel size, which is not true in this ALMA dataset. Hence we cannot include the continuum by setting the default {\ratran} \mbox{\textit{central}} parameter, which describes the radius and blackbody temperature of the central source, in the straightforward manner. Instead, in our input physical model, we have created a \emph{pseudo}-continuum in the innermost three grid cells of the 1-D input model by setting (1) the outer radius of the third grid cell to be the physical radius of the radio continuum, (2) the ``kinetic temperature'' to be the brightness temperature of the continuum, (3) the outflow velocity to be zero, (4) the turbulence velocity to be $100\,\kms$ to get an effectively flat continuum spectrum within the velocity range of interest, and (5) the molecular abundance to be exceedingly high to get an optically thick core which blocks all the line emission from behind it. The exact number of grid cells representing the \emph{pseudo}-continuum does not affect the results. The velocity range of the {\ratran} image cubes was selected to be $\pm 25\,\kms$ from the systemic velocity, which is the same as for our ALMA image products.

In our modelling, the continuum level and spectral line absorption/emission were fitted from independent sets of parameters. The radius and effective temperature of the radio continuum were determined by fitting the modelled continuum levels to the ones in the observed spectra extracted from the centre, from 32\,mas and from 64\,mas. Beyond these distances the continuum level is effectively zero. The derived radius and effective temperature of the \emph{pseudo}-continuum is $R_{\rm continuum} = 3.60 \times 10^{13}\,{\rm cm}$ (21.8\,mas) and $2600\,{\rm K}$, respectively. These values are comparable to the mean radii and brightness temperatures of the elliptical disks fitted by us (Appendix \ref{sec:appendix_cont}), by \citet{mrm2015}, and by \citet{vro2015}.



\subsection{Modelling results}
\label{sec:model_results}

In the models of Mira's extended atmosphere and its inner wind, power-laws are adopted for the H$_2$ gas density and kinetic temperature profiles such that the density and temperature attain their maximum values at the outer surface of the radio photosphere, $R_{\rm continuum}$. The profiles of the physical parameters are expressed as functions of the radial distance from the continuum centre, which is defined as ``radius'' in the following discussion and in the plots of the input physical models. In order to reproduce the intensity of the spectra extracted from the centre and different projected distances, SiO abundance (relative to molecular hydrogen abundance) has to decrease with radius. We assume a simple two-step function for the SiO abundance, where the outer abundance is ${\sim}1\%$ of the inner abundance. The radius at which SiO abundance drops significantly is assumed to be $r_{\rm cond} = 1.0 \times 10^{14}\,{\rm cm} \approx 5\,R_{\star}$ in our preferred model. As we will discuss in Sect. \ref{sec:discuss-dust}, the observed spectra can still be fitted if $r_{\rm cond} \gtrsim 4\,R_{\star}$ or if the outer SiO abundance is ${\sim}10\%$ of the inner value (i.e., a degree of condensation of 90\%). The depletion of SiO molecule represents the dust condensation process in the transition zone between the inner dynamical atmosphere and the outer, fully accelerated circumstellar envelopes. For the H$_2$O molecule, however, condensation onto dust grains or solid ice is not expected in the modelled region where the gas temperature is at least a few hundred Kelvin. Furthermore, in the non-equilibrium chemical modelling of \citet{gobrecht2016}, the H$_2$O abundance in the inner winds of the oxygen-rich Mira variable IK Tau remains roughly constant with radius at a given stellar pulsation phase. So we assume the H$_2$O abundance (relative to H$_2$) near Mira to be constant at $5.0 \times 10^{-6}$ throughout the modelled region (out to $5.0 \times 10^{14}\,{\rm cm}\approx 25\,R_{\star} \approx 0{\farcs}3$). For the reason discussed in Section \ref{sec:model_preferred}, we have also considered an alternative H$_2$O abundance profile with a sharp increase in H$_2$O abundance near the radio photosphere.

In our radiative transfer modelling, the expansion/infall velocity, gas density, and gas kinetic temperature are the crucial parameters in the input physical model. We have empirically explored different types of profiles that are plausible in the inner winds and circumstellar envelopes of evolved stars. To improve the readability of the article, we present in Appendix \ref{sec:appendix_model} various plausible models that fail to reproduce the observed spectra. In this section, we only discuss our preferred model--Model 3--in which both infall and outflow layers coexist in the extended atmosphere of Mira. In Sect. \ref{sec:discuss-hydrodyn}, we compare the velocity, density, and temperature profiles in our preferred model with the those predicted by current hydrodynamical models of pulsating stellar atmospheres. We also model the line radiative transfer with the atmospheric structures derived from those hydrodynamical models.

\subsubsection{Preferred model: mixed infall and outflow}
\label{sec:model_preferred}

Our modelling shows that pure infall would produce too much emission in the blueshifted velocities of the spectra than is observed (Appendix \ref{sec:appendix_model}). The excess emission component, as we have discussed in Sect. \ref{sec:result_spec}, originates from the far-side of the innermost layer (beyond the radio photosphere) of Mira's extended atmosphere that is not blocked by the radio continuum disk. In our preferred model, we introduce a thin expanding layer (${\sim}5 \times 10^{11}\,{\rm cm} \approx 0.03\,R_{\star}$) in the innermost radii between the radio photosphere and the globally infalling layer. Alternating outflow and infall velocity profiles have been calculated numerically by \citet{bowen1988a,bowen1988b} for Mira-like variables, and subsequently adopted by \citet{h96sio,h01h2o} to simulate the SiO and H$_2$O masers from a Mira-like M-type variable star at a single stellar phase. The infall velocity immediately above this expanding layer is about $7.3\,\kms$, and the expansion velocity below this layer is about $4.0\,\kms$. The outer infalling gas and the inner expanding layer produce a shocked region, with the shock velocity of $\Delta V \lesssim 12\,\kms$, near the radio photosphere of Mira. The maximum gas infall speed of ${\sim}7\,\kms$ is consistent with the proper motions of SiO maser spots around another oxygen-rich Mira variable TX Cam, which lie in the velocity range of $5$--$10\,\kms$ \citep{diamond2003}. The emission from the far-side of the expanding layer would appear at redshifted velocities and the absorption from the near-side would be in the blueshifted part (i.e., the usual P Cygni profile). The excess emission from the pure infall models is therefore reduced to a level that fits the observed spectra.

\begin{figure*}[!htbp]
\centering
\includegraphics[width=\modelwidth]{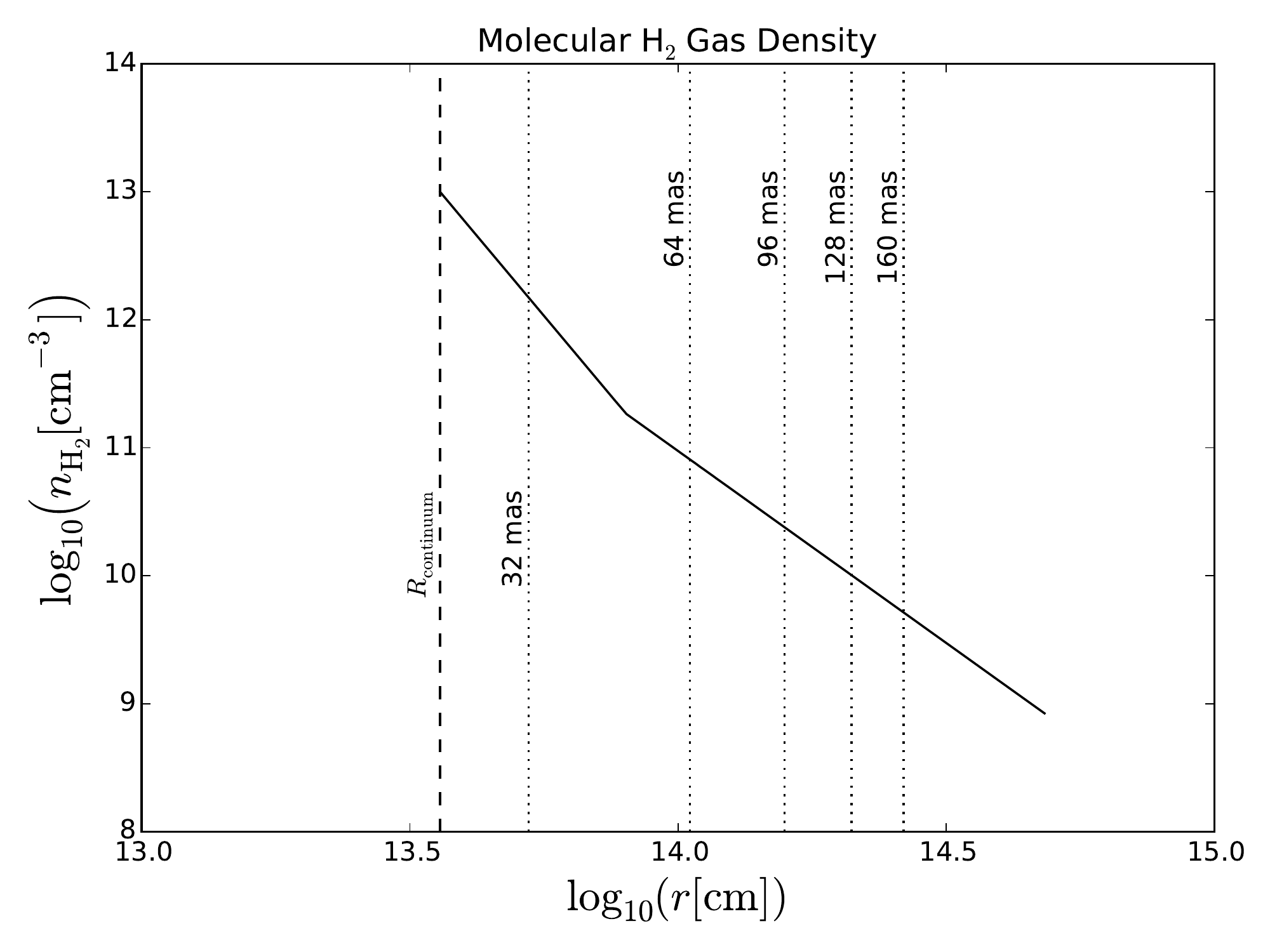}
\includegraphics[width=\modelwidth]{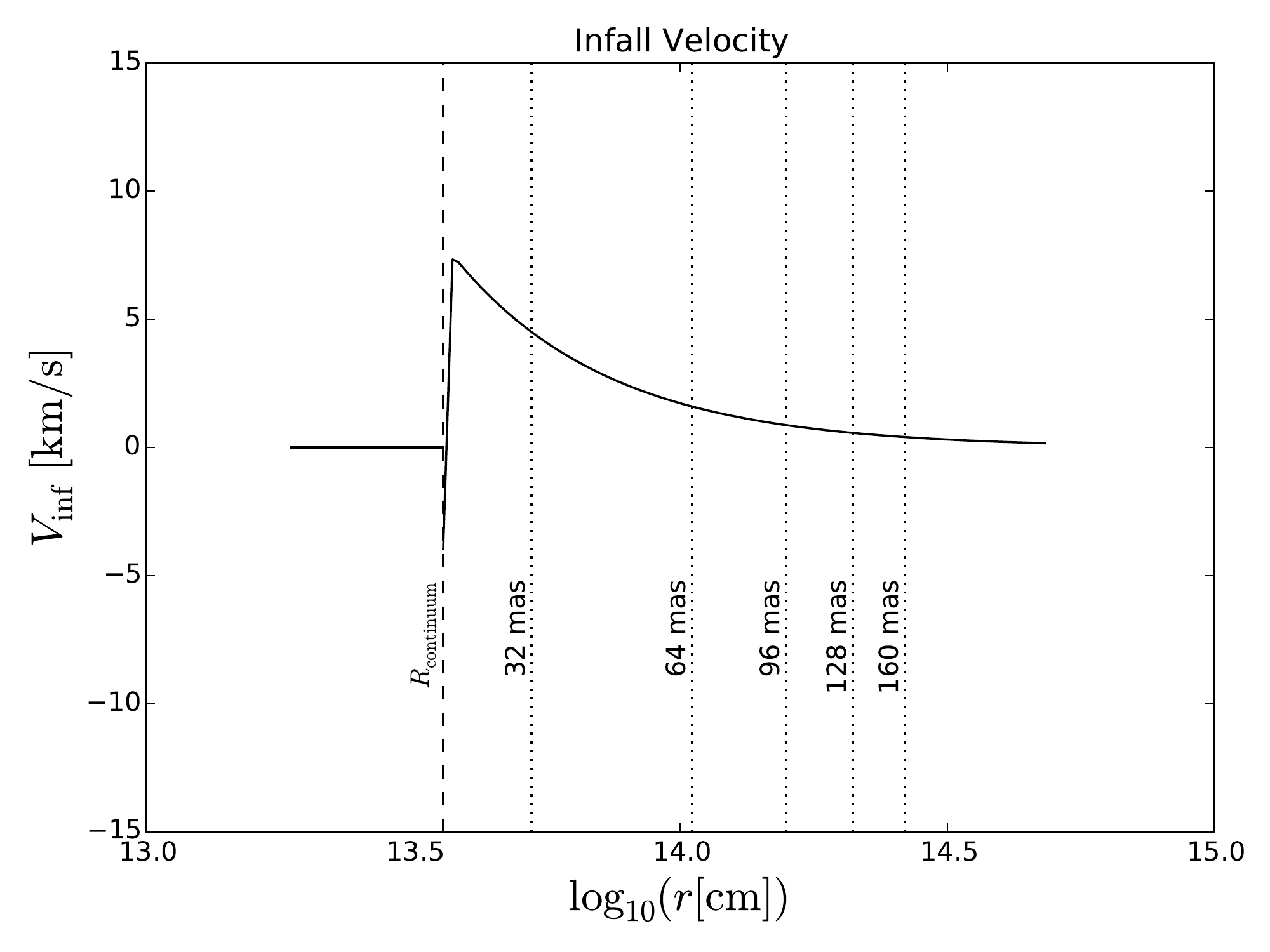}\\
\includegraphics[width=\modelwidth]{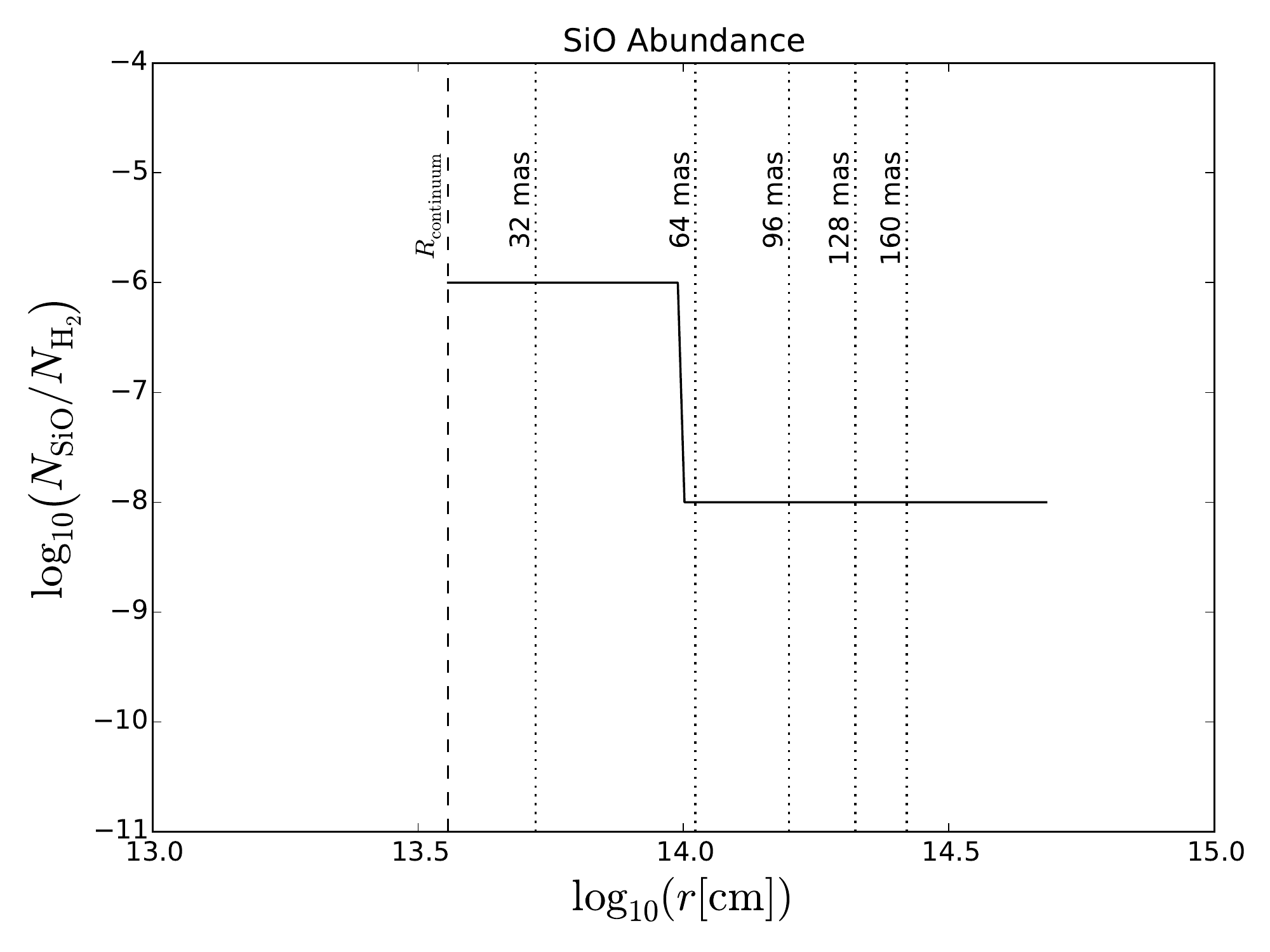}
\includegraphics[width=\modelwidth]{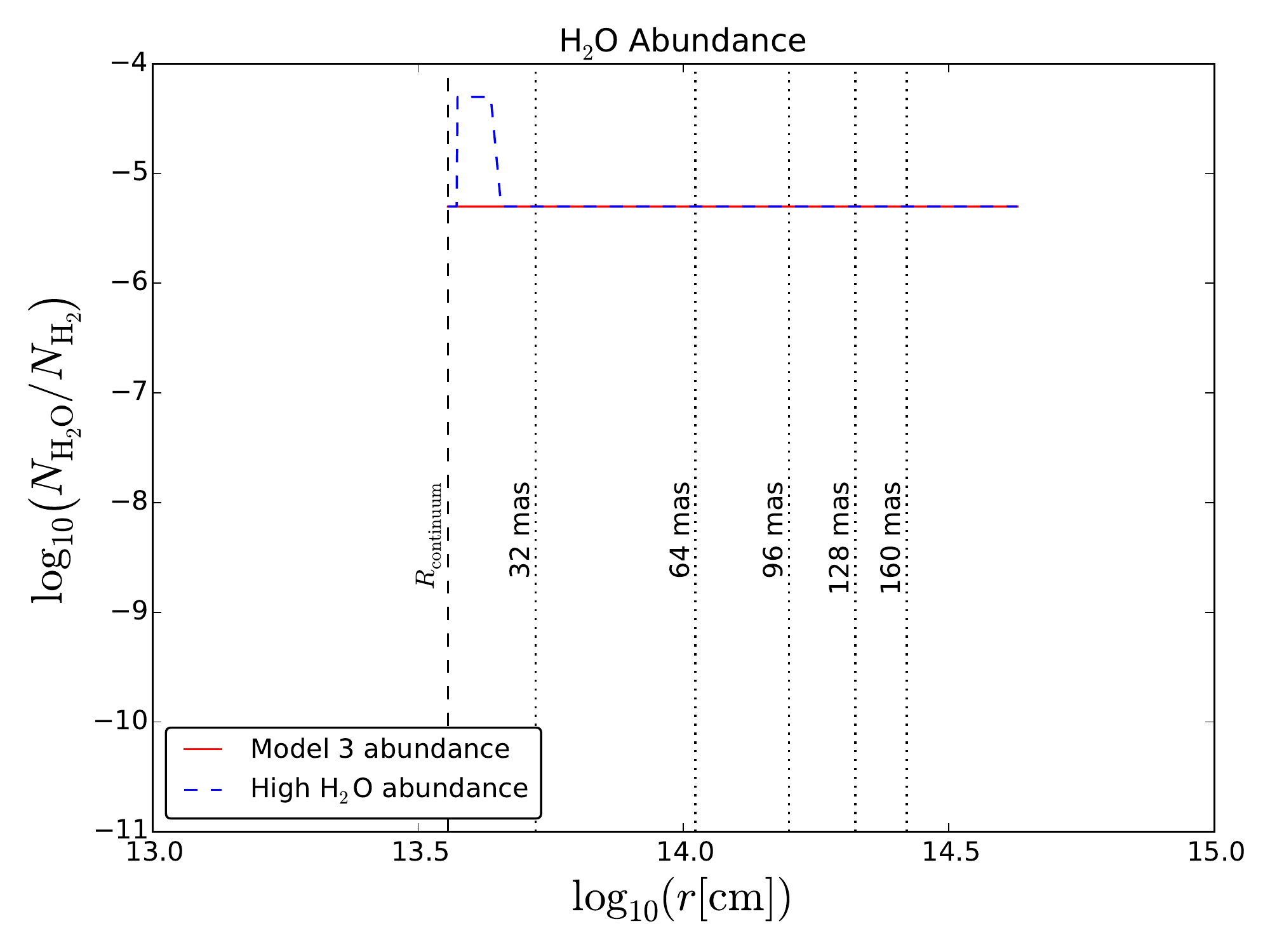}\\
\includegraphics[width=\modelwidth]{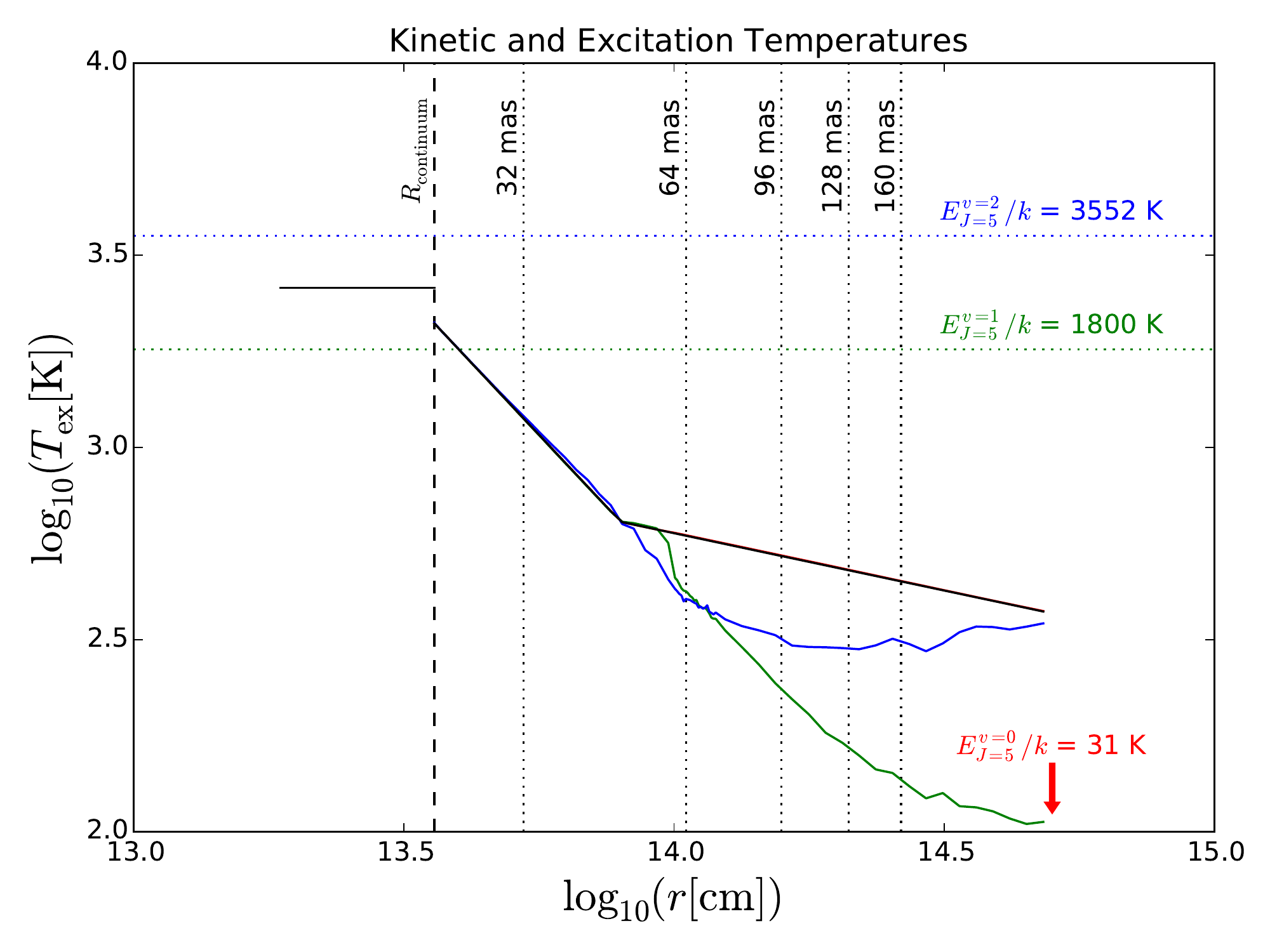}
\includegraphics[width=\modelwidth]{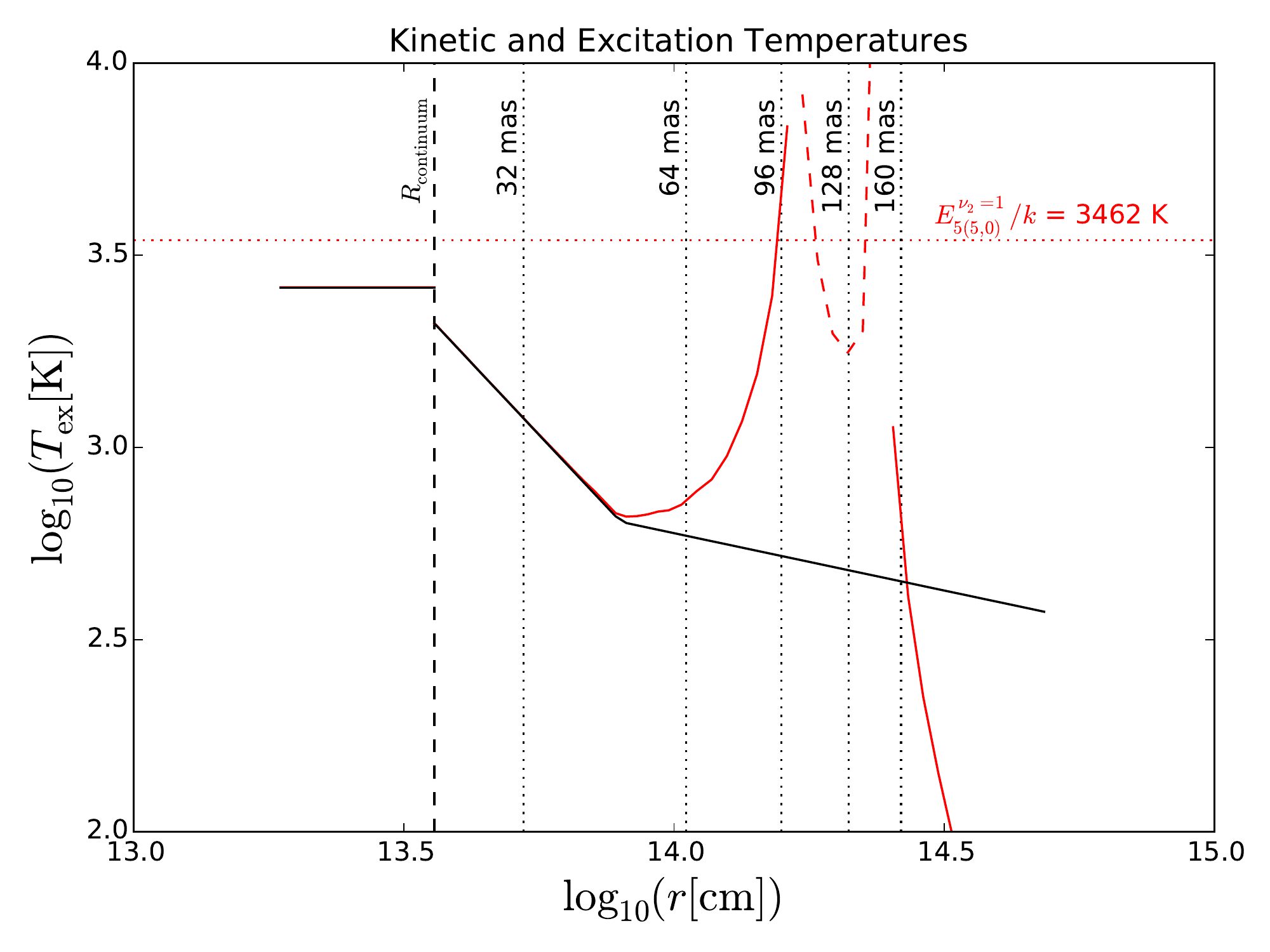}
\caption{Inputs of our preferred model. Shown in the panels are the H$_2$ gas density (\textbf{top-left}) , infall velocity (negative represents expansion) (\textbf{top-right}), $^{28}$SiO abundance (\textbf{middle-left}), H$_2$O abundance (\textbf{middle-right}), and the kinetic temperature (in black) and excitation temperatures (in colours) of the three $^{28}$SiO transitions (\textbf{bottom-left}) and the H$_2$O transition (\textbf{bottom-right}). In the bottom-right panel, solid red line indicates positive excitation temperature (i.e., non-maser emission) of the H$_2$O transition, and the dashed red line indicates the absolute values of the negative excitation temperature (i.e., population inversion) between $1.7 \times 10^{14}$ and $2.4 \times 10^{14}\,{\rm cm}$. Small negative values for the excitation temperature would give strong maser emission. Vertical dotted lines mark the radii at which the spectra were extracted; coloured horizontal dotted lines in the bottom panels indicates the upper-state energy ($E_{\rm up}/k$) of the respective transitions. The innermost layer within $R_{\rm continuum}$ represents the grid cells for the \emph{pseudo}-continuum, in which the input values for H$_2$ gas density and molecular abundances are above the range of the plots.}
\label{fig:model3}
\clearpage
\end{figure*}

To properly fit the line profiles, the radius of peak infall velocity is adopted to be $3.75 \times 10^{13}\,{\rm cm}$, where the gas density is almost $10^{13}\,{\rm cm}^{-3}$. Figure \ref{fig:model3} shows the important input parameters in our model, including the molecular H$_2$ gas density (top-left), infall velocity (top-right), molecular SiO and H$_2$O abundances (middle) and the gas kinetic temperature (bottom). The bottom row of Fig. \ref{fig:model3} also shows the excitation temperatures of the SiO and H$_2$O transitions (in colour).

\begin{figure*}[!htbp]
\centering
\includegraphics[trim=1.0cm 2.0cm 2.0cm 1.5cm, clip, width=\multispecwidth]{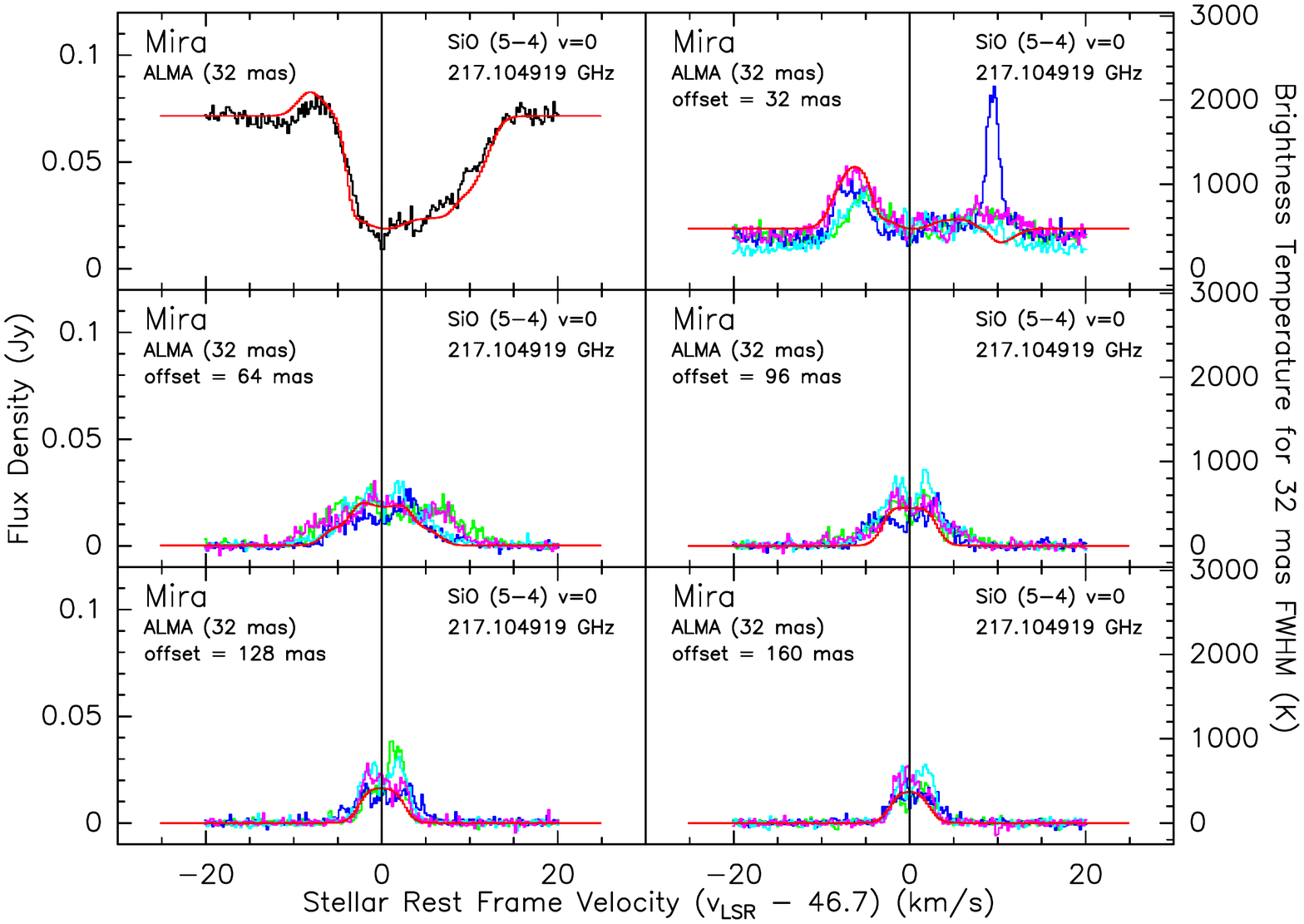}
\caption{Preferred model: spectra of SiO ${\varv} = 0$ $J=5-4$ at various positions. The black histogram is the observed spectrum at the centre of continuum, green, blue, cyan and magenta histograms are the observed spectra along the eastern, southern, western and northern legs, respectively, at various offset radial distances as indicated in each panel. The red curves are the modelled spectra predicted by {\ratran}. Our model does not produce the population inversion (i.e., negative excitation temperature) required for maser emission in this SiO transition, so we do not expect our modelled spectra to show any maser emission, as seen in the spike in the upper-right panel (see text for the discussion of the spike).}
\label{fig:m3siov0spec}
\end{figure*}

\begin{figure*}[!htbp]
\centering
\includegraphics[trim=1.0cm 7.3cm 2.0cm 1.5cm, clip, width=\multispecwidth]{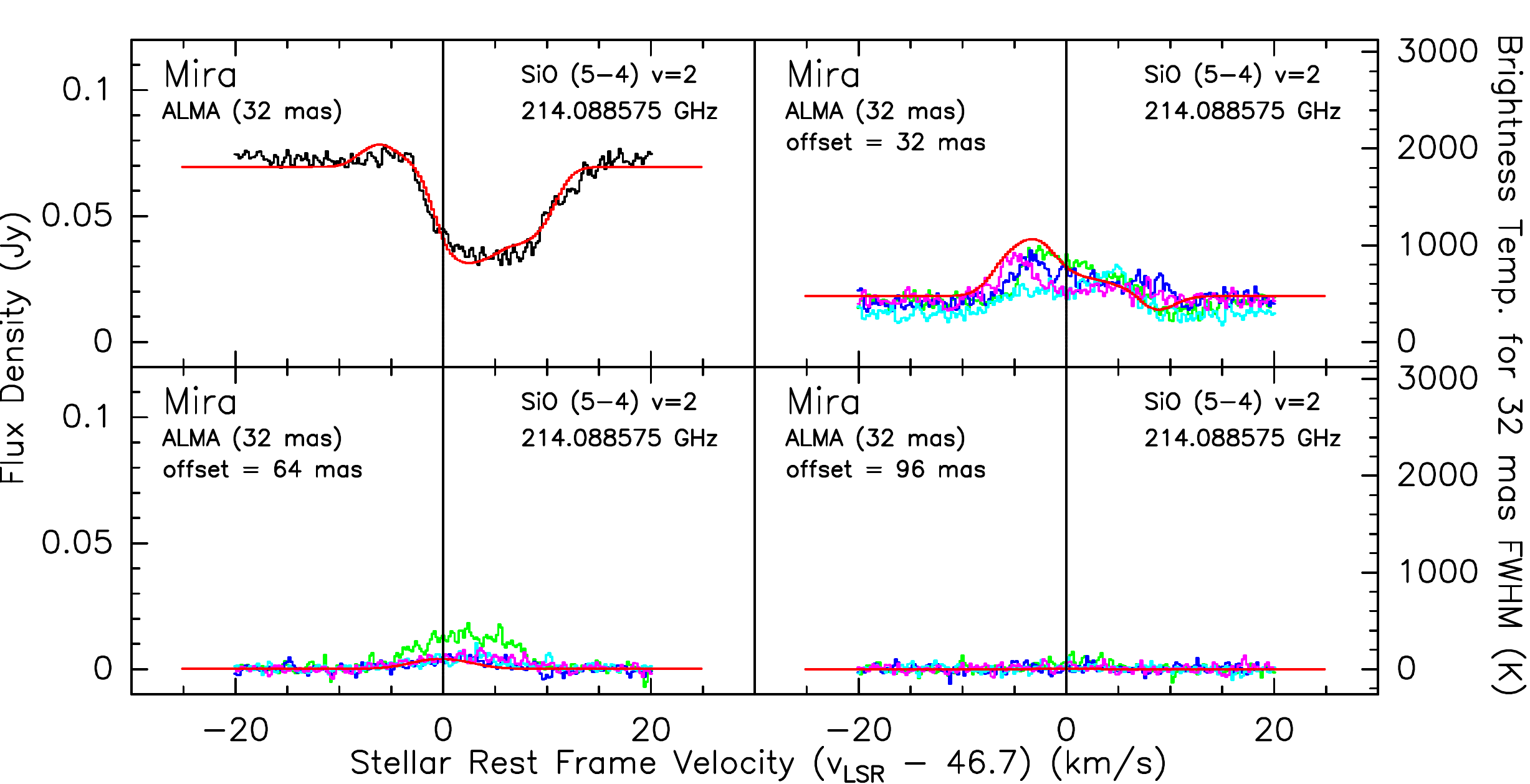}\caption{Preferred model: spectra of SiO ${\varv} = 2$ $J=5-4$ at various positions. The black histogram is the observed spectrum at the centre of continuum, green, blue, cyan and magenta histograms are the observed spectra along the eastern, southern, western and northern legs, respectively, at various offset radial distances as indicated in each panel. The red curves are the modelled spectra predicted by {\ratran}.}
\label{fig:m3siov2spec}
\end{figure*}

\begin{figure*}[!htbp]
\centering
\includegraphics[trim=1.0cm 7.3cm 2.0cm 1.5cm, clip, width=\multispecwidth]{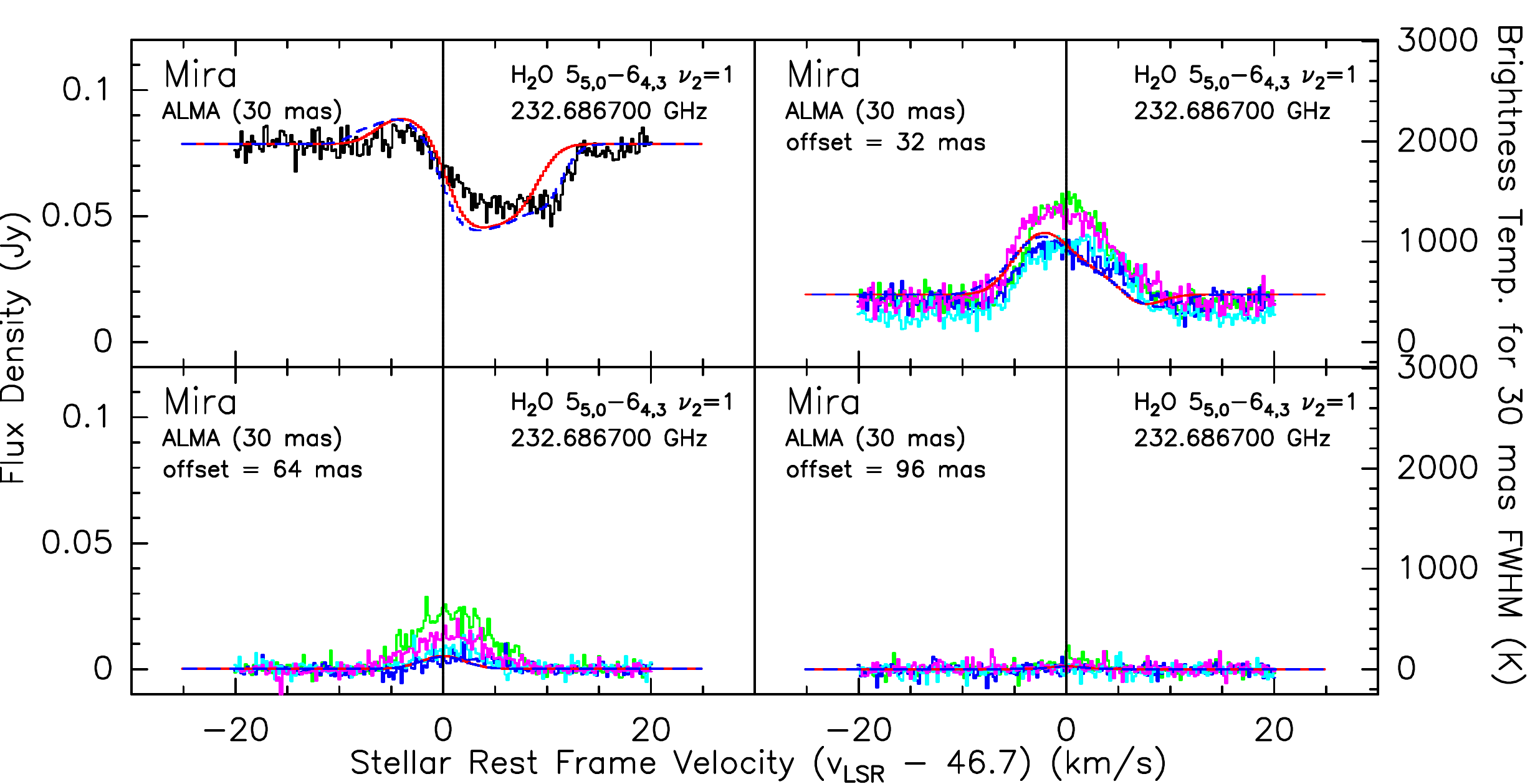}
\caption{Preferred model: spectra of H$_2$O $v_2=1$ $J_{K_a,K_c}=5_{5,0}-6_{4,3}$ at various positions. The black histogram is the observed spectrum at the centre of continuum, green, blue, cyan and magenta histograms are the observed spectra along the eastern, southern, western and northern legs, respectively, at various offset radial distances as indicated in each panel. The red curves are the modelled spectra predicted by {\ratran}, and the blue dashed curves are the same model adopting a high H$_2$O abundance (see the middle-right panel of Fig. \ref{fig:model3}).}
\label{fig:m3h2ov1spec}
\end{figure*}

Figures \ref{fig:m3siov0spec}, \ref{fig:m3siov2spec}, and \ref{fig:m3h2ov1spec} show the comparison of our modelled and observed spectra of SiO ${\varv} = 0$ $J=5-4$, SiO ${\varv} = 2$ $J=5-4$, and H$_2$O $v_2=1$ $J_{K_a,K_c}=5_{5,0}-6_{4,3}$, respectively. The top-left panel of these figures show the spectra extracted from the line-of-sight towards the continuum centre.

The top-right panel of Fig. \ref{fig:m3siov0spec} shows the modelled and observed SiO spectra at 32\,mas. In our modelled spectrum, there is a small absorption feature near the redshifted velocity of $+10\,\kms$, which is not seen in the data. This spectral feature is indeed part of the broad absorption as seen along the line-of-sight towards the radio continuum, which appears in the spectra at 32\,mas due to beam convolution. Hence, we may have introduced too much absorption, in particular near the peak infall velocities, to the model. Inhomogeneities in the images may have introduced additional emission features to the spectra, but this is not the case here. For example, there is a sharp spike in the observed spectra extracted from the southern position (in blue). This feature is due to an intensely emitting SiO clump at ${\sim}26\,{\rm mas}$ to the south of the continuum centre. The maximum brightness temperature of this clump in the map is ${\sim}2300\,{\rm K}$. The intense emission from this clump is probably due to maser action, because, if it were of thermal nature, then one would also expect the corresponding $^{29}$SiO line to be detected with intense emission from this clump. However, this clump is too far away from the other positions from which the spectra were extracted to contribute significant emission. Another possible explanation is that the infall velocity in our model may decrease too quickly with radius. For example, at the offset of 64\,mas, our modelled SiO ${\varv} = 0$ spectrum appears to be narrower than the observed spectra (middle-left panel of Fig. \ref{fig:m3siov0spec}). We have tried including a constant velocity layer of $10^{13}\,{\rm cm}$ at the peak infall velocity, but we still could not eliminate the absorption feature near $+10\,\kms$. If we adopt a much higher temperature, up to about $2600\,{\rm K}$, in the immediate proximity of the radio photosphere, then we would introduce too much blueshifted emission to the resultant spectra. Also, our spherically symmetric and homogeneous model obviously fails to reproduce the features arising in individual clumps.

In Fig. \ref{fig:m3h2ov1spec}, we present two different models of the H$_2$O spectral fitting using different input abundance profiles, which are plotted in the middle-right panel of Fig. \ref{fig:model3}. As shown in the top-left panel of Fig. \ref{fig:m3h2ov1spec}, the modelled spectra using constant abundance profile (``Model 3 abundance''; in red) do not fit well to the observed H$_2$O absorption spectra (in black) along the line-of-sight towards the continuum centre. In particular, the modelled spectrum does not show the strong observed absorption in the extreme redshifted velocities $>10\,\kms$. Hence, we have to introduce a sharp rise in the input H$_2$O abundance by about 10 times, to $5.0 \times 10^{-5}$, within the innermost region where the infall velocity peaks (``High H$_2$O abundance''; in blue) in order to reproduce the strong redshifted absorption feature in the spectrum.

Overall, considering the complexity of Mira's extended atmosphere and inner wind, we believe that Model 3 can satisfactorily reproduce most of the features in the observed SiO and H$_2$O spectra in ALMA Band 6. We therefore adopt it as our preferred model and use it as the base model of our further tests in Section \ref{sec:discussion}.


\section{Discussion}
\label{sec:discussion}

\subsection{Caveat in the interpretation of the gas density}
\label{sec:limitations}

In our modelling, we assume that the gas in Mira's extended atmosphere is composed of purely neutral, molecular hydrogen (H$_2$) in its rotationally ground state, $J=0$. At radii close to the radio photosphere of evolved stars, atomic hydrogen \emph{could be} the dominant species in terms of number density \citep{glassgold1983,doel1990,rm1997}. \citet{glassgold1983} have demonstrated that, the atmosphere of an evolved star with the effective temperature of about 3000\,K would be essentially atomic, and those of a star of about 2000\,K would be essentially molecular. Since the effective temperature of the star is expected to be higher than the brightness temperature of the radio photosphere \citep[e.g.][]{rm1997}, there should be significant amount of atomic hydrogen present in the regions being modelled. In fact, intense hydrogen Balmer series emission lines have long been detected in the atmosphere of Mira \citep[e.g.][]{joy1926,joy1947,joy1954,gillet1983,fabas2011}. The hydrogen emission is thought to be the results of dissociation and recombination of the atom due to shock waves propagating through the partially ionized hydrogen gas in the atmospheres of Mira variables \citep[e.g.][]{fox1984,fadeyev2004}. In addition, molecular hydrogen could well be excited to higher rotational levels (see our discussion in Appendix \ref{sec:appendix_h2rotation}).

We note that the collisional rate coefficients between SiO molecule and atomic hydrogen (H) and electrons (e$^{-}$) have already been computed by \citet{palov2006} and \citet{varambhia2009}, respectively. However, in this study, we did not attempt to calculate the fractional distribution of atomic/molecular hydrogen, or to consider the collisions between SiO molecule and atomic hydrogen, helium or electrons. Hence, the derived H$_2$ gas density from our {\ratran} modelling is just a proxy of the densities of all possible collisional partners of SiO, including rotationally excited molecular hydrogen, atomic hydrogen, helium, and even electrons, in the extended atmosphere of Mira.

\begin{figure*}[!htbp]
\centering
\includegraphics[trim=1.0cm 12.6cm 2.0cm 1.5cm, clip, width=\multispecwidth]{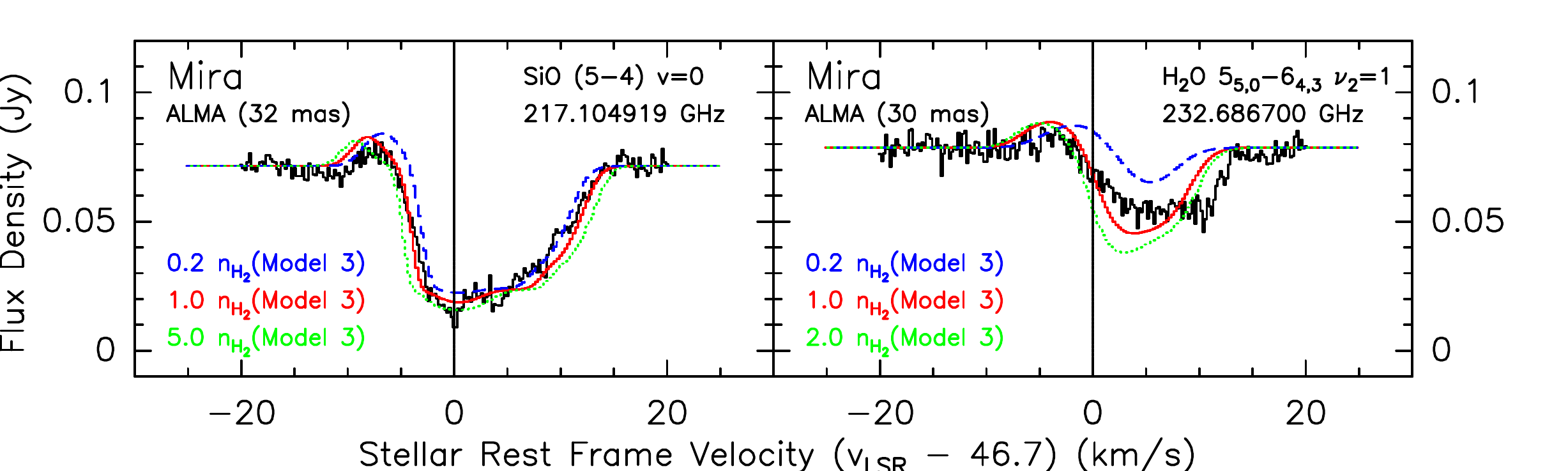}
\caption{Spectra of SiO ${\varv} = 0$ $J=5-4$ (left) and H$_2$O $v_2=1$ $J_{K_a,K_c}=5_{5,0}-6_{4,3}$ (right) extracted from the centre of the continuum. The black histogram is the observed spectrum, and the red curves are the modelled spectra from our preferred model, Model 3. The blue dashed curves are the modelled spectra by reducing the input H$_2$ gas density by a factor of 5; and the green dotted curves are the modelled spectra by increasing the input gas density by a factor of 5 for the modelling of SiO, and a factor of 2 for H$_2$O.}
\label{fig:densitytest}
\end{figure*}

In order to examine how well the H$_2$ gas density in our preferred model is constrained, we have modelled the SiO and H$_2$O spectra by scaling the gas density by various factors. Figure \ref{fig:densitytest} shows the results of these sensitivity tests on the input gas density. We have found that the SiO spectra, extracted from the line-of-sight towards the centre of the continuum, does not vary too much, even if the gas density is varied by about an order of magnitude. On the other hand, the H$_2$O spectra extracted from the centre shows significant change in the absorption depth even if the gas density is changed by a factor of ${\sim}2$. Hence, assuming other input parameters (particularly the molecular abundances and gas temperature) of Model 3 are fixed, our derived gas density is tightly constrained. The gas density reaches $10^{12}$--$10^{13}\,{\rm cm}^{-3}$ just beyond the radio photosphere. This is consistent with other models that explain the radio continuum fluxes from Mira's radio photosphere \citep{rm1997}, and the near-infrared H$_2$O spectrum \citep{yamamura1999}. On the other hand, the derived gas density is much higher (by 2--4 orders of magnitude) than those predicted from hydrodynamical models (see Sect. \ref{sec:discuss-codex}).

\subsection{Structure of the extended atmospheres}
\label{sec:discuss-atm}

Our modelling results of the molecular emission and absorption of SiO and H$_2$O gas allow us to compare the structure of extended atmosphere of Mira as inferred from previous observations in various frequencies. We will first briefly summarise the relevant observations of Mira in Sect. \ref{sec:discuss-atm-intro}, then discuss our interpretation on Mira's molecular layer in Sect. \ref{sec:discuss-layer} and dust condensation zone in Sect. \ref{sec:discuss-dust}.


\subsubsection{Previous observations}
\label{sec:discuss-atm-intro}

Combining their centimetre-wavelength observations and millimetre/infrared fluxes in the literature, \citet{rm1997} have demonstrated that long-period variables have a ``radio photosphere'' with a radius about twice that of the optical/infrared photosphere. The latter is determined from the line-free regions of the optical or infrared spectrum and is defined as the stellar radius, $R_{\star}$. In the following discussion, we adopt the value of $R_{\star}$ to be $12.3\,{\rm mas}$ (or $292\,R_{\astrosun}$) as determined by \citet{perrin2004}. The spectral index in the radio wavelengths is found to be $1.86 ({\approx} 2)$, close to the Rayleigh-Jeans law at low frequencies of an optically thick blackbody \citep{rm1997}. \citet{mrm2015} and \citet{planesas2016} found that this spectral index can also well-fit the submillimetre flux densities of $o$ Cet at 338\,GHz in ALMA Band 7 and at 679\,GHz in ALMA Band 9, respectively.

The ``radio photosphere'' encloses a hot, optically thick molecular layer (${\sim} 2 \times 10^3$\,K) predominantly emitting in the infrared. Observations have revealed that this molecular layer lies between radii of ${\sim} 1 $ and $2\,R_{\star}$. \citet{haniff1995} found that, for $o$ Cet, the derived radius of the strong TiO absorption near 710\,nm with a uniform disk model is about $1.2 \pm 0.2\,R_{\star}$. \citet{perrin2004} derived, by fitting a model consisting of an infrared photosphere and a thin, detached molecular (H$_2$O+CO) layer to the infrared interferometric data, that the radius of the molecular layer around $o$ Cet is about $2.07 \pm 0.02\,R_{\star}$. Alternatively, \citet{yamamura1999} have modelled the H$_2$O spectral features in the near-infrared (${\sim}2\text{--}5\,\mu{\rm m}$) spectrum of $o$ Cet with a stack of superposing plane-parallel layers: the star, an assumed hot SiO (2000\,K) layer, a hot H$_2$O (2000\,K), and a cool H$_2$O (1200\,K) layers. Assuming the hot SiO layer has a radius of $2.0\,R_{\star}$, they derived the radii of the hot and cool H$_2$O layers to be $2.0\,R_{\star}$ and $2.3\,R_{\star}$, respectively. \citet{ohnaka2004} employed a more realistic model for the extended molecular layer with two contiguous spherical shells, a hotter and a cooler H$_2$O shells, above the mid-infrared photosphere. By fitting to the $11\,\mu{\rm m}$ spectrum, \citet{ohnaka2004} derived the radii of $1.5\,R_{\star}$ and $2.2\,R_{\star}$ for the hot (1800\,K) and cool (1400\,K) H$_2$O shells, respectively.

Beyond the molecular layer and the ``radio photosphere'', there is a ring-like region of SiO maser emission at the radius between $2\,R_{\star}$ and $3\,R_{\star}$. Maser emission naturally arises from a ring-like structure because the maser requires a sufficiently long path length of similar radial velocity in order to be tangentially amplified to a detectable brightness \citep{diamond1994}. Such a SiO maser ring has been imaged in detail at various stellar phases towards the oxygen-rich Mira variable TX Cam \citep[e.g.][]{diamond2003,yi2005}. For $o$ Cet, \citet{rm2007} have directly imaged the radio photosphere and the SiO $J=1$--$0$ maser emission at 43\,GHz and found that the radii of the radio photosphere and the SiO maser ring are about $2.1\,R_{\star}$ and $3.3\,R_{\star}$, respectively, with $R_{\star} = 12.29 \pm 0.02\,\rm mas$ being the radius of the infrared photosphere as model-fitted by \citet{perrin2004}.

Further out, beyond the SiO maser emission region, dust grains start to form. The major types of dust around oxygen-rich AGB stars are corundum (Al$_2$O$_3$) and silicate dust. Using the hydrodynamical model from \citet{ireland2004a} and \mbox{\citet{ireland2004b}}, \citet{gray2009} have modelled SiO maser emission in Mira variables and found that the presence of Al$_2$O$_3$ dust may either enhance or suppress SiO maser emission. From interferometric observations of various Mira variables at near-infrared ($2.2\,\mu$m), mid-infrared ($8$--$13\,\mu$m) and radio (43\,GHz, 7\,mm) wavelengths, \citet{perrin2015} fitted the visibility data with models similar to \citet{perrin2004} (stellar photosphere + detached shell of finite width) and found that Al$_2$O$_3$ dust predominantly forms between $3\,R_{\star}$ and $4.5\,R_{\star}$, while silicate dust forms in $12\,R_{\star}$--$16\,R_{\star}$, which is significantly beyond the radius of SiO maser emission and the silicate dust formation radius derived from previous observations \citep[e.g.][]{danchi1994}.


\subsubsection{Molecular layer}
\label{sec:discuss-layer}

From our visibility analysis (see Appendix \ref{sec:appendix_cont}), we determine the mean radius of the 1.3\,mm-photosphere, to be $R_{\rm 229\,GHz} = 22.90 \pm 0.05\,{\rm mas}$ ($543\,R_{\astrosun}$). This is about 1.9 times the size of the near-infrared photosphere ($R_{\star} = 12.3\,{\rm mas}$; $292\,R_{\astrosun}$) as determined by \citet{perrin2004}. As we have summarised, previous visibility modelling of near- (2--5\,$\mu$m) and mid-infrared (11\,$\mu$m) interferometric data has suggested the existence of an optically thick, hot molecular H$_2$O+SiO layer with a maximum radius of $2.3\,R_{\star}$ (${\sim}30\,{\rm mas}$) \citep{yamamura1999,ohnaka2004}. Thus, this ALMA SV observation has the sufficient angular resolution to resolve the hot molecular layer in the millimetre-wavelength regime and in addition allows probing its velocity structure.

Modelling the spectral lines of H$_2$O and SiO molecules at various projected radial distance from the star, we have determined that the kinetic gas temperature within the mid-infrared molecular layer ($30\,{\rm mas} \sim 5 \times 10^{13}$\,cm) has to be about 1400--2100\,K. The temperature range is consistent with those previously modelled by \citet{rm1997} from their centimetre-wavelength observations of Mira's radio photosphere, and by \citet{yamamura1999} and \citet{ohnaka2004} from infrared observations using simple models of contiguous, uniform molecular H$_2$O+SiO layers.

In our maps, the emission from the vibrationally excited ($E_{\rm up}/k > 3500\,{\rm K}$) SiO ${\varv} = 2$ and H$_2$O $v_2=1$ lines has an extent of ${\lesssim} 100$\,mas (${\lesssim} 8\,R_{\star}$). The core emission region of the SiO 5--4 ${\varv} = 0$  vibrational ground state line (i.e., excluding the extended filamentary or arc-like emission feature to the west/south-west) that is detected at $\ge 3\sigma$ has radii between 200\,mas ($3.3 \times 10^{14}\,{\rm cm}$; to the south-east) and 600\,mas ($9.9 \times 10^{14}\,{\rm cm}$; to the west), and the size of the half-maximum emission is roughly 100--150\,mas (see Fig. \ref{fig:siov0chan_csub}). Hence SiO emits rotational emission up to a radius of ${\sim}50\,R_{\star}$, far beyond the radius of the molecular layer  probed by infrared interferometers. \citet{perrin2015}, assuming that the mid-infrared $N$-band visibilities between 7.80\,$\mu$m and 9.70\,$\mu$m as the only signature of gas-phase SiO emission, concluded that the SiO can only be found in gas phase within $3\,R_{\star}$. We suggest that this discrepancy is due to the excitation effect of SiO molecules. The ground state SiO line in the ALMA SV observation has an energy above the ground of only ${\sim}30\,{\rm K}$, and therefore is excited throughout the region within the silicate dust condensation zone. While the ALMA images indicate a significant amount of gas-phase SiO molecules, the gas temperature beyond the molecular layer (${\lesssim}1000\,{\rm K}$) is insufficient to collisionally excite the SiO molecules to higher vibrational states. Thus, the SiO molecule does not produce detectable infrared emission beyond ${\sim}3\,R_{\star}$ even if it is abundant there.


\subsubsection{Dust shells and the sequence of dust condensation}
\label{sec:discuss-dust}

The radii of dust shells around Mira have been measured with infrared interferometry at 11\,$\mu$m by \citet{danchi1994} and \citet{lopez1997}. A single silicate dust shell from 60--2500\,mas with a dust temperature of 1063\,K at the inner radius was adopted in the model of \citet{danchi1994}. \citet{lopez1997} used a two-shell model, composed purely of silicate grains, at radii of 50\,mas and 200\,mas. The dust temperature of the inner radius of the inner dust shell is about 1160--1380\,K. These results suggest that dust grains start to form around Mira at a temperature above 1000\,K and a radius of ${\sim}2$--$3\,R_{\star}$, where $R_{\star}$ was determined to be $19.3$--$23.6\,{\rm mas}$. If we use our adopted value of 12.3\,mas for $R_{\star}$, then the inner dust formation radius would be ${\sim}4$--$5\,R_{\star}$. Compared to the recent model of \citet{perrin2015}, this range is significantly smaller than the silicate formation radii, which is at least $12\,R_{\star}$, but is consistent with the radii of corundum formation.

Studies of silicate dust formation suggest that efficient condensation occurs only when the gas temperature drops to below 600\,K \citep[e.g.][]{gail1998}. This allows the SiO gas to emit for a much larger radius in the extended atmosphere of Mira, as compared to the radii of its dust shells derived previously and described above. The discussion of higher silicate dust condensation temperatures has been recently revived by \citet{gail2013}. Their new measurements of the vapour pressure of solid SiO suggest that gas-phase SiO molecules may first nucleate into SiO clusters, and then condense onto dust grains \citep{gail2013}. The gas temperature (assumed to be also the dust temperature at the inner boundary of the dust shell) at which SiO gas start to deplete is estimated to be about 600\,K, for a mass-loss rate, \Mdot, of ${\sim}10^{-6}\,M_{\astrosun}\,{\rm yr}^{-1}$, increasing to 800\,K for $\Mdot = 10^{-4}\,M_{\astrosun}\,{\rm yr}^{-1}$. The SiO nucleation process thus allows the depletion of SiO gas to begin at a higher gas temperature, i.e., at a smaller inner radius, than previously thought. However, the result of \citet{gail2013} still cannot explain the high dust temperature ($>1000\,{\rm K}$) as derived from visibility fitting by \citet{danchi1994} and \citet{lopez1997}.

\begin{figure*}[!htbp]
\centering
\includegraphics[width=0.65\textwidth]{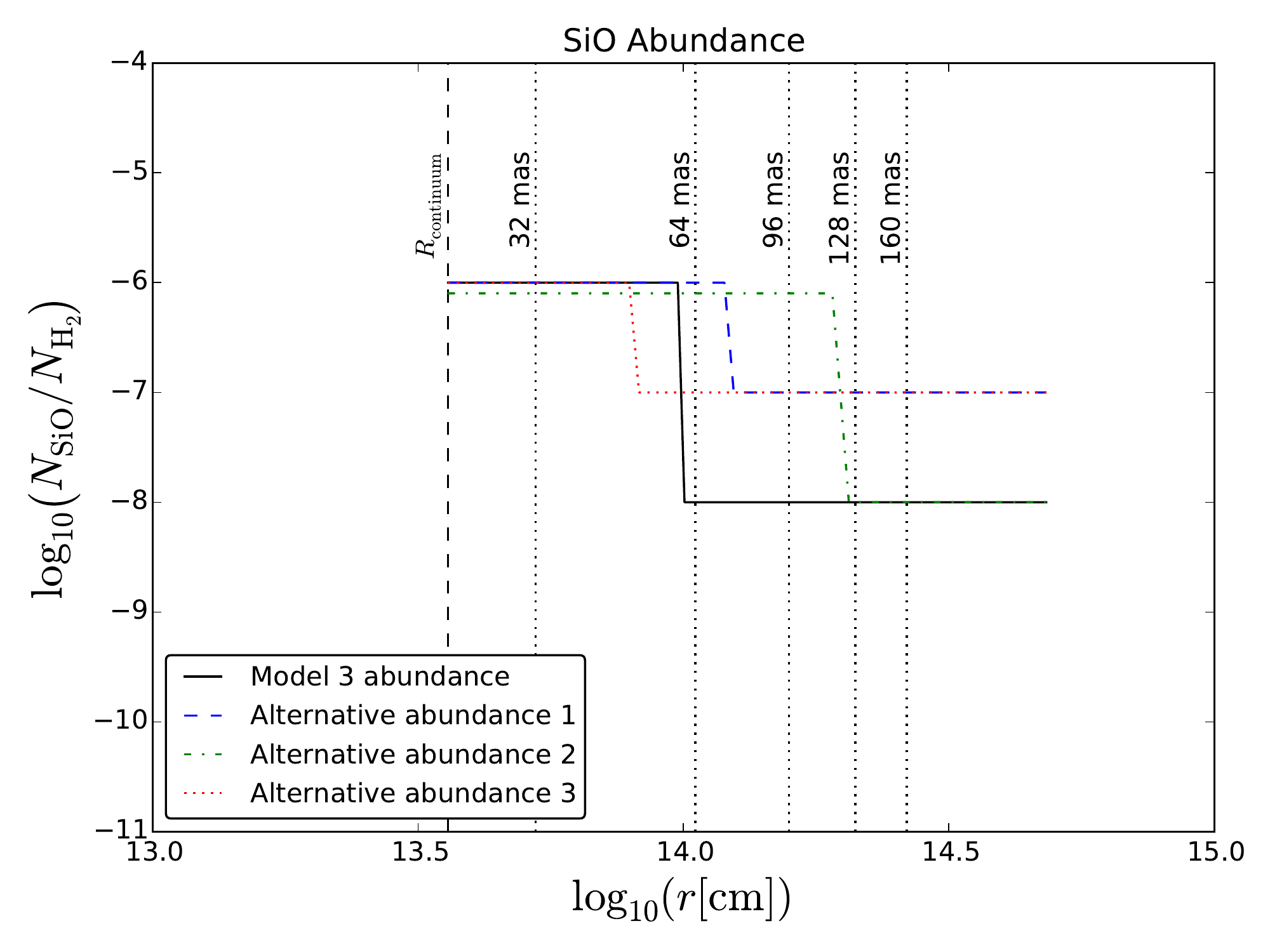}
\caption{Alternative input $^{28}$SiO abundance profiles for Model 3 which produce similar modelled spectra. The solid black curve is the same 2-step abundance profile as in Model 3, which is close to the minimum possible abundance to fit the ALMA spectra. The three other coloured curves are alternative abundance profiles using also 2-step functions. The abundance profile 1 (blue, dashed) has an inner abundance of $1 \times 10^{-6}$ up to the radius of ${\sim}6\,R_{\star}$, and an outer abundance of $1 \times 10^{-7}$; profile 2 (green, dashed-dotted) has an inner abundance of $8 \times 10^{-7}$ up to ${\sim}10\,R_{\star}$, and an outer abundance of $1 \times 10^{-8}$; and profile 3 (red, dotted) has an inner abundance of $1 \times 10^{-6}$ up to ${\sim}4\,R_{\star}$, and an outer abundance of $1 \times 10^{-7}$.}
\label{fig:test-abun}
\end{figure*}

In our Model 3, the radius at which the $^{28}$SiO abundance decreases significantly is adopted to be ${\sim}60\,{\rm mas}$, which corresponds to $1.0 \times 10^{14}\,{\rm cm}$ or ${\sim}5\,R_{\star}$. The modelled gas temperature at this radius is ${\sim}600\,{\rm K}$. Besides this SiO abundance profile, we have also tested with other 2-step functions in order to find out the maximum and minimum amount of SiO molecules, and the possible range of SiO depletion radii, required to reproduce the observed ALMA spectra. Figure \ref{fig:test-abun} shows three examples of alternative SiO abundance models. All these models are as good as our Model 3 in reproducing the spectra. Our experiments show that, at the very least, the SiO abundance should be ${\sim}1 \times 10^{-6}$ within ${\sim}4\,R_{\star}$ and ${\sim}10^{-8}$--$10^{-7}$ beyond that. In other words, SiO molecules cannot deplete onto dust grains in a significant amount within ${\sim}4\,R_{\star}$. This radius is consistent with the inner radius of the silicate dust shells derived in the literature. Our tests, however, have shown that the synthesised SiO spectra in the outer radii are not sensitive to a higher value of the SiO abundance, or the exact shape of the abundance profile. The actual radius where gas-phase SiO molecule condenses onto dust grains therefore may be much further from the star than $4\,R_{\star}$. Moreover, the actual degree of SiO gas depletion, through silicate dust condensation, nucleation of molecular clusters, or other gas-phase chemical reactions \citep[e.g.][]{gail2013,gobrecht2016}, may not be as high as assumed in our preferred model.

The gas temperature at which SiO gas start to deplete in our models is about $490$--$600\,{\rm K}$, much below the dust temperature at the inner dust shells as derived observationally by \citet{danchi1994} and \citet{lopez1997}. This temperature is also somewhat lower, by about $100\,{\rm K}$, than the gas temperature at which SiO gas starts to nucleate into clusters \citep{gail2013}. However, we note that the visibility models derived by \citet{danchi1994} and \citet{lopez1997} have assumed that the dust around Mira is composed of pure silicate grains. The derived parameters are therefore based on the adopted optical properties of silicate dust grains. Other possible compositions of dust grains around oxygen-rich stars, such as corundum or the mixture of corundum and silicate, cannot be excluded but have not been explored in those models.

Corundum, the crystalline form of aluminium oxide (Al$_2$O$_3$), has a high condensation temperature of ${\sim}1700\,{\rm K}$ \citep[e.g.][]{grossman1974,lorenzmartins2000} and is the most stable aluminium-containing species at a temperature below 1400\,K \citep{gail1998}. \citet{lml1990} have classified the circumstellar dust shells of oxygen-rich AGB stars into several groups according to the spectral features found in their mid-infrared spectral energy distributions (SEDs). These SED groups show (a) broad emission feature from 9 to 15\,$\mu$m, (b) multiple components with peaks near 10, 11.3 and 13\,$\mu$m, and (c) strong, well-defined characteristic silicate peaks at 9.8 and 18\,$\mu$m. \citet{lml1990} have also suggested that the circumstellar dust shells follow an evolutionary sequence starting from the class showing the broad feature, then multiple components and finally silicate features in the SED. From a survey of O-rich AGB stars, most of them being Mira variables, \citet{lorenzmartins2000} have successfully fitted the SEDs showing (a) broad features, (b) multiple components (or the intermediate class), and (c) silicate features and with corundum grains, a mixture of corundum and silicate grains, and pure silicate grains, respectively. Their results show that the inner radius of the modelled dust shells increases from the broad class, the intermediate class, and the silicate class. The fitted dust temperature of the hottest grains also follows the same sequence. In addition, they also found that the optical depths of the corundum-dominated emission are much smaller than those of the silicate-dominated emission. \citet{lorenzmartins2000} thus concluded that their results were consistent with the evolutionary sequence suggested by \citet{lml1990}. Corundum grains are the first species to form in the circumstellar dust shells, at a small radius of ${\sim}2$--$3\,R_{\star}$ and a temperature of ${\sim}1400$--$1600\,{\rm K}$. At a later stage, silicate grains start to form and dominate the emission features in the SED. The inner radius of the silicate dust shells and temperature of the hottest silicate grains are ${\sim}5$--$20\,R_{\star}$ and ${\sim}500$--$1000\,{\rm K}$, respectively.

Our modelling results have shown that gas-phase SiO starts to deplete at a radius of \emph{at least} $4\,R_{\star}$ and a gas temperature of ${\lesssim}600\,{\rm K}$. In addition, the observed spectra show that SiO molecules survive in the gas-phase well below 1000\,K. This is apparently inconsistent with the fitting by \citet{danchi1994} and \citet{lopez1997}, that the silicate dust shells form at a temperature above 1000\,K. We therefore suggest that the inner hot dust shells around Mira may indeed be composed of other grains types, possibly corundum, instead of silicate grains as previously assumed. Although no prominent spectral features of corundum have been reported \citep[e.g.][]{lopez1997,lobel2000}, we note that the corundum grains may be coated with silicates when the temperature becomes low further out in the dust shell \citep[e.g.][Sect. 6.1 and references therein]{karovicova2013}. The optical depth of pure corundum grains, which only exist close to the star, may also be much lower than that of silicate grains \citep[e.g.][]{lorenzmartins2000} and therefore the corundum features may not be easily distinguished in the SED from the silicate features.


\subsubsection{Maser emission}
\label{sec:discuss-maser}

Among all the spectral lines covered in this ALMA SV observation, only SiO ${\varv} = 1$ $J=5-4$ and ${\varv} = 1$ $J=2-1$ (in ALMA Band 3; not included in this article) exhibit strong maser emission (Fig. \ref{fig:band6lines}). In the images, isolated SiO ${\varv} = 1$ $J=5-4$ maser spots are seen outside Mira's radio photosphere, primarily at the radial distances between ${\sim}30$ and $120\,{\rm mas}$ from the fitted position of the radio continuum. The (relative) spatial distribution of the SiO ${\varv} = 1$ maser is consistent with those previously reported by \citet{boboltz2004} and \citet{cotton2008} (and references therein) in other lower-$J$ maser transitions. The presence of maser emission indicates that the SiO gas in those maser-emitting spots \emph{must not} be in local thermodynamic equilibrium (LTE). In our preferred (1-D and smooth) model, the gas density is uniformly high throughout the maser-emitting region and hence all possible maser action is quenched. The predicted excitation temperature does not show any negative values throughout the modelled region (Fig. \ref{fig:model3}) and therefore population inversion, a prerequisite for maser emission to take place, of the SiO ${\varv} = 1$ transition cannot be predicted by our simple model. We note that our model does not include any external infrared radiation field, in particular the infrared pumping band of SiO molecule near 8.1\,$\mu$m ($\Delta{\varv} = 1$, fundamental) and 4.0\,$\mu$m ($\Delta{\varv} = 2$, first overtone). The only input radiation field in our model is a blackbody of $2600\,{\rm K}$, representing the radio photosphere of Mira. This temperature is lower than the typical infrared effective temperature of Mira, which is about 3200\,K \citep{woodruff2004}. So the radiation field does not realistically approximate the radiative excitation of SiO molecule to higher vibrational states. Because our modelling aims to explain the \emph{general} physical conditions of Mira's extended atmosphere, we did not attempt to construct a sophisticated model to explain both maser and non-maser emission.

The water line covered in this SV observation, H$_2$O $v_2=1$ $J_{K_a,K_c}=5_{5,0}-6_{4,3}$, near 232.7\,GHz does not show any maser emission. \citet{gray2016} have conducted extensive radiative transfer modelling to explore the physical conditions under which the modelled H$_2$O lines (including all possible lines covered by ALMA) would exhibit maser emission in the envelopes of evolved stars. Slab geometry and silicate dust, which is optically thin in the millimetre-wavelengths and optically thick in the radiative pumping bands of H$_2$O's vibrational states (e.g., $v_2$ band at $6.27\,\mu{\rm m}$), have been assumed in their modelling \citep{gray2016}. The 232.7-GHz H$_2$O emission is seen from the radio photosphere up to about 80\,mas (Fig. \ref{fig:h2ov1chan_csub}). Hence, the H$_2$O-emitting region corresponds to the kinetic temperature of ${\sim}550$--$2100\,{\rm K}$, the gas density of $n_{{\rm H}_2} {\sim} 4 \times 10^{10}$--$1 \times 10^{13}\,{\rm cm}^{-3}$, and hence the H$_2$O molecular density of $n_{{\rm H}_2{\rm O}} {\sim} 2 \times 10^{5}$--$5 \times 10^{7}\,{\rm cm}^{-3}$ in our preferred model. Our derived H$_2$O molecular density lies well within the range in the model of \citet{gray2016} that is predicted to exhibit strong maser emission in the 232.7-GHz transition \emph{if} the dust temperature is high enough. The absence of the 232.7-GHz H$_2$O maser is consistent with a (silicate-type) dust temperature lower than approximately $900$, $1000$, and $1600\,{\rm K}$ for the respective gas kinetic temperature of about $500$, $1000$, and $1500\,{\rm K}$ \citep[see Fig. 10 of][]{gray2016}. These comparisons, although not a conclusive proof, suggest that hot dust grains that are optically thick at the $v_2$ band ($6.27\,\mu{\rm m}$) did not exist in Mira's extended atmosphere during the ALMA SV observation.


\subsection{Comparison with current hydrodynamic models}
\label{sec:discuss-hydrodyn}

\subsubsection{Early hydrodynamic models for stellar pulsation}
\label{sec:discuss-old-hydro-models}


There are many numerical hydrodynamical calculations on Mira variables to simulate the variation of pulsation velocity, number density, kinetic temperature as functions of the stellar phase and/or radial distance from the star. Pioneering work includes the studies of \citet{willsonhill1979}, \citet{hillwillson1979}, \citet{wood1979}, \citet{willson1987}, and \citet{beach1988} and \citet{bowen1988a,bowen1988b,bowen1989} (hereafter, Bowen's models). \citet{wood1979}, \citet{willson1987}, and \citet{bowen1988a,bowen1988b,bowen1989} have compared the effect of radiation pressure on dust on the mass-loss rate and the velocities of the stellar outflows. The outflow/infall velocity profiles as a function of radius as derived from these models are qualitatively similar. These authors all predict alternating outflow and infall layers in close proximity to the star, within about $4$--$6 \times 10^{13}\,{\rm cm}$. Beyond that radius, the dust-driven winds exhibit accelerating outflow. In the region where material expands and falls back, large-scale shocks are produced at the interface between the outer, infall layer and the inner outflow.

Infall motions in the extended atmosphere of the star can be observed as inverse P Cygni profiles in the spectra. The material at the near side of the star will show redshifted absorption and the material at the far side that is not blocked by the continuum will show blueshifted emission. These emission features are present even if there is no hot material (perhaps shock-heated) with a temperature higher than the continuum brightness temperature, and are also known as the ``nebular'' effect \citep[e.g.][]{bessell1996,scholz2000}, in which the large volume of the highly extended atmosphere, albeit only weakly emitting per unit volume, adds up to produce significant emission. For example, the redshifted absorption in the CO second overtone ($\Delta {\varv} = 3$) lines from $o$ Cet indicates infall motions in the deep photospheric layers of the star \citep{hinkle1984}. Results from the early spectroscopy by \citet{joy1926,joy1954} also suggest infall motion in the extended atmosphere of $o$ Cet, based on modern information on the systemic (centre-of-mass) velocity of the star \citep[see also the interpretation by][]{gabovits1936}.

Bowen's models have been adopted by \citet{h96sio} and \citet{h01h2o} to simulate the SiO (${\varv} = 1$ and $2$) and H$_2$O (ground state and $v_2=1$) masers from a template M-type Mira (parameters were based on $o$ Cet) at a single stellar phase, and by \citet{gh00} and \citet{h02} to simulate the variability of SiO (${\varv} = 1$) masers at various epochs of a stellar cycle of the model Mira. \citet{gray2009} have comprehensively reviewed the success and limitations of these precursor models for SiO maser simulations. One major drawback of Bowen's hydrodynamical solutions is that the pulsation phase in the model (with phase 0 defined as the moment when the inner boundary of the model atmosphere, or the ``piston'', is moving outwards at the maximum speed) is disconnected from the stellar phase as determined from the optical or infrared brightness variations \citep{h96sio}. In addition, the assumption of a constant infrared radiation field by dust and the stellar photosphere was also too simplistic for Mira variables \citep{gray2009}.

Solving the hydrodynamical equations as presented in \citet[Chap. 12]{rm1967}, who adopt the von Neumann-Richtmyer pseudo-viscosity method as the artificial shock dissipation mechanism \citep{vNR1950}, the authors of early hydrodynamic models have derived the dynamical structures around Mira-like variables. Bowen's models considered also the thermal relaxation of shocks via radiative cooling from neutral hydrogen atoms at high temperatures (${\gtrsim}6000\,{\rm K}$) and from other species (which is represented by an assumed cooling coefficient) at low temperatures \citep[see Sects. II(b) and II(c)(iii) of][]{bowen1988a}. Bowen's models predicted the existence of an extended (${\sim}10^{14}\,{\rm cm}$) post-shock region of elevated gas temperatures (${\sim}10\,000\,{\rm K}$) near the optical/infrared photosphere.

On the other hand, using the hydrogen Balmer series emission lines as the signatures of shock-heated region, \citet{fox1984} and \citet{fox1985} developed a theoretical model of shock waves in the atmosphere of Mira variables and predicted that these pulsation-driven shocks dissipate within a very thin region in the extended atmospheres, of a several orders of magnitude smaller extent than the stellar radius (i.e., $\ll 10^{12}\,{\rm cm}$). The circumstellar shock models by \citet{willacy1998} and \citet{gobrecht2016} predict that the cooling length of H$_2$ dissociation, depending on the shock velocities, is typically $<10^8\,{\rm cm}$. The very narrow post-shock region suggests that the relaxation of the shocked material towards radiative equilibrium and local thermodynamic equilibrium is essentially instantaneous and therefore the post-shock heating might be neglected. \citet{woitke1996} argued that Bowen's cooling rate was underestimated by a few orders of magnitude, thus resulting in an atmosphere with highly elevated gas temperature. Furthermore, based on the observational constraints from Mira's radio continuum emission at 8\,GHz, \citet{rm1997,rm1997b} have shown that the amplitude of gas temperature disturbance, probably due to shocks or pulsations, can only be about $300\,{\rm K}$ (assuming a shock propagation speed of $7.3\,\kms$). These are in contrast with Bowen's non-equilibrium models which expect a rather extended shock-heated region with highly elevated gas temperature. \citet{bessell1989} have compared the synthesised spectra from Bowen's non-equilibrium model with the observed spectrum between 600\,nm and 4\,$\mu$m, and found that they did not match at all. Moreover, the hydrogen spectra (e.g., H$\alpha$) predicted by \citet{luttermoser1990,luttermoser1992} using Bowen's model were much broader, and have different line profiles, than what was actually observed \citep{woodsworth1995}. Hence, the non-equilibrium models with extended high-temperature regions are not suitable for the extended atmospheres of Mira variables \citep[see also the discussions in][]{woitke1998,willson2000}.

\begin{figure*}[!htbp]
\centering
\includegraphics[height=0.21\textheight]{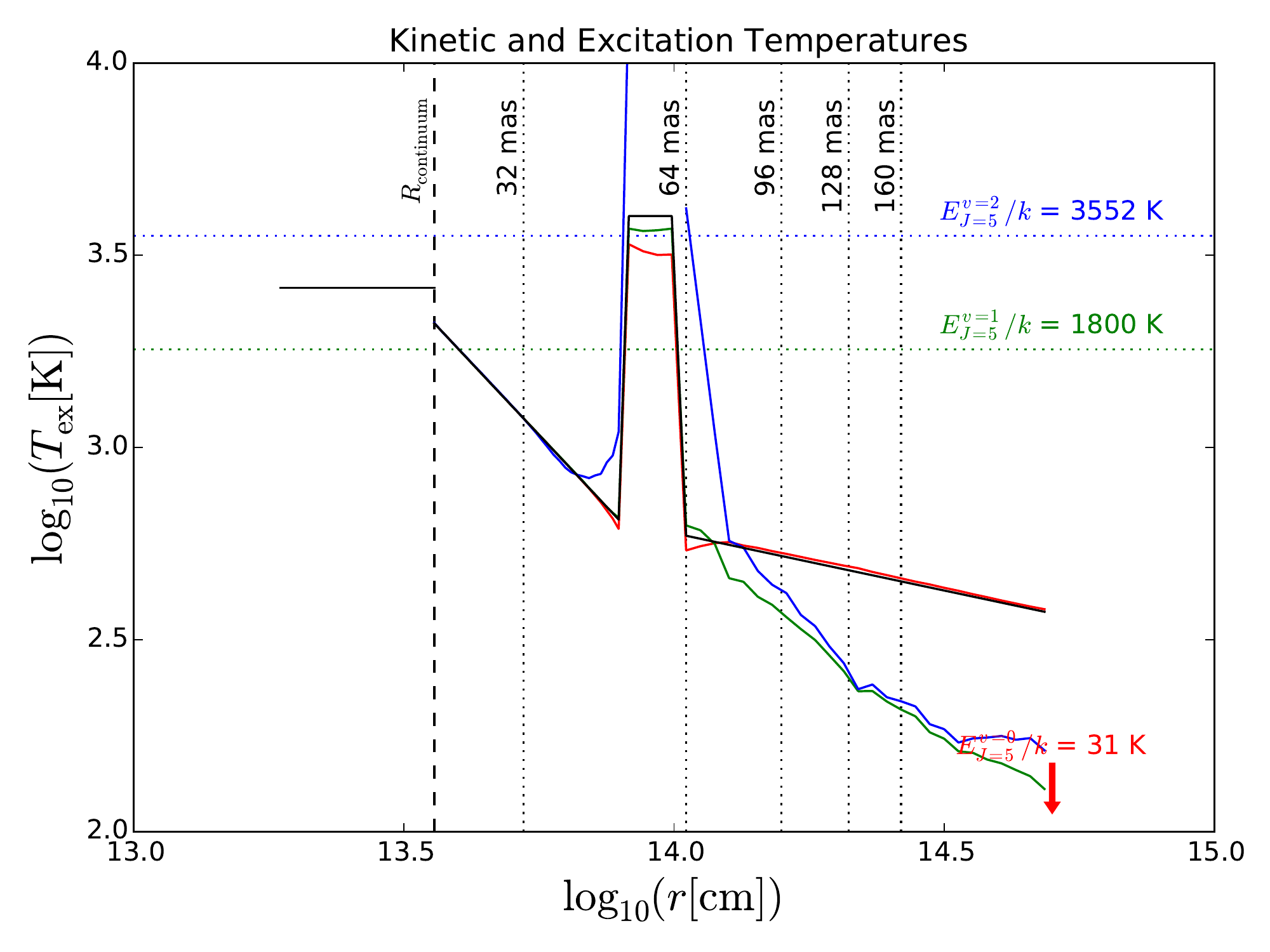}
\includegraphics[trim=1.0cm 7.3cm 2.0cm 1.5cm, clip, height=0.21\textheight]{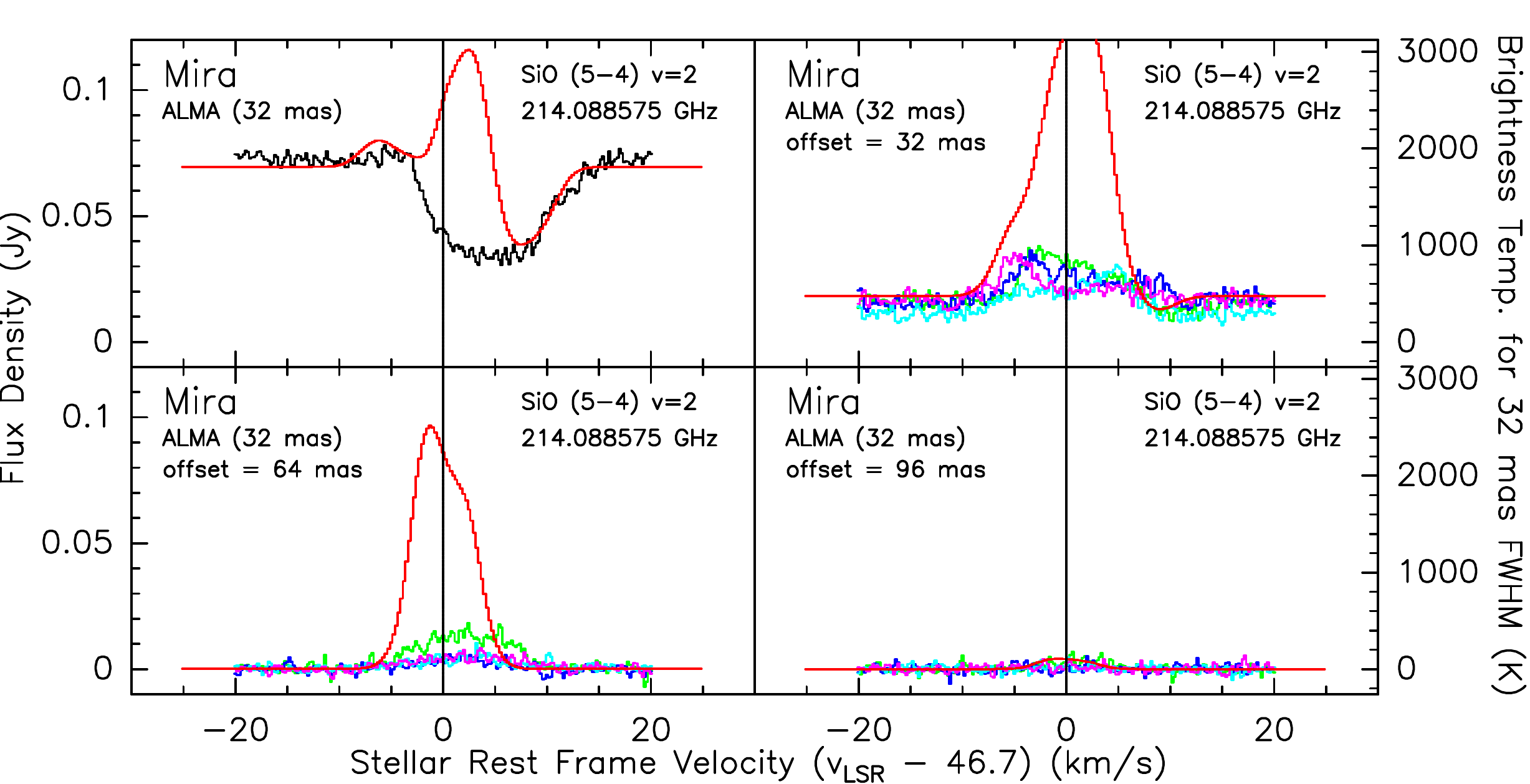} \\
\caption{The preferred model with a layer of an elevated gas temperature. The layer has a width of $2.0 \times 10^{13}\,{\rm cm}$ and an arbitrarily chosen temperature of $4000\,{\rm K}$. The left panel shows the elevated gas temperature profiles and the right panel shows the observed and modelled SiO ${\varv} = 2$ $J=5-4$ spectra as examples. All other parameters are the same as our preferred model (Model 3).}
\label{fig:test-hightemp}
\end{figure*}

For the purpose of verification, we have also conducted similar tests to \citet{bessell1989} by introducing to our preferred model (Model 3) an arbitrary extended layer (${\gtrsim}10^{12}\,{\rm cm}$) of elevated gas temperature (${\sim}4000\,{\rm K}$) at various radial distances from Mira. The elevated temperature is significantly higher than the brightness temperature of the stellar radio continuum (${\sim}2600\,{\rm K}$). Figure \ref{fig:test-hightemp} shows an example of our tests. Even though the gas temperature is elevated for just a relatively thin layer ($2 \times 10^{13}\,{\rm cm} \sim R_{\star}$) as compared to Bowen's non-equilibrium model (${\sim}10^{14}\,{\rm cm}$), the synthesised spectra exhibit very strong emission features that are absent in the data. In fact, if we further extend the zone of elevated gas temperature, the emission spikes only become more prominent. Therefore our tests suggest that during this particular ALMA SV observation, the extended atmosphere of Mira did not contain any extended (${\gtrsim}10^{12}\,{\rm cm}$) shock-heated (above the brightness temperature of the radio continuum) region as predicted from the non-equilibrium model. This is consistent with those shock wave models that suggest that the shock relaxation in Mira atmosphere takes place within a very thin zone compared to the stellar radius \citep[e.g.][]{fox1984,fox1985}. In addition, we do not expect the existence of stellar chromosphere, as suggested to explain the H$\alpha$ absorption lines from some semi-regular variables \citep{luttermoser1994,wood2004}, beyond the radio photosphere of Mira during this ALMA SV observation. Indeed, a stellar chromosphere may not exist around Mira at all because the H$\alpha$ line from the star has only been seen in emission, not absorption \citep[e.g.][]{joy1947,gillet1983,gillet1985}.


\subsubsection{More recent hydrodynamic models}
\label{sec:discuss-new-hydro-models}

Based on the assumption of negligible post-shock heating, \citet{bessell1996} and \citet{hofmann1998} have made series of pulsation models to theoretically calculate the photospheric structure (e.g., density, temperature, outflow/infall velocities) of Mira variables. Applying the predictions from these model series, \citet{bessell1996} and \citet{scholz2000} have predicted the near-infrared spectra of selected atomic and molecular lines, which exhibit both the normal and inverse P Cygni profiles, and show significant variations with the stellar phase.

New hydrodynamical solutions have also been derived by \citet{ireland2004a,ireland2004b} and adopted by \citet{gray2009} to simulate SiO maser emission in Mira variables. The model series are based on self-excited pulsation as in \citet{hofmann1998}, instead of ``piston''-generated pulsation in other hydrodynamical models. Simulations have shown that SiO masers would form a ring of typically ${\sim}2.2\,R_{\star}$ around the star \citep{gray2009}, which is close to but beyond the radius of the radio photosphere. The results are consistent with the previous observations of SiO maser shells around Mira by \citet{rm1997,rm2007}. Hence, strong SiO masers within the radio photosphere are prohibited \citep{rm1997,gray2009}. Comparing the hydrodynamical solutions of \citet{ireland2004a,ireland2004b} with observations, the strongest inner shocks (with $\Delta V > 20 \kms$) are enclosed within the radio photosphere, and the shocks beyond that only cause a velocity change of ${\sim}7 \kms$. The low shock velocity beyond the radio continuum is also consistent with the proper motion velocities of SiO masers observed in TX Cam \citep{diamond2003} and the nearly constant radio light curves observed in several Mira variables \citep{rm1997,rm1997b}. In addition, \citet{gray2009} have found that the presence of corundum (Al$_2$O$_3$) dust grains can both enhance or suppress SiO maser emission.

More recently, \citet{ireland2008,ireland2011} and \citet{scholz2014} have developed the Cool Opacity-sampling Dynamic EXtended ({\codex}) atmosphere models, based on an improved self-excited pulsation code by \citet{keller2006} to determine the pressure and luminosity, and an opacity-sampling method to derive the gas and dust temperatures \citep[see, also, the review by][]{ireland2011b}. The mass density and velocity profiles in the extended atmosphere can also be computed. Thus far, the {\codex} models have only been compared to the spectro-interferometric and spectro-photometric observations of Mira variables in the infrared wavelengths \citep[e.g.][]{woodruff2009,wittkowski2011,hillen2012}. Our radiative transfer modelling of the submillimetre molecular transitions from Mira as observed by the ALMA long baselines, including the vibrational ground state lines that are less affected by excitation effects, can provide an alternative probe to the detailed physical conditions in Mira's extended atmosphere at a spatial scale of a few $R_{\star}$. We will present in Sects. \ref{sec:discuss-codex} and \ref{sec:discuss-acceleration} our modelling results using the predicted atmospheric structures from {\codex} and alternative velocity profiles in Mira's extended atmosphere.


\subsubsection{Kinematics}
\label{sec:discuss-codex}

As described above, our high angular-resolution ALMA long baseline data clearly resolve the line emission/absorption from the extended atmosphere of Mira. Employing radiative transfer modelling we can test the atmospheric structures predicted by the latest hydrodynamical models. In particular, we consider the hydrodynamical model series, {\codex}, developed by \citet{ireland2008,ireland2011} and \citet{scholz2014}. We use the \texttt{o54} $5400\,L_{\astrosun}$ model series, which is based on the stellar parameters of our target source, Mira ($o$ Cet). From the {\codex} developers we have obtained the atmospheric structure information of the \texttt{o54} model series, including the total mass density, kinetic temperature \citep[which is not the computed non-grey equilibrium temperature in the final model atmospheres, see][]{ireland2008}, and the radial expansion or infall velocity of the gas at different radii. Among the 6 complete stellar pulsation cycles computed by the developers, the models at the stellar phases closest to this 2014 ALMA SV dataset (near phase 0.45) are models 250420\footnote{These model numbers correspond to various phases of the models in the \texttt{o54} series in chronological order. Models 248480--251160 are the models in the compact atmospheric cycles, during which most of the mass in the model atmosphere is contained within a relatively small radial extent; models 260820--263740 are those in the extended atmospheric cycles; and models 285180--291860 cover almost four consecutive pulsation cycles of the model atmosphere, with intermediate extents of the mass-zones \citep[see Fig. 1 and Tables 2, 3, and 4 of][]{ireland2011}.} (phase 0.38), 261740 (0.40), 286060 (0.41), 287880 (0.40), 289440 (0.40), and 291820 (0.41) \citep{ireland2008,ireland2011}.

Because of the assumption of instantaneous shock dissipation, there is no extended zone of elevated gas kinetic temperature above the stellar brightness temperature in the extended atmosphere. The gas temperature profiles of the {\codex} models are, respectively, lower than (by $<500\,{\rm K}$) and higher than (by $<700\,{\rm K}$) than what we have empirically derived in Sect. \ref{sec:model_results} for the inner radii (${\lesssim}5 \times 10^{13}\,{\rm cm} \approx 2.5\,R_{\star}$) and outer radii. The temperature in Mira's extended atmosphere is always $<2000\,{\rm K}$, except at radii very close to the radio continuum.

The {\codex} code computes the mass density of all particles in the atmosphere. In our {\ratran} input models, we convert the mass density to number density by division with the average mass of the particles in the wind, $m_{\rm part} = 0.7 m_{{\rm H}_2} + 0.3 m_{{\rm He}} = 4.3372 \times 10^{-24}\,{\rm g}$, assuming the typical helium mass fraction of $Y \approx 0.3$ \citep[e.g.][]{wood1977,lattanzio2003,ireland2008,ventura2013}. In our modelling, however, we only consider the collision of H$_2$O and SiO with rotational ground-state H$_2$ molecules only.

However, if we apply this conversion to the input models, the number density in Mira's extended atmosphere will be about $10^{8}$--$10^{10}\,{\rm cm}^{-3}$, which is too low to excite the molecules and to produce significant emission or absorption in the synthesised spectra. The deep absorption features observed towards the continuum disk and the emission profiles towards various radial distances from the star can only be reproduced by arbitrarily increasing the converted number density by a large factor. In our modelling, therefore, we scale the number density of all {\codex} models by a factor of $10^4$ such that the density outside the radio photosphere reaches at least $10^{12}\,{\rm cm}^{-3}$. This density is similar to the one adopted in our empirical modelling (Sect. \ref{sec:modelling}), and is also consistent with that derived by \citet{rm1997}, who modelled the centimetre-wavelength radio fluxes from the radio photospheres of Mira variables (including Mira). Furthermore, the density is also compatible with the lower limit of $10^{11}\,{\rm cm}^{-3}$, which is estimated by modelling the near-infrared H$_2$O spectrum and assuming a relatively high H$_2$O abundance of ${\sim}3 \times 10^{-4}$ \citep{yamamura1999}. If the adopted H$_2$O abundance is similar to our assumed value of ${\sim}10^{-5}$, then the lower limit of the gas density derived by \citet{yamamura1999} would also be close to the values in our modelling. We, however, note that the gas density should be interpreted with caution -- see our discussion in Sect. \ref{sec:limitations}.

In our initial test, we have replaced our preferred model (Model 3) with the {\codex}-predicted gas number density (scaled by $10^4$), kinetic temperature and radial velocity profiles. All other parameters, namely the SiO abundance profile, the local velocity dispersion, and the radio continuum are the same as Model 3. The outer radius of our model is greater than the outer boundaries of the {\codex} models. The outer boundaries depend on individual {\codex} models. For each model, we extrapolate the number density and kinetic temperature in power-laws near the outer boundary, i.e., linear extrapolation in log-log relation. We also assume that the infall or expansion velocity at and beyond the {\codex} model boundary to be constant because we have no information of the kinematics beyond that. Since the gas density and molecular SiO abundance are usually very low to cause significant excitation in the outer regions, the precise extrapolation method has little effect on the synthesised spectra.

Figure \ref{fig:test-codexvel-a} shows the results of our initial test of the velocity profiles from these 6 {\codex} models. Remarkably, all the 6 models near the observed stellar phase are able to qualitatively reproduce the general spectral features, including (1) the strong absorption profile in the line-of-sight towards the radio continuum of Mira, and (2) gradually decreasing emission flux and line width in the lines-of-sight towards increasing radial offsets from the star. Closer inspection reveals that all the 6 {\codex} models produce an extra high-velocity absorption wings in either the redshifted or blueshifted parts of the spectra towards the centre of the continuum. The velocities of the absorption wings correspond to the high-velocity gas at $10$--$20\,\kms$ near the radio continuum of Mira, i.e., $3.6 \times 10^{13} \,{\rm cm}$ ($1.8\,R_{\star}$). Because there is no sign of any absorption wings broader than ${\pm} 10\,\kms$ in the observed spectra, the strong velocity variation as seen in these models, in particular models 261740, 286060, 287880, and 289440, cannot explain the specific atmosphere structure of Mira during the time of the 2014 ALMA SV observation.

The velocity profiles of {\codex} models 250420 and 291820 are qualitatively similar to our Model 3 (see Fig. \ref{fig:model3}, top-right panel). First, the extended atmosphere exhibits slowly varying, infall motion over a large range of radii. Second, there is a sharp change in velocity, representing a strong shock front, with $\Delta V \gtrsim 10\,\kms$ just outside the radio continuum ($3.60 \times 10^{13} {\rm cm}$). The strong shock front in model 250420 is located at $3.64 \times 10^{13} {\rm cm}$ and that in model 291820 is at $3.83 \times 10^{13} {\rm cm}$ \citep{ireland2011}. In our second test, we increase the radius of our radio \emph{pseudo}-continuum to engulf the strong shock fronts in models 250420 and 291820. Figure \ref{fig:test-codexvel-c} shows the modelled spectra as a result of hiding the strong shock fronts. The high-velocity absorption wings in the blueshifted part of the SiO spectra extracted from the continuum centre has disappeared, and hence the synthesised spectra can now better fit the observed ALMA spectra. We therefore conclude that, at the time of the 2014 ALMA SV observation (stellar phase ${\sim}0.45$), there does not exist any strong shock with a high velocity of $\Delta V \gtrsim 20\,\kms$ in the extended atmosphere of Mira \emph{beyond} the radius of its 229-GHz radio photosphere. Furthermore, the infall and outflow velocities of the gas beyond the radio photosphere of Mira are bounded by ${\sim}7\,\kms$ and ${\sim}4\,\kms$, respectively.

Our finding is consistent with previous observations of SiO maser emission from other oxygen-rich Mira variables. For example, \citet{wittkowski2007} found that the expansion velocity of the SiO maser-emitting shell around S Ori is about $10\,\kms$ by fitting the projected radii of the maser spots and their line-of-sight velocities; \citet{diamond2003} also found that the infall velocities of the SiO maser emission around TX Cam range from $5$--$10\,\kms$ by tracing the proper motions of the maser spots. In fact, based on their shock damping model, \citet{rm1997b} have excluded the shock propagation velocities significantly (by a few $\kms$) higher than $7\,\kms$ in order to explain the multi-epoch radio flux variation of Mira at 8\,GHz, assuming that the amplitude of temperature disturbance due to shocks is ${\gtrsim}300\,{\rm K}$ or the factor of density compression is ${\gtrsim}2$. Hence, any gas infall or outflow motion with speed above $10\,\kms$ is unlikely beyond the radio photosphere of Mira, and therefore the very high-velocity shocks of $\Delta V \gtrsim 20\,\kms$ as predicted by the {\codex} models, in particular models 285180 (phase 0.80) and 287980 (phase 0.61) in the same \texttt{o54} model series \citep{ireland2011}, are not expected.

The above exercises have demonstrated that the atmospheric structures predicted by the {\codex} models can qualitatively reproduce the general spectral features of the molecular transitions originating from the extended atmospheres and the inner wind of Mira. As already suggested by the authors of {\codex}, the derived structures of the model atmosphere, such as the stellar radius, mass density, gas temperature, and velocity, exhibit significant cycle-to-cycle variations and appear to be chaotic \citep{ireland2011,ireland2011b}. The combination of the expansion/infall velocity profile and the locations of shock fronts are different in each cycle. It may take tens of stellar cycles (over a decade) for similar atmospheric structures to reappear \citep[e.g. Fig. 1 of][]{ireland2011}. We therefore expect that the radio and (sub)millimetre spectra of the molecular transitions would also exhibit significant cycle-to-cycle variation, in addition to rapid variation within a single cycle. Long-term (multi-cycle and multi-epoch) monitoring of Mira variables with the ALMA long baseline is therefore necessary to fully test the predictions from hydrodynamical models, especially the amplitudes of pulsation-driven shocks above the radio photospheres of these stars.


\begin{figure*}[!htbp]
\centering
\raisebox{0.1\codexspecheight}{\includegraphics[height=\codexmodelheight]{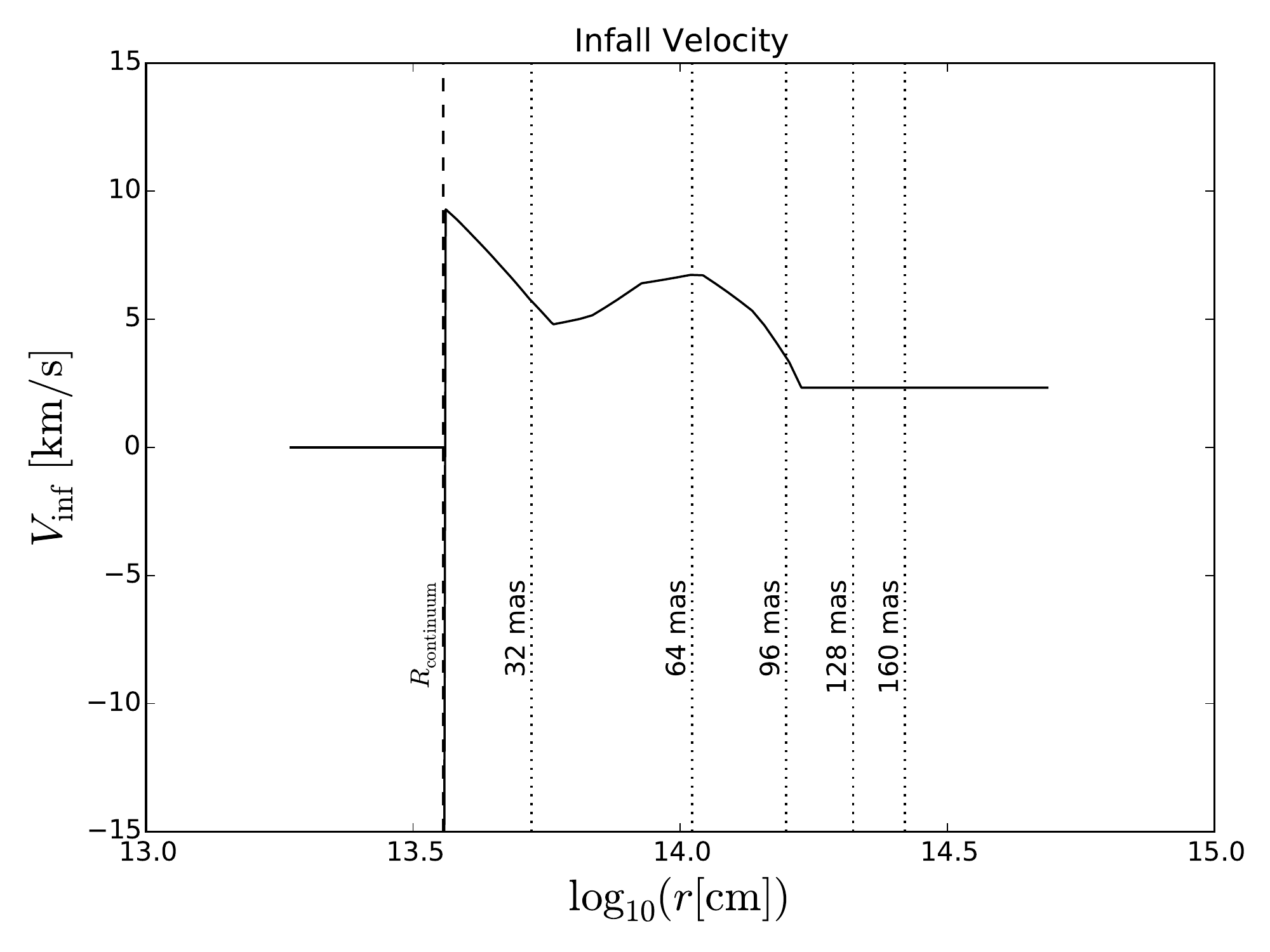}}
\put(-190,160){{\parbox{1.3cm}{{\codexplt} \\ 250420}}}
\includegraphics[trim=1.0cm 2.0cm 2.0cm 1.5cm, clip, height=\codexspecheight]{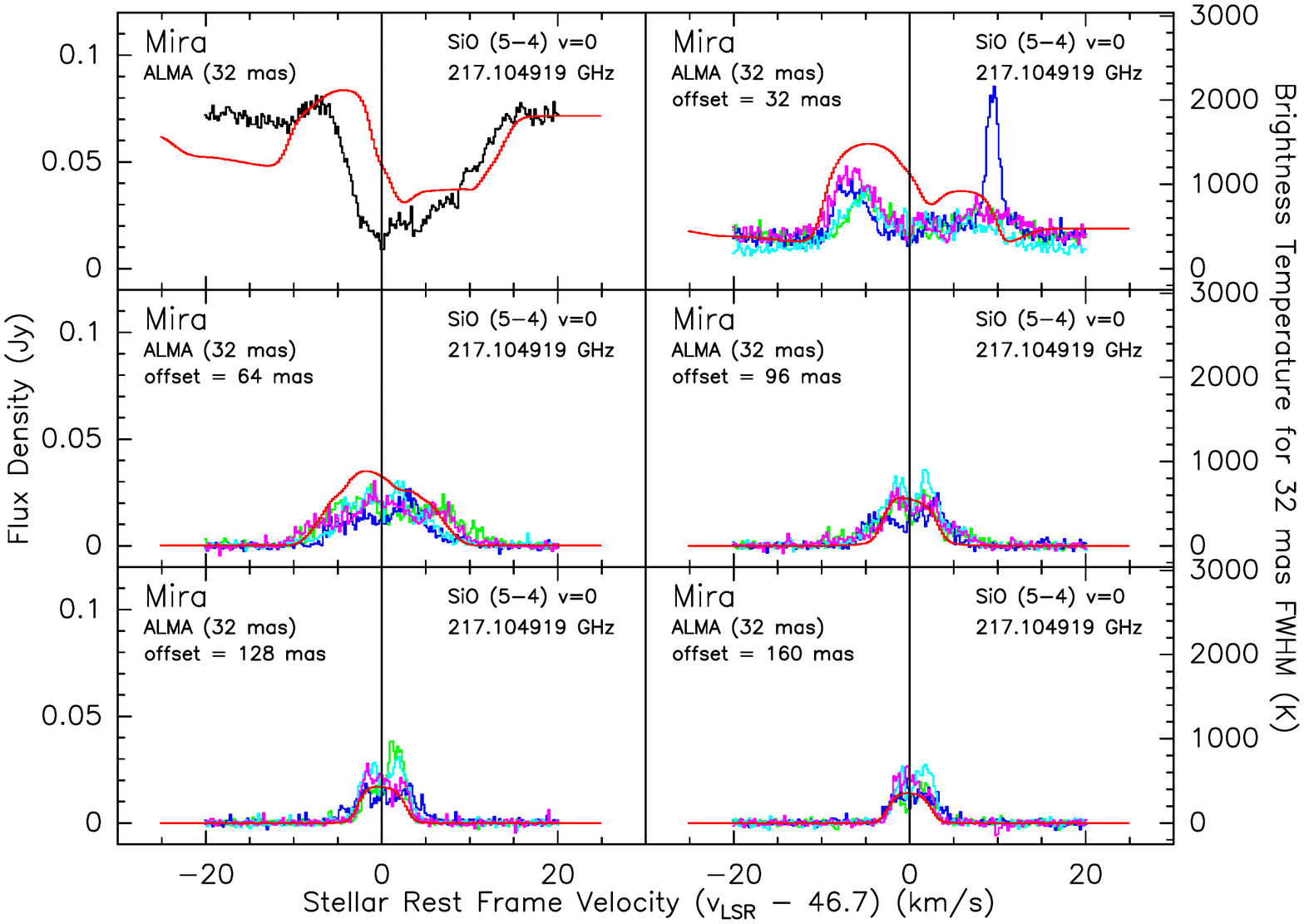} \\
\raisebox{0.1\codexspecheight}{\includegraphics[height=\codexmodelheight]{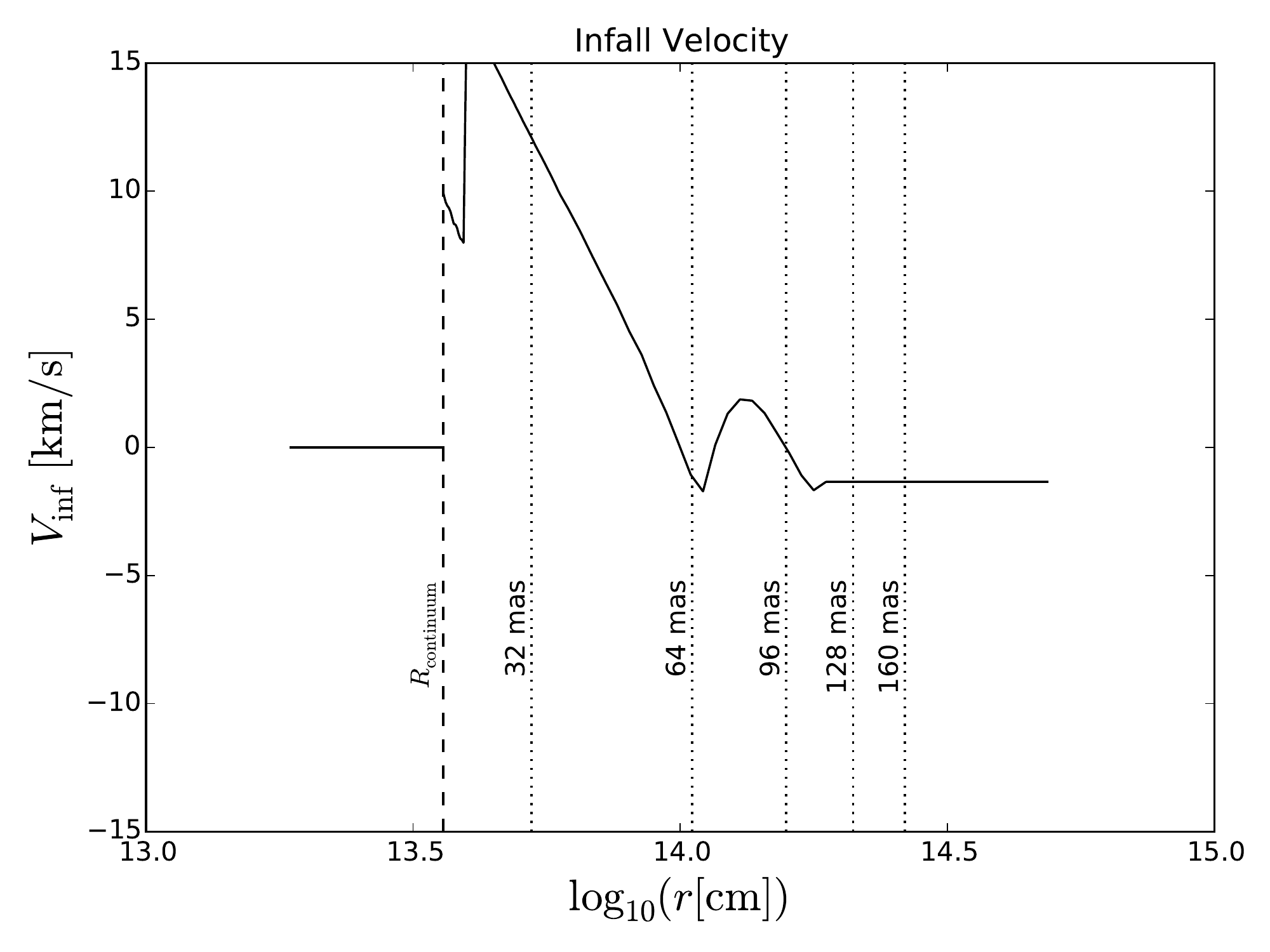}}
\put(-190,160){{\parbox{1.3cm}{{\codexplt} \\ 261740}}}
\includegraphics[trim=1.0cm 2.0cm 2.0cm 1.5cm, clip, height=\codexspecheight]{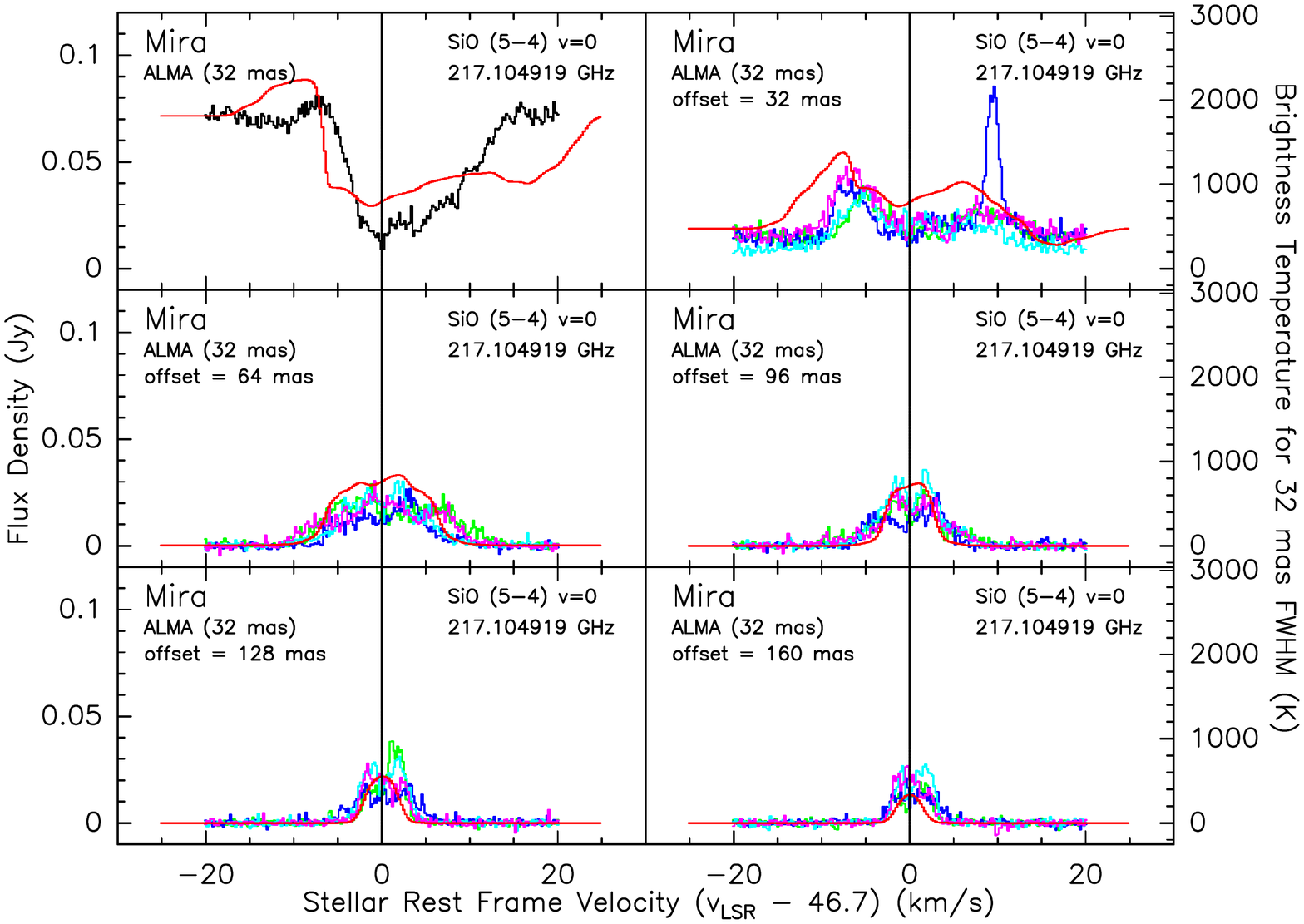} \\
\raisebox{0.1\codexspecheight}{\includegraphics[height=\codexmodelheight]{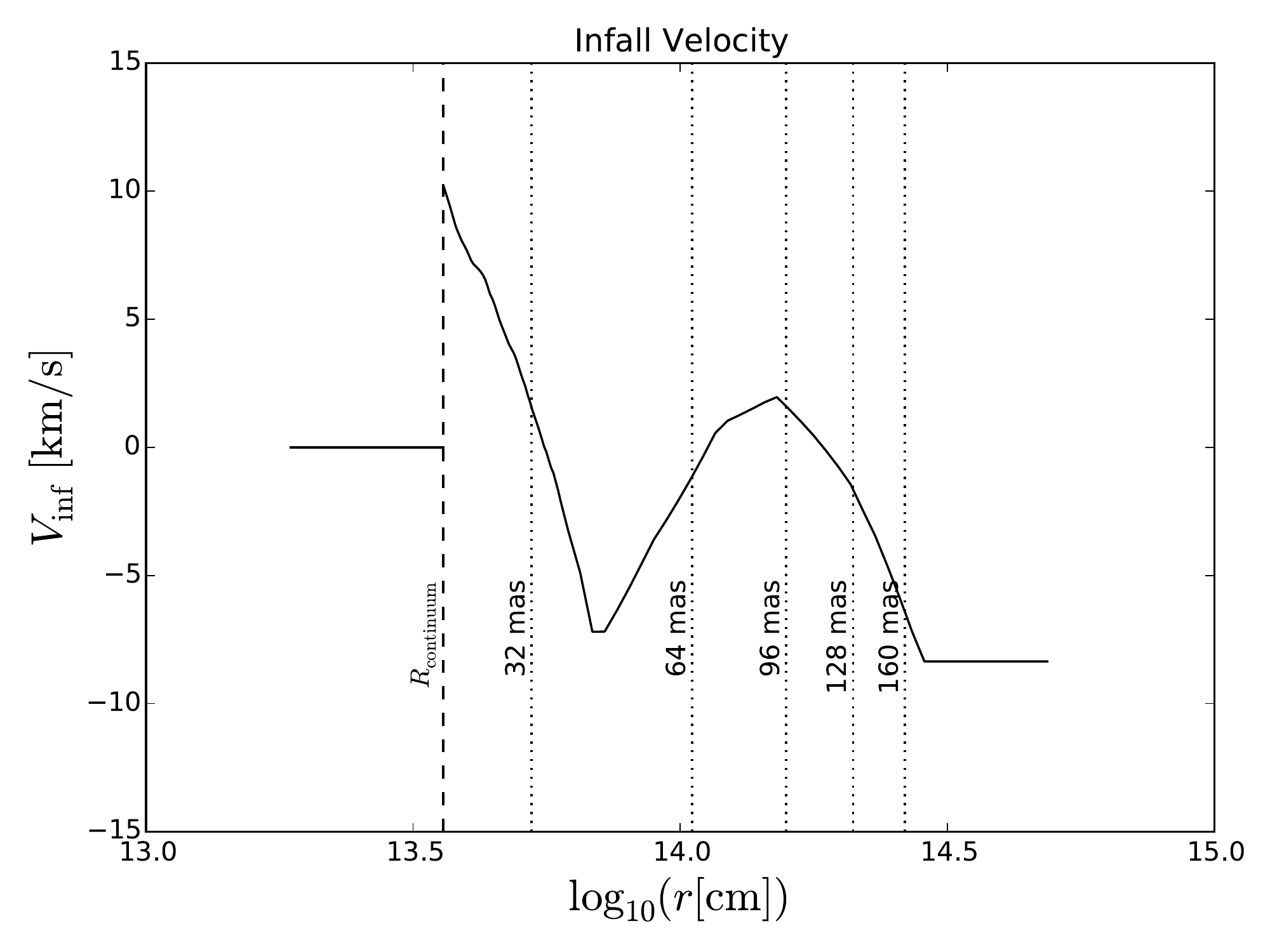}}
\put(-190,160){{\parbox{1.3cm}{{\codexplt} \\ 286060}}}
\includegraphics[trim=1.0cm 2.0cm 2.0cm 1.5cm, clip, height=\codexspecheight]{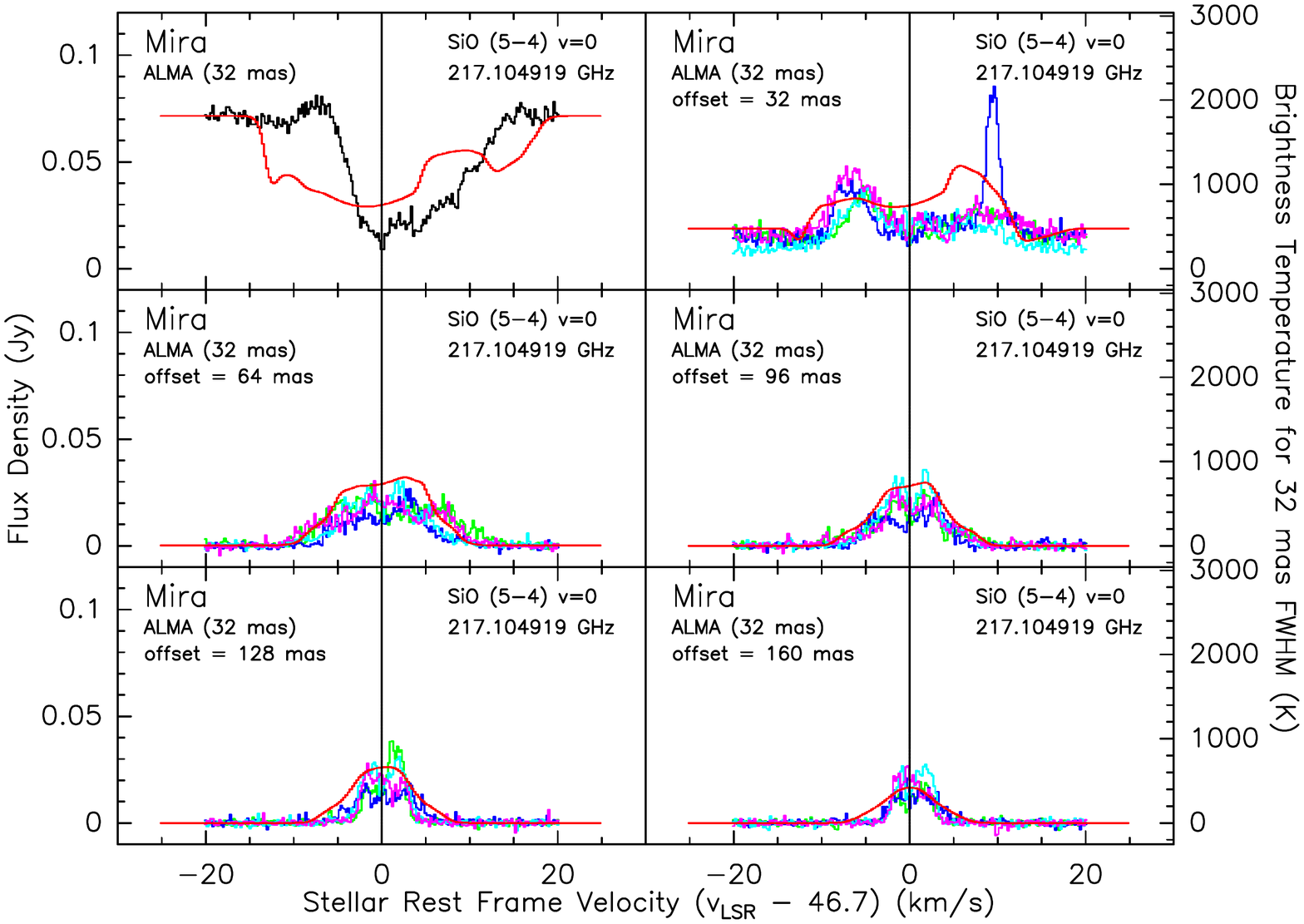}
\caption{Modified Model 3 with input gas density, kinetic temperature and velocity profiles from {\codex} models 250420 (top), 261740 (middle) and 286060 (bottom). Special scaling has been applied to the input gas density (see text). The infall velocity profiles are plotted on the left and the selected resultant spectra are shown on the right. We applied constant extrapolation of the infall velocity beyond the outer boundary of {\codex} models.}
\label{fig:test-codexvel-a}
\end{figure*}

\begin{figure*}[!htbp]
\ContinuedFloat
\centering
\raisebox{0.1\codexspecheight}{\includegraphics[height=\codexmodelheight]{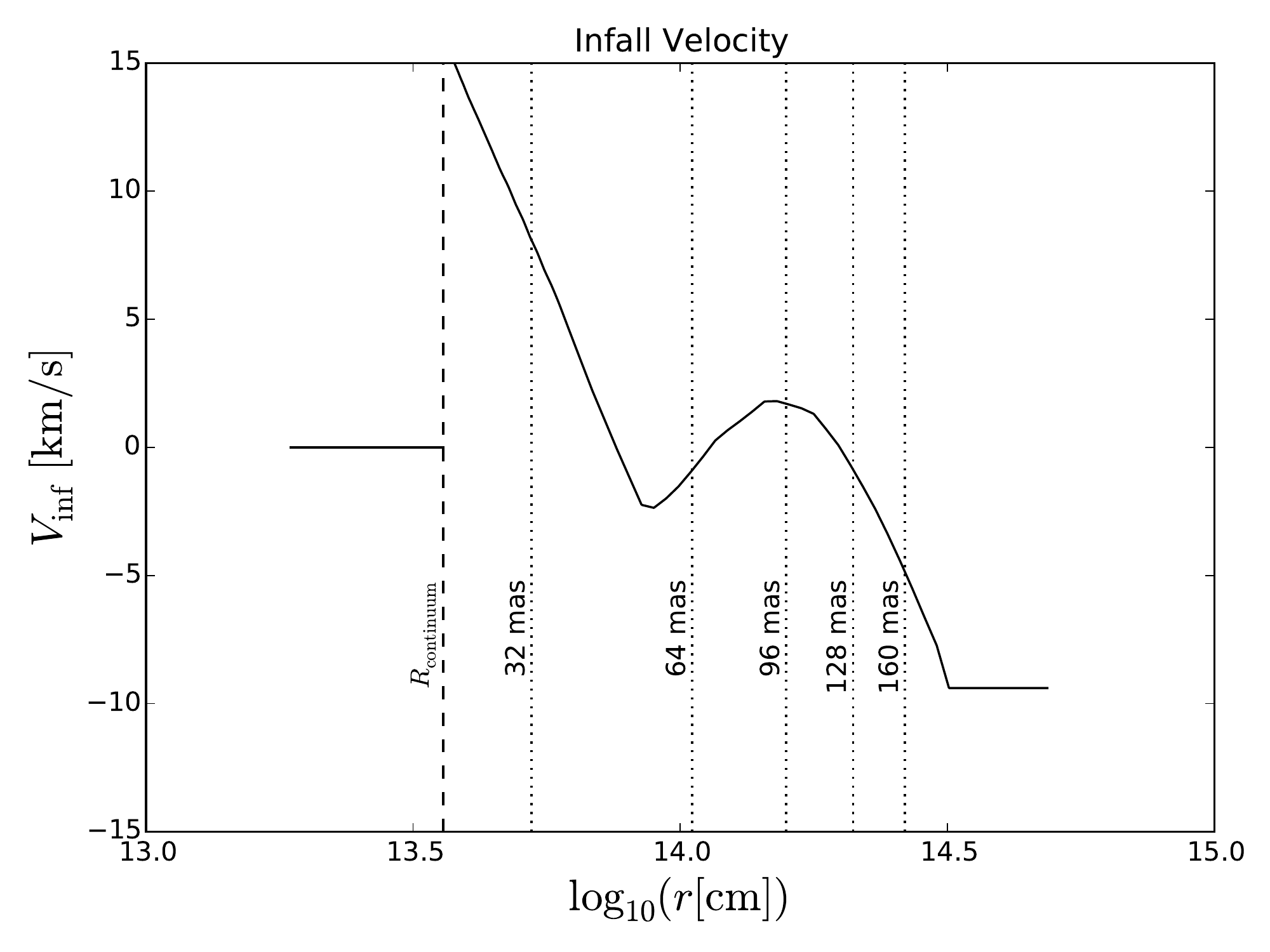}}
\put(-190,160){{\parbox{1.3cm}{{\codexplt} \\ 287880}}}
\includegraphics[trim=1.0cm 2.0cm 2.0cm 1.5cm, clip, height=\codexspecheight]{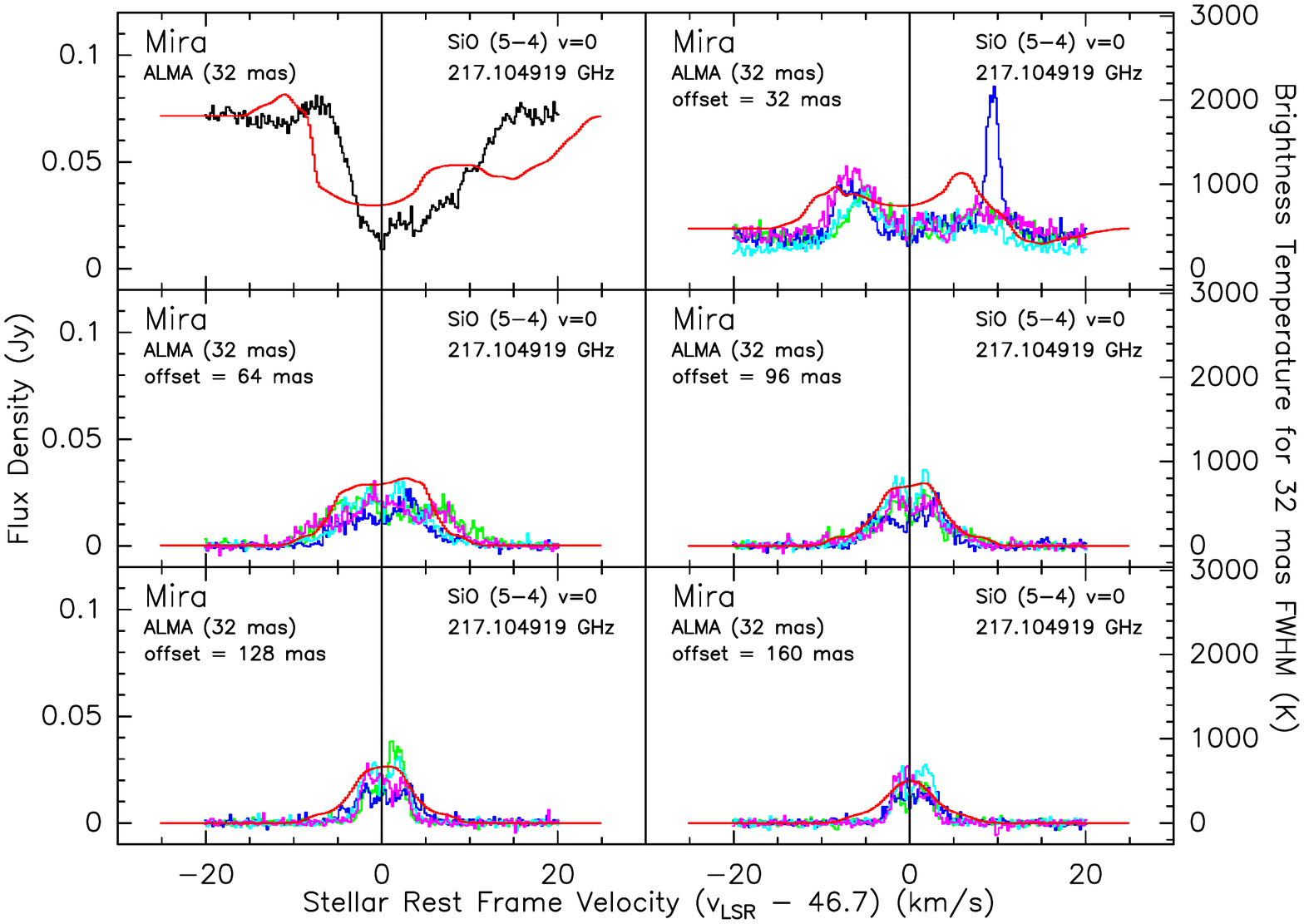} \\
\raisebox{0.1\codexspecheight}{\includegraphics[height=\codexmodelheight]{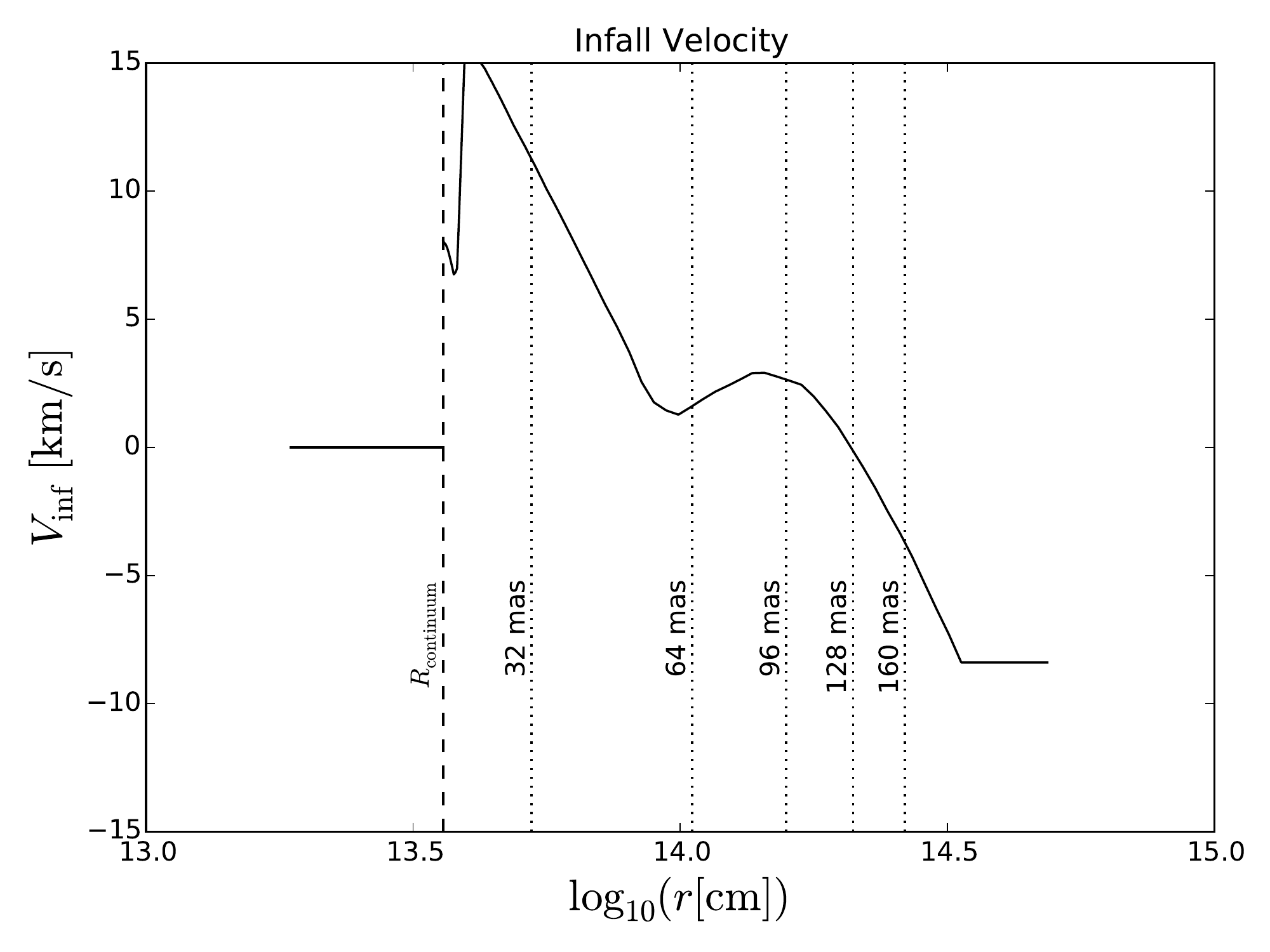}}
\put(-190,160){{\parbox{1.3cm}{{\codexplt} \\ 289440}}}
\includegraphics[trim=1.0cm 2.0cm 2.0cm 1.5cm, clip, height=\codexspecheight]{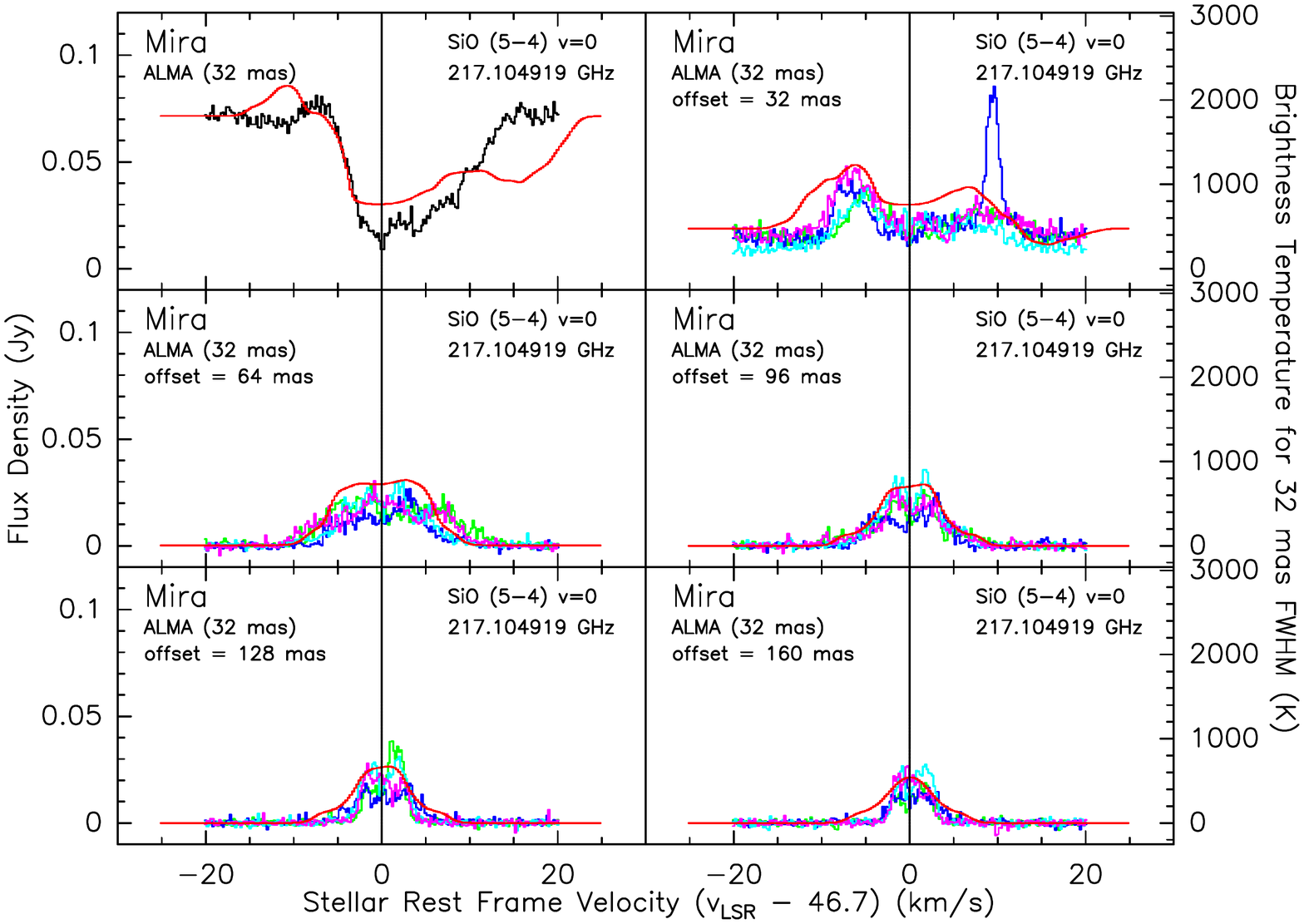} \\
\raisebox{0.1\codexspecheight}{\includegraphics[height=\codexmodelheight]{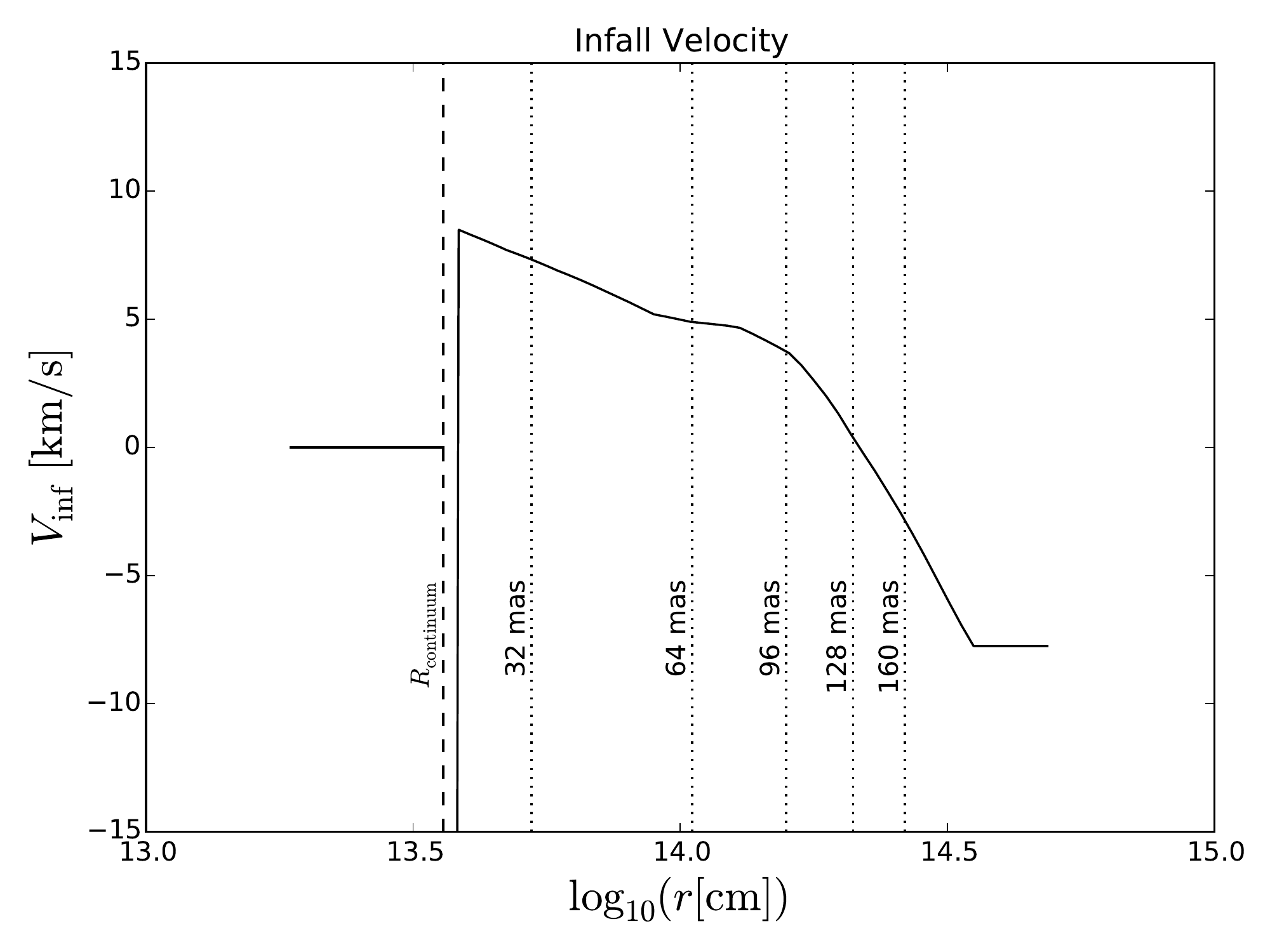}}
\put(-190,160){{\parbox{1.3cm}{{\codexplt} \\ 291820}}}
\includegraphics[trim=1.0cm 2.0cm 2.0cm 1.5cm, clip, height=\codexspecheight]{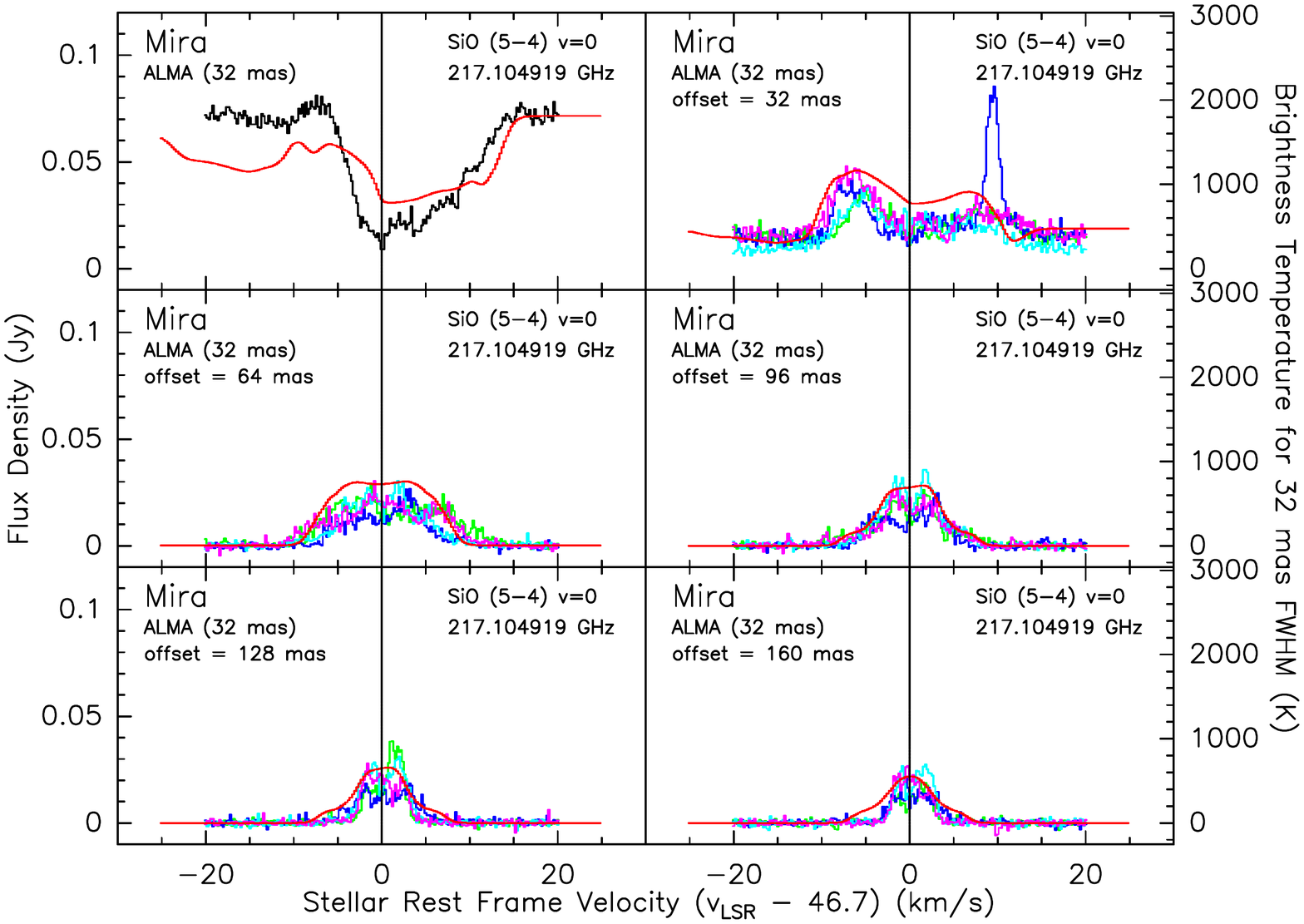}
\caption[]{Continued for {\codex} models 287880 (top), 289440 (middle) and 291820 (bottom).}
\label{fig:test-codexvel-b}
\end{figure*}

\begin{figure*}[!htbp]
\centering
\includegraphics[width=\modelwidth]{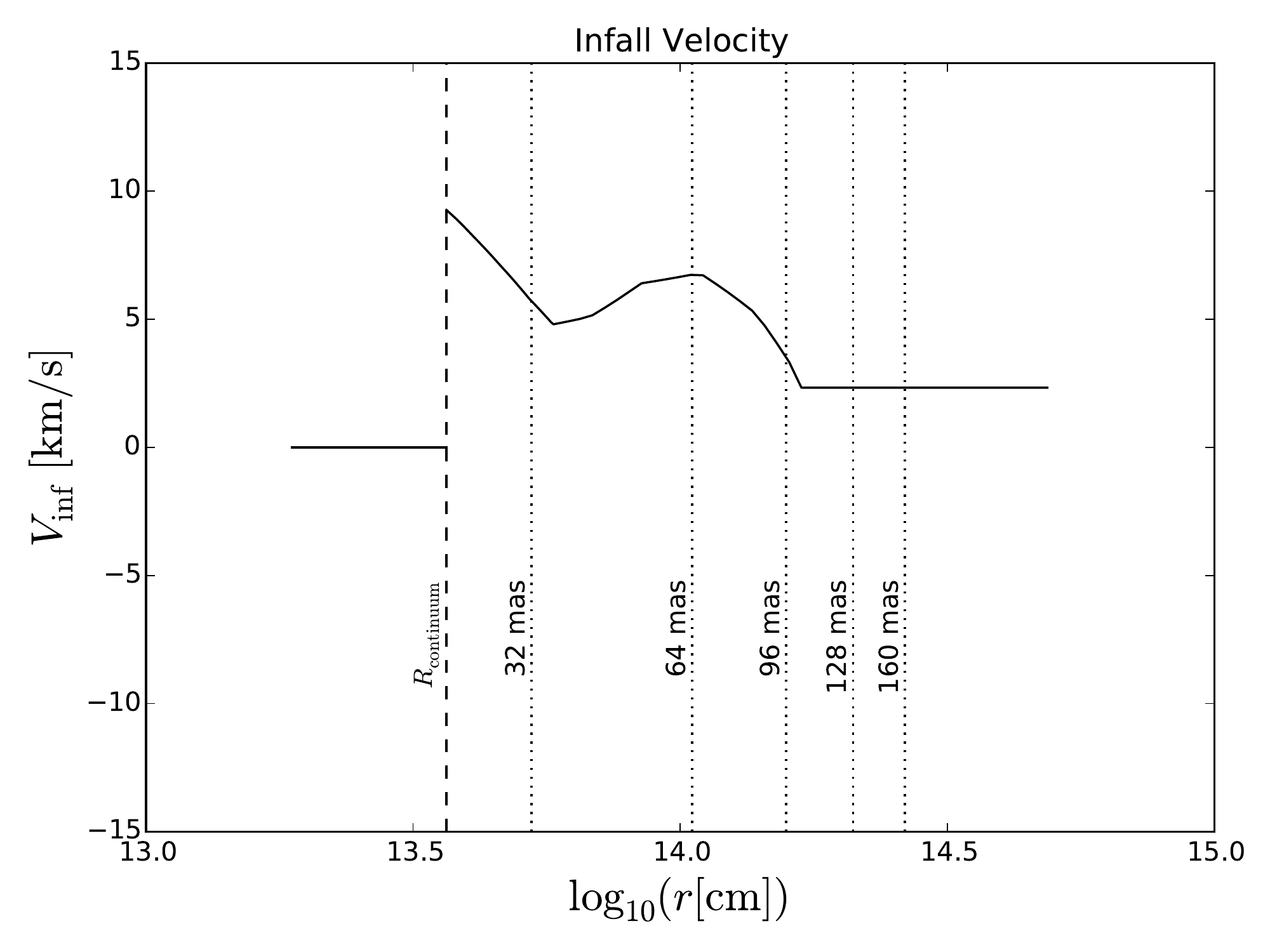}
\put(-220,155){{\parbox{1.5cm}{{\codexplt} \\ 250420 \\ (modified)}}}
\includegraphics[width=\modelwidth]{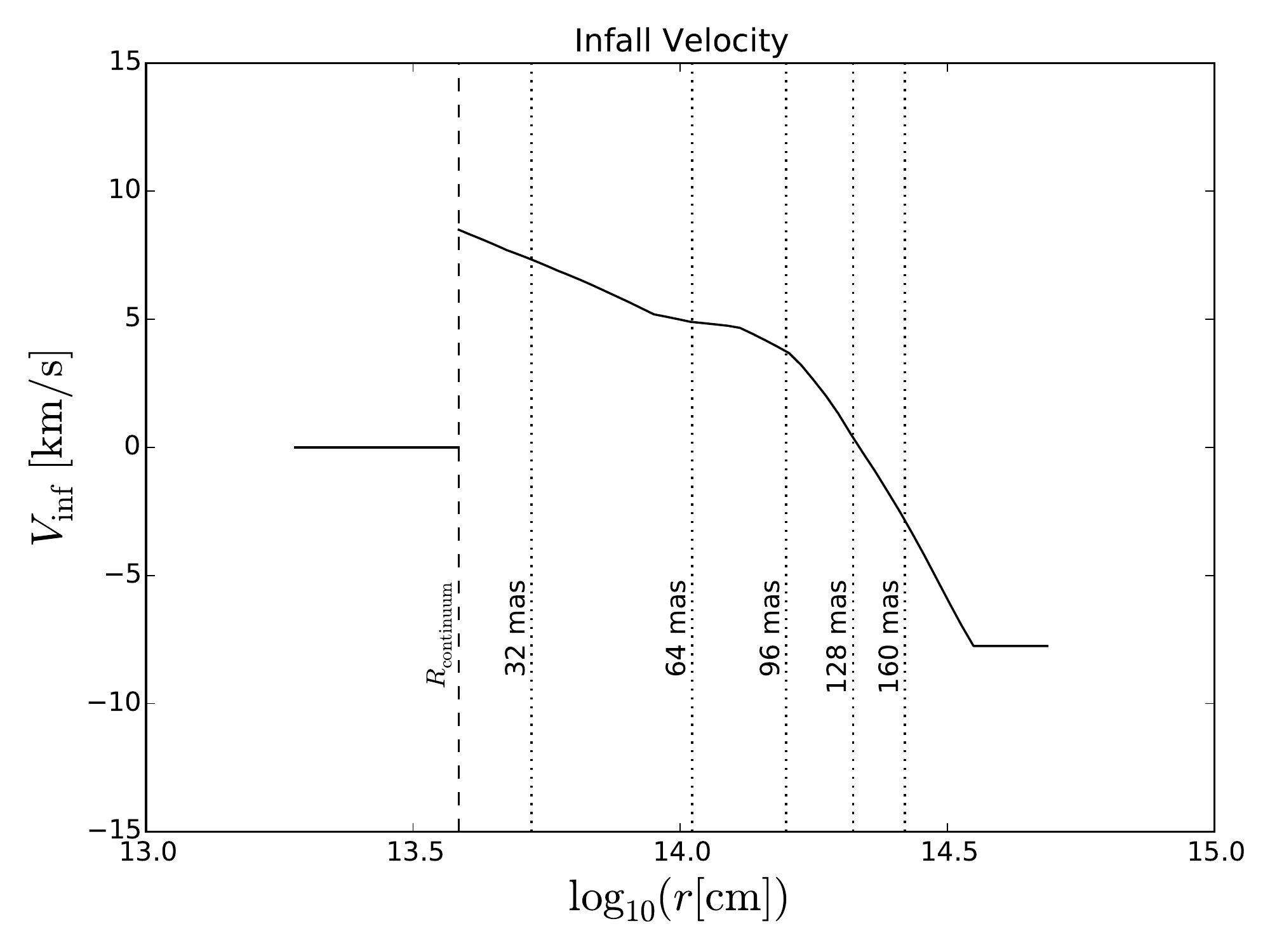} 
\put(-220,155){{\parbox{1.5cm}{{\codexplt} \\ 291820 \\ (modified)}}} \\
\includegraphics[trim=1.0cm 2.0cm 2.0cm 1.5cm, clip, width=\modelwidth]{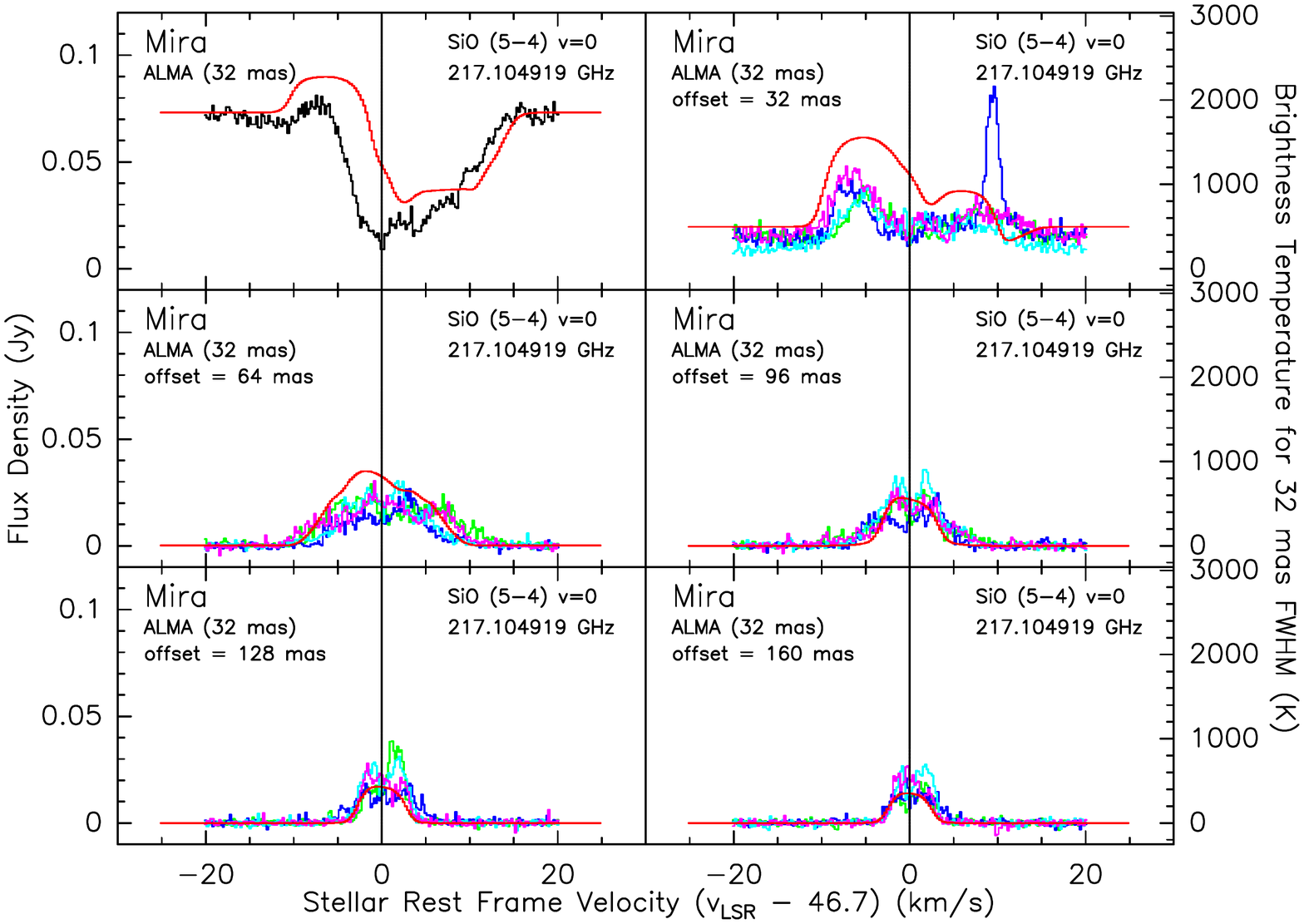}
\includegraphics[trim=1.0cm 2.0cm 2.0cm 1.5cm, clip, width=\modelwidth]{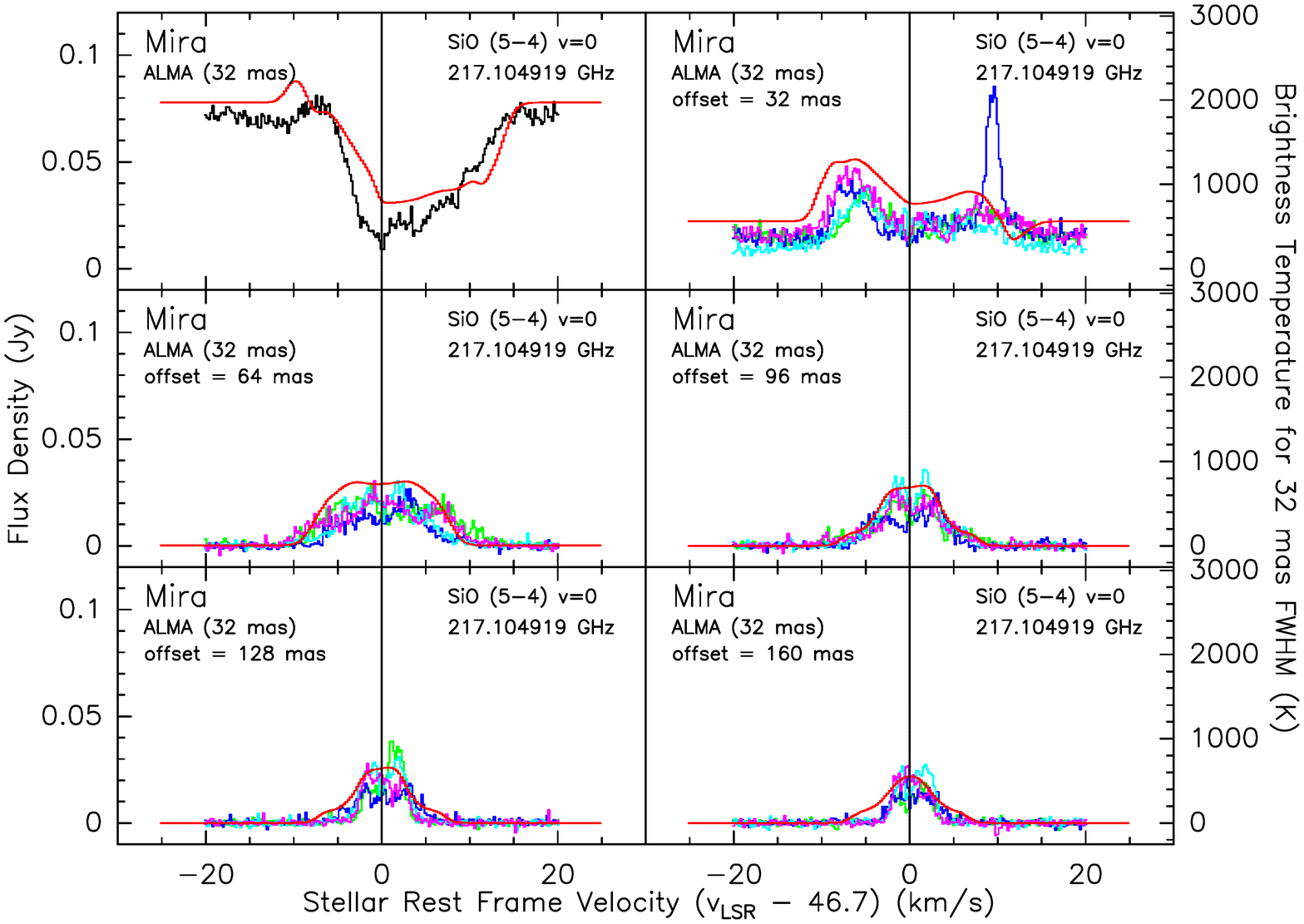} \\
\includegraphics[trim=1.0cm 7.3cm 2.0cm 1.5cm, clip, width=\modelwidth]{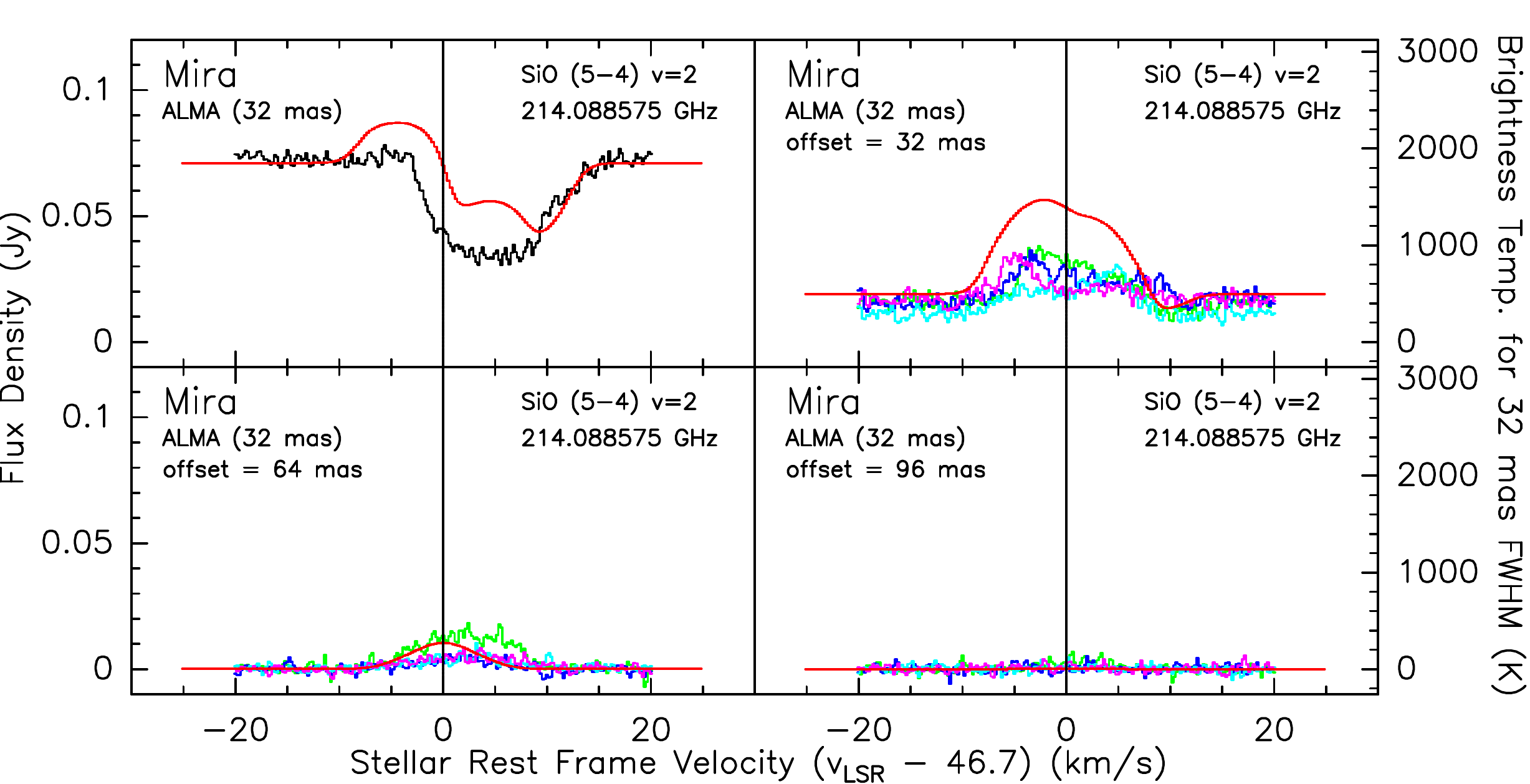}
\includegraphics[trim=1.0cm 7.3cm 2.0cm 1.5cm, clip, width=\modelwidth]{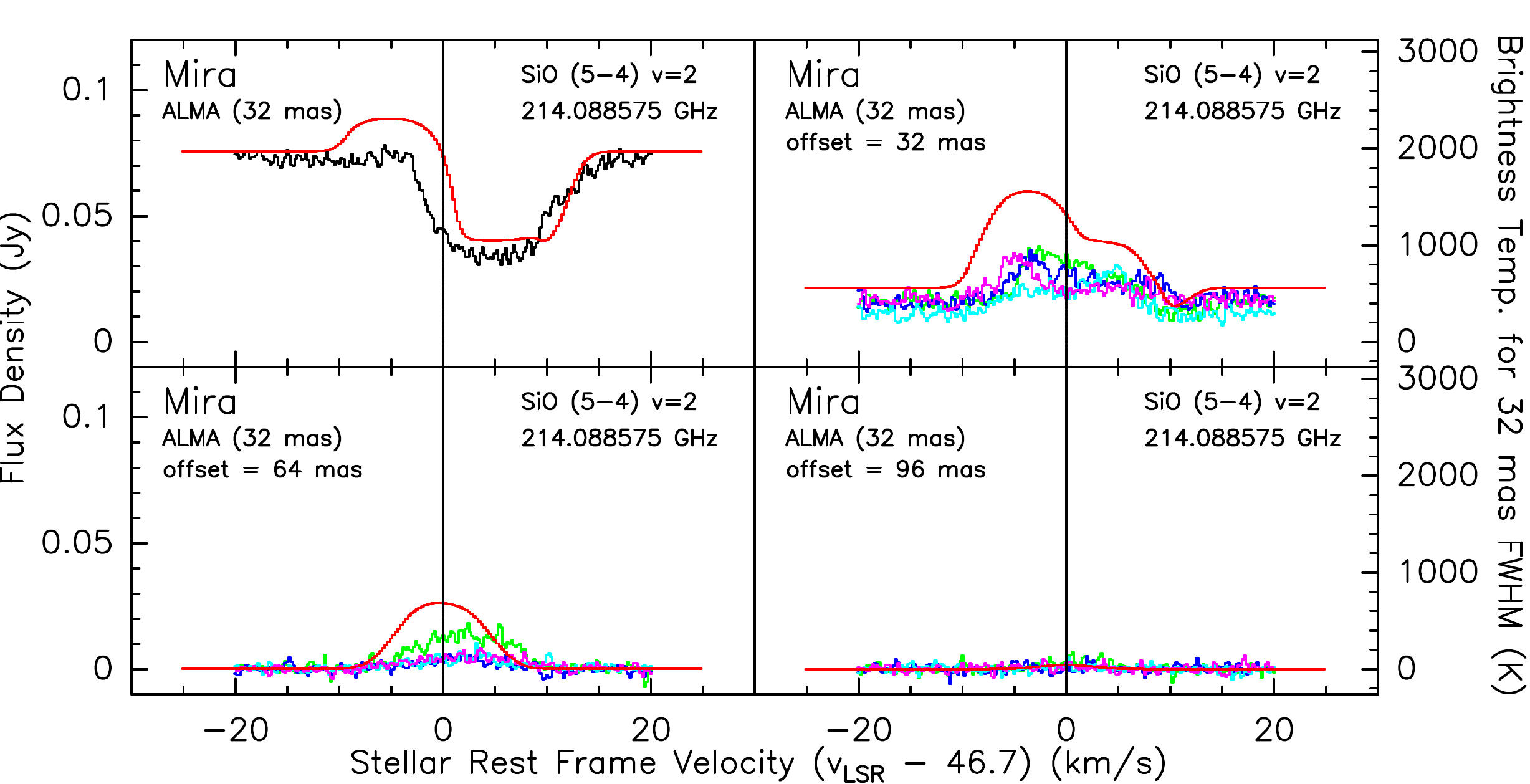}
\caption{Similar to Fig. \ref{fig:test-codexvel-a}, except that the radius of the \emph{pseudo}-continuum in the model, $R_{\rm continuum}$ is changed to $3.65 \times 10^{13} {\rm cm}$ for model 250420 (left column) and $3.85 \times 10^{13} {\rm cm}$ for model 291820 (right column). The modelled continuum levels are therefore slightly higher than the ones adopting $R_{\rm continuum} = 3.60 \times 10^{13} {\rm cm}$. The strong shock fronts predicted in these two models are hidden inside the radio continuum. The middle and bottom rows show the selected resultant SiO spectra from the corresponding models.}
\label{fig:test-codexvel-c}
\end{figure*}


\subsubsection{Wind acceleration}
\label{sec:discuss-acceleration}

\citet{hofner2003} also developed models of the dynamical atmospheres by solving time-dependent dynamic equations and frequency-dependent radiative transfer equations (i.e. non-grey dynamical model), including a time-dependent description of dust formation for carbon-rich atmospheres \citep[see, also, the review by][]{hofner2015}. Based on H\"{o}fner's hydrodynamical model, \citet{gautschyloidl2004} were able to reproduce the infrared spectra of carbon-rich stars in the wavelength range of 0.5--25\,$\mu$m, observed at various stellar phases, with a single consistent model atmosphere for each star. \citet{nowotny2005a,nowotny2005b,nowotny2010} have also compared the synthesised spectra of CO and CN in the infrared wavelengths with the observed spectra of carbon-rich stars and found that the general line profiles and radial velocities of the observed photospheric lines can be explained by the model atmospheres derived from the hydrodynamical code of \citet{hofner2003}. On the other hand, in the case of oxygen-rich model atmospheres, \citet{hofner2003} adopted a simple parametrised description for the dust opacity as in Eq. (5) of \citet{bowen1988a}. If non-grey radiative transfer was considered in the dynamical models, however, there would be too little radiative pressure to drive the winds in oxygen-rich stars \citep[e.g.][]{woitke2006,hofner2007a}. Both \citet{woitke2006} and \citet{hofner2007b} have concluded that the iron content of the wind-driving dust grains must be low, otherwise the dust condensation radius would be too far away from the star ($>10\,R_{\star}$) and hence dust-driven winds could not form. \citet{hofner2007a} have considered the possibility that the winds being driven by a small amount of carbon-bearing grains in the oxygen-rich atmospheres. By varying the grain properties in their dynamical models, \citet{hofner2008} has predicted that the size of the wind-driving (iron-free) grains must be in the narrow range between 0.1 and 10\,$\mu$m. \citet{scicluna2015} have reported the detection of large grains with an average size of ${\sim}0.5\,\mu$m in the oxygen-rich circumstellar envelope of the red supergiant (RSG) VY Canis Majoris based on optical polarimetric imaging. However, the implication of their results on AGB stars, which are the low-mass counterparts of RSGs, is unclear. In contrast, \citet{ireland2011} and \citet{ireland2011b} have found that both large iron-poor grains and small ($<70\,{\rm nm}$) iron-rich grains may drive the winds in oxygen-rich atmospheres, although the material still has to be lifted to a radius of at least $3$--$5\,R_{\star}$, where dust grains start to condensate efficiently.

One notable difference between the results of the {\codex} code and that developed by H\"{o}fner is that {\codex} models exhibit large-scale velocity variations at radii up to ${\sim}10\,R_{\star}$, while the models based on H\"{o}fner's code only show large-scale velocity variation of $\Delta V > 10\,\kms$ within ${\sim}2$--$3\,R_{\star}$. In H\"{o}fner's model atmospheres, the winds are efficiently accelerated by (Fe-free) dust grains that condense at ${\sim}3\,R_{\star}$ \citep[e.g.][]{hofner2009,hofner2015} and therefore the velocity profile becomes a generally continuous outflow with small-amplitude pulsation ($\Delta V \lesssim 2\,\kms$) due to previous shock episodes \citep[e.g.][]{hofner2003,nowotny2005a,nowotny2010}. 

\begin{figure*}[!htbp]
\centering
\includegraphics[width=\modelwidth]{./fig_models/sio1195-lv-vinf.pdf}
\includegraphics[width=\modelwidth]{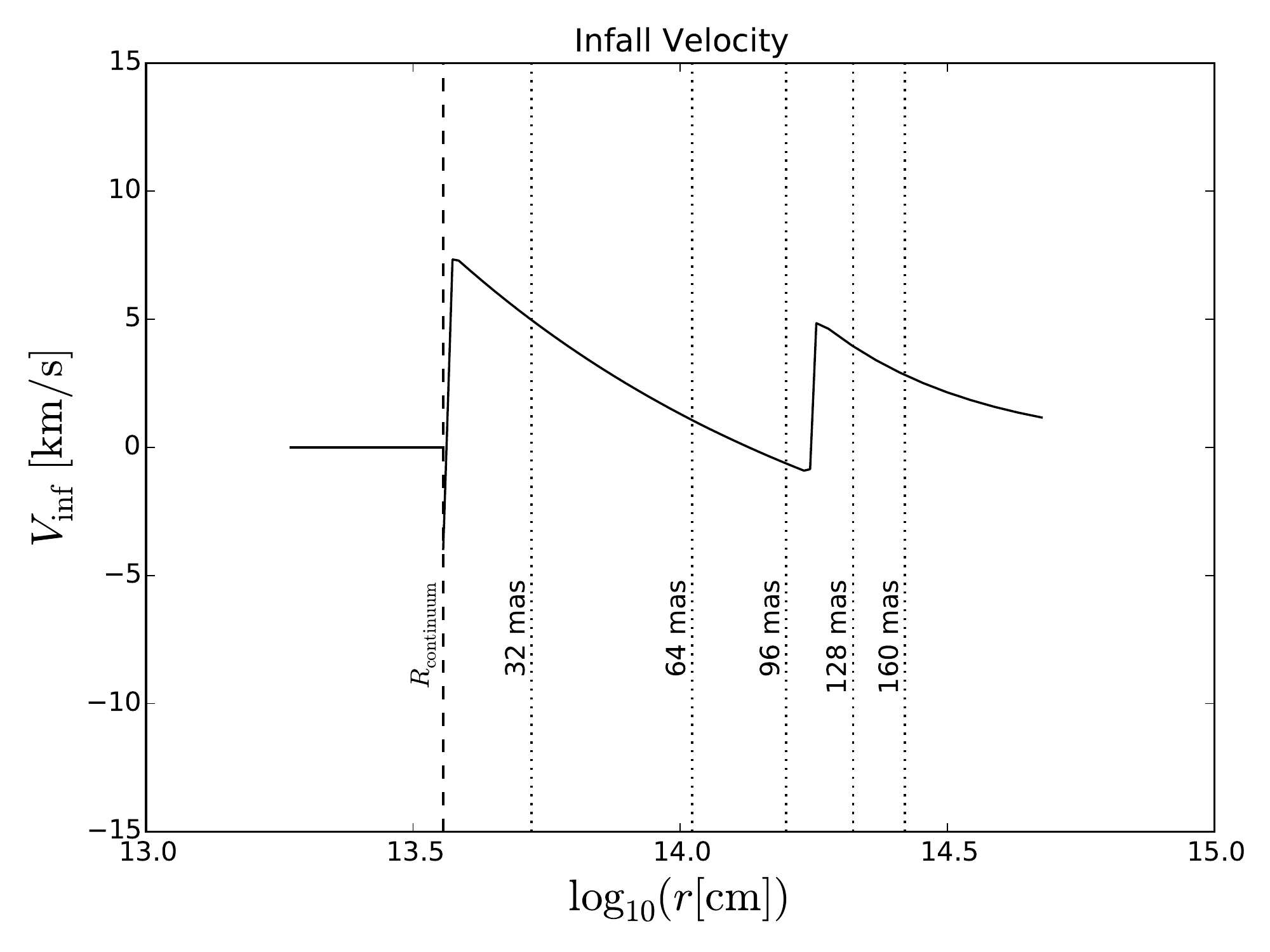} \\
\includegraphics[trim=1.0cm 2.0cm 2.0cm 1.5cm, clip, width=\modelwidth]{./fig_models/sio1195-spec_sio54v0_all.pdf}
\includegraphics[trim=1.0cm 2.0cm 2.0cm 1.5cm, clip, width=\modelwidth]{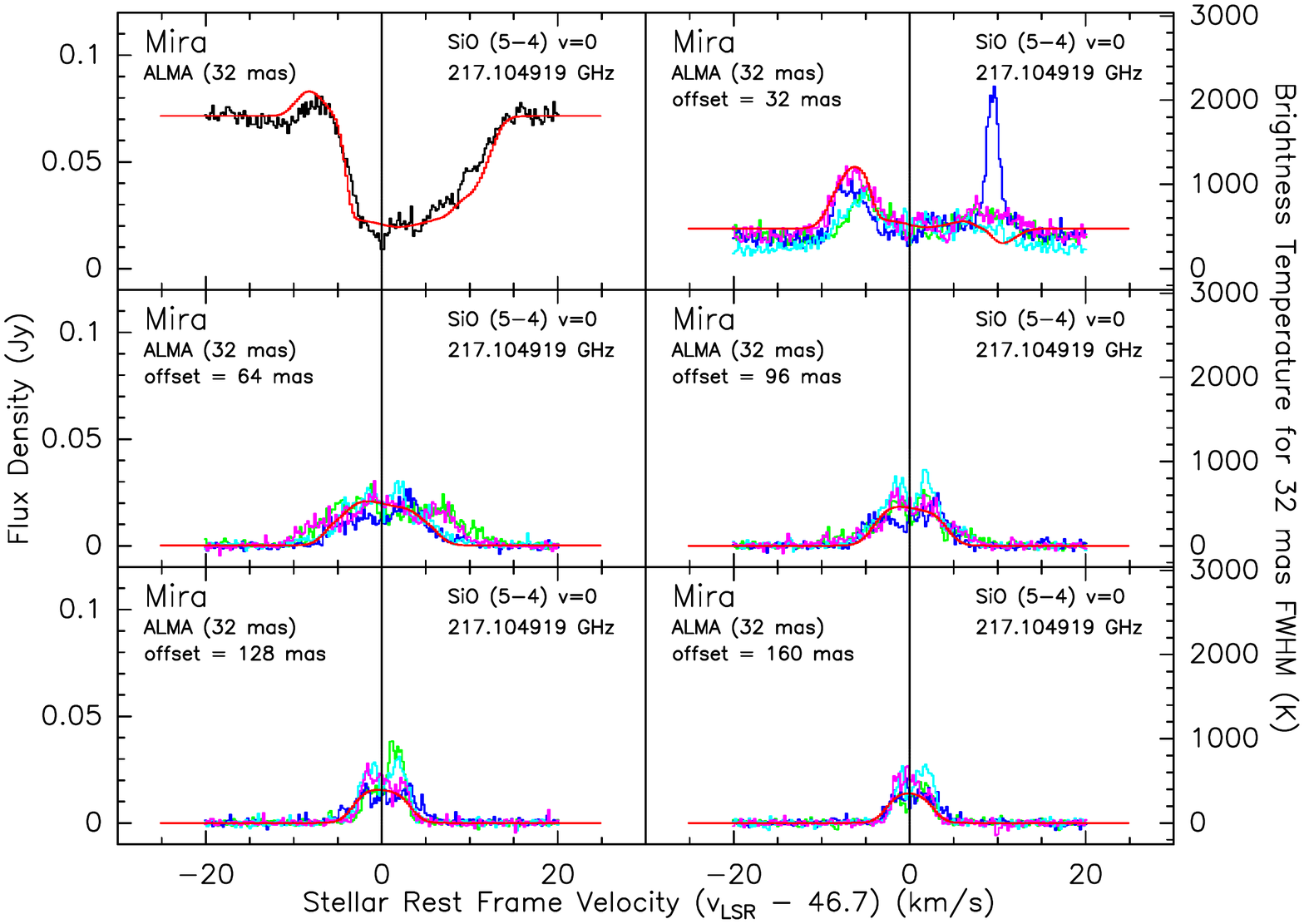} \\
\includegraphics[trim=1.0cm 7.3cm 2.0cm 1.5cm, clip, width=\modelwidth]{./fig_models/sio1195-spec_sio54v2_all.pdf}
\includegraphics[trim=1.0cm 7.3cm 2.0cm 1.5cm, clip, width=\modelwidth]{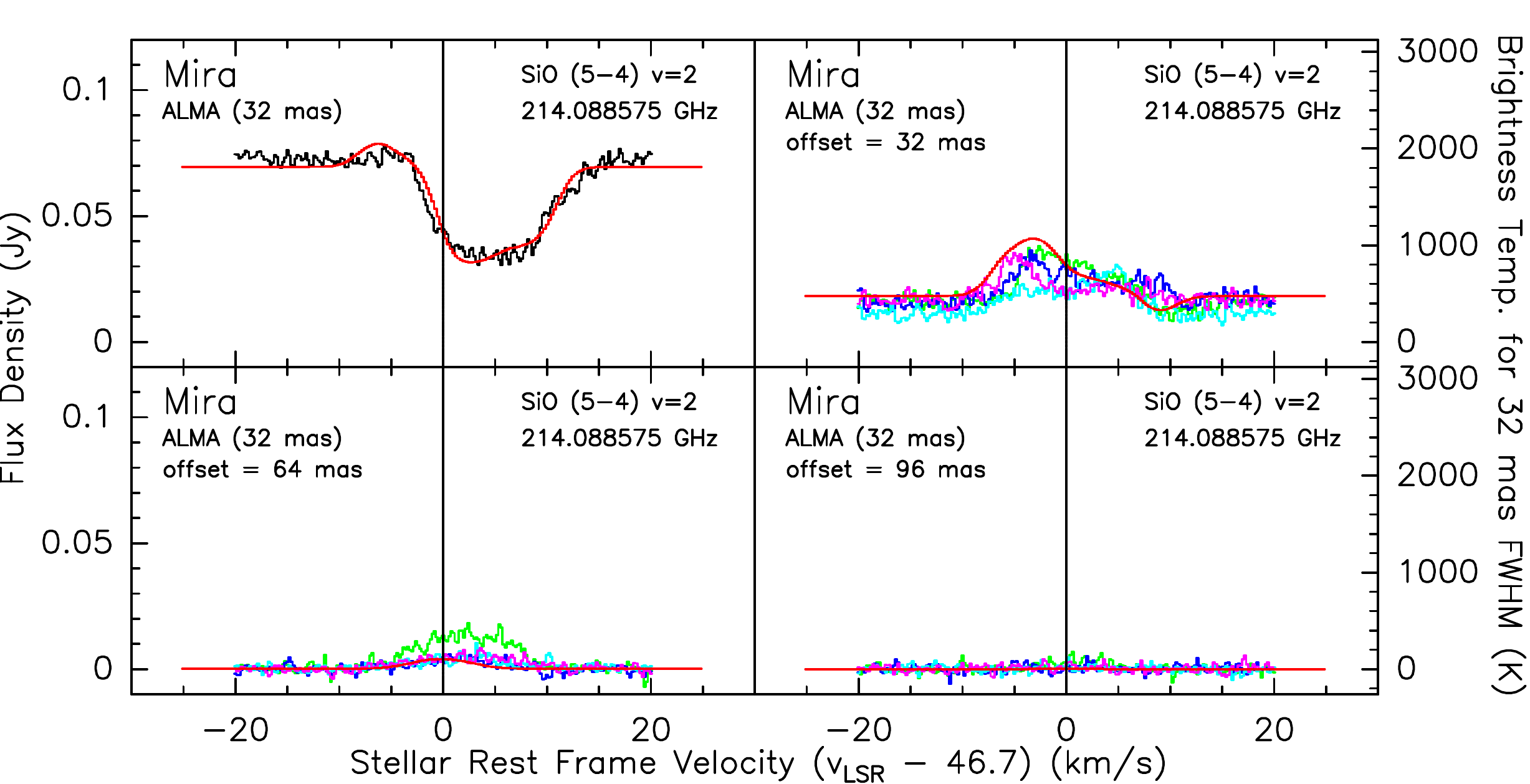}
\caption{The input radial velocity profile (top row) and the modelled SiO (middle and bottom rows) spectra of two nearly identical models, except that for the model on the left, there is only one large-scale velocity variation close to the radio continuum, while for the model on the right, there is an additional strong velocity variation near $1.8 \times 10^{14}\,{\rm cm} = 109\,{\rm mas} \sim 5\,R_{\star}$.}
\label{fig:test-velovariation}
\end{figure*}

We have conducted another test to examine whether large-scale velocity variations of $\Delta V {\sim} 5\,\kms$ at a large radial distance ${\sim}10\,R_{\star}$ from the star are possible. Instead of employing H\"{o}fner's model atmospheres directly, we have constructed a model nearly identical to our preferred Model 3 with a modified input velocity profile. Figure \ref{fig:test-velovariation} shows the results of this test. The model on the left column presents the input velocity profile (top panel), the modelled and observed SiO ${\varv} = 0$ (middle) and SiO ${\varv} = 2$ (bottom) spectra of Model 3 for comparison. The model on the right column shows the input and results of the alternative velocity variation. This alternative model exhibits significant increase in the infall velocity, with $\Delta V \approx 6\,\kms$ at ${\sim}9\,R_{\star}$, which can only be seen in Ireland's {\codex} atmospheres but not in H\"{o}fner's. As long as the infall velocity does not exceed ${\sim} 5\,\kms$, the modelled spectra would not show significant difference regardless of the radial distance where the gas from the extra shock episode is located. Based on the molecular spectra of Mira at this particular stellar phase alone, we can neither distinguish whether such an extra episode of shocked gas exists at all nor determine its possible distance from the star. On the other hand, our test concerning the SiO molecular abundance in Sect. \ref{sec:discuss-dust} shows that the radius at which SiO starts to condense onto dust grains is at least $4\,R_{\star}$. This suggests that the actual wind acceleration may occur beyond ${\sim}4\,R_{\star}$ and hence some velocity variations could still be possible beyond ${\sim}2$--$3\,R_{\star}$.


\section{Conclusions}
\label{sec:concl}


\begin{enumerate}
\item With the long ALMA baselines of ${\sim}15\,{\rm km}$, we are now able to probe the physical conditions in the extended atmospheres and the inner winds (within the dust condensation zone) of AGB stars in unprecedented detail. Mira ($o$ Cet) has been observed as a Science Verification target in the 2014 ALMA Long Baseline Campaign. The angular resolution of the long baseline is ${\sim}30\,{\rm mas}$ at 220\,GHz, which is high enough to resolve the radio continuum of Mira. For the first time, spectral line absorption against the stellar radio continuum has been clearly imaged in the millimetre wavelengths (1.3\,mm) in the transitions of SiO ${\varv} = 0, 1, 2$ $J=5-4$ and H$_2$O $v_2=1$ $J_{K_a,K_c}=5_{5,0}-6_{4,3}$.

\item Through radiative transfer modelling, we are able to reconstruct the detailed physical conditions of Mira's extended atmosphere, namely the gas density, kinetic temperature, abundance of SiO and H$_2$O molecules, and the expansion/infall velocity as functions of radial distance from the star. We fit the SiO and H$_2$O spectra along the lines-of-sight towards the stellar continuum, and towards positions at various sky-projected radii and position angles from the star. In our preferred model which successfully reproduces the spectra of SiO ${\varv} = 0, 2$ and H$_2$O $v_2=1$, the extended atmosphere shows infall motion in general. A shock of velocity $\Delta V \sim 12\,\kms$ is found above Mira's 229\,GHz radio photosphere. The SiO abundance drops significantly from $1 \times 10^{-6}$ to $1 \times 10^{-8}$--$1 \times 10^{-7}$ at the radius of about $1.0 \times 10^{14}\,{\rm cm} = 5\,R_{\star}$, where $R_{\star} = 12.3\,{\rm mas} = 292\,R_{\astrosun}$ is our adopted radius of Mira's infrared photosphere. However, we have also shown that the SiO depletion radius may indeed be anywhere from $4\,R_{\star}$ outwards. In addition, the H$_2$O spectra may be better fitted by adopting an abundance distribution that show a sharp rise in abundance (by about 10 times) near the radio photosphere.

\item We have also tested the predictions from current hydrodynamical models, in particular the {\codex} model series that are tailored for Mira. We have used the predicted atmospheric structures as the inputs of our line radiative transfer modelling. The models successfully reproduce, qualitatively, the absorption features against the continuum. After fine-tuning the radial distances and the magnitudes of the major shock front(s), which are chaotic in nature, the synthesised spectra from the {\codex} models can then reasonably well-fit the observed spectra in this SV observation. Considering the chaotic nature of Mira's extended atmosphere, the modelled spectra from {\codex}'s atmospheres fit remarkably well to the observed ALMA spectra. In addition, we also demonstrated that some other models of Mira's circumstellar environment (e.g. the presence of a chromosphere) do not have support in this ALMA data.

\item We have carried out model fitting of Mira's radio continuum emission at 229.6\,GHz and compared with two other independent results published in mid-2015. Our continuum models for Mira A and B are consistent with those fitted by \citet{mrm2015}. The single uniform disk model for Mira A is consistent with a radio photosphere with a brightness temperature of $2611 \pm 51\,{\rm K}$. On the other hand, we have not found any sign of a compact (${\sim}4.7\,{\rm mas}$) hotspot ($T_b \sim 10\,000\,{\rm K}$) in Mira A's continuum as suggested by \citet{vro2015}, even though we have adopted essentially the identical fitting procedure as them and cross-checked with another visibility fitting software.

\item The long ALMA baselines have demonstrated its capability to produce high-angular resolution and high-sensitivity spectral line images in the extended atmospheres of evolved stars. In order to test the validity of current hydrodynamical models and address the long-standing puzzle of dust condensation and wind-driving mechanism in oxygen-rich evolved stars, long-term (multi-cycle and multi-epoch) monitoring of Mira variables and AGB stars are necessary.
\end{enumerate}



\begin{acknowledgements}
This paper makes use of the following ALMA data: ADS/JAO.ALMA{\#}2011.0.00014.SV. ALMA is a partnership of ESO (representing its member states), NSF (USA) and NINS (Japan), together with NRC (Canada) and NSC and ASIAA (Taiwan), and KASI (Republic of Korea), in cooperation with the Republic of Chile. The Joint ALMA Observatory is operated by ESO, AUI/NRAO and NAOJ. We thank M. Scholz, M.~J. Ireland and P.~R. Wood for kindly providing their model atmospheres from the \texttt{o54} series of the Cool Opacity-sampling Dynamic EXtended ({\codex}) atmosphere model. We thank M.~J. Reid for a careful reading of the manuscript and for his highly useful comments. We also thank the anonymous referee, L.~D. Matthews, and E.~W. Greisen for their insightful comments and suggestions, which have inspired some of our tests to address the imaging issues. We acknowledge with thanks the variable star observations from the AAVSO International Database contributed by observers worldwide and used in this research. This research has made use of NASA's Astrophysics Data System Bibliographic Services. This research has made use of the SIMBAD database, operated at CDS, Strasbourg, France. PyFITS is a product of the Space Telescope Science Institute, which is operated by AURA for NASA. K.~T. Wong was supported for this research through a stipend from the International Max Planck Research School (IMPRS) for Astronomy and Astrophysics at the Universities of Bonn and Cologne and also by the Bonn-Cologne Graduate School of Physics and Astronomy (BCGS). K.~T. Wong also acknowledges the BCGS Honors Budget which covers the digitisation cost of the Ph.D. thesis of R.~C. Doel.
\end{acknowledgements}


\bibliographystyle{aa}
\bibliography{aa-2015-27867}


\begin{appendix}


\section{Continuum analysis of Mira A and B}
\label{sec:appendix_cont}

The size and flux density of the elliptical disk component fitted by \citet{mrm2015} are $52.2\,{\rm mas} \times 40.7\,{\rm mas}$ ($d \equiv \sqrt{\theta_{\rm maj} \theta_{\rm min}} = 7.6 \times 10^{13}\,{\rm cm}$) and $S_{229.6\,{\rm GHz}} = 92 \pm 30 \,{\rm mJy}$, respectively, and those of the additional Gaussian component are $26.0\,{\rm mas} \times 23.9\,{\rm mas}$ ($d = 4.1 \times 10^{13}\,{\rm cm}$) and $S_{229.6\,{\rm GHz}} = 59\,{\rm mJy}$, respectively. However, significantly different results have been reported by \citet{vro2015}. The elliptical disk component fitted by \citet{vro2015} has a size of ${\sim}43.4\,{\rm mas} \times 38.5\,{\rm mas}$ ($d = 6.7 \times 10^{13}\,{\rm cm}$) and a flux density of $S_{229.6\,{\rm GHz}} = 139 \pm 0.2 \,{\rm mJy}$, while the additional Gaussian, which was assumed circular in their model, has $d = 4.7 \,{\rm mas}$ ($0.8 \times 10^{13}\,{\rm cm}$) and $S_{229.6\,{\rm GHz}} = 9.6 \pm 0.1 \,{\rm mJy}$. At 229.6\,GHz, the brightness temperature of the 4.7-mas Gaussian component is approximately 10\,000\,K \citep{vro2015}, which is much higher than that of the uniform elliptical disk component of Mira A. \citet{vro2015} therefore argue the existence of a hotspot, or a cluster of unresolved spots, and speculate that its origin may be connected to the stellar magnetic activity of Mira A.

We have also conducted similar continuum analysis of the Mira AB system in ALMA Band 6. Apart from the data in the continuum windows of this SV dataset ($\texttt{spw}=0, 7$; at 229.6\,GHz) as already analysed by \citeauthor{mrm2015} and \citeauthor{vro2015}, we also include the continuum data from the six spectral line windows ($\texttt{spw}=1$--$6$ and $8$--$13$). The continuum data in these spectral line windows were prepared by averaging and splitting out the line-free channels in the windows corresponding to each distinct frequency range (see the first two columns of Table \ref{tab:continuum}), and then exporting the visibility data to Miriad. The visibilities of Mira A and B were fitted simultaneously using the similar model adopted by \citet{mrm2015} and \citet{vro2015}, i.e., a uniform disk plus an additional Gaussian component for Mira A, and a single Gaussian model for Mira B. We fit the visibilities with the task \texttt{uvfit} (Revision 1.10, 2013 August) in Miriad. Miriad/\texttt{uvfit} is preferred to the \texttt{uvmodelfit} task in CASA because it allows multiple model components to be fitted simultaneously. We have cross-checked the $uv$-fitting results from the respective tasks in Miriad and CASA with single-component models, and both tasks return similar results. To compare the results of \citet{mrm2015} and \citet{vro2015}, the visibilities of the 229-GHz continuum windows ($\texttt{spw}=0, 7$) have been fitted in the similar manner.

\begin{figure*}[!htbp]
\centering
\includegraphics[width=0.42\textwidth]{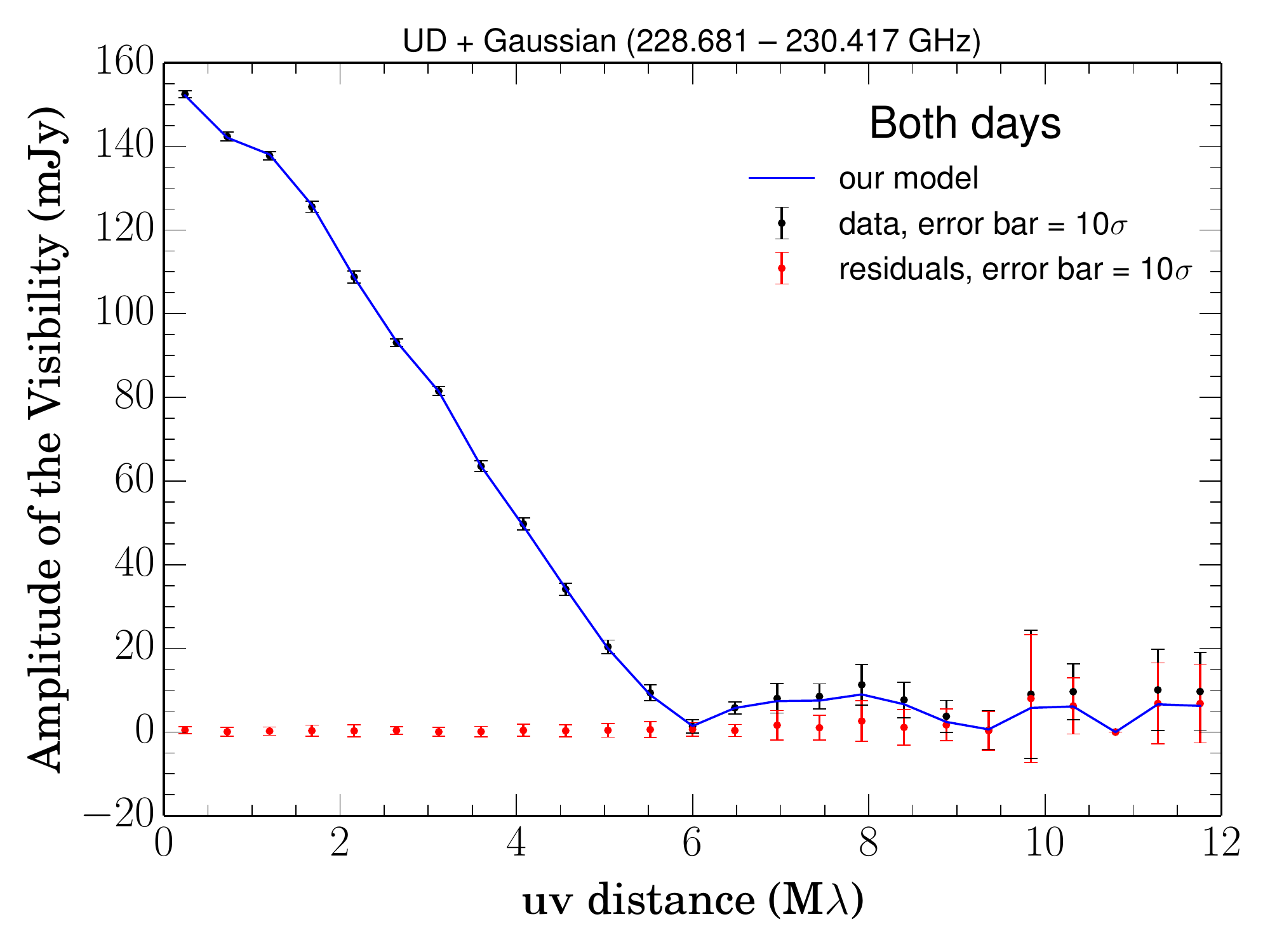}\\[-6pt]
\includegraphics[width=0.42\textwidth]{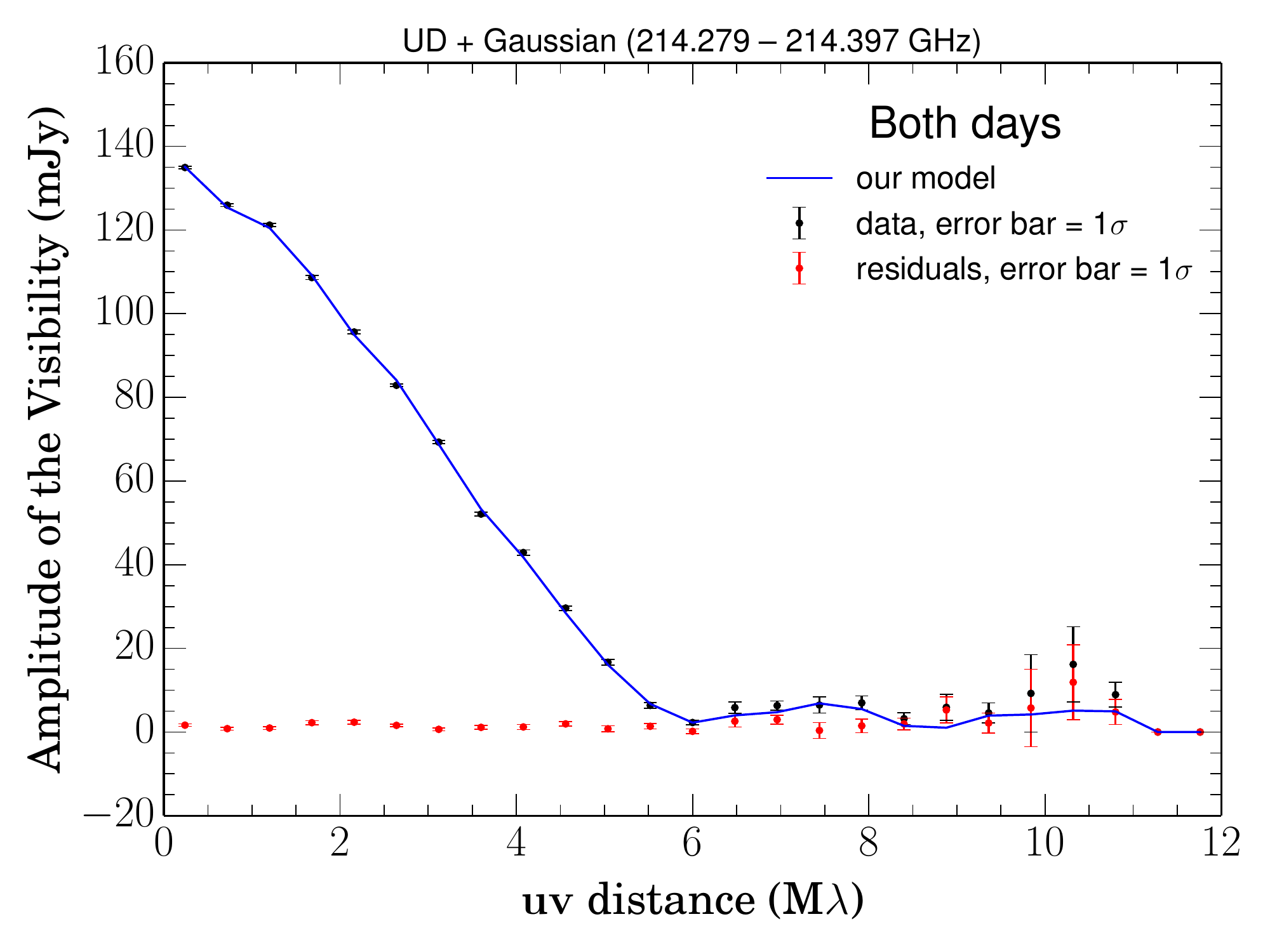}
\includegraphics[width=0.42\textwidth]{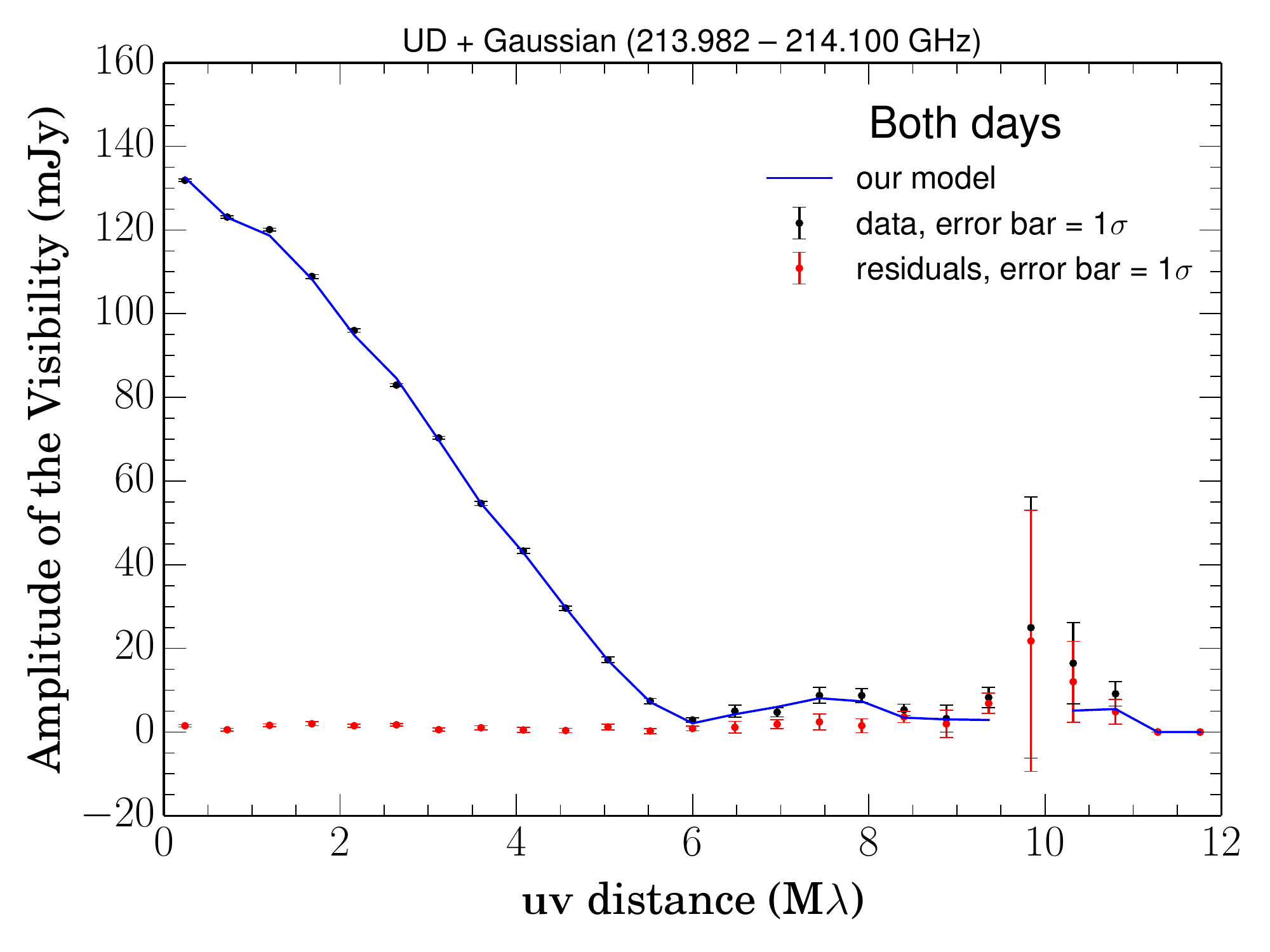}\\[-6pt]
\includegraphics[width=0.42\textwidth]{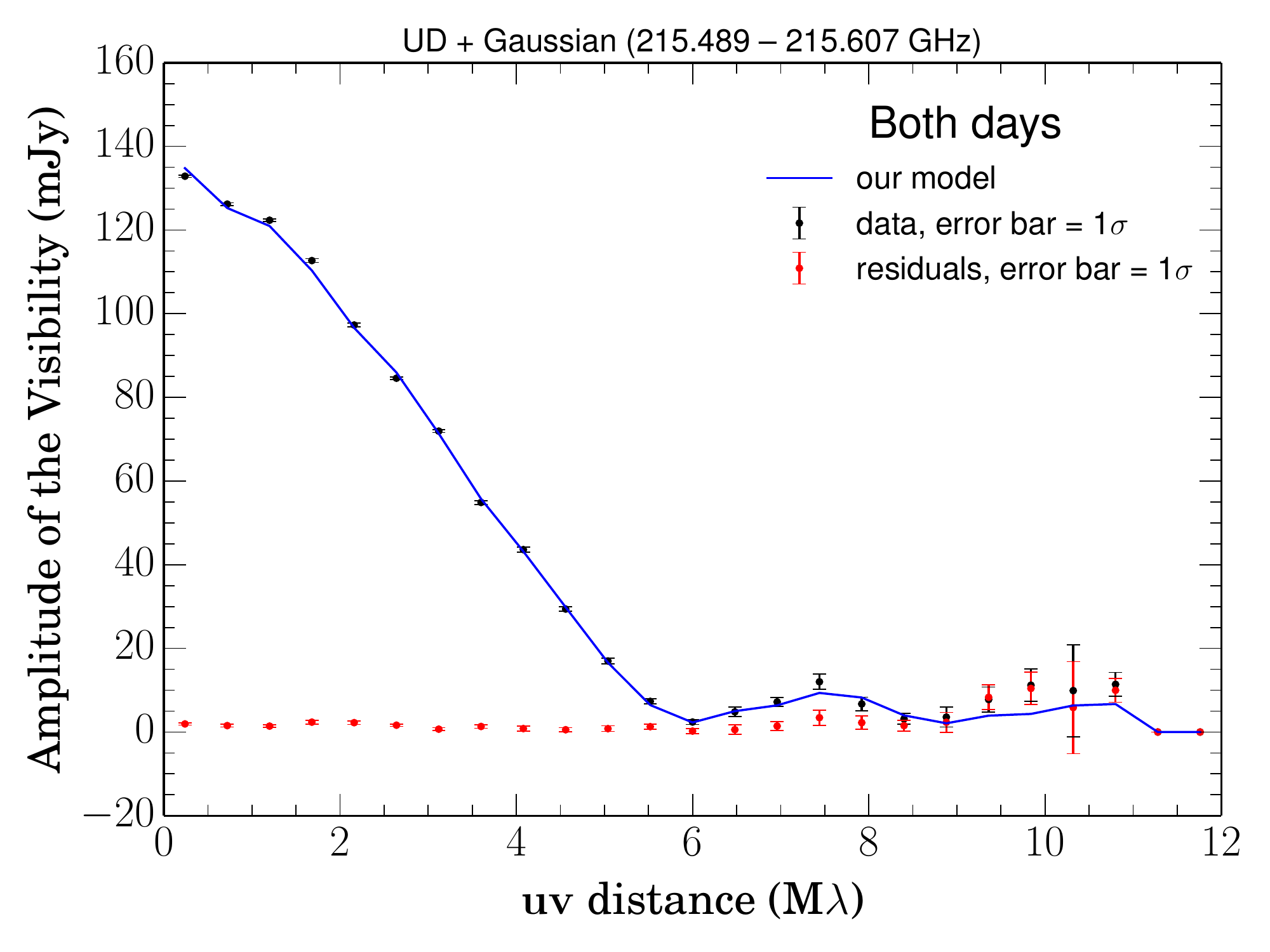}
\includegraphics[width=0.42\textwidth]{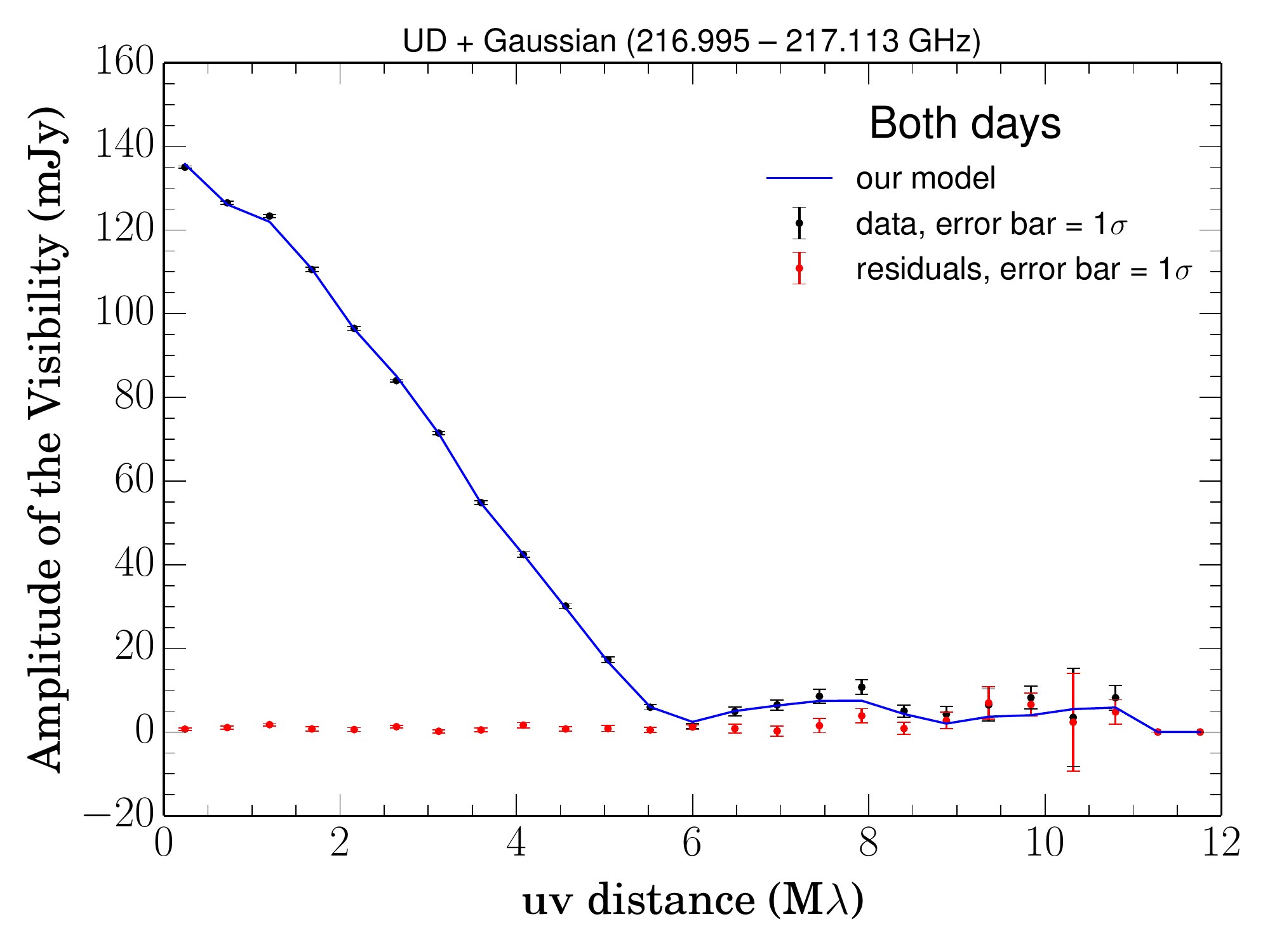}\\[-6pt]
\includegraphics[width=0.42\textwidth]{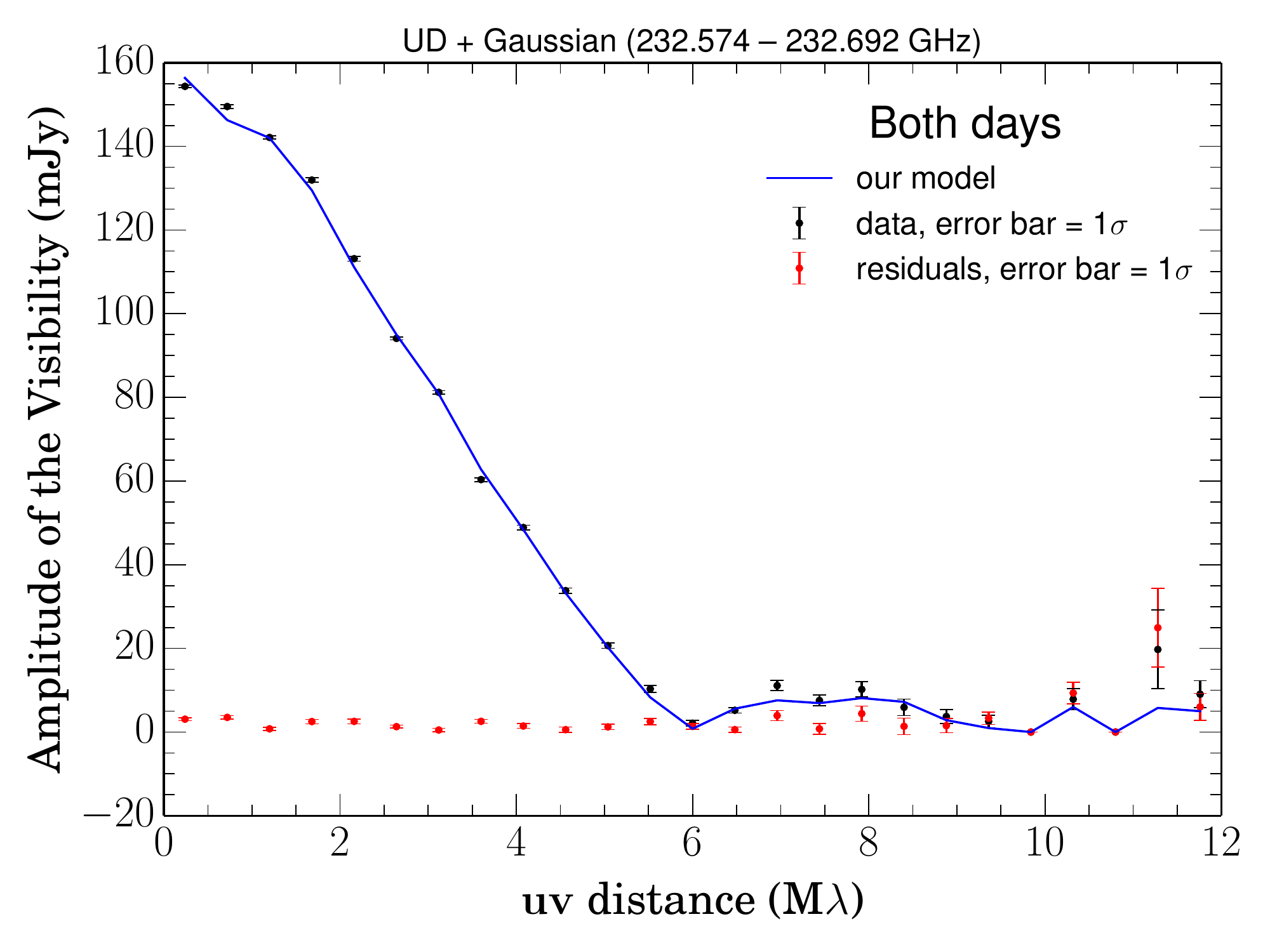}
\includegraphics[width=0.42\textwidth]{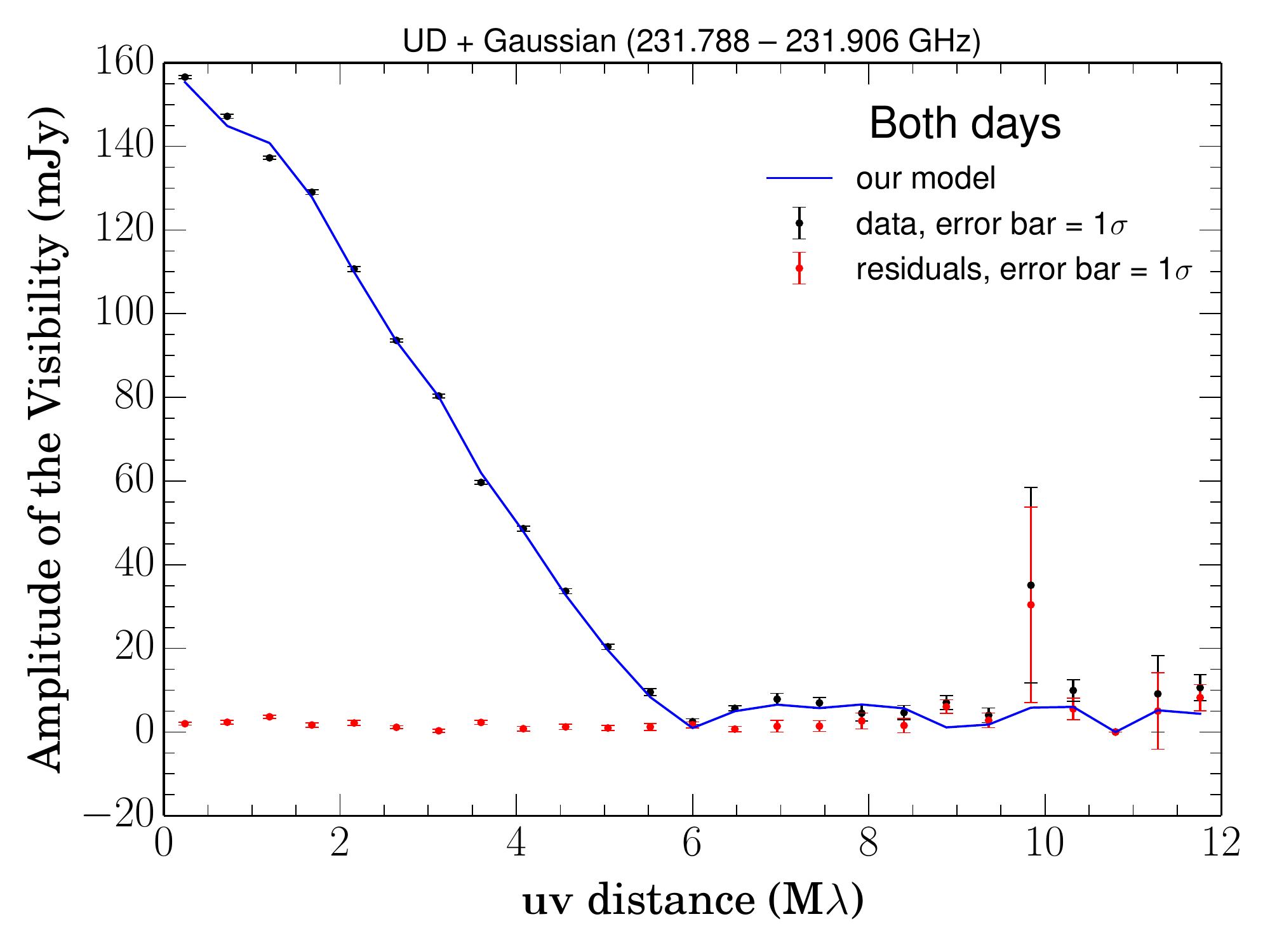}
\caption{Visibility plots of the continuum fitting with the task \texttt{uvfit} in Miriad. The plot on the top is the continuum window at 229.6\,GHz, and the six below are (from top to bottom, and from left to right) the spectral line windows in ascending order of frequency: 214.0\,GHz, 214.3\,GHz, 215.5\,GHz, 271.1\,GHz, 231.8\,GHz, 231.6\,GHz as listed in Table \ref{tab:continuum}. The error bars in the continuum window and spectral line windows are 10 times and 1 time, respectively, the standard deviation in the mean of the amplitude of visibilities in the respective bin of $uv$-distance. The large error bar near $10\,{\rm M}\lambda$ is due to small number of data points.}
\label{fig:cont-uvfit-vis}
\end{figure*}

\begin{figure*}[!htbp]
\centering
\includegraphics[trim=2.0cm 0.0cm 3.0cm 0.0cm, clip, width=1.3\miriadmapwidth]{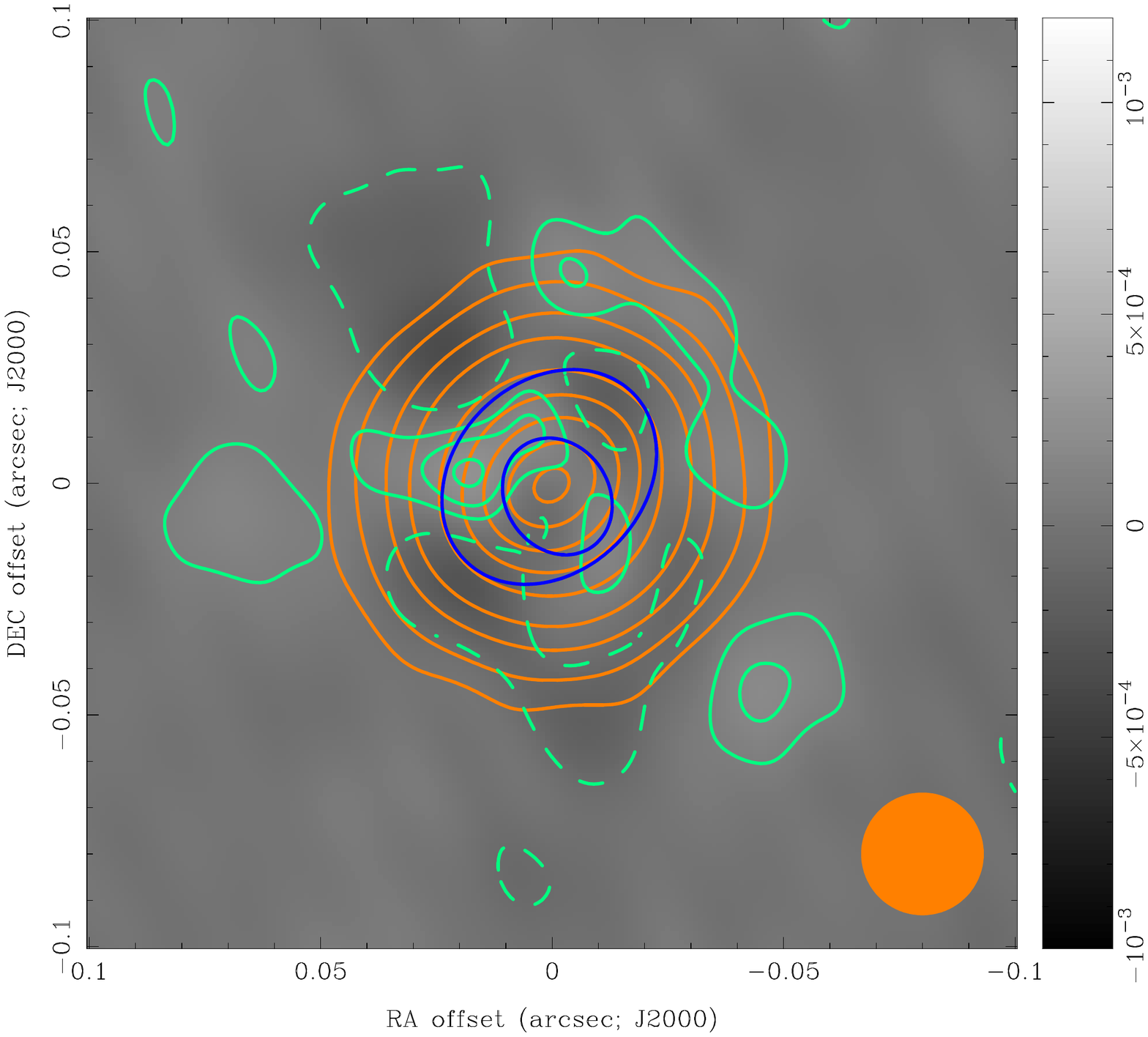}
\put(-145,152){\Large{228.681--230.417\,GHz}}\\
\includegraphics[trim=2.0cm 0.0cm 3.0cm 0.0cm, clip, width=\miriadmapwidth]{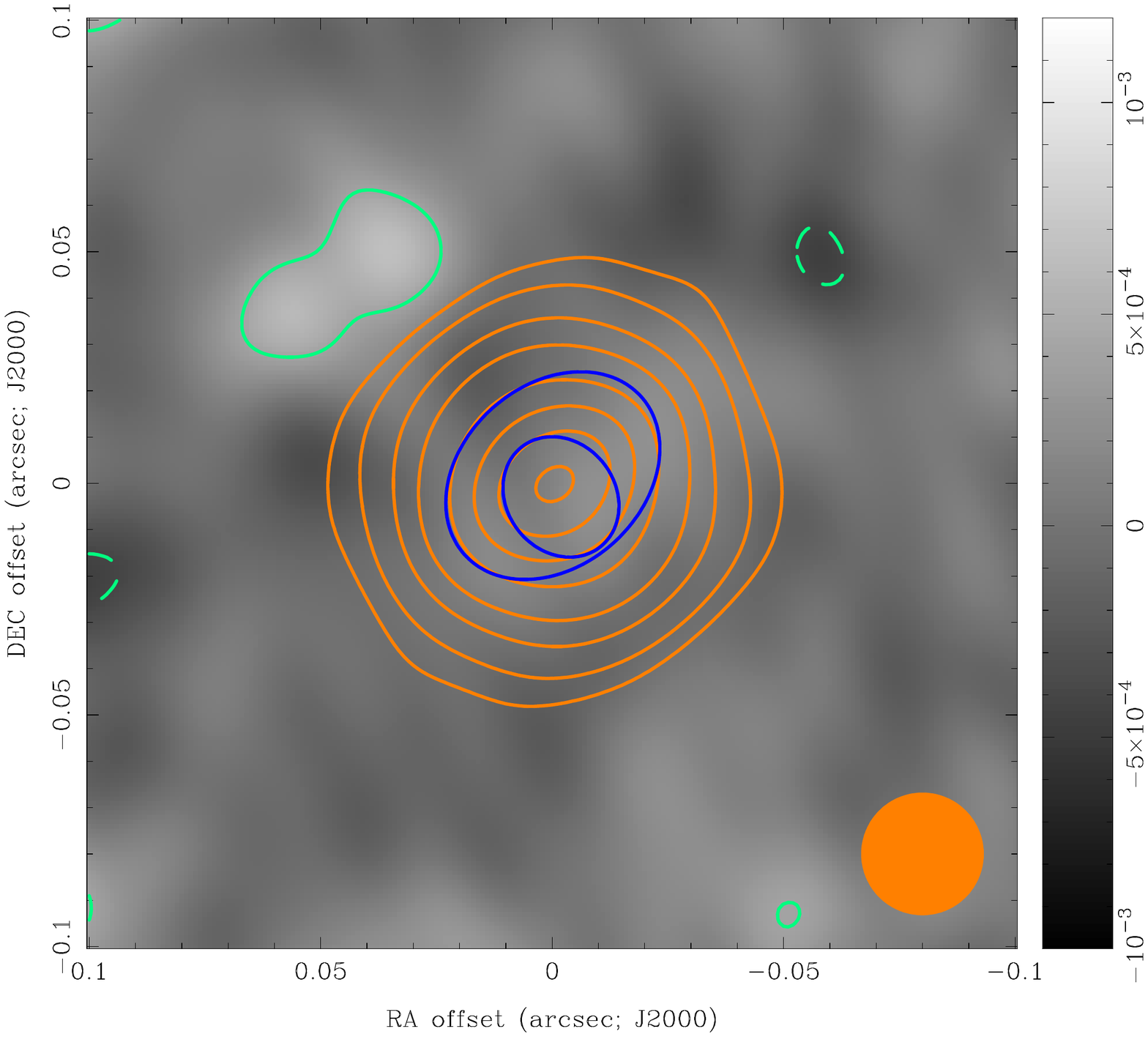}
\put(-110,117){{213.982--214.100\,GHz}}
\includegraphics[trim=2.0cm 0.0cm 3.0cm 0.0cm, clip, width=\miriadmapwidth]{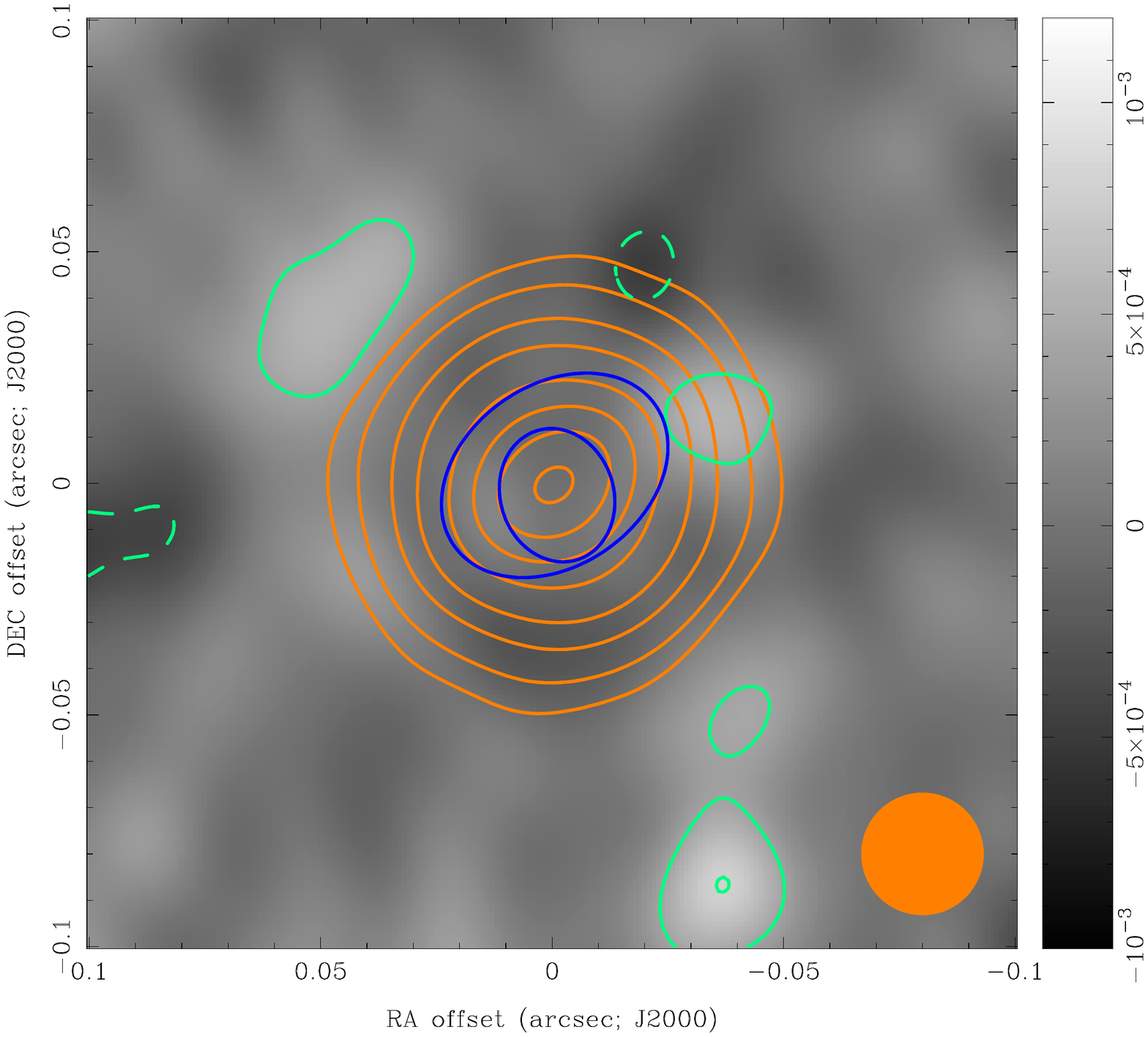}
\put(-110,117){{214.279--214.397\,GHz}}
\includegraphics[trim=2.0cm 0.0cm 3.0cm 0.0cm, clip, width=\miriadmapwidth]{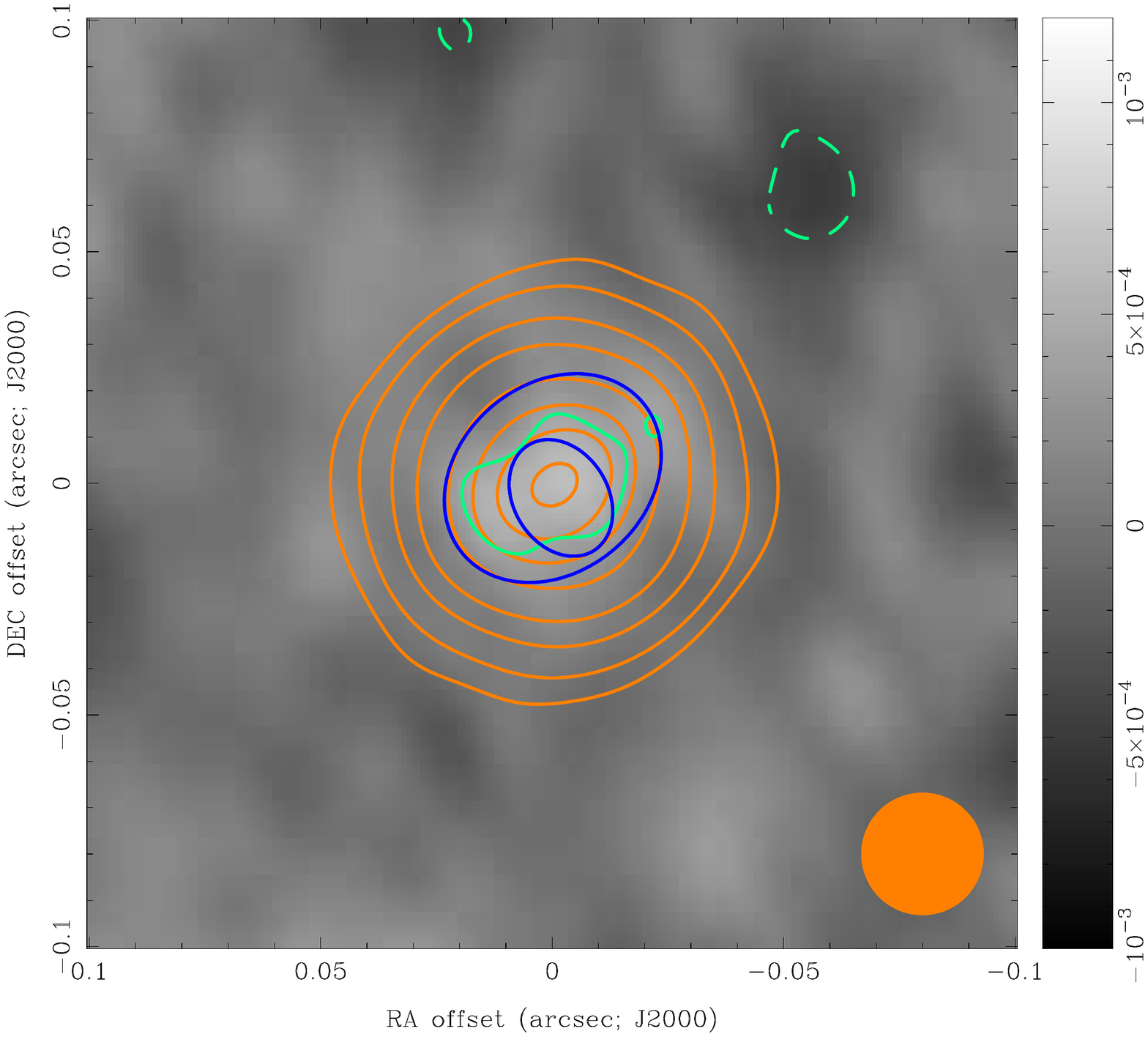}
\put(-110,117){{215.489--215.607\,GHz}}\\
\includegraphics[trim=2.0cm 0.0cm 3.0cm 0.0cm, clip, width=\miriadmapwidth]{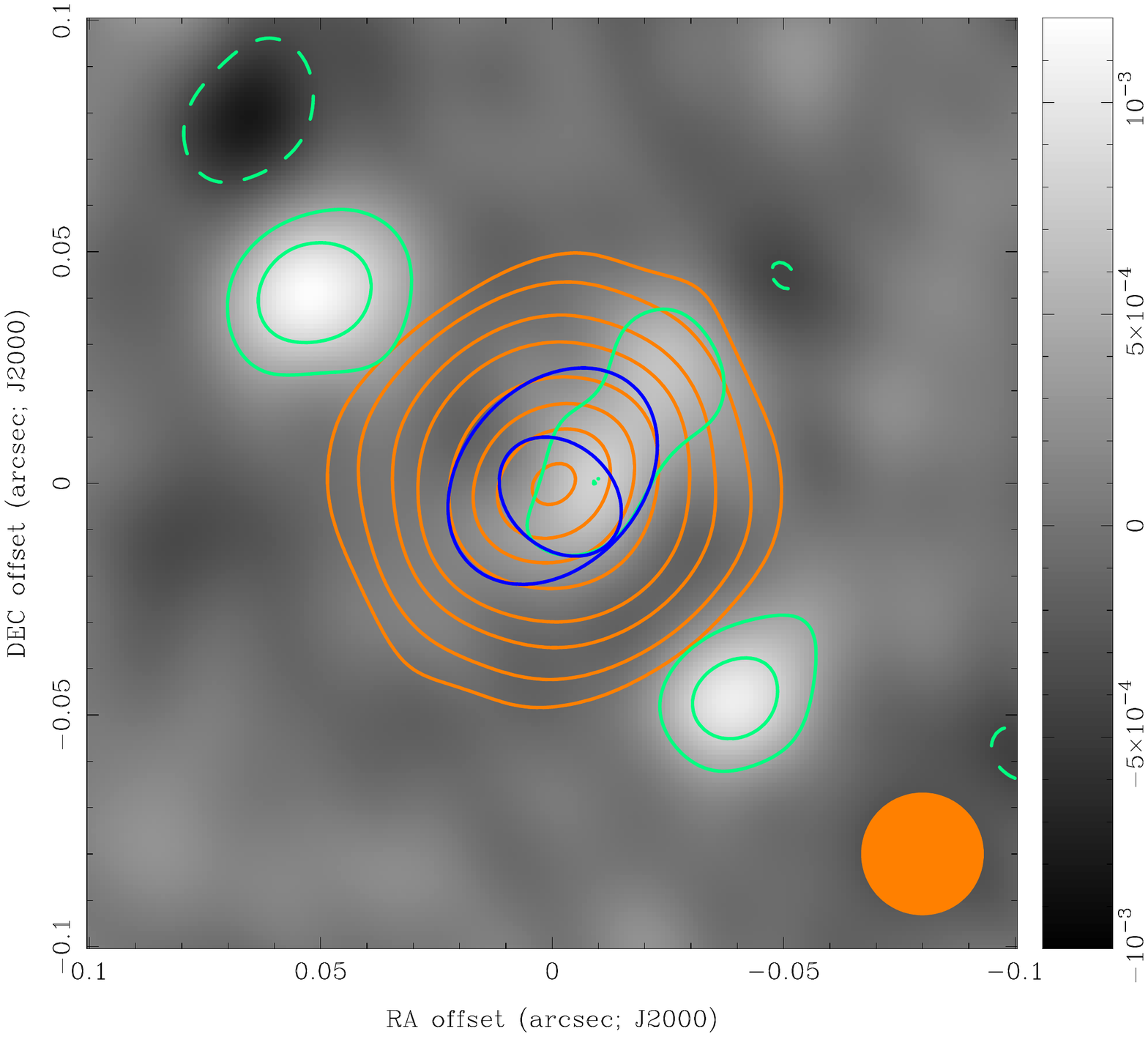}
\put(-110,117){{216.995--217.113\,GHz}}
\includegraphics[trim=2.0cm 0.0cm 3.0cm 0.0cm, clip, width=\miriadmapwidth]{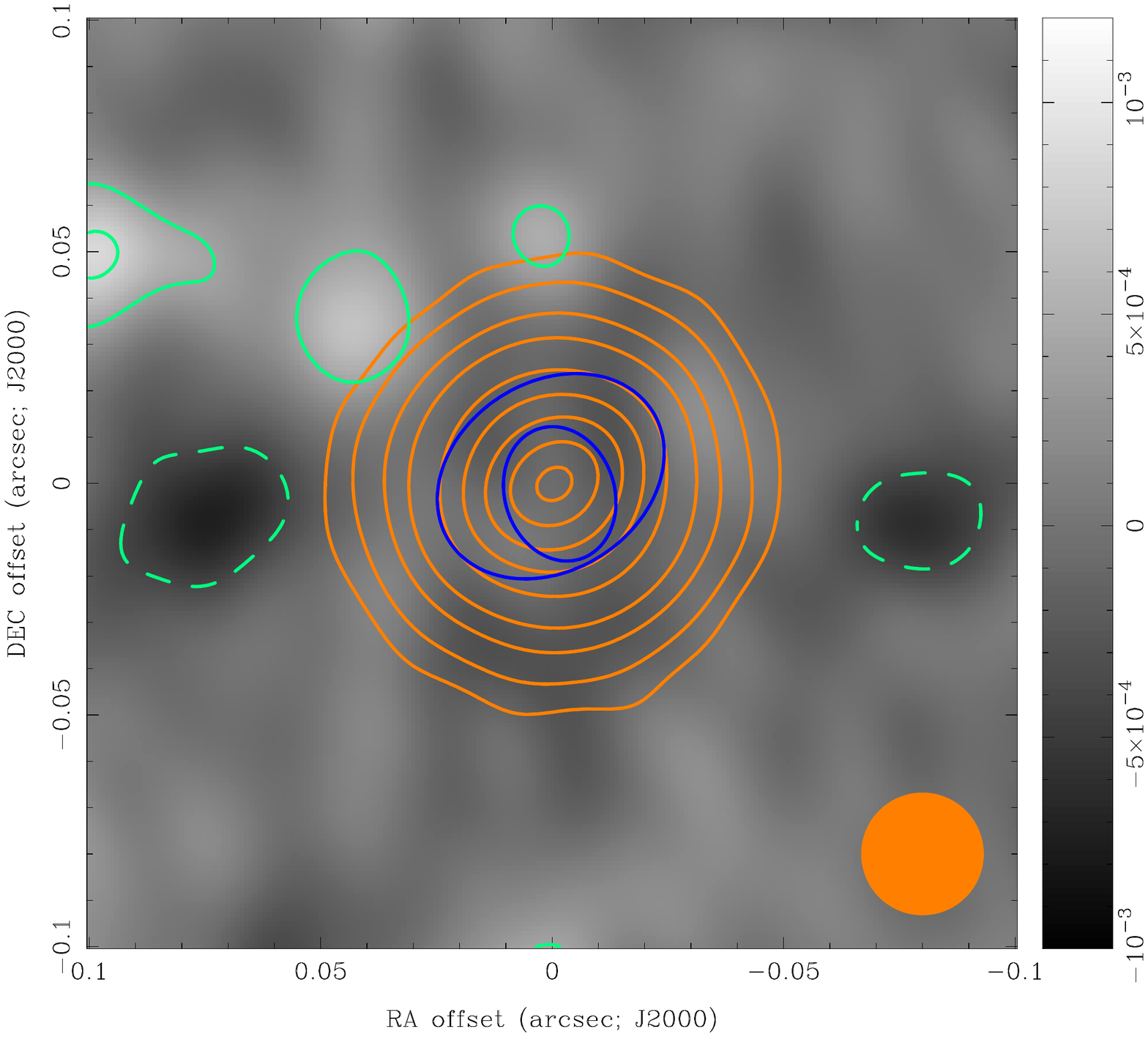}
\put(-110,117){{231.788--231.906\,GHz}}
\includegraphics[trim=2.0cm 0.0cm 3.0cm 0.0cm, clip, width=\miriadmapwidth]{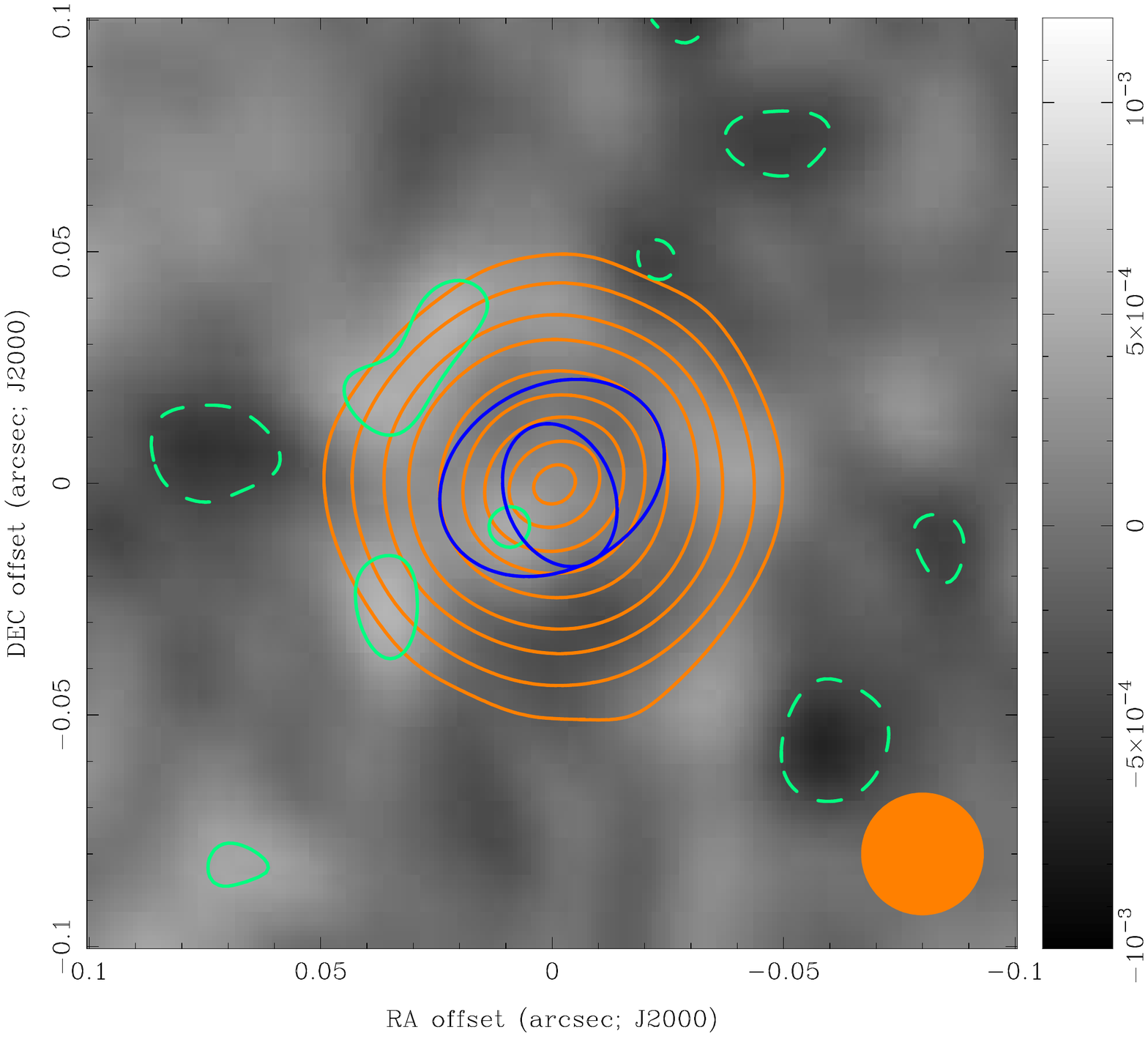}
\put(-110,117){{232.574--232.692\,GHz}}
\caption[]{Image showing the continuum fitting results with the task \texttt{uvfit} in Miriad. The large map on the top is the continuum window at 229.6\,GHz, the six smaller maps below are (from top to bottom, and from left to right) the spectral line windows in ascending order of frequency: 214.0\,GHz, 214.3\,GHz, 215.5\,GHz, 271.1\,GHz, 231.8\,GHz, and 231.6\,GHz as listed in Table \ref{tab:continuum}. In all the panels, the orange contours are the continuum models for Mira A, with parameters listed in Table \ref{tab:continuum}. The sizes of the (larger) uniform disk and (smaller) Gaussian models for the respective spectral windows are also drawn in blue open ellipses. The continuum contour levels are $-3,3,10,30,60,120,180,240,300,340 \times 0.20\,{\rm mJy}$. The green contours and the greyscale maps are the residual images (difference between the data and modelled continuum) of the respective spectral windows. The residual contour levels are $-2,2,4,6 \times \sigma$, where $\sigma=0.04\,{\rm mJy}$ for the continuum window (top) and $\sigma=0.20\,{\rm mJy}$ for the six line windows (bottom two rows). The restoring beam of $0{\farcs}027$ FWHM is indicated in orange at the bottom-right corner in each panel.}
\label{fig:cont-uvfit}
\end{figure*}

Table \ref{tab:continuum} shows the $uv$-fitting results to the continuum data in the ascending order of the range of rest frequencies. For Mira A, whilst the major and minor axes of the elliptical components appear to be consistent among all spectral windows, the fluxes shared by the uniform disk and the Gaussian component show great variation. The formal errors of the fluxes reported by \texttt{uvfit} are ${\sim}9$\,mJy for the continuum window and $35$--$40$\,mJy for the six spectral line windows. Hence, the fitted fluxes of the individual components of Mira A in the line windows are not as reliable as those in the continuum window at 229.6\,GHz. Figures \ref{fig:cont-uvfit-vis} and \ref{fig:cont-uvfit} show the visibility plots and the maps, respectively, of the fitted continuum models and residuals for all seven continuum and spectral line windows in the Band 6 data. In the maps, there are some ${\sim}6\sigma$ residuals in the 229-GHz continuum window. However, our experiment shows that adding one or two more Gaussian components to the $uv$-fitting does not significantly improve the results. The residuals in the spectral line windows appear to be less significant than the continuum, probably because the map rms noises are much higher (by a factor of about four) than that in the map of the continuum window.


\begin{table*}[!htbp]
\caption{Photospheric parameters from fitting the continuum visibility data, with the \texttt{uvfit} task in the Miriad software, of Mira A and B in different continuum and spectral line windows of ALMA Band 6.}
\label{tab:continuum}
\centering
\begin{tabular}{lccccccc}
\hline\hline
SPWs & Freq. Range & \multicolumn{2}{c}{Mira A Uniform Disk Model} & \multicolumn{2}{c}{Mira A Gaussian Model} & \multicolumn{2}{c}{Mira B Gaussian Model} \\
 &  & $S_{\nu} (\Delta S_{\nu})$ & $\theta_{\rm maj} \times \theta_{\rm min}$, P.A. & $S_{\nu} (\Delta S_{\nu})$ & $\theta_{\rm maj} \times \theta_{\rm min}$, P.A. & $S_{\nu} (\Delta S_{\nu})$ & $\theta_{\rm maj} \times \theta_{\rm min}$, P.A. \\
 & (GHz) & (mJy) & (mas $\times$ mas, $\degr$) & (mJy) & (mas $\times$ mas, $\degr$) & (mJy) & (mas $\times$ mas, $\degr$) \\
\hline
2, 9 & 213.982--214.100 & 84(36) & $51.2 \times 39.1$, $-48.3$ & 45(37) & $27.4 \times 23.5$, $38.2$ & 9.4(1.8) & $25.7 \times 19.0$, $-81.4$ \\
1, 8  & 214.279--214.397 & 76(37) & $53.1 \times 39.1$, $-55.1$ & 56(37) & $29.2 \times 24.5$, $17.6$ & 9.4(1.9) & $25.6 \times 19.0$, $-60.1$ \\
3, 10 & 215.489--215.607 & 97(40) & $50.8 \times 40.6$, $-50.0$ & 35(41) & $26.8 \times 20.4$, $31.9$ & 9.2(1.7) & $23.4 \times 19.1$, $82.3$ \\
4, 11 & 216.995--217.113 & 90(35) & $52.2 \times 38.8$, $-41.8$ & 43(36) & $29.1 \times 22.6$, $48.0$ & 9.8(1.9) & $24.7 \times 19.7$, $-89.6$ \\
0, 7  & 228.681--230.417 & 102(9) & $51.2 \times 41.0$, $-45.0$ & 47(9) & $26.4 \times 22.4$, $34.0$ & 11.3(0.5) & $25.5 \times 22.5$, $72.7$ \\
6, 13 & 231.788--231.906 & 91(40) & $52.1 \times 40.6$, $-57.1$ & 61(41) & $29.5 \times 23.7$, $17.9$ & 11.3(2.2) & $28.1 \times 22.4$, $12.3$ \\
5, 12 & 232.574--232.692 & 95(36) & $51.1 \times 39.5$, $-60.4$ & 59(36) & $31.8 \times 23.5$, $21.3$ & 11.1(2.1) & $25.8 \times 24.7$, $43.1$ \\
\hline
\end{tabular}
\tablefoot{The columns are (from left to right): ALMA Band 6 spectral windows (SPWs), rest frequency range of the continuum data (GHz), and the total flux ($S_{\nu}$, mJy) and the formal error of the fitting in the parenthesis ($\Delta S_{\nu}$, mJy), major and minor axes ($\theta_{\rm maj}$, $\theta_{\rm min}$, mas) and the position angle (P.A., degrees) of each component adopted in the visibility fitting. Mira A was fitted with an elliptical uniform disk together with an elliptical Gaussian component, and Mira B with an elliptical Gaussian model. The formal errors for the major and minor axes are typically ${\lesssim}1\,{\rm mas}$ for the models of Mira A, and ${\lesssim}2\,{\rm mas}$ for Mira B, except for the continuum windows ($\texttt{spw}=0, 7$) in which the errors are ${\lesssim}0.2\,{\rm mas}$ for Mira A and ${\sim}0.3\,{\rm mas}$ for Mira B. The choices of model components are the similar to those adopted by \citet{mrm2015} and \citet{vro2015}.}
\end{table*}


The size and flux density of our elliptical disk model for Mira A in the continuum window are $51.2\,{\rm mas} \times 41.0\,{\rm mas}$ ($d = 7.6 \times 10^{13}\,{\rm cm}$) and $S_{229.6\,{\rm GHz}} = 102 \pm 9 \,{\rm mJy}$, respectively. This corresponds to a brightness temperature of $1630 \pm 175\,{\rm K}$. The uncertainly takes into account the formal errors of the flux (which is the dominant formal error), major and minor axes of the disk model (${\sim}0.1\,{\rm mas}$), and the bandwidth of the continuum spectral window (${\sim}1.7\,{\rm GHz}$). The additional Gaussian component for Mira has a size of $26.4\,{\rm mas} \times 22.4\,{\rm mas}$ ($d = 4.0 \times 10^{13}\,{\rm cm}$) and a flux density of $S_{229.6\,{\rm GHz}} = 47 \pm 9 \,{\rm mJy}$, respectively. These values are consistent with those from \citet{mrm2015} within uncertainties, but significantly different from the fitting of \citet{vro2015}.

\begin{figure*}[!htbp]
\centering
\includegraphics[width=0.42\textwidth]{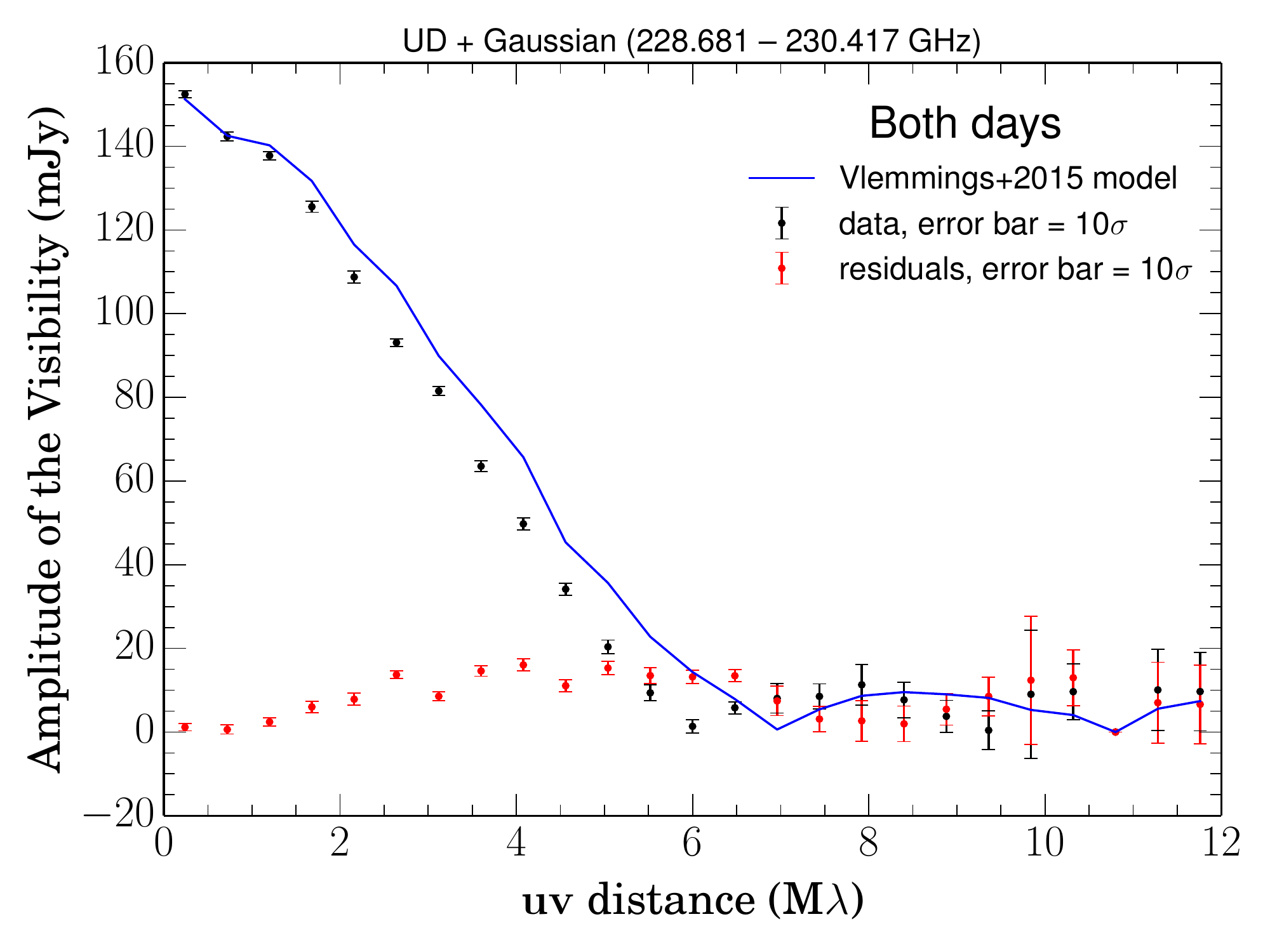}
\includegraphics[width=0.42\textwidth]{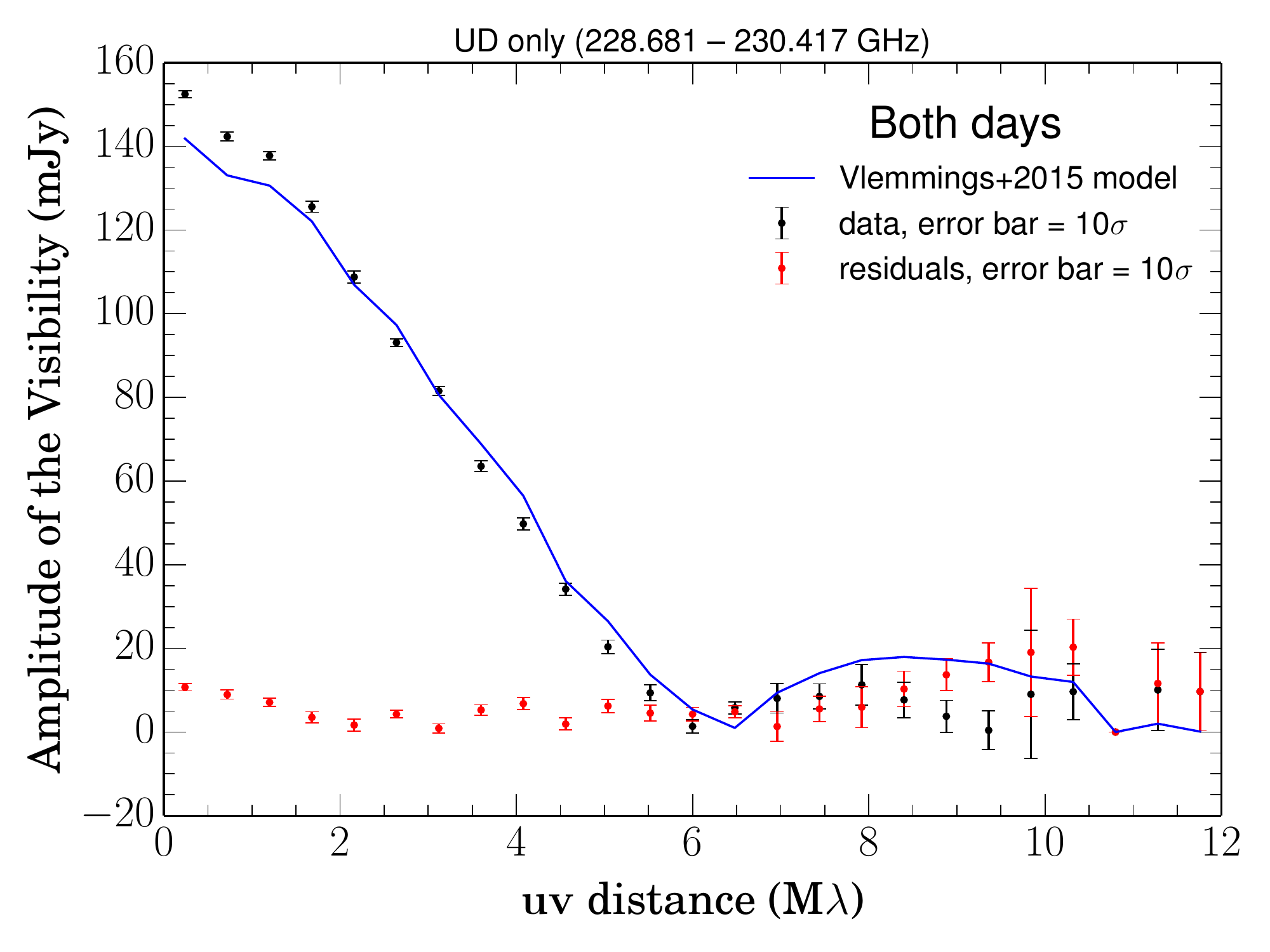}\\[-6pt]
\includegraphics[width=0.42\textwidth]{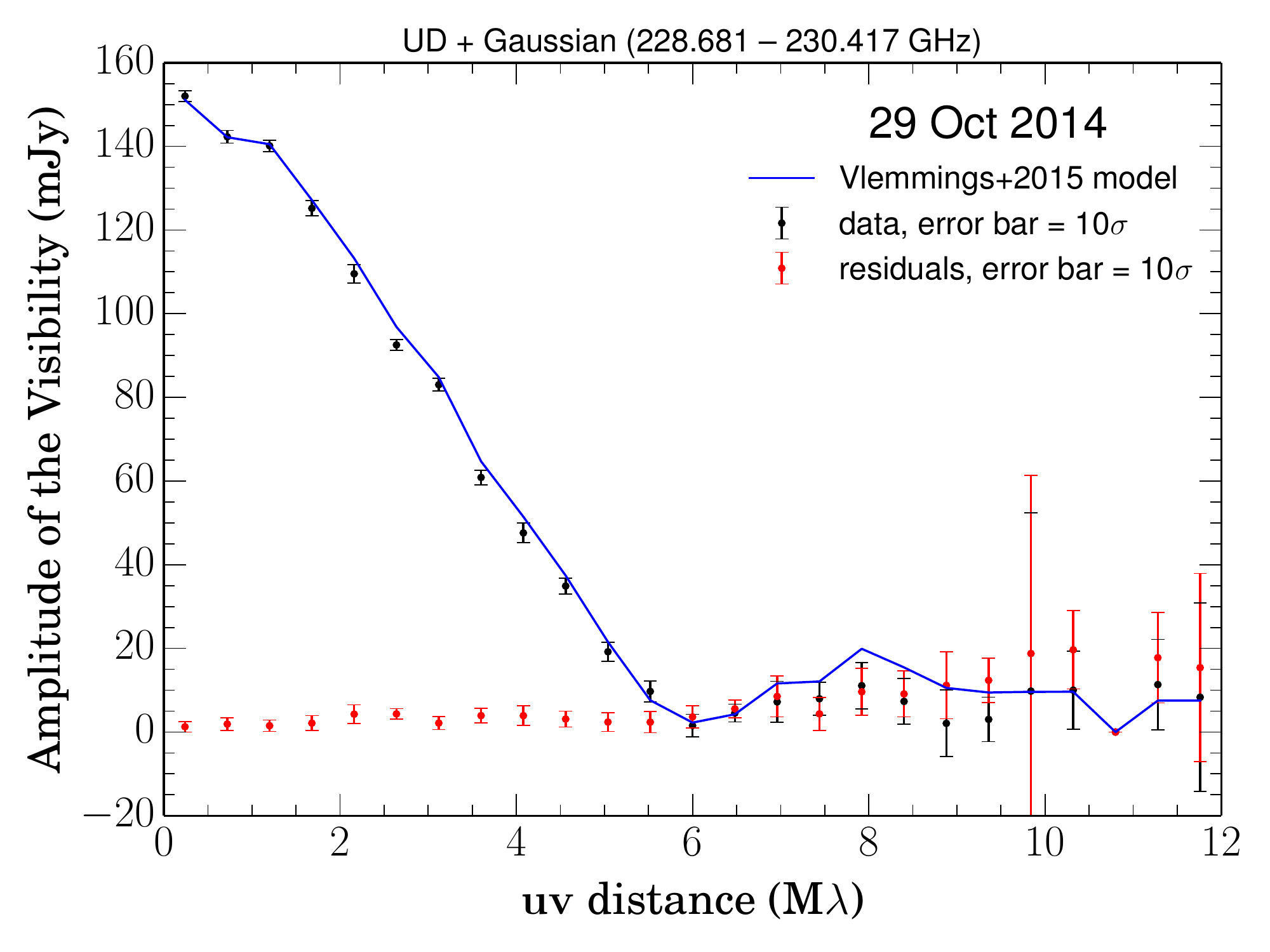}
\includegraphics[width=0.42\textwidth]{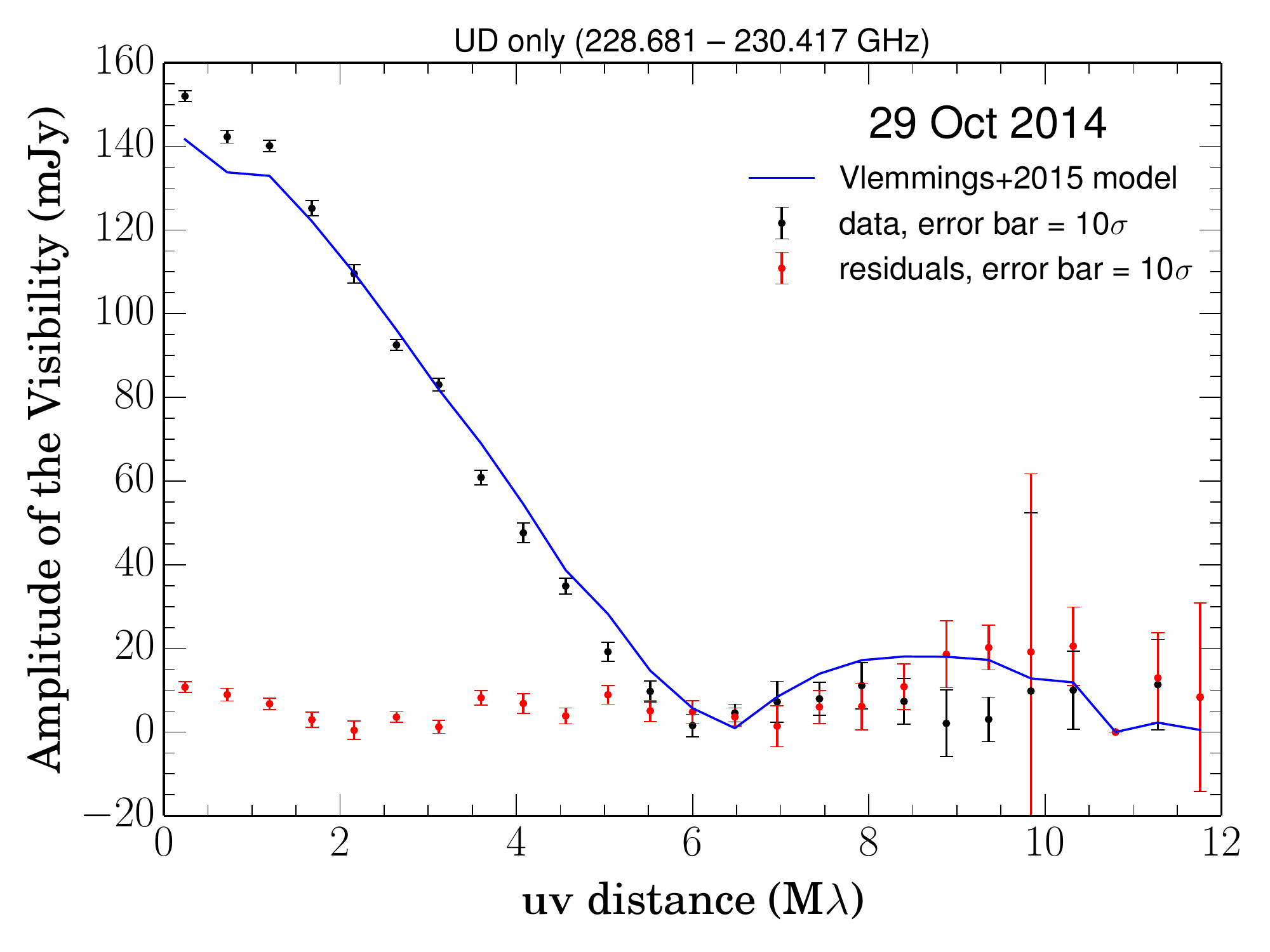}\\[-6pt]
\includegraphics[width=0.42\textwidth]{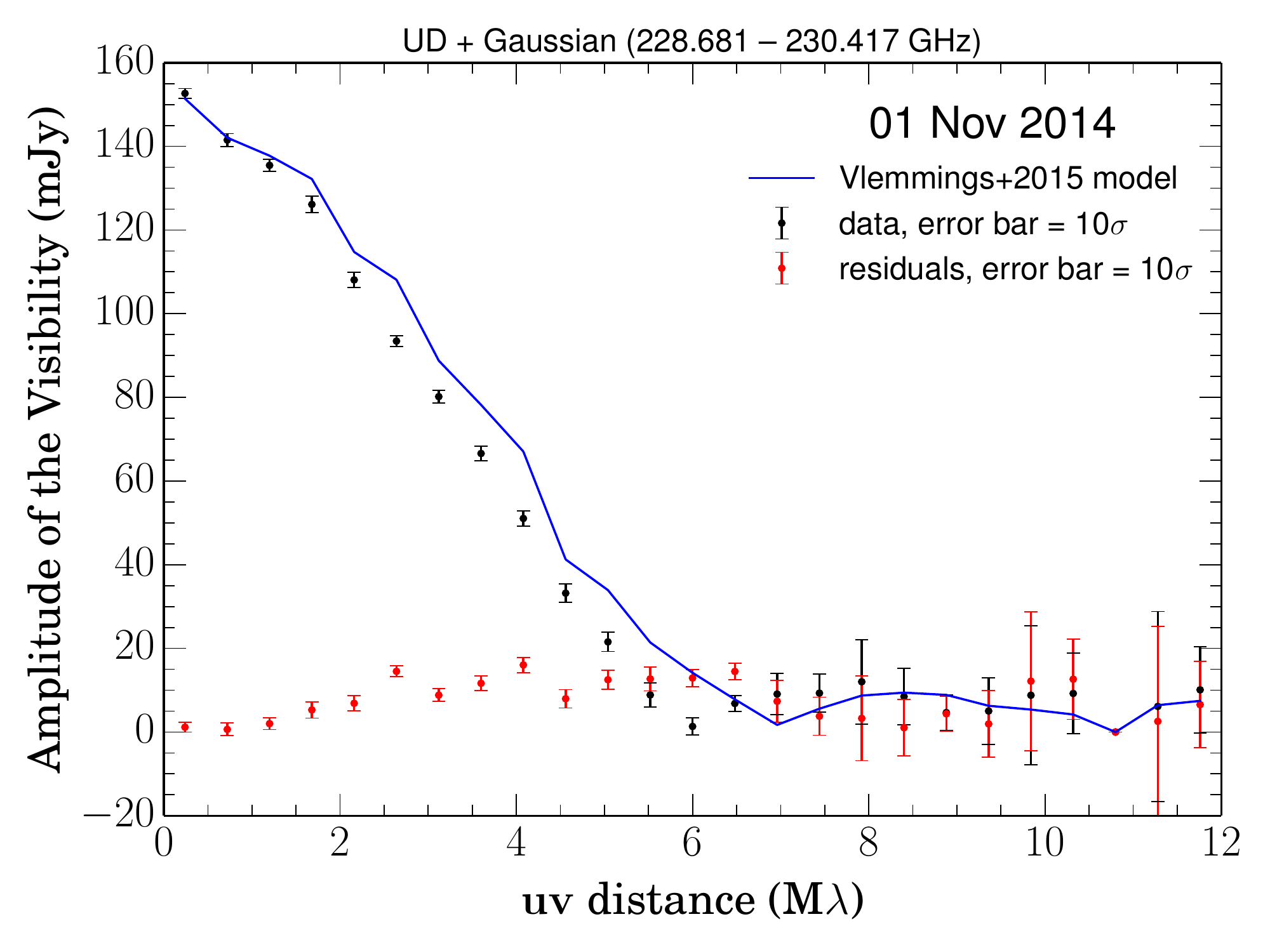}
\includegraphics[width=0.42\textwidth]{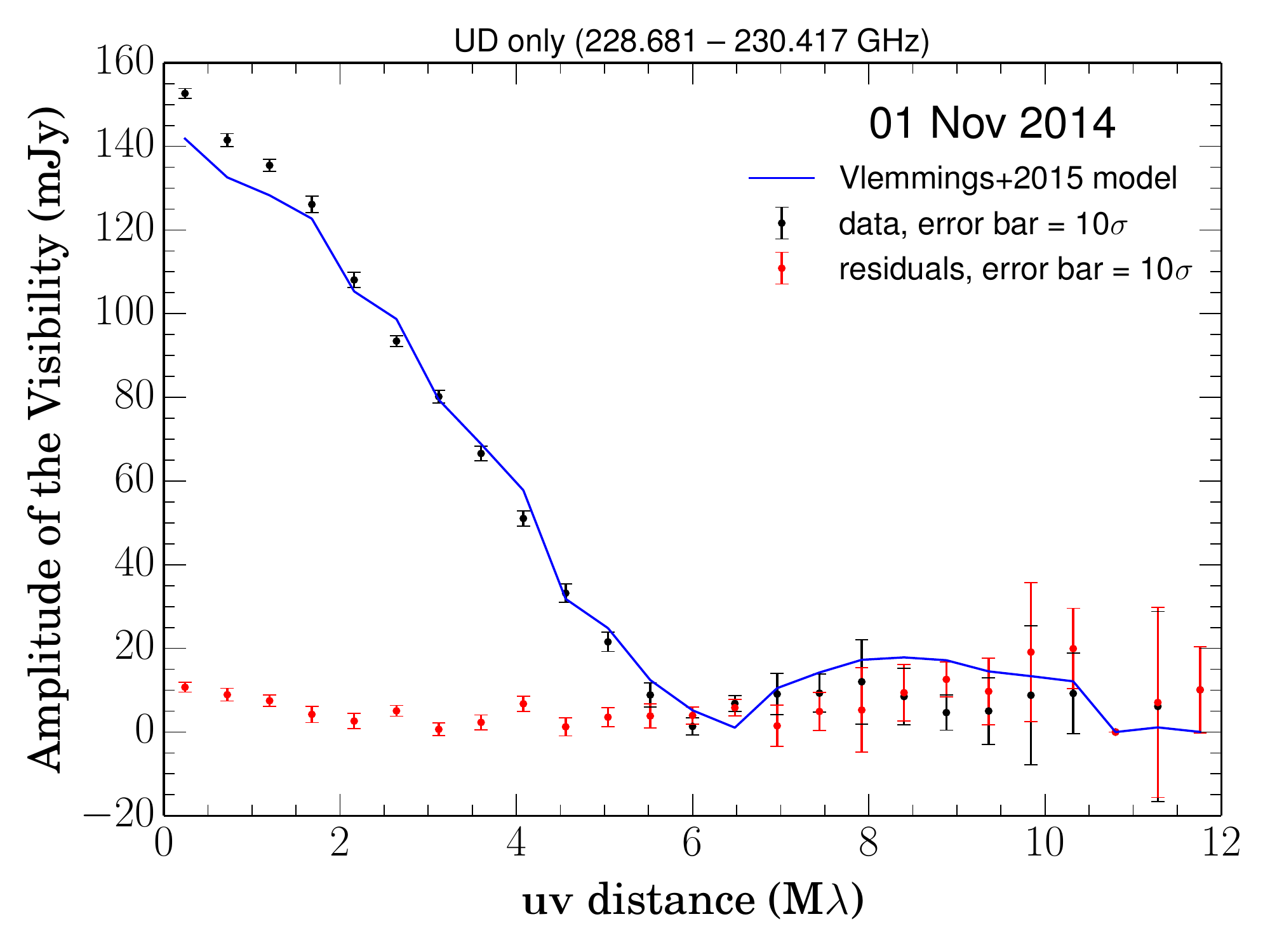}
\caption{Visibility plots of the continuum fitting with the task \texttt{uvfit} in Miriad using the parameters obtained by \citet{vro2015}. All the plots are the results of the continuum window at 229.6\,GHz. The top panels are the results without splitting the datasets into individual days. The middle and bottom panels are the results of split data for 2014 October 29 and 2014 November 01, respectively. The 3 panels on the left are results using uniform disk and a Gaussian to fit Mira A, and the 3 on the right are results using only a uniform disk for Mira A. The error bars in the continuum window and spectral line windows are 10 times and 1 time, respectively, the standard deviation in the mean of the amplitude of visibilities in the respective bin of $uv$-distance. The large error bar near $10\,{\rm M}\lambda$ is due to small number of data points.}
\label{fig:cont-vro-vis}
\end{figure*}

\begin{figure*}[!htbp]
\centering
\includegraphics[trim=2.0cm 0.0cm 3.0cm 0.0cm, clip, width=1.3\miriadmapwidth]{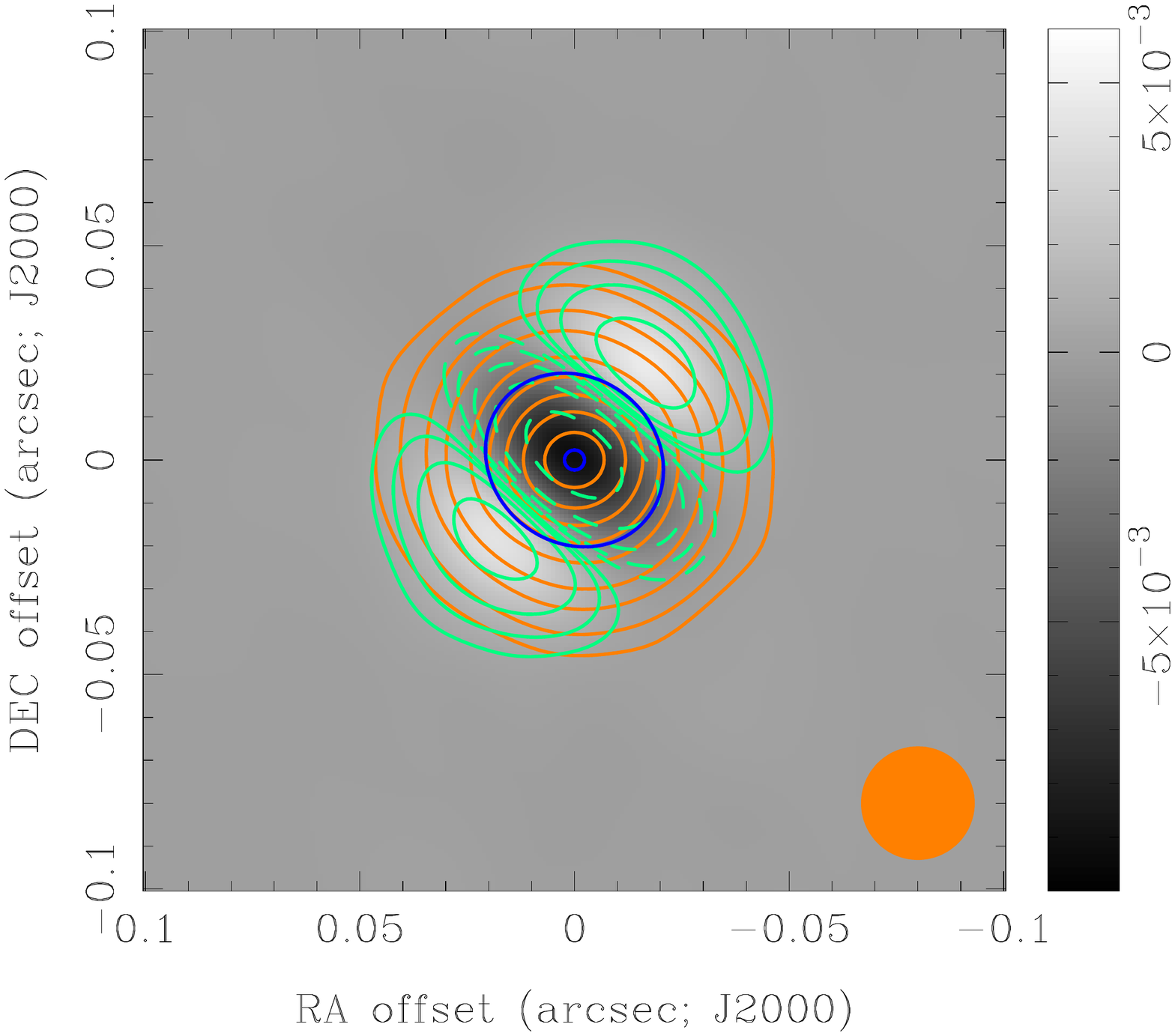}
\put(-115,150){{UD + Gaussian}}
\put(-140,135){\footnotesize{Both days \hspace{0.30\miriadmapwidth} Mira A}}
\includegraphics[trim=2.0cm 0.0cm 3.0cm 0.0cm, clip, width=1.3\miriadmapwidth]{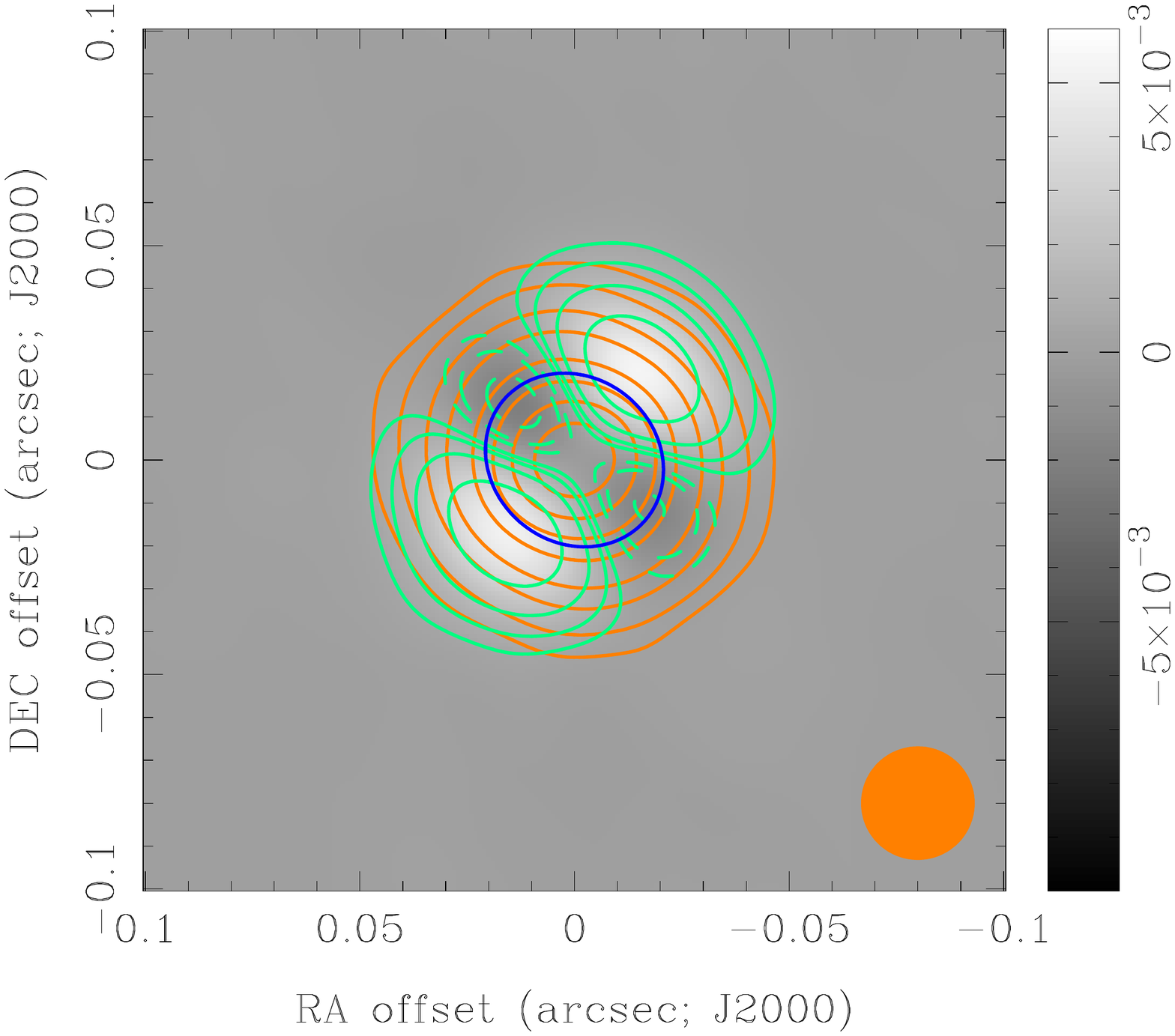}
\put(-105,150){{UD-only}}
\put(-140,135){\footnotesize{Both days \hspace{0.30\miriadmapwidth} Mira A}}\\
\includegraphics[trim=2.0cm 0.0cm 3.0cm 0.0cm, clip, width=1.3\miriadmapwidth]{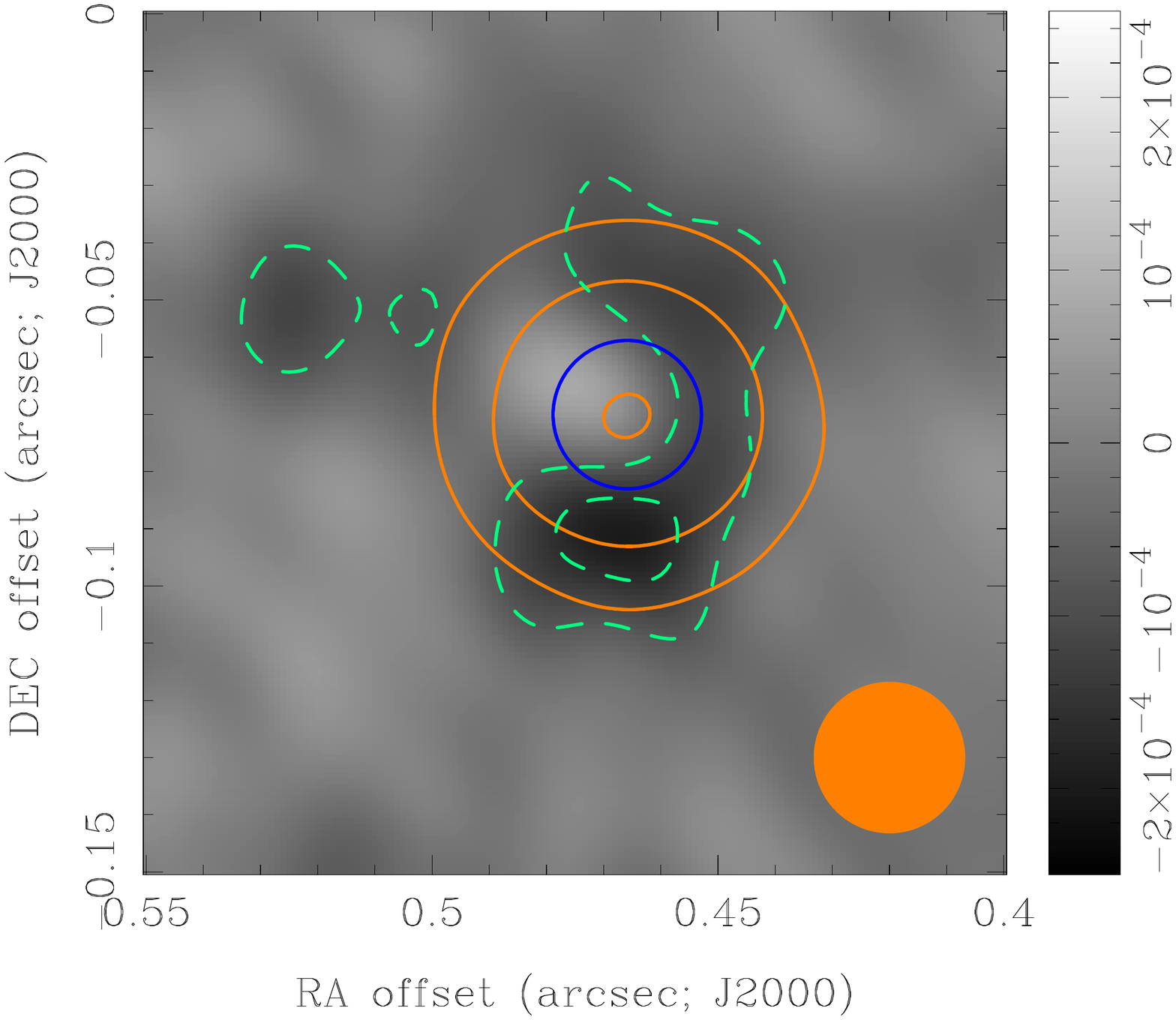}
\put(-115,150){{UD + Gaussian}}
\put(-140,135){\footnotesize{Both days \hspace{0.30\miriadmapwidth} Mira B}}
\includegraphics[trim=2.0cm 0.0cm 3.0cm 0.0cm, clip, width=1.3\miriadmapwidth]{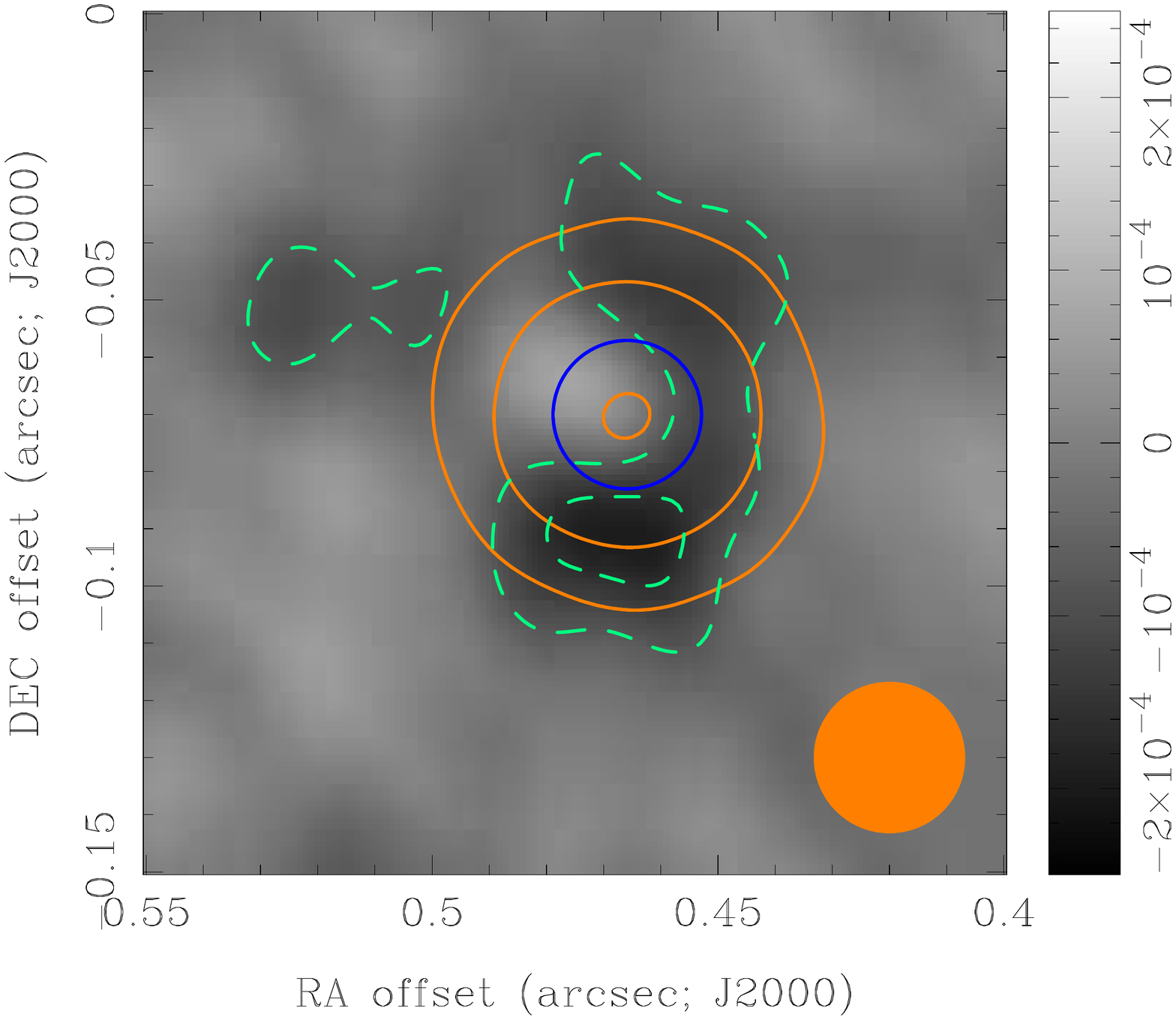}
\put(-105,150){{UD-only}}
\put(-140,135){\footnotesize{Both days \hspace{0.30\miriadmapwidth} Mira B}}
\caption[]{Results of continuum fitting with the task \texttt{uvfit} in Miriad without splitting the data into two observing days. The top and bottom rows show the results for Mira A and Mira B, respectively. The left column shows the results when Mira A was fitted with a uniform disk and a Gaussian (UD + Gaussian), while the right shows the results when Mira was fitted with the same uniform disk only (UD-only). The parameters for Mira A and B are fixed as those from \citet{vro2015}, except the position of the additional Gaussian component for the UD + Gaussian model of Mira A. The sizes of the (larger) uniform disk and (smaller) Gaussian models for the respective spectral windows are drawn in blue open ellipses/circles. In all the panels, the orange contour levels for the continuum model are $3, 10, 30, 60, 120, 180, 240, 300, 360, 420$, and $540 \times 0.20\,{\rm mJy}$. The green contours and the greyscale maps are the residual images produced after the modelled continuum visibilities have been subtracted from the respective windows. The residual contour levels for the images of Mira A (top row) are $-160, -80, -40, -20, -10, 10, 20, 40$, and $80 \times \sigma$, and for Mira B (bottom row) are $-4, -2, 2$, and $4 \times \sigma$, where $\sigma=0.04\,{\rm mJy}$. The restoring beam of $0{\farcs}027$ FWHM is indicated in orange at the bottom right corner in each panel.}
\label{fig:cont-vro}
\end{figure*}

We have tested the $uv$-fitting model of \citet{vro2015} in Miriad/\texttt{uvfit} by setting their model parameters for Mira A and B as the initial inputs. Without fixing any parameters, all the fitting trials with Miriad/\texttt{uvfit} eventually converge to our values as presented in Table \ref{tab:continuum}. The derived major and minor axes of the additional Gaussian component of Mira A are always in the range of ${\sim}20$--$30$\,mas, which are much larger than 4.7\,mas as reported by \citet{vro2015}. On the other hand, even if we fixed the sizes of the uniform disk and Gaussian components to be the same as their values, we still could not obtain the same results as in \citet{vro2015}. Figures \ref{fig:cont-vro-vis} (top two panels) and \ref{fig:cont-vro} show the visibility plots and the maps, respectively, in the 229-GHz continuum window, of the continuum model and residuals of both Mira A and B using the parameters fitted by \citet{vro2015}. In the model where Mira A was fitted with two components, a uniform disk (UD) and a 4.7-mas Gaussian, we obtain huge residuals that appear to be completely different to the residual image presented in Fig. 1 (right) of \citet{vro2015}. Moreover, in the UD-only model for Mira A (using the best-fit disk model by \citeauthor{vro2015}), we do not obtain the single, compact and bright hotspot near the phase centre as shown in Fig. 1 (middle) of \citet{vro2015}. Instead, we find a pair of strong positive residual features along the SE-NW axis and strong negative features along the SW-NE axis, which may be indicative of an inappropriate shape of the subtracted uniform disk component. We also note the UD-only residual image of \citet{vro2015} may show a negative spot in the north-east of the central compact hotspot (their Fig. 1 middle). However, \citet{vro2015} did not show any negative contour lines or negative colour scale in their figures.


\citet{vro2015} fitted the continuum visibilities differently from \citet{mrm2015} and from us in two ways. First, they fitted the continuum dataset of each observing day independently, instead of combing both days together. Second, \citet{vro2015} used a different $uv$-fitting tool, \texttt{UVMULTIFIT}, which has been developed by the Nordic ALMA Regional Center Node\footnote{\url{http://www.nordic-alma.se/support/software-tools}} and can be implemented in CASA \citep{uvmultifit}. 

\begin{figure*}[!htbp]
\centering
\includegraphics[trim=2.0cm 0.0cm 3.0cm 0.0cm, clip, width=\miriadmapwidth]{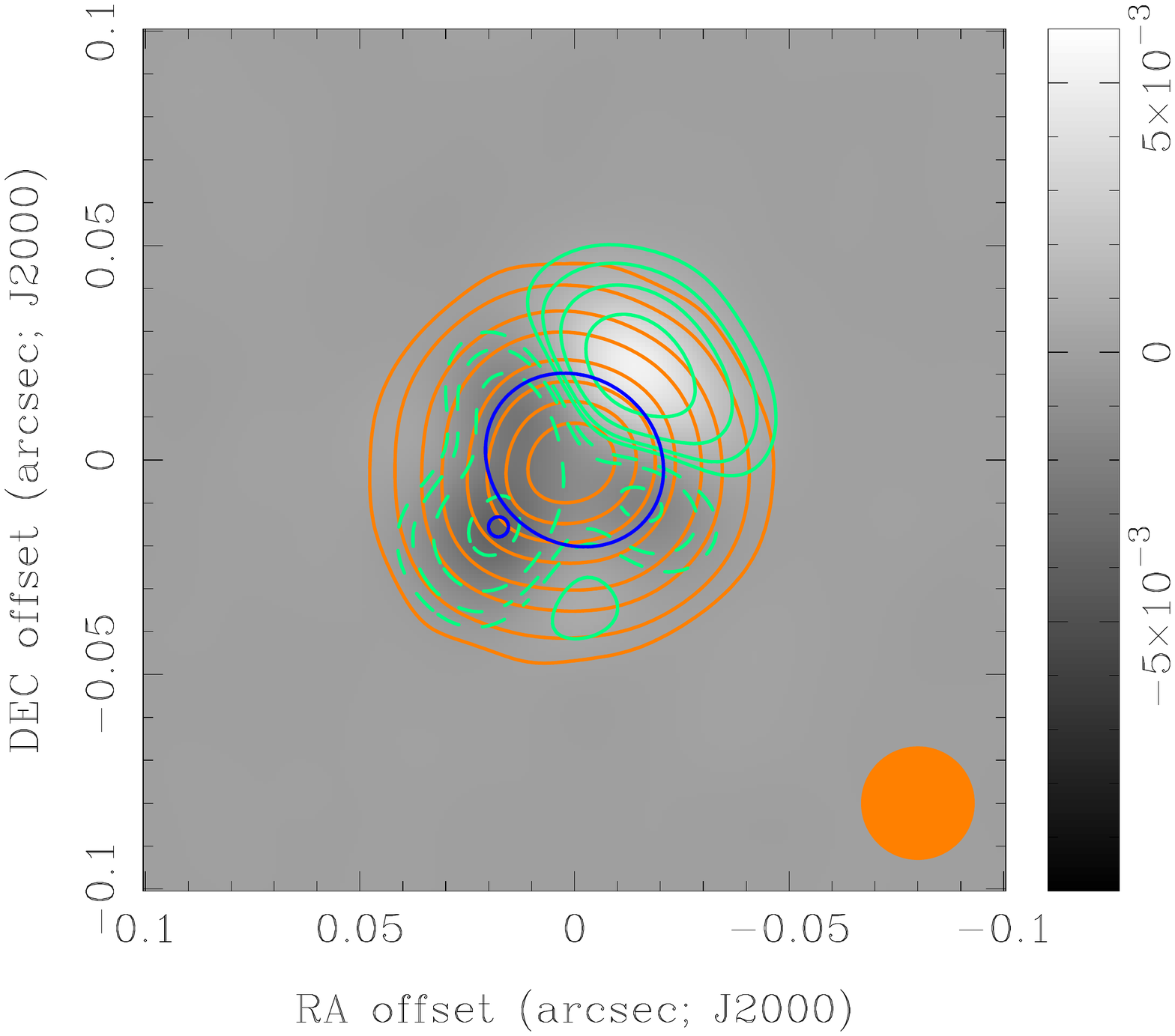}
\put(-95,115){{UD + Gaussian}}
\put(-110,102){\footnotesize{2014 Oct 29 \hspace{0.07\miriadmapwidth} Mira A}}
\includegraphics[trim=2.0cm 0.0cm 3.0cm 0.0cm, clip, width=\miriadmapwidth]{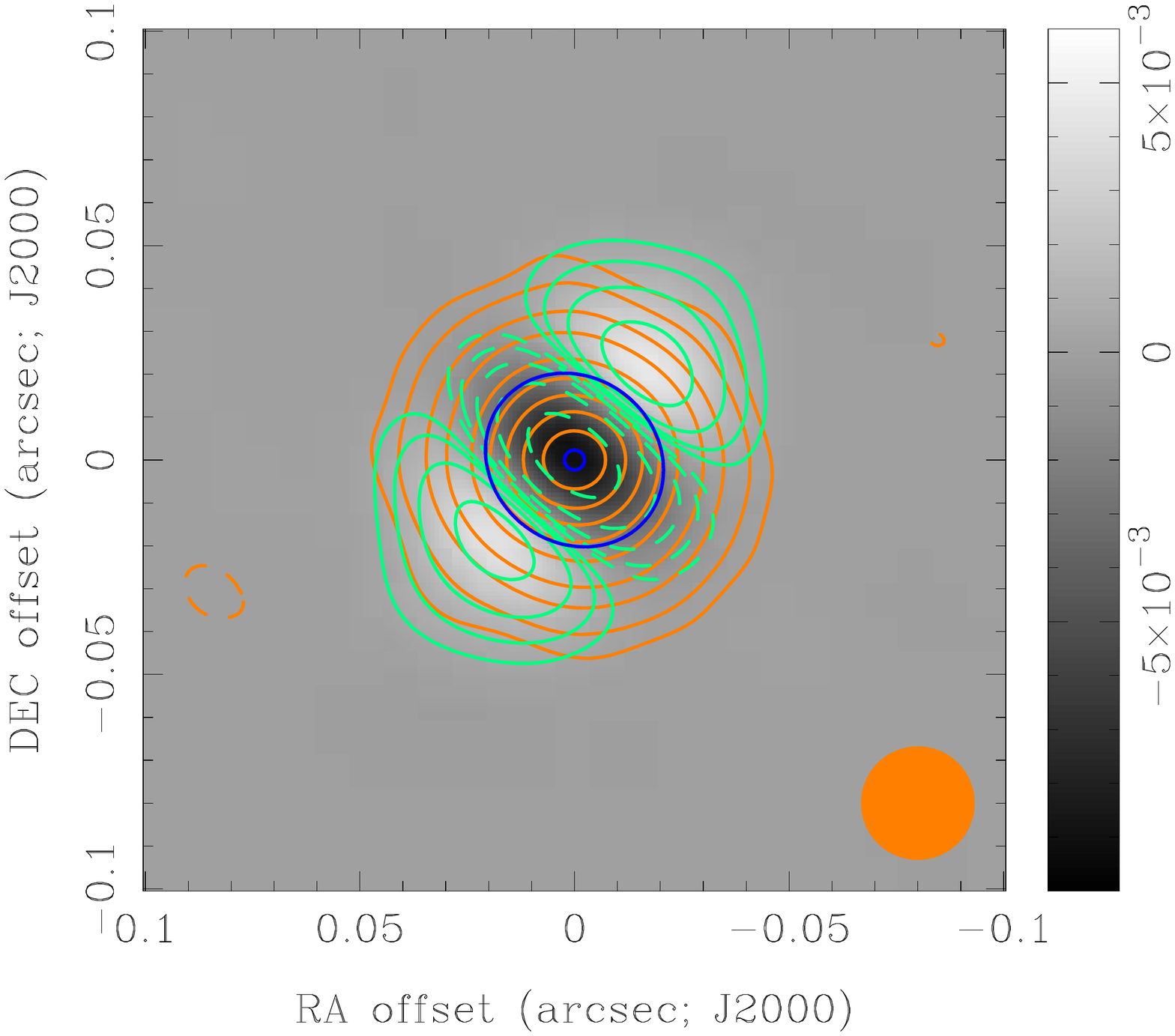}
\put(-95,115){{UD + Gaussian}}
\put(-110,102){\footnotesize{2014 Nov 01 \hspace{0.07\miriadmapwidth} Mira A}}
\includegraphics[trim=2.0cm 0.0cm 3.0cm 0.0cm, clip, width=\miriadmapwidth]{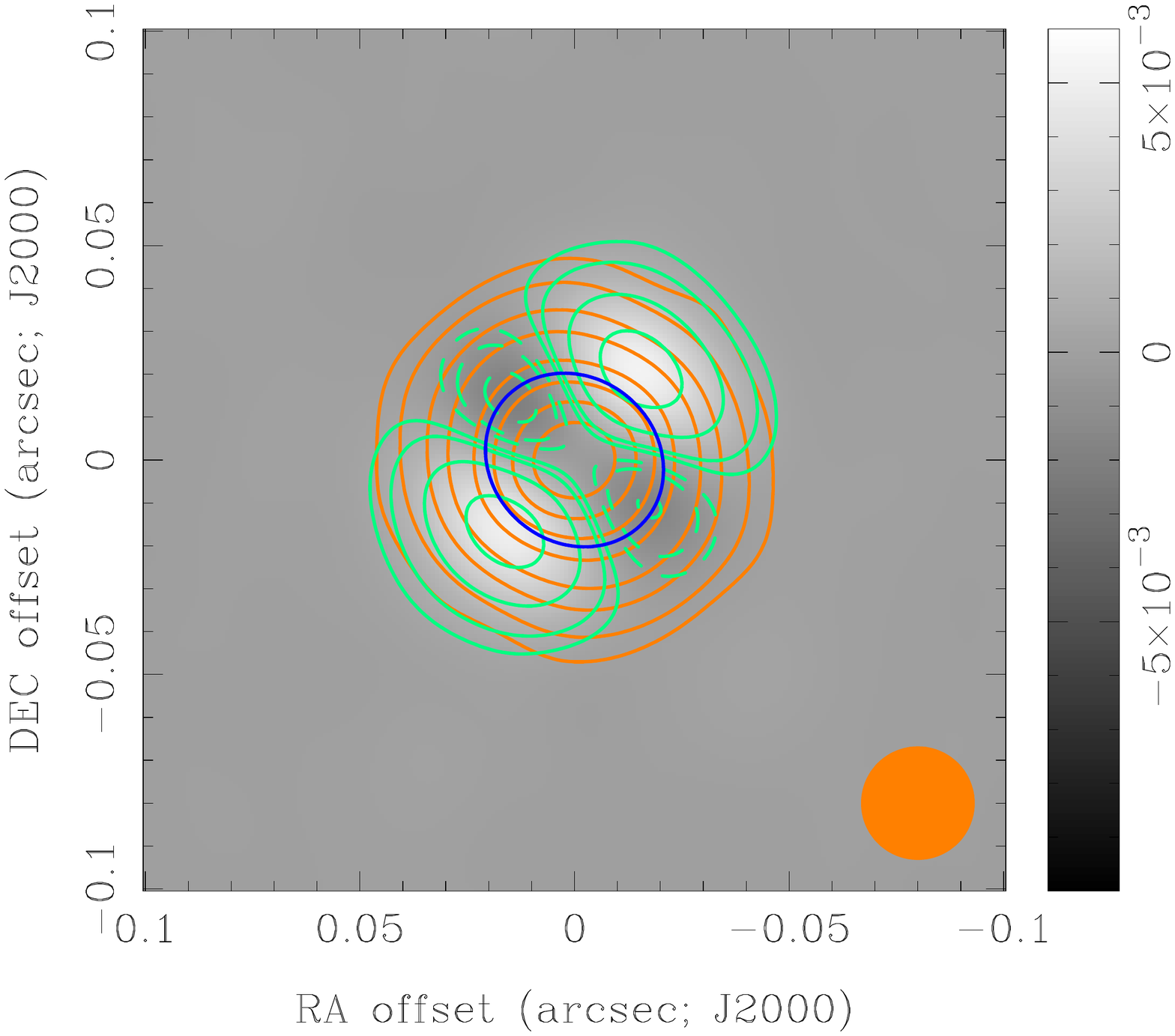}
\put(-85,115){{UD-only}}
\put(-110,102){\footnotesize{2014 Oct 29 \hspace{0.07\miriadmapwidth} Mira A}}
\includegraphics[trim=2.0cm 0.0cm 3.0cm 0.0cm, clip, width=\miriadmapwidth]{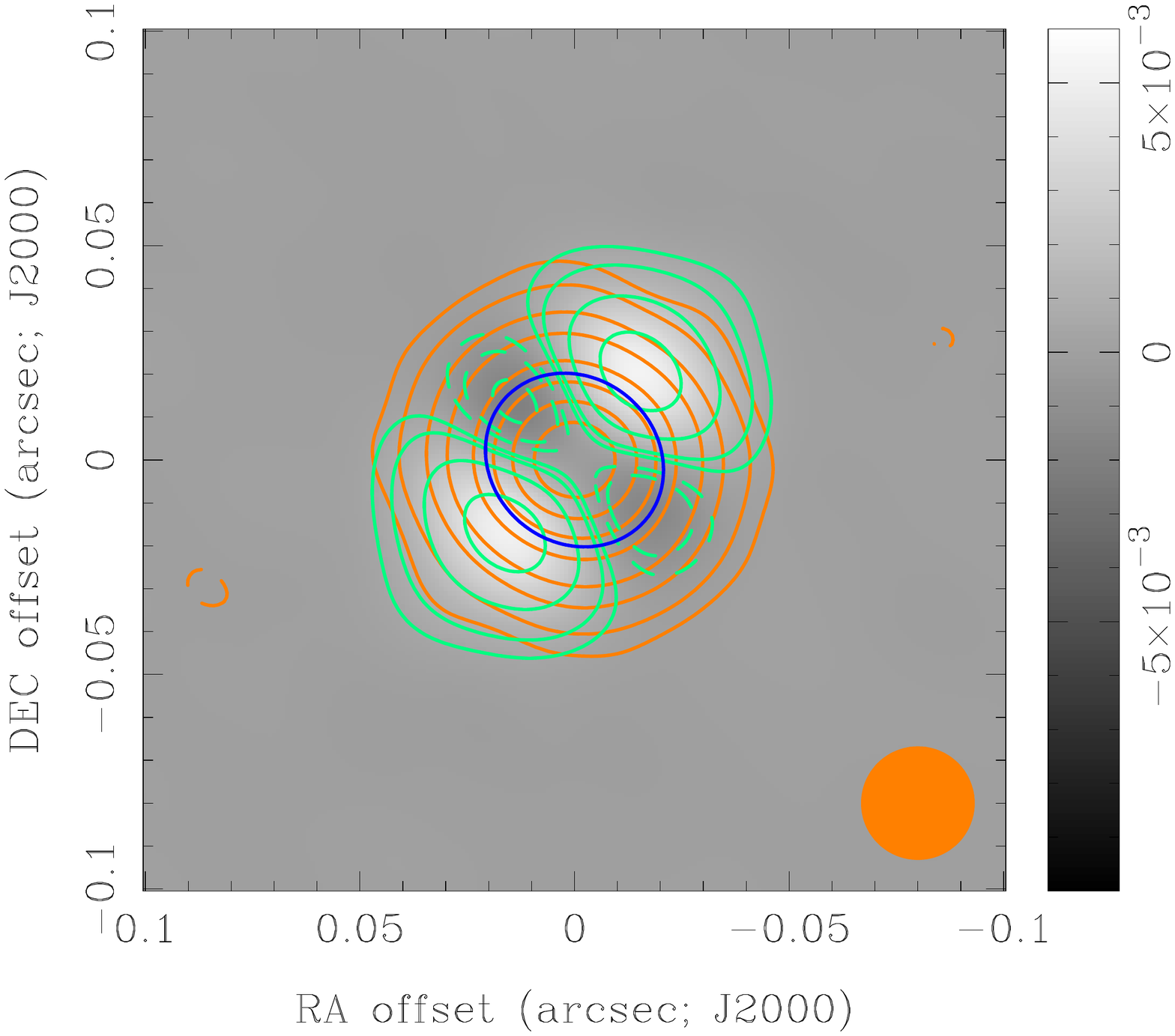}
\put(-85,115){{UD-only}}
\put(-110,102){\footnotesize{2014 Nov 01 \hspace{0.07\miriadmapwidth} Mira A}}\\
\includegraphics[trim=2.0cm 0.0cm 3.0cm 0.0cm, clip, width=\miriadmapwidth]{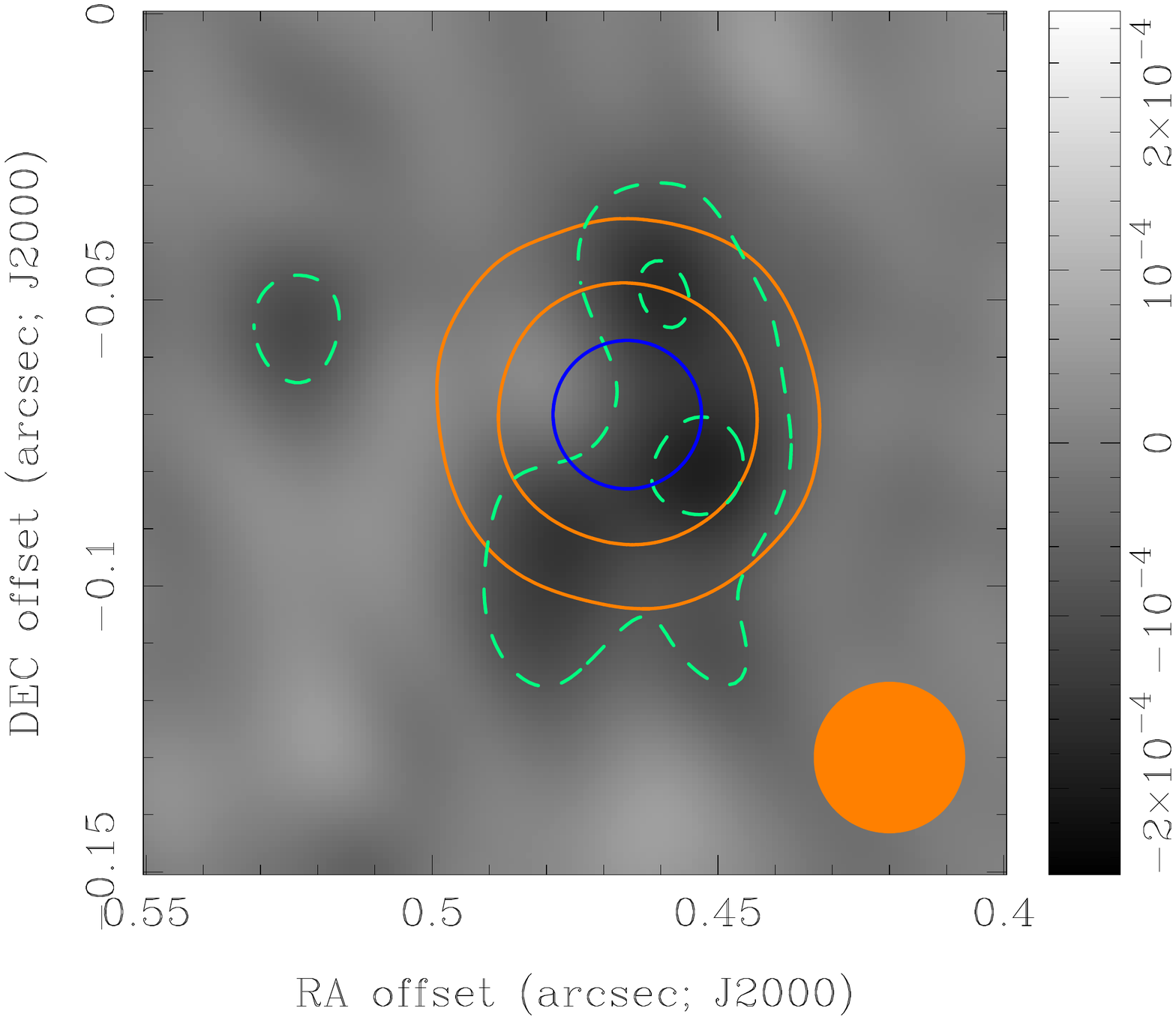}
\put(-95,115){{UD + Gaussian}}
\put(-110,102){\footnotesize{2014 Oct 29 \hspace{0.07\miriadmapwidth} Mira B}}
\includegraphics[trim=2.0cm 0.0cm 3.0cm 0.0cm, clip, width=\miriadmapwidth]{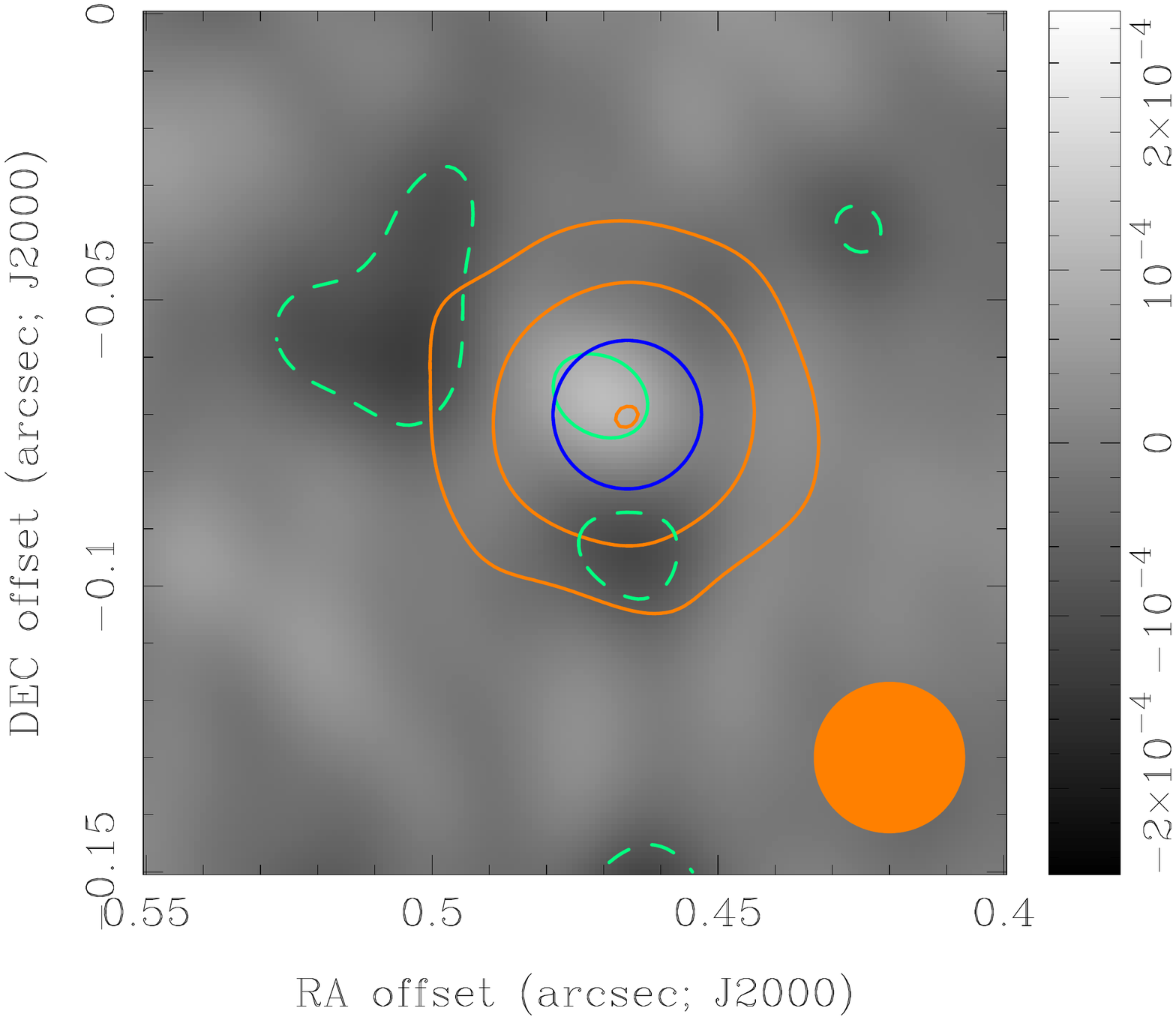}
\put(-95,115){{UD + Gaussian}}
\put(-110,102){\footnotesize{2014 Nov 01 \hspace{0.07\miriadmapwidth} Mira B}}
\includegraphics[trim=2.0cm 0.0cm 3.0cm 0.0cm, clip, width=\miriadmapwidth]{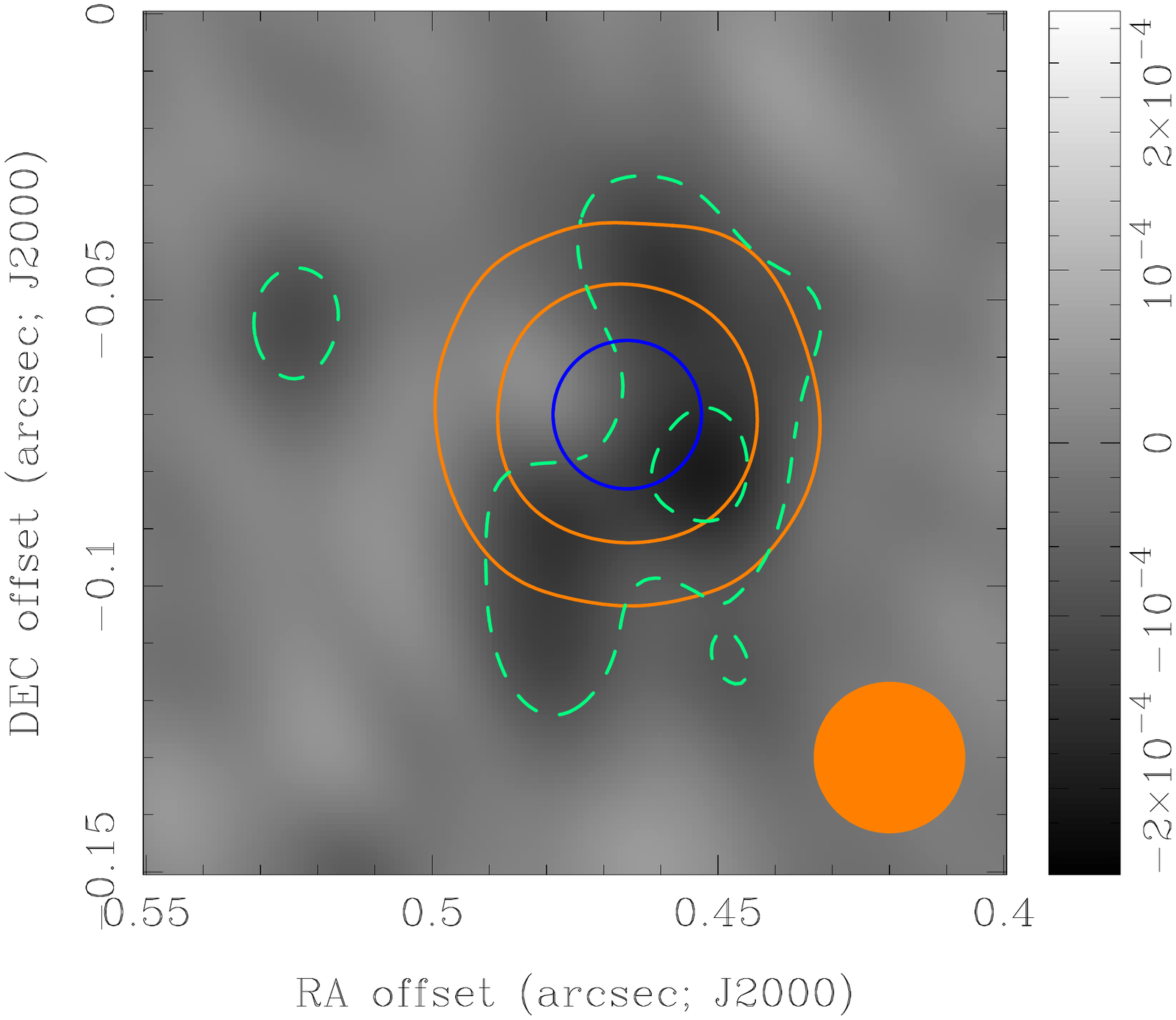}
\put(-85,115){{UD-only}}
\put(-110,102){\footnotesize{2014 Oct 29 \hspace{0.07\miriadmapwidth} Mira B}}
\includegraphics[trim=2.0cm 0.0cm 3.0cm 0.0cm, clip, width=\miriadmapwidth]{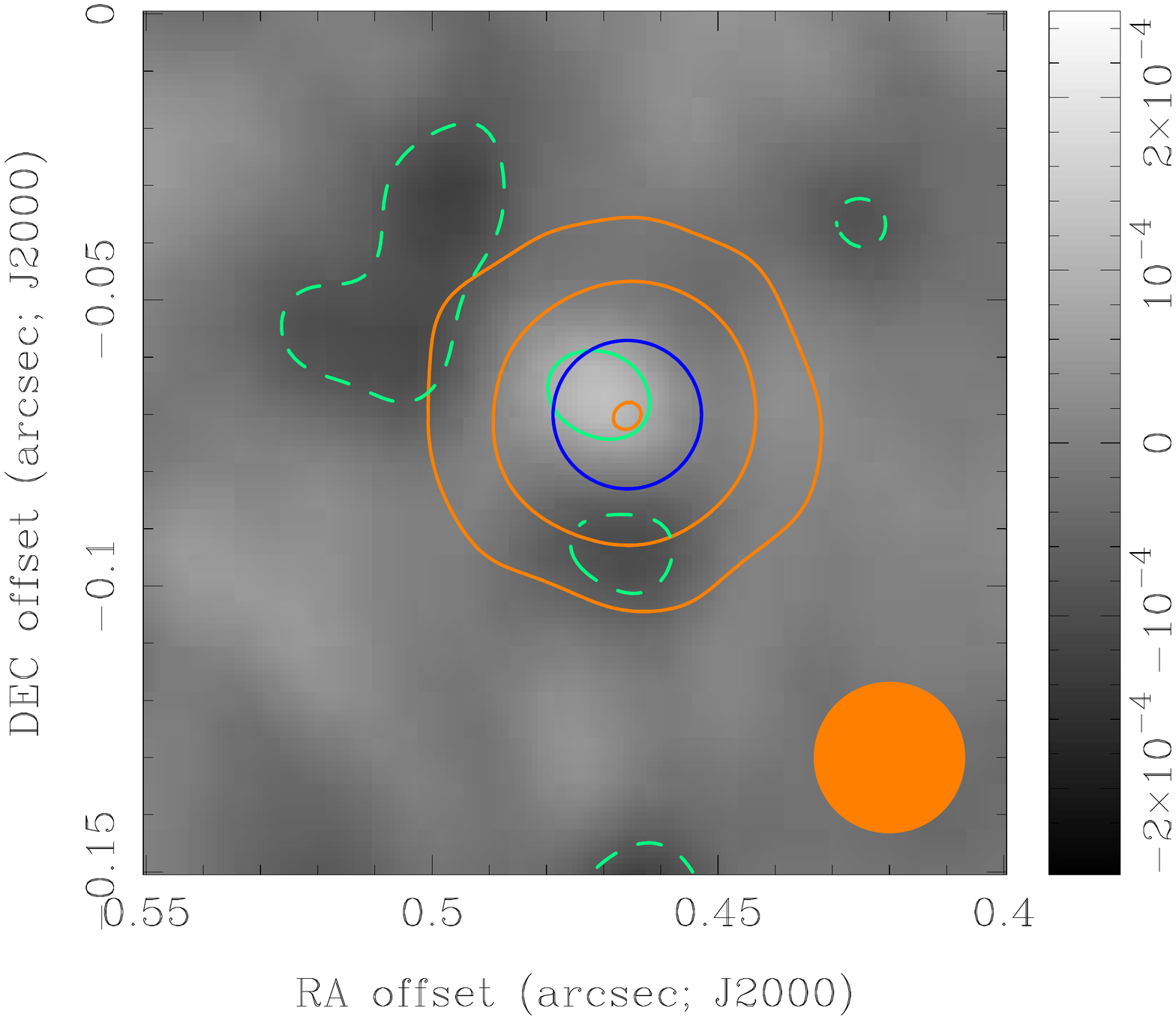}
\put(-85,115){{UD-only}}
\put(-110,102){\footnotesize{2014 Nov 01 \hspace{0.07\miriadmapwidth} Mira B}}
\caption[]{Same as Fig. \ref{fig:cont-vro} for the fitting results on each day of observation.}
\label{fig:cont-vro2d}
\end{figure*}

We have conducted additional tests to see if we may reproduce the fitting results of \citet{vro2015} by following any one, or both aspects of their approach. Figures \ref{fig:cont-vro-vis} (middle two and bottom two panels) and \ref{fig:cont-vro2d} show the visibility plots and maps, respectively, of the fitting results using Miriad/\texttt{uvfit}, by splitting the continuum data into individual days of observation. In the fitting, we fixed the sizes and flux densities of the model components of Mira A and B to be the same as \citet{vro2015}. The positions of the uniform disk (UD) component of Mira A and the Gaussian component of Mira B were also fixed at the respective stellar positions. The only free parameter in the UD + Gaussian model is therefore the position of the additional Gaussian component of Mira A. Although there are some significant differences between the residual maps among the two observing days, we did not obtain any satisfactory fitting with the parameters obtained by \citet{vro2015}, neither did we find any evidence for the bright spot in the UD-only model. In the UD + Gaussian model for the first day of observation (2014 Oct 29), we even found that the best-fit position of the purported hotspot to be outside the uniform disk component of Mira A, which is far-away from the best-fit position as reported by \citet{vro2015}.


Using CASA/\texttt{UVMULTIFIT} (version 2.1.4-r3), we also obtained the similar results as reported above. Assuming the additional Gaussian component is circular, which is the exact same model as \citet{vro2015}, the size and flux density of the elliptical disk component for Mira A are found to be $50.3\,{\rm mas} \times 41.0\,{\rm mas}$ and $S_{229.6\,{\rm GHz}} = 92.9 \pm 0.1 \,{\rm mJy}$, respectively; and those of the additional Gaussian component are $25.4\,{\rm mas} \times 25.4\,{\rm mas}$ and $S_{229.6\,{\rm GHz}} = 56.6 \pm 0.1 \,{\rm mJy}$, respectively.

We have therefore failed to reproduce any evidence of the compact ($<5$\,mas) Gaussian component in Mira A as suggested by \citet{vro2015}, even if we carried out the fitting in the exact same approach with the same set of model components (elliptical uniform disk + circular Gaussian). Self-calibration may be a possible cause for the discrepancy. However, from their Sect. 2, it is unclear whether \citet{vro2015} have derived the self-calibration solutions on their own or obtained the solutions from the ALMA Science Portal (like what we did). Although \citet{vro2015} discovered a compact hotspot from their fitting in Band 6, they could not find similar structure in the continuum data of Band 3 (94.2\,GHz) in this ALMA SV dataset with a rms sensitivity of ${\sim}40\,\mu{\rm Jy}\,{\rm beam}^{-1}$. Assuming the size of 4.7\,mas and Gaussian brightness distribution, \citet{vro2015} estimated in their Sect. 4.1.3 the upper limit of the hotspot brightness temperature in Band 3 to be $T_b \lesssim 17\,500\,{\rm K}$. However,  by repeating the same calculation, we have obtained an upper limit of $T_b \lesssim 250\,{\rm K}$, which is a factor of 70 lower than their estimate. Hence we cast doubt on the existence of any detectable compact hotspot near the centre of Mira A's radio continuum disk near 229\,GHz in this ALMA SV data.


\section{A discussion on continuum subtraction in spectral line imaging}
\label{sec:appendix_contsub}


It is a standard practice to subtract continuum emission from spectral line data (either from the visibility domain or from the image domain) before image deconvolution \citep[e.g.][Sect. 11.9]{vanlangevelde1990,thompsonbook2001}. Otherwise, the image deconvolution algorithm (e.g., CLEAN) has to reconstruct a continuum image for each individual, narrowband channel. Because of the non-linear nature of image deconvolution, small differences (such as the frequency-dependent $uv$-coverage or the noise) among channels may lead to very different convergence of the continuum image \citep[e.g.][Sect. 6]{cornwell1992,rupen1999}. Further discussions on the necessity of continuum subtraction in spectral line imaging can be found in Sect. 1 of \citet{sault1994}, Sect. 16.9 of the Miriad User Guide\footnote{\url{http://www.atnf.csiro.au/computing/software/miriad/}} (version 14 Sep 2015), and Sect. 8.3 of the $\cal AIPS$\footnote{NRAO Astronomical Image Processing System.} ${\cal C}ook{\cal B}ook$\footnote{\url{http://www.aips.nrao.edu/cook.html}} (version 31-DEC-2015).

\subsection{The imaging problem}
\label{sec:appendix_imagingproblem}

\begin{figure*}[!htbp]
\centering
\includegraphics[height=\spectraheight]{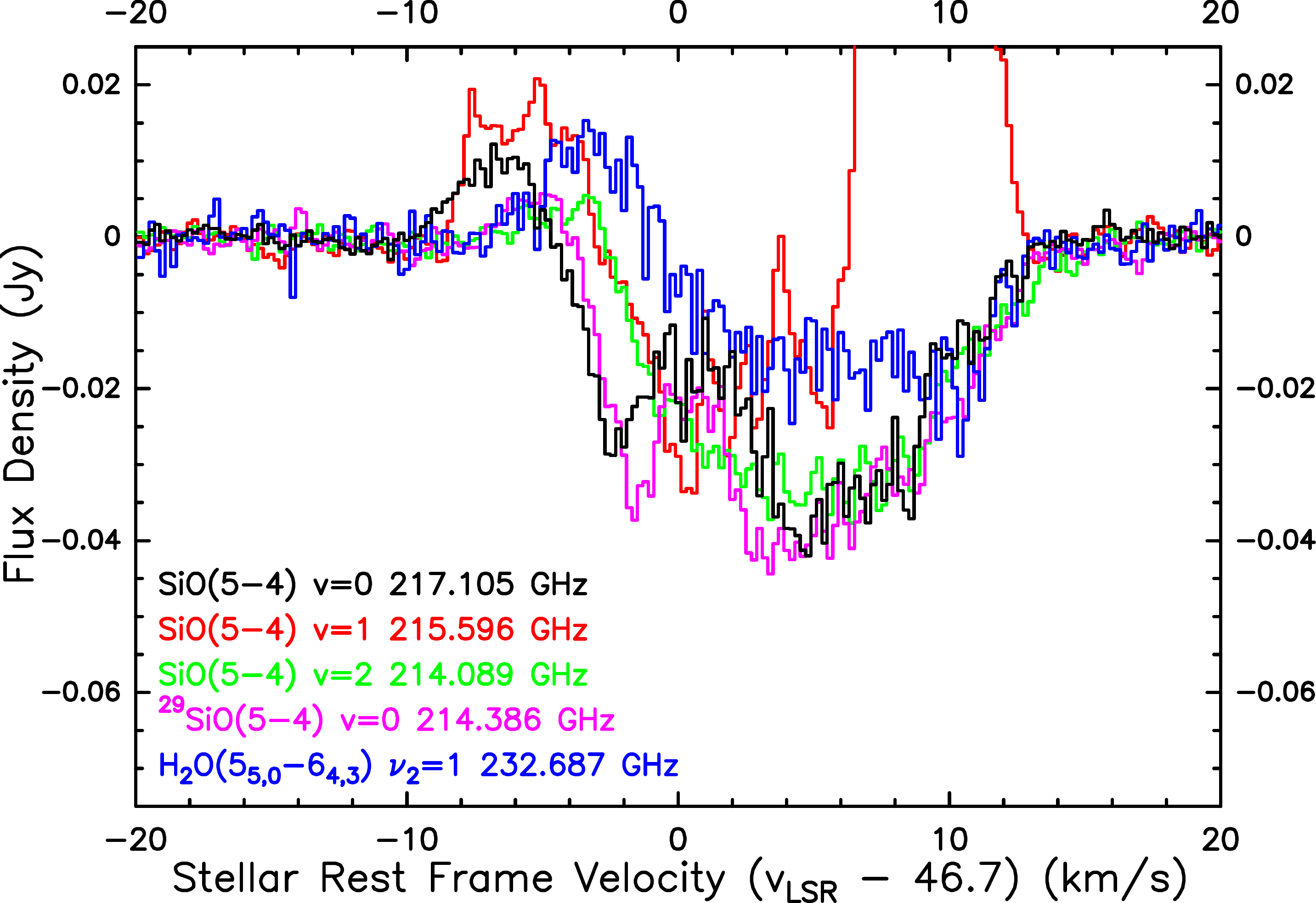}
\caption{Spectral lines in ALMA Band 6 extracted from the line-of-sight towards the centre of Mira's continuum from the continuum-subtracted maps. The channel maps are presented in Figs. \ref{fig:siov0chan_csub}--\ref{fig:h2ov1chan_csub}. The maser emission from SiO ${\varv} = 1$ $J=5-4$ transition (red colour) above $0.025\,{\rm Jy}$ is not shown in this figure.}
\label{fig:band6lines_csub}
\end{figure*}

In our imaging exercises, we find that the images produced from the continuum-subtracted visibility data (hereafter, ``continuum-subtracted images/spectra'') and those from the full (line and continuum) data (hereafter, ``full data images/spectra'') are significantly different. Figure \ref{fig:band6lines_csub} shows the continuum-subtracted spectra of all the observed SiO and H$_2$O lines in ALMA Band 6 extracted from the centre of Mira's radio continuum. Comparing these spectra to the full data spectra shown in Fig. \ref{fig:band6lines}, we expected the only differences being the trivial constant offsets of about $70$--$80\,{\rm mJy/beam}$, which are the peak radio continuum fluxes of Mira in the respective frequency ranges. In reality, however, the depths of absorption near the systemic velocity channels in the continuum-subtracted spectra are significantly shallower than expected, especially for the vibrationally ground state transitions of $^{28}$SiO and $^{29}$SiO. The effects appear as perplexing ``bumps'' near the systemic velocity in the continuum-subtracted spectra. In these velocity channels, bright SiO-emitting clumps and large-scale ($\gtrsim 0{\farcs}1$) SiO emission are present around Mira's radio photosphere. The discrepancy also appears in the nearby pixels which show SiO line absorption against the continuum of Mira's radio photosphere.

In the line-free channels (positive and negative velocity ends of the spectra), the continuum-subtracted spectra have much lower noise than the full data spectra. This is consistent with the fact that CLEAN would introduce additional noise to the image areas where real sources (in our case, the radio continuum) exist (see, for example, Sect. 8.3 of the $\cal AIPS$ ${\cal C}ook{\cal B}ook$).


\subsection{Continuum subtraction in the visibility and the image domains}
\label{sec:appendix_uvim_domain}

We have first compared the imaging results of continuum subtraction in the visibility domain and in the (DIRTY) image domain, using the CASA tasks \texttt{uvcontsub} (then \texttt{clean}) and \texttt{imcontsub} (then \texttt{deconvolve}, which deconvolves the images purely in the image domain), respectively. Both approaches produce the same ``bumps'' in the absorption spectra. Hence, we conclude that the discrepancy between the continuum-subtracted and the full data images arises from the image deconvolution (CLEAN) process, but not continuum subtraction.

The following are the two most probable causes of the discrepancy:
\begin{enumerate}
\item the full data imaging has failed to recover all SiO line emission, possibly extended, around Mira's radio photosphere (i.e., missing emission) or, conversely,
\item the continuum-subtracted imaging has failed to recover the correct amount of SiO line absorption against Mira's radio continuum (i.e., missing absorption).
\end{enumerate}

\subsection{Natural weighting versus robust weighting}
\label{sec:appendix_weighting}

The first possibility depends on the scales of emission structures and the antenna configuration. In the full data images, all real astrophysical signals would have non-negative fluxes because the continuum emission is included in the data. The strongest line absorption can at most attenuate all the radio flux (i.e., down to zero, but not negative) from the continuum behind the column of gas along the line-of-sight. In the continuum-subtracted images, on the other hand, strong negative signals would appear in the pixels of Mira's radio photosphere---representing the line absorption against the continuum---and, strong positive signals would appear around the continuum disk---representing the line-emitting gas outside the radio photosphere. Because of the negative ``hole'' amidst the positive pixels, any real emission or absorption structures in the continuum-subtracted images would appear to be less extended than those in the full data images. Since the visibility plane sampling ($uv$-sampling) of ALMA in the long baseline configuration is rather sparse, full data imaging may not be able to recover all possible extended SiO emission from the inner wind of Mira.


Our imaging (Sect. \ref{sec:obs}) has adopted robust weighting with the parameter $\mathcal{R}_{\rm Briggs}=0.5$. If the ``missing emission'' scenario is true, then the discrepancy between the continuum-subtracted and the full data spectra should be alleviated under natural weighting, which gives relatively higher weightings to data of short baselines and hence is more sensitive to extended structures. Natural weighting can therefore recover more missing flux (if it exists at all) from the extended SiO emission around the radio photosphere of Mira. Since natural weighting gives slightly larger synthesised beam than robust weighting, we only compare the maps of CLEAN component (hereafter, \texttt{CC}) models under the two different weighting schemes to eliminate the effect of the beam sizes. We integrate the total flux of the \texttt{CC} models within circles of various spatial scales, centred at Mira's radio continuum, and obtain the \texttt{CC} spectra of all the images. In contrary to what is expected from the ``missing emission'' scenario, we find that the full data \texttt{CC} spectra (not presented) are essentially identical under both weighting schemes. The fact that a higher weighting in short baselines does not recover more fluxes suggests that the ALMA antenna configurations in the SV observation were not missing any significant amount of extended SiO emission from Mira's inner wind. The absorption profiles of the continuum-subtracted \texttt{CC} spectra, on the other hand, show ``bumps'' that are even more prominent under natural weighting than under robust weighting.


In addition to the above test, we also do not expect any missing flux from extended emission structure for two reasons. First, the shortest baseline of this ALMA SV observation is 15\,m, corresponding to a maximum recoverable scale\footnote{See footnote \ref{footnote:mrs}.} of over $11\arcsec$ for the observed SiO transitions, which is at least one order of magnitude larger than the expected SiO depletion radius. Gas density and kinetic temperature also drop rapidly beyond the inner wind of Mira. So we expect the excitation of SiO molecules (both isotopologues) beyond the radius of the detected emission to be both varying and very weak. It is unlikely that a smooth and large-scale SiO emission exists, and that the missing flux of which could significantly affect the absorption spectra. Second, negative ``bowls'' around smaller-scale emission structures, which are characteristic features if smooth extended emission is missing \citep[e.g.][]{bajaja1979,braun1985}, are not seen in any of our images.

\subsection{Number of CLEAN iterations}
\label{sec:appendix_iterations}

\begin{figure*}[!htbp]
\centering
\includegraphics[height=\spectraheight]{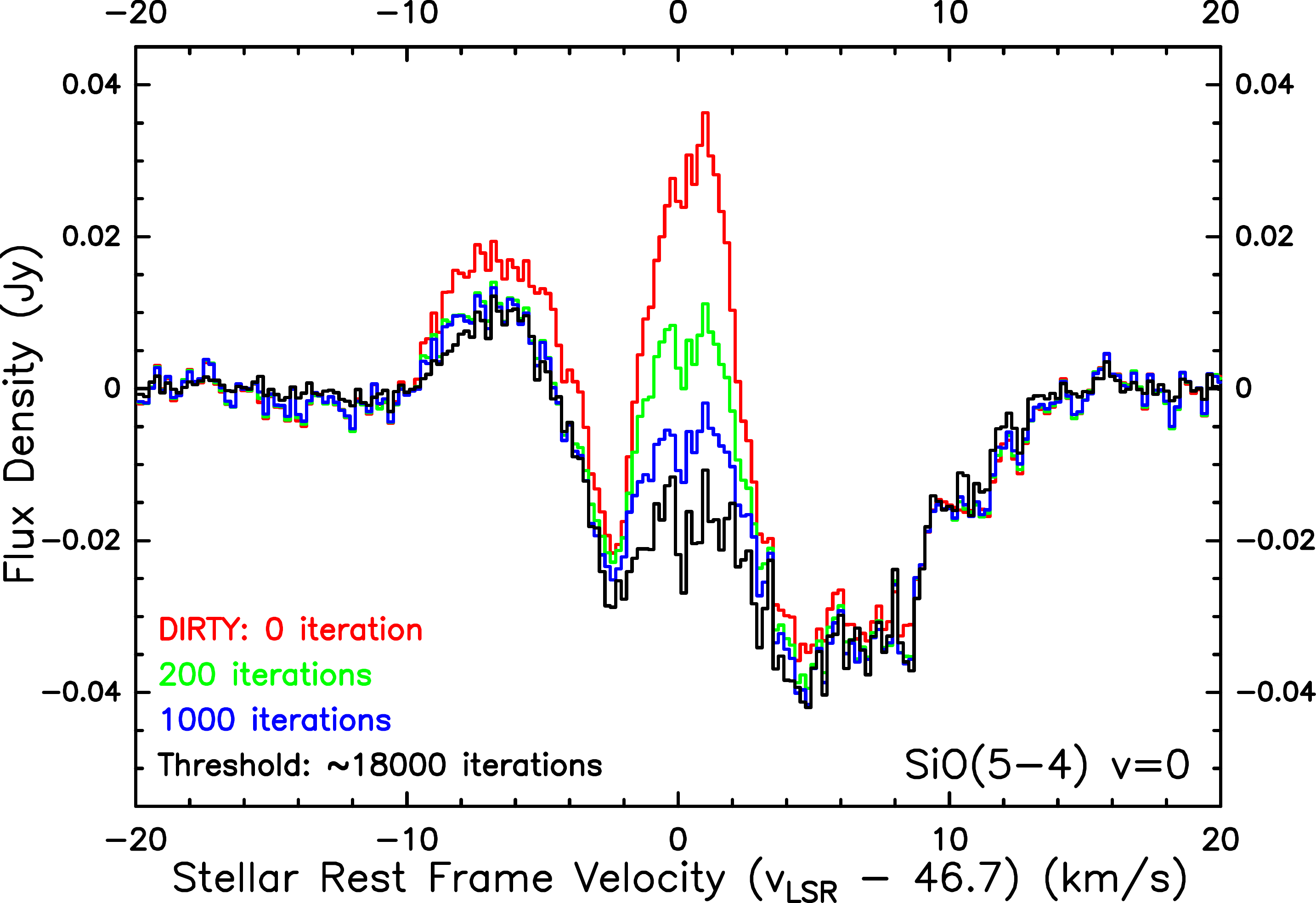}
\caption{Continuum-subtracted absorption profiles of the SiO ${\varv} = 0$ $J=5-4$ transition along the line-of-sight towards the centre of Mira's continuum extracted from images produced by different CLEAN stopping criteria. Images containing the spectra from top to bottom are CLEANed with: 0 iteration (i.e., DIRTY images), 200, 1000, and approximately 18\,000 (fully CLEANed to the threshold, see Sect. \ref{sec:obs}) iterations, respectively.}
\label{fig:band6lines_iter}
\end{figure*}

We therefore consider the second possibility in which the CLEAN process of the continuum-subtracted images may have underestimated the amount of foreground SiO line absorption against the background radio continuum. In order to trace the process of CLEAN iterations and the first batch of \texttt{CC} models selected by the algorithm, we make two additional sets of images, each using a small fixed number of CLEAN iterations instead of an rms noise-dependent threshold as the stopping criteria. Figure \ref{fig:band6lines_iter} shows the spectra extracted from the images of SiO ${\varv} = 0$ $J=5-4$ CLEANed with 200 and 1000 iterations, as well as the spectra from the DIRTY images (no CLEAN iteration) and the fully CLEANed (down to the threshold value as specified in Sect. \ref{sec:obs}) images. In Fig. \ref{fig:band6lines_iter}, the ``bump'' feature is seen in all spectra. Its amplitude reaches the maximum in the spectrum extracted from the DIRTY images, and gradually reduces as the number of iterations increases. In general, the DIRTY images are the most severely ``corrupted'' by the sidelobes of ALMA's point-spread function (i.e., the DIRTY beam). As CLEAN proceeds to more iterations, the effect of the sidelobes should diminish and eventually vanish below the noise when a low-enough threshold is reached. The apparent resemblance of the ``bump'' in the fully CLEANed spectra to that in the DIRTY spectra and the decreasing trend of the ``bump'' amplitude in the continuum-subtracted spectra with an increasing number of CLEAN iterations seem to suggest that the ``bump'', or excess emission in individual channels, may indeed be the artefacts introduced by the sidelobes of the array's point-spread function that have not been completely removed by the CLEAN algorithm.

A closer inspection of the 200- and 1000-iteration \texttt{CC} model maps (not presented) shows that the first 1000 \texttt{CC} models selected by the CLEAN algorithm essentially all come from the strong SiO emission in the immediate proximity of Mira's radio photosphere. Only a handful of (negative) \texttt{CC} models were found within the radio continuum disk of Mira. We therefore speculate that, the initial deconvolution of the strong, extended (scales of the order of $0{\farcs}1$) SiO emission outside the radio continuum may have impaired the subsequent deconvolution of the line absorption within the continuum disk.

\subsection{Long-baseline image deconvolution}
\label{sec:appendix_longbaseline}

\begin{figure*}[!htbp]
\centering
\includegraphics[height=\spectraheight]{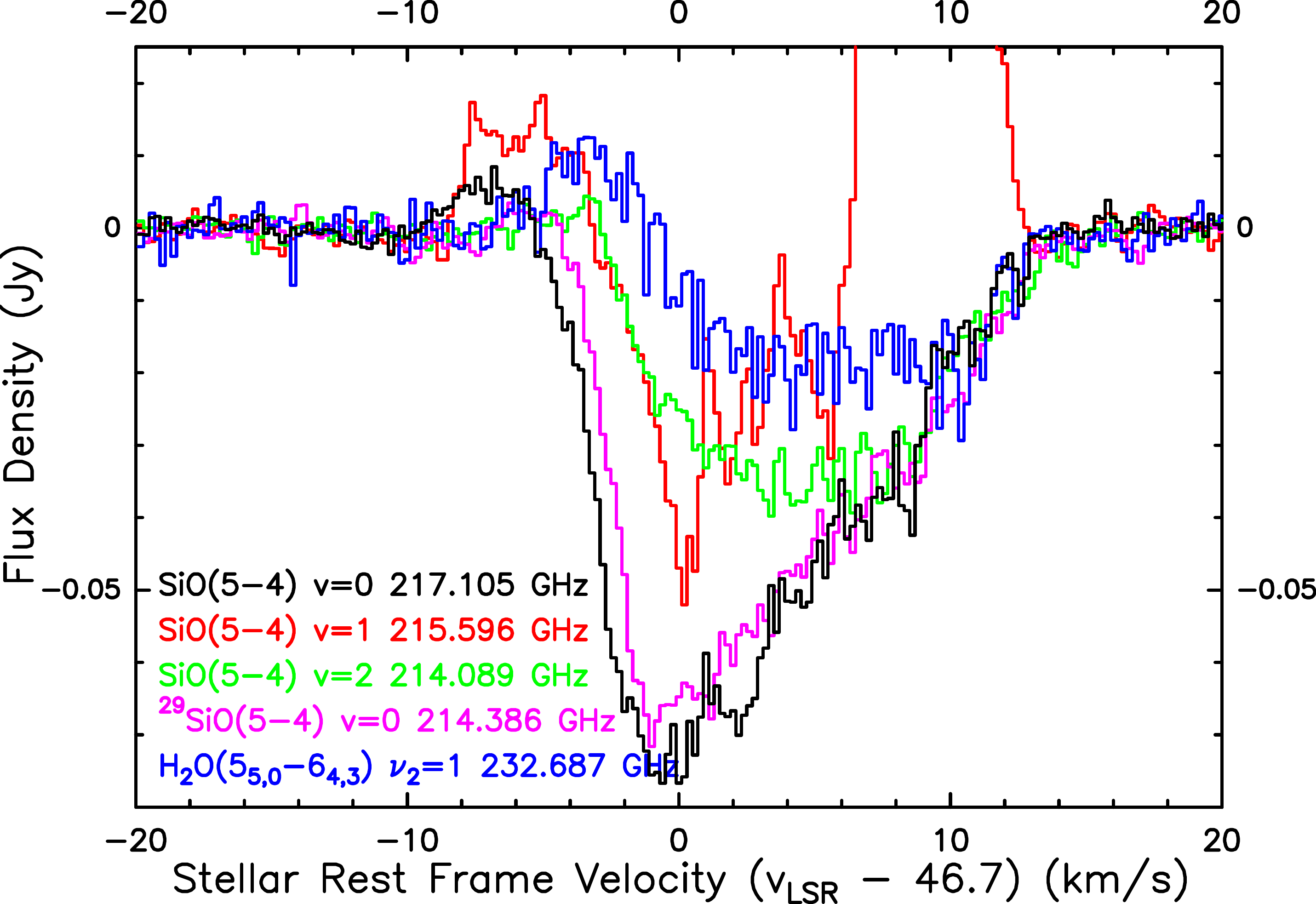}
\caption{Spectral lines in ALMA Band 6 extracted from the line-of-sight towards the centre of Mira’s continuum from the continuum-subtracted maps, imaged with baselines longer than $500\,{\rm m}$ only. The maser emission from SiO ${\varv} = 1$ $J=5-4$ transition (red colour) above $0.025\,{\rm Jy}$ is not shown in this figure.}
\label{fig:band6lines_uv500}
\end{figure*}

To examine the performance of image deconvolution when we exclude the contribution from the extended (${\sim}0{\farcs}1$) SiO emission outside Mira's radio photosphere, another set of continuum-subtracted images are produced for all the observed SiO and H$_2$O lines in Band 6 using only baselines longer than $500\,{\rm m}$. The minimum baseline of $500\,{\rm m}$ corresponds to maximum recoverable scales of about $342$--$347\,{\rm mas}$ for the SiO lines, and about $319\,{\rm mas}$ for the H$_2$O line. In these long-baseline images, the detected flux of the SiO emission around the radio photosphere is significantly reduced or even eliminated. Figure \ref{fig:band6lines_uv500} shows the spectra extracted from the centre of the radio continuum in the long-baseline, continuum-subtracted images. The long-baseline spectra do not show any ``bumps'' in the original continuum-subtracted spectra (Fig. \ref{fig:band6lines_csub}), and resemble the similar inverse P Cygni absorption profiles in the full data spectra (Fig. \ref{fig:band6lines}). However, the maximum depths of absorption in the long-baseline spectra of $^{28}$SiO and $^{29}$SiO ${\varv} = 0$ transitions appear to be even larger, by about $5\,{\rm mJy}$, than the flux level of the radio continuum being absorbed (cf. Fig. \ref{fig:band6lines}). We do not have a definitive answer for this ``over-absorption'' problem. In the long-baseline images, negative ``bowls'' are seen at radial distances larger than the SiO emission region. These are the expected characteristic features due to the lack of short baselines ($<500\,{\rm m}$). The region of the continuum disk amidst the SiO emission is exterior to the ``hollow'' emission region, and therefore may indeed contain the artefactual negative signals due to the lack of short baselines in addition to the real absorption from the foreground SiO gas.

\subsection{CASA simulation of the ALMA SV observation}
\label{sec:appendix_casasimulation}

Finally, we also simulate the ALMA SV observation of Mira to test if our preferred model (Sect. \ref{sec:model_preferred}), which apparently produces satisfactory fits to the observed full data spectra, can also reproduce the observed ``bumps'' in the continuum-subtracted spectra. Positive results would imply that our preferred model can consistently reproduce both the full data and the continuum-subtracted spectra, regardless of the (still uncertain) causes of the discrepancy in the spectra; negative results, on the other hand, would suggest that our model is still far from approximately describing the structures and physical conditions of Mira's extended atmosphere and inner wind.

In the CASA simulation task \texttt{simobserve}, we use the \textsc{sky} output model for the $^{28}$SiO ${\varv} = 0$ $J=5-4$ transition from {\ratran} modelling (Sect. \ref{sec:modelling}) as the input \texttt{skymodel}. For simplicity, we regard the two days of Mira's SV observation as one continuous observing block in the simulation, and adopt approximate values for the observation parameters. The hour angle at the mid-point of the simulated observation (\texttt{hourangle}) and the total on-source time (\texttt{totaltime}) are set to be $-0.8\,{\rm h}$ and $4000\,{\rm s}$, respectively. We also use the actual ALMA antenna configuration in Mira's SV observation as the input \texttt{antennalist}. Table \ref{tab:antenna_pads} lists all 39 ALMA antenna pads (stations) deployed in the SV observation. Note, however, that only 35 or 36 antennae out of 39 were used in each day of observation \citep{lbc2014}. Moreover, we did not consider data flagging during the calibration process. These would slightly affect the array's point-spread function and the map rms noise, but we do not expect a significant deviation from the reality. The longest and shortest baselines also resemble those of the real visibility data. \texttt{simobserve} produces a simulated visibility data, containing both the continuum and line data (as in our radiative transfer model). We then perform continuum subtraction (with the task \texttt{uvcontsub}) and imaging (with \texttt{clean}) in the exact same ways as the real data, which are described in Sect. \ref{sec:obs}.

\begin{table}[!htbp]
\caption{ALMA antenna pads deployed in Mira's SV observation.}
\label{tab:antenna_pads}
\centering
\begin{tabular}{cccc|ccc}
\hline\hline
\multicolumn{4}{c}{Nominal 2014 LBC array\tablefootmark{a}} & \multicolumn{3}{c}{Additional antennae\tablefootmark{b}} \\
\hline
A113 & A131 & W207 & P401 & A001 & A029 & A078 \\
A118 & A132 & W210 & P402 & A004 & A035 & A082 \\
A121 & A133 & S301 & P404 & A005 & A058 & T701 \\
A122 & W201 & S303 & P405 & A006 & A072 & T703 \\
A124 & W204 & S306 & P410 & A011 & A075 &      \\
A127 & W206 & S309 &      & A024 & A076 &      \\
\hline
\end{tabular}
\tablefoot{
\tablefoottext{a}{23 antennae in the nominal 2014 ALMA Long Baseline Campaign (LBC) configuration with a minimum baseline of about 400\,m \citep[cf. Fig. 1 of][]{lbc2014}.}
\tablefoottext{b}{16 additional antennae of short spacings that were available during either day, or both days, of the Science Verification observation of Mira in ALMA Band 6.}}
\end{table}

\begin{figure*}[!htbp]
\centering
\includegraphics[height=\spectraheight]{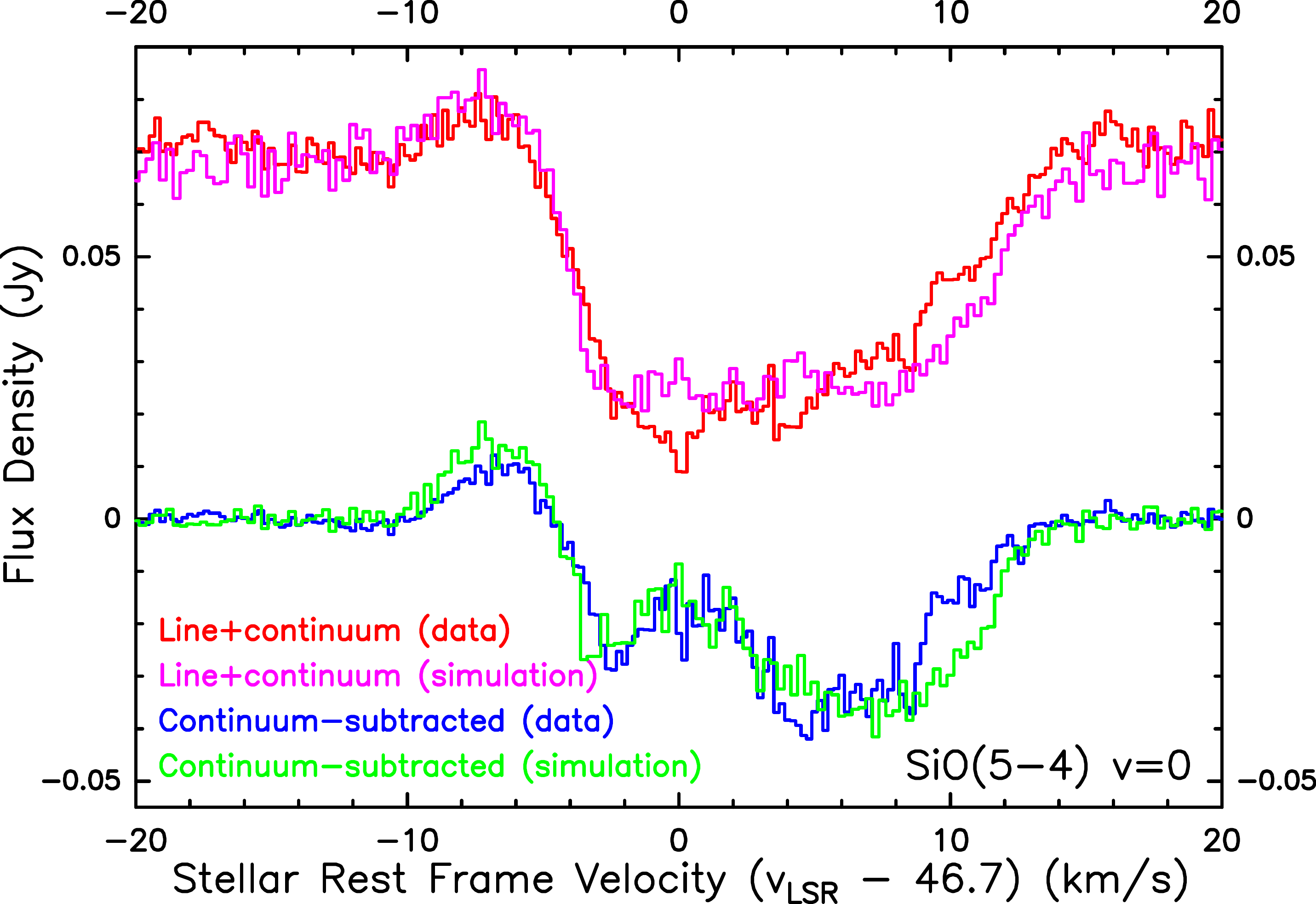}
\caption{$^{28}$SiO ${\varv} = 0$ $J=5-4$ spectra extracted from the line-of-sight towards the centre of Mira's continuum from the full data (line+continuum) images (red and magenta at the top) and from the continuum-subtracted images (blue and green at the bottom). The red and blue spectra correspond to the images from the actual data of Mira's ALMA SV observation, and the magenta and green spectra correspond to the simulated data using CASA task \texttt{simobserve}. The imaging procedures for the observed and simulated data are identical.}
\label{fig:band6lines_sim}
\end{figure*}

Figure \ref{fig:band6lines_sim} shows the resultant spectra of the simulation, overlaid with the real spectra as already presented in Figs. \ref{fig:band6lines} (full data) and \ref{fig:band6lines_csub} (continuum-subtracted). Our simulation successfully reproduces the same ``bump'', with the same amplitude (within the noise) and in the same velocity range, in the absorption profile of the continuum-subtracted spectrum. Because this ``bump'' is not present in (1) the full data spectrum, (2) the spectrum extracted from modelled images convolved with a circular beam (i.e., the modelled spectra in the top-left panel of Fig. \ref{fig:m3siov0spec}), or (3) our input sky brightness model, we conclude that the ``bump'' in the continuum-subtracted spectrum is indeed a spurious feature introduced during the imaging (CLEAN) process. We also suggest that the ``bump'' is caused by the ``missing absorption'' scenario as described in Appendix \ref{sec:appendix_uvim_domain}. Based on the analyses in the previous sections (Appendices \ref{sec:appendix_weighting}--\ref{sec:appendix_longbaseline}), we speculate that the image deconvolution of the strong SiO emission surrounding the continuum disk, which corresponds to the first batch of the CLEAN component models, has probably impaired the deconvolution of the line absorption region in the images. However, it is still unclear how this has happened despite the fact that CLEAN should be able to handle positive and negative signals equally well. In view of the possible limitations of the CLEAN algorithm, it may be worthwhile to perform imaging tests for this SV data with novel aperture synthesis techniques in radio astronomy \citep[e.g.][]{sutter2014,carrillo2014,lochner2015,junklewitz2016}.

There are a few other important features in the simulated images and spectra. The noise in the line-free channels of the simulated full data spectrum is comparable to the observed one, and both are significantly higher than the observed/simulated continuum-subtracted spectra. This further supports our earlier claim in Appendix \ref{sec:appendix_imagingproblem} that CLEANing the radio continuum in individual channels could introduce additional noise to the image products.

We also notice that the simulated spectra do not fit well to the observed ones near the velocity of $+10\,\kms$. We attribute this to the imperfection of our physical model rather than issues in the data processing (see Sect. \ref{sec:model_preferred}).

\begin{figure*}[!htbp]
\centering
\includegraphics[trim=0.0cm 0.2cm 0.0cm 0.2cm, clip, width=\singlemapwidth]{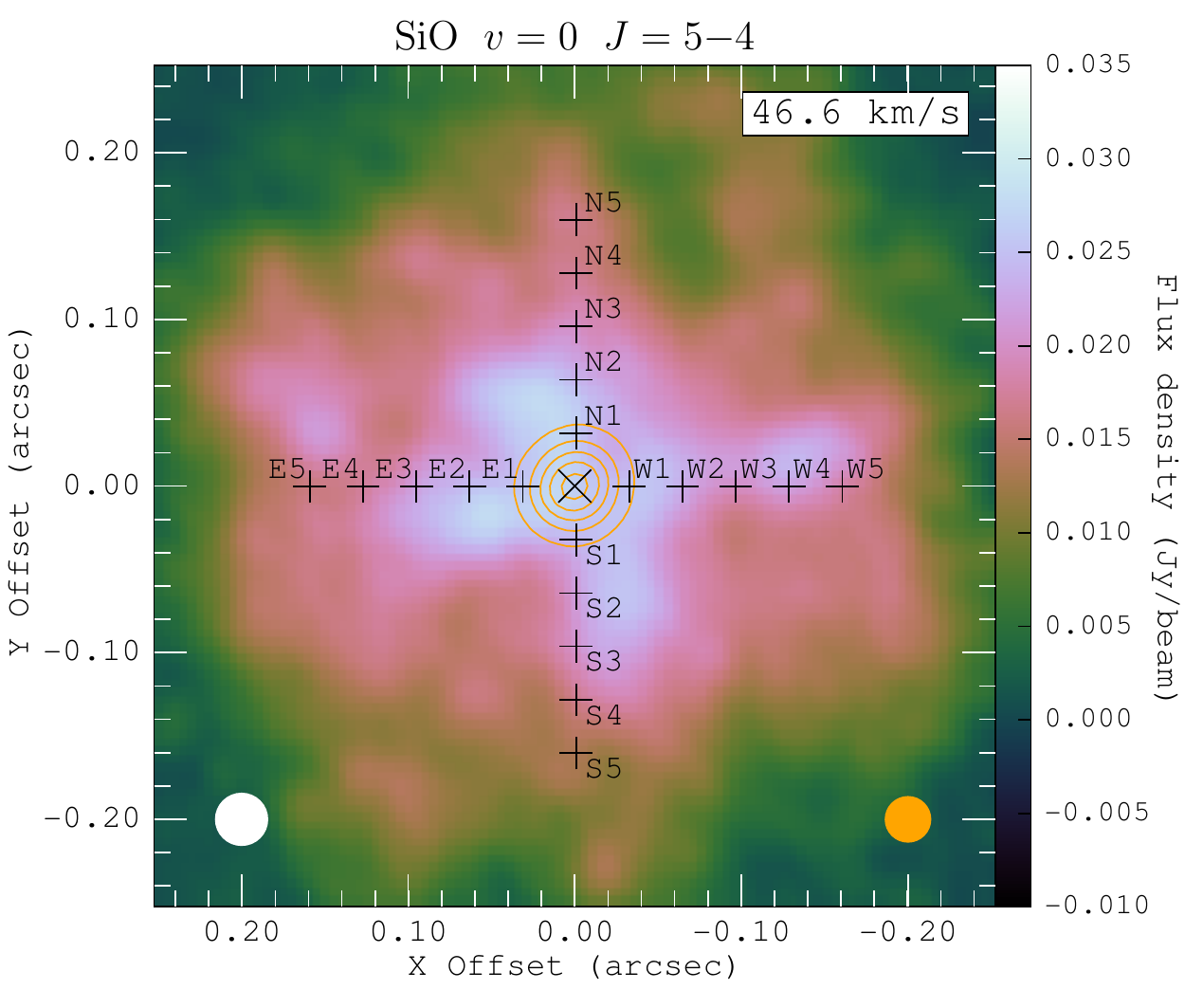}
\caption[]{Same as Fig. \ref{fig:array} for the \emph{simulated} map of SiO ${\varv} = 0$ $J=5-4$ (with the continuum) at the channel of velocity $46.6 \kms$ with a channel width of $1\,\kms$. The velocity is slightly different from the systemic velocity of $46.7 \kms$ because of the different velocity grid in our model. The orange contours represent the real data of Mira's radio continuum at 229\,GHz.}
\label{fig:array_sim}
\end{figure*}

\begin{figure*}[!htbp]
\centering
\includegraphics[trim=0.0cm 1.1cm 0.0cm 0.9cm, clip, width=\channelmapwidth]{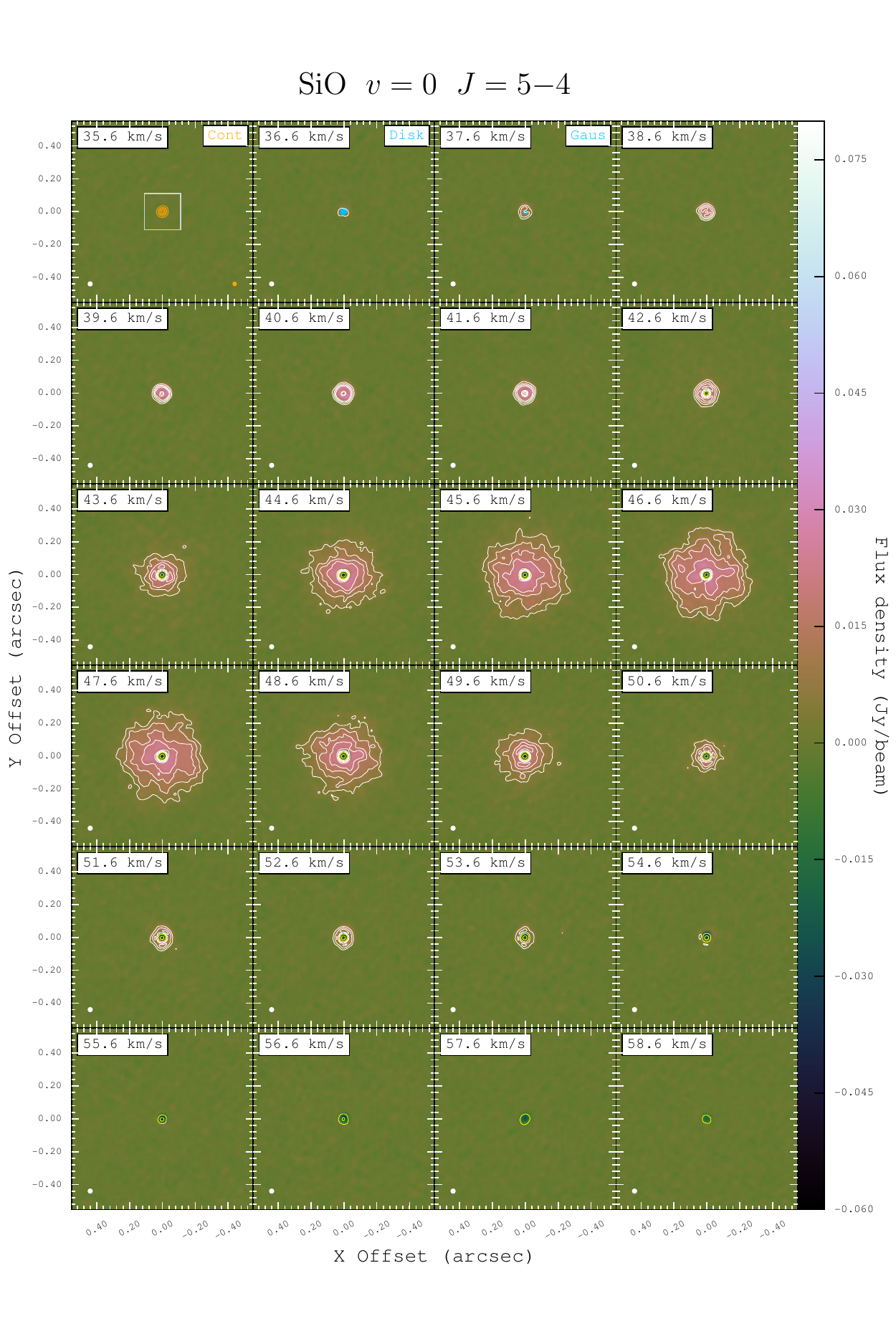}
\caption[]{Same as Fig. \ref{fig:siov0chan_csub} for the \emph{simulated} channel maps of post-imaging continuum-subtracted SiO ${\varv} = 0$ $J=5-4$. 
The white contours represent 6, 12, 18, 24, 48, and $72\sigma$ and yellow contours represent $-72$, $-60$, $-48$, $-36$, $-24$, $-12$, and $-6\sigma$, where $\sigma = 0.80\,{\rm mJy}\,{\rm beam}^{-1}$ is the map rms noise in the real data. The LSR velocities are slightly different from those in Fig. \ref{fig:siov0chan_csub} because of the different velocity grid in our model.}
\label{fig:siov0chan_csub_sim}
\end{figure*}

Our input sky brightness model, which only models the region around Mira up to $0{\farcs}3$ and ignores the remote, arc-like emission feature, is spherically symmetric. The modelled intensity varies smoothly with radius. So one would expect that the simulated images to behave in a similar manner. The simulated images, however, do contain a lot of clumpy structures that are very similar to what we see in Figs. \ref{fig:siov0chan_csub}, \ref{fig:siov0chanzoomed_csub}, and \ref{fig:array}. Figure \ref{fig:array_sim} shows the simulated full data image of the SiO ${\varv} = 0$ line near the systemic velocity, i.e. the channel in which the scale of the SiO emission is the largest. Similar to the same map from the actual observation (Fig. \ref{fig:array}), the simulated map also shows structures of enhanced brightness all over the region being modelled. Figure \ref{fig:siov0chan_csub_sim} shows the simulated channel maps of the images, with the continuum subtracted \emph{after} imaging. These maps also resemble the real data counterpart as shown in Fig. \ref{fig:siov0chan_csub}: the core SiO line-emitting region appears to be globally spherically symmetric, but it also contains some irregular structures of the similar spatial scales. Qualitatively speaking, the simulated maps are more symmetric and smoother than the real observed maps. The presence of clumpy structures in the simulated images may be due to the sparse $uv$-sampling of the ALMA long baseline configuration, which makes image deconvolution difficult. In addition, the signal-to-noise ratio may not be sufficient to clearly detect the SiO emission in Mira's extended atmosphere. Hence, the simulated ALMA observation suggests that the clumpy structures in the extended atmosphere and inner wind of Mira may not be all real.

\section{Full data channel maps (without continuum subtraction)}
\label{sec:appendix_maps}

\begin{figure*}[!htbp]
\centering
\includegraphics[trim=0.0cm 1.1cm 0.0cm 0.9cm, clip, width=\channelmapwidth]{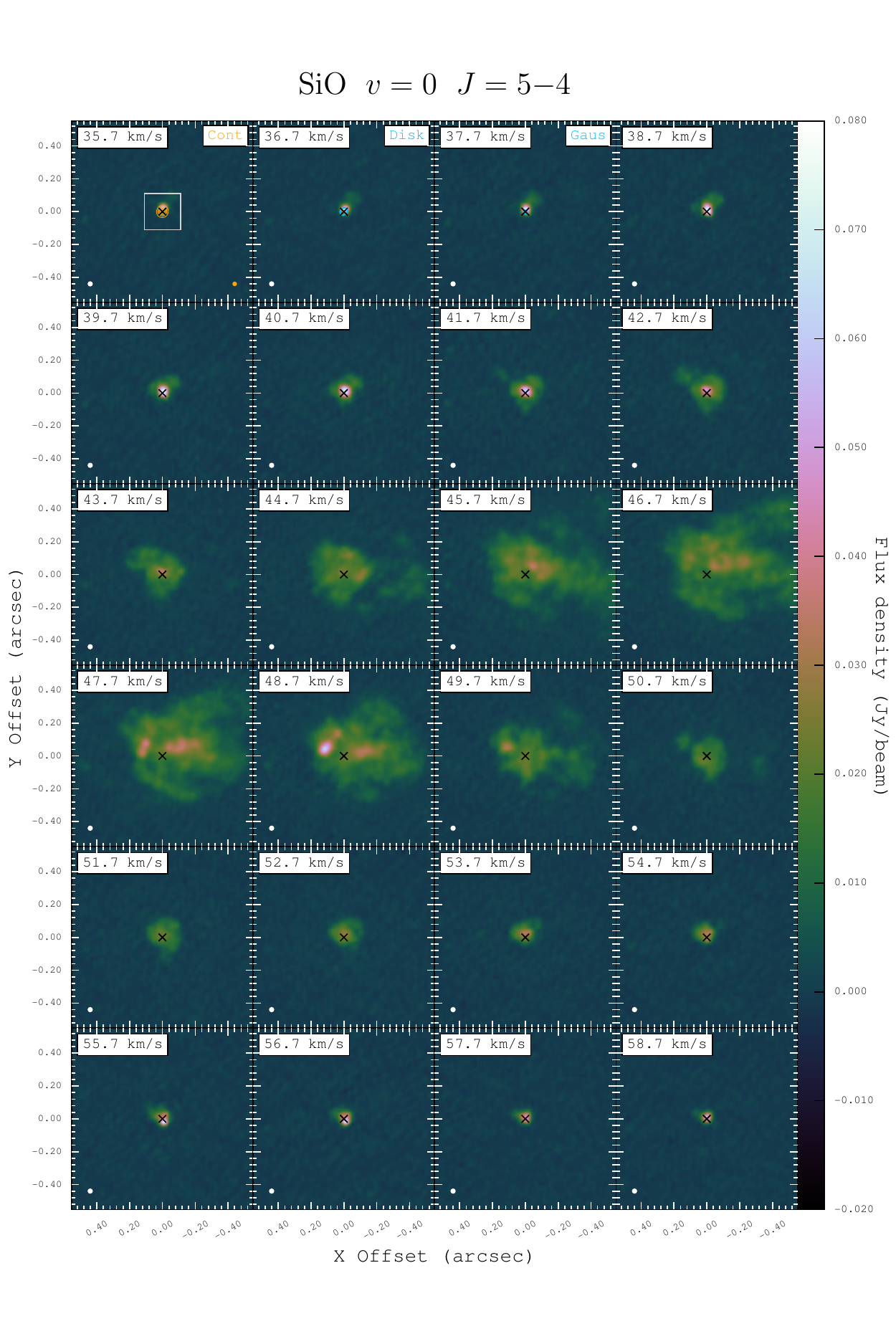}
\caption[]{Same as Fig. \ref{fig:siov0chan_csub} for the channel maps, including continuum emission from Mira A, of SiO ${\varv} = 0$ $J=5-4$. 
The absolute position of Mira A is denoted by a black cross. 
The map rms noise is $0.80\,{\rm mJy}\,{\rm beam}^{-1}$. 
In the first panel of the top row, the white box centred at Mira A indicates the $0{\farcs}22 \times 0{\farcs}22$ region of the zoomed maps of SiO ${\varv} = 0$ (Fig. \ref{fig:siov0chanzoomed}), ${\varv} = 2$ (Fig. \ref{fig:siov2chan}) and H$_2$O $v_2=1$ (Fig. \ref{fig:h2ov1chan}).}
\label{fig:siov0chan}
\end{figure*}

\begin{figure*}[!htbp]
\centering
\includegraphics[trim=0.0cm 1.1cm 0.0cm 0.9cm, clip, width=\channelmapwidth]{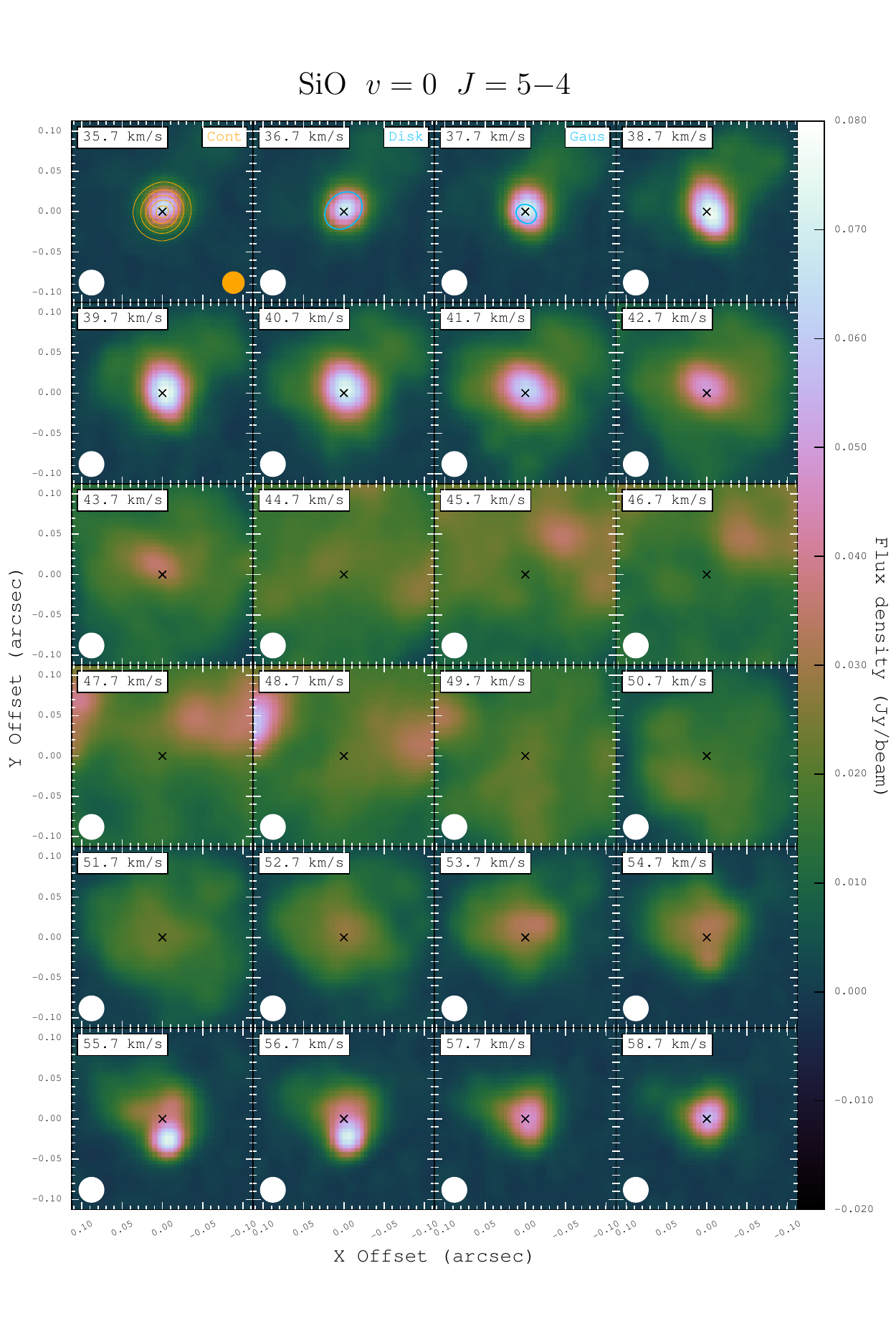}
\caption[]{Same as Fig. \ref{fig:siov0chan} for the zoomed ($0{\farcs}22 \times 0{\farcs}22$) channel maps of SiO ${\varv} = 0$ $J=5-4$ including continuum emission from Mira A.}
\label{fig:siov0chanzoomed}
\end{figure*}

\begin{figure*}[!htbp]
\centering
\includegraphics[trim=0.0cm 1.1cm 0.0cm 0.9cm, clip, width=\channelmapwidth]{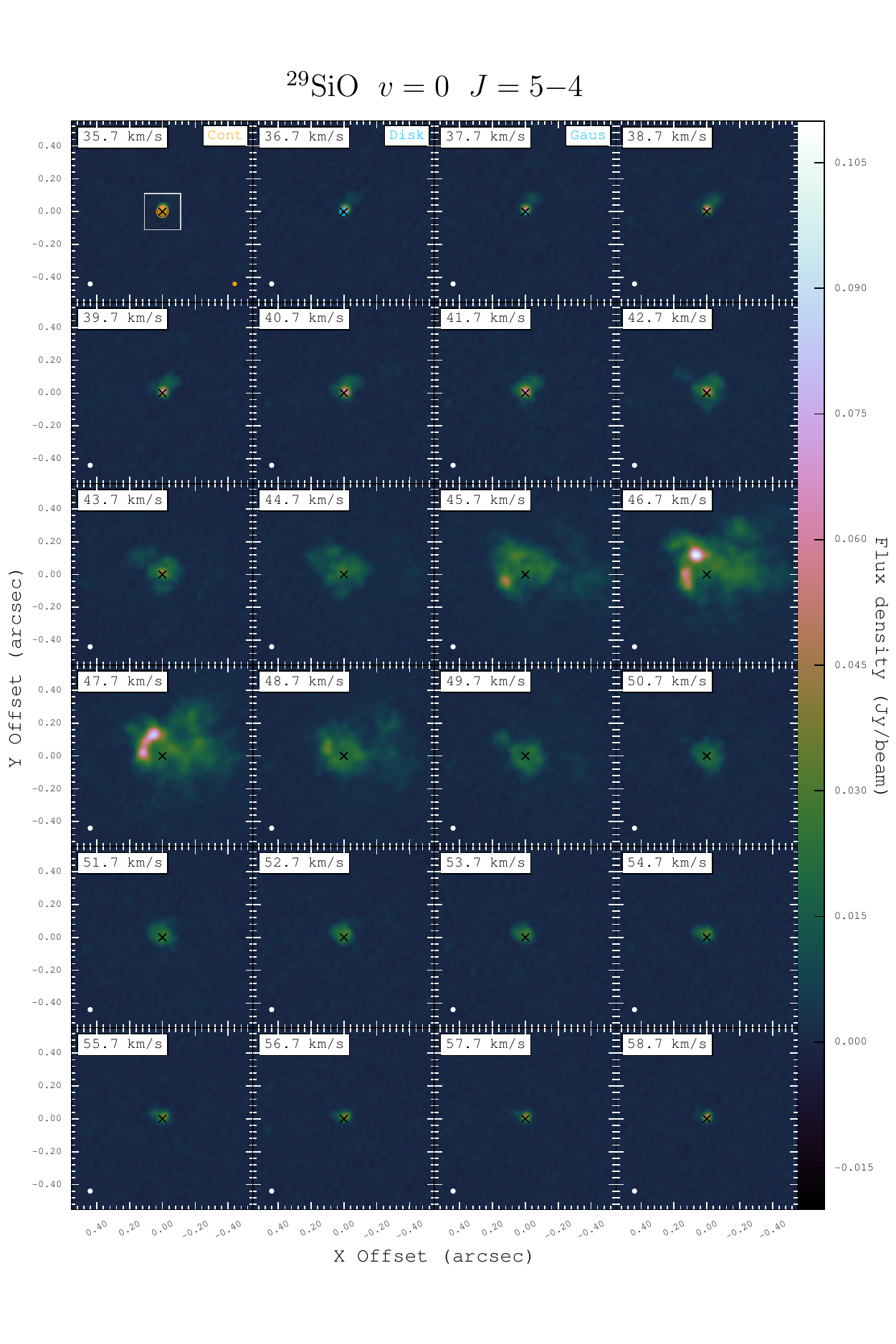}
\caption[]{Same as Fig. \ref{fig:siov0chan} for the channel maps of $^{29}$SiO ${\varv} = 0$ $J=5-4$ including continuum emission from Mira A. 
The map rms noise is $0.65\,{\rm mJy}\,{\rm beam}^{-1}$.}
\label{fig:29siochan}
\end{figure*}

\begin{figure*}[!htbp]
\centering
\includegraphics[trim=0.0cm 1.1cm 0.0cm 0.9cm, clip, width=\channelmapwidth]{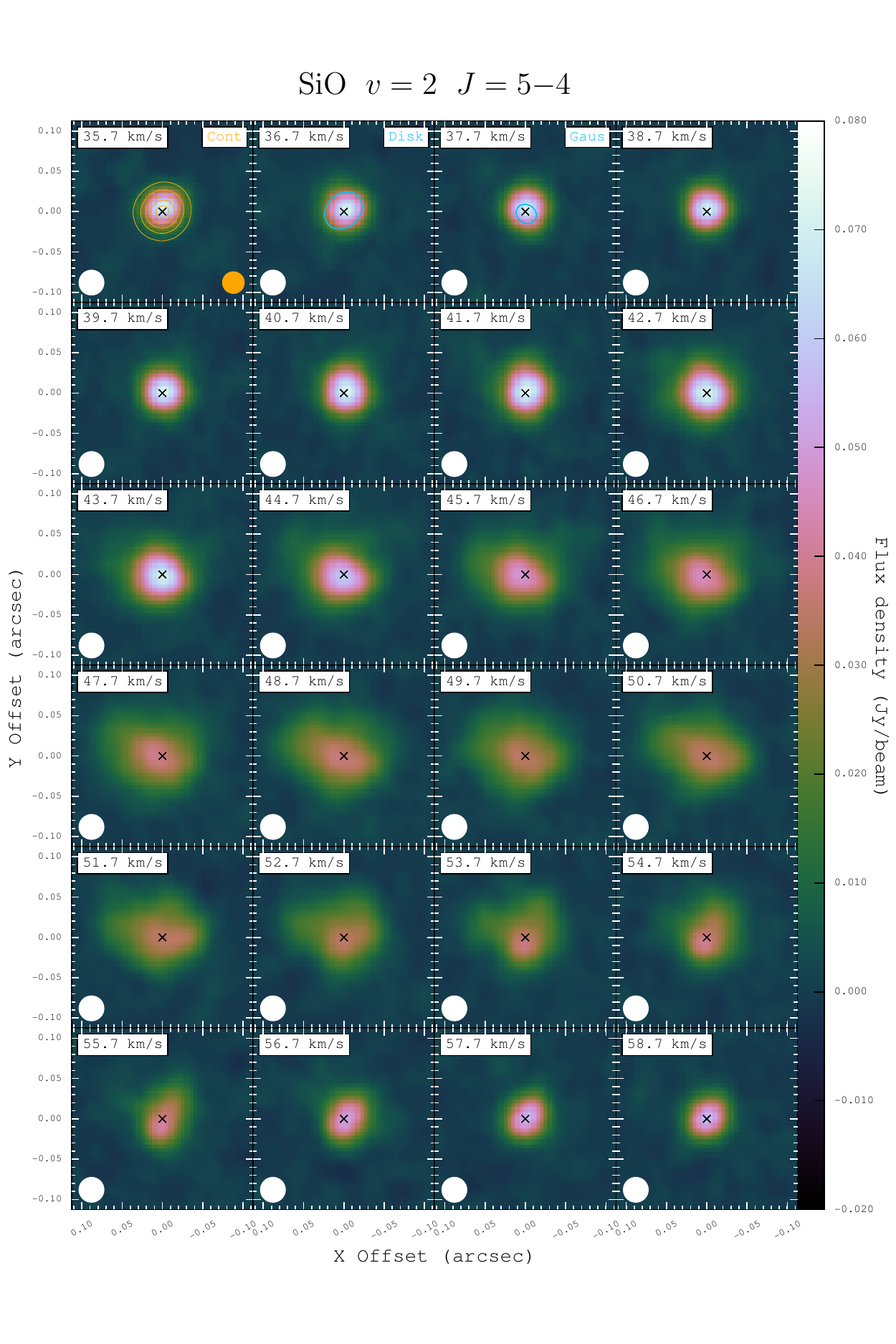}
\caption[]{Same as Fig. \ref{fig:siov0chan} for the zoomed ($0{\farcs}22 \times 0{\farcs}22$) channel maps of SiO ${\varv} = 2$ $J=5-4$ including continuum emission from Mira A. 
The map rms noise is $0.72\,{\rm mJy}\,{\rm beam}^{-1}$.}
\label{fig:siov2chan}
\end{figure*}

\begin{figure*}[!htbp]
\centering
\includegraphics[trim=0.0cm 1.1cm 0.0cm 0.9cm, clip, width=\channelmapwidth]{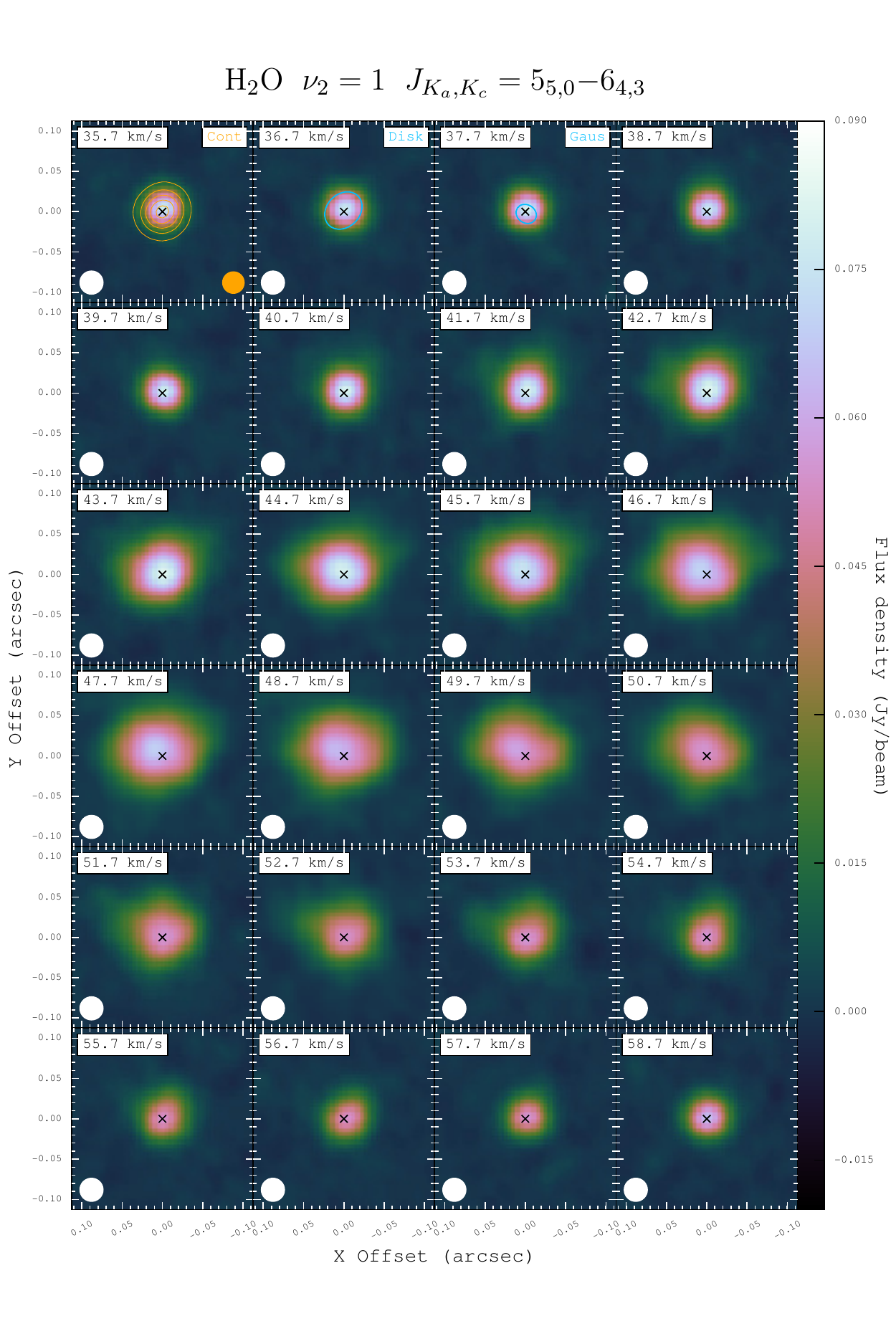}
\caption[]{Same as Fig. \ref{fig:siov0chan} for the zoomed ($0{\farcs}22 \times 0{\farcs}22$) channel maps of SiO ${\varv} = 2$ $J=5-4$ including continuum emission from Mira A. 
The circular restoring beam of $0{\farcs}030$ FWHM for the H$_2$O images is indicated in white at the bottom-left in each panel. 
The map rms noise is $0.72\,{\rm mJy}\,{\rm beam}^{-1}$.}
\label{fig:h2ov1chan}
\end{figure*}

Figures \ref{fig:siov0chan}--\ref{fig:h2ov1chan} show the channel maps without continuum subtraction. As explained in Sects. \ref{sec:obs} and \ref{sec:result_maps}, we extract from these maps the spectra for our radiative transfer modelling and discussion in Sects. \ref{sec:result_spec}, \ref{sec:modelling}, and \ref{sec:discussion}.


\section{Extrapolation of SiO--H$_2$ collisional rate coefficients}
\label{sec:appendix_siorates}

\subsection{Pure rotational transitions in the vibrational ground state ${\varv} = 0$}
\label{sec:appendix_purerotrates}

Collisional rate coefficients between SiO and para-H$_2$ ($J=0$) are often approximated by scaling the rate coefficients of the SiO--He system by the square root of the ratio of the reduced masses of the two collisional partners, i.e., $\displaystyle\sqrt{\frac{\mu_{\rm Si-He}}{\mu_{{\rm Si-H}_2}}} = 1.38$ \citep[e.g.][]{lamda}. Based on the Gordon-Kim electron gas model \citep{gordonkim1972}, \citet{bg1981} have derived the potential energy surface (PES) of the SiO--He system, from which the potential energy of the system in terms of the orientation of the two molecules can be derived \citep[e.g.][Chap. 2]{comchembook2011}. \citet{bg1983a,bg1983b} then computed the collisional cross sections and hence the corresponding rate coefficients for ${\varv} \le 2$, $J \le 39$ and $\Delta {\varv} = 0, 1$ for the H$_2$ gas temperature range between 1000\,K and 3000\,K. \citet{turner1992} have derived the rate coefficients, for ${\varv} = 0$ and $J \le 20$ only, in a lower temperature range of $20$--$300$\,K by extrapolating the PES of \citet{bg1981} to shorter collision distances between SiO and He. With the \emph{ab initio} method as described in their Section 2, \citet{dayou2006} have derived a new PES for the SiO--He system which differs significantly from that derived by \citet{bg1981}. The resultant rate coefficients from \citet{dayou2006} are different from those from \citet{turner1992} by up to an order of magnitude, with the odd/even-$\Delta J$ propensity rule reversed. The differences are qualitatively similar to the finding by \citet{thomas1980} that the rate coefficients derived by an electron gas model could be underestimated or overestimated by about 2 to 3 times in the odd- or even-$\Delta J$ transitions, respectively, relative to \emph{ab initio} calculation. Hence, in our SiO datafile, we adopt the rate coefficients from \citet{dayou2006} (${\varv} = 0$, $J \le 26$) for the collisional transitions of SiO molecule in the vibrational ground state. The rate coefficients for higher $J$ and temperatures are extrapolated by empirically derived analytical approximations as described in the following. For all the function fitting in the following discussion, we use the Levenberg-Marquardt algorithm, as implemented in the Python/SciPy module \texttt{scipy.optimize}, as the non-linear least squares fitting method.

\subsubsection{Extrapolation to high $J$ ($J_u > 26$)}

At the gas temperature $T$, the collisional rate coefficients, $k_{J_uJ_l}$, from the initial upper rotational level ($J_u$) to the final lower level ($J_l$) in the vibrational ground state (${\varv} = 0$) can be computed with the infinite-order sudden approximation (IOS) \citep[e.g.][]{goldflam1977}:
\begin{equation}
\label{eq:ratecoeff}
k_{J_uJ_l} (T) = \left( 2J_l+1 \right) \sum_{j=|J_u-J_l|}^{J_u+J_l} 
\begin{pmatrix}
J_u & J_l & j \\
0 & 0 & 0 \end{pmatrix}^2
(2j+1) \, k_{j0} (T) \cdot A^{J_u}_j(T) , 
\end{equation}
where 
$ \begin{pmatrix}
j_i & j_f & j \\
0 & 0 & 0 \end{pmatrix} $ is the Wigner 3-$j$ symbol. For the special cases in which the bottom three entries of the 3-$j$ symbol equal 0, it has been shown that \citep[e.g.][]{edmondsbook1960,messiahbook1962}, if $S \equiv j_i + j_f + j$ is odd, then
\begin{align}
\begin{pmatrix}
j_i & j_f & j \\
0 & 0 & 0 \end{pmatrix} &= 0, 
\end{align}
and if $S \equiv j_i + j_f + j$ is even, then
\begin{align}
\notag \begin{pmatrix}
j_i & j_f & j \\
0 & 0 & 0 \end{pmatrix} = &\left(-1\right)^{S/2} \sqrt{\frac{(S-2j_i)!(S-2j_f)!(S-2j)!}{(S+1)!}} \\
&\times \frac{(S/2)!}{(S/2-j_i)!(S/2-j_f)!(S/2-j)!}.
\end{align}
The IOS approximation requires that the difference in the rotational energy levels is insignificant to the kinetic energy of the system, which may not be the case for high $\Delta J$ transitions. An adiabaticity factor for purely rotational transitions (i.e., $\Delta {\varv} = 0$), $A^{J_u}_j(T)$, is therefore required to account for this effect \citep{ramaswamy1979,depristo1979a}. The adiabaticity factor at the gas temperature $T$ is given by
\begin{align}
\label{eq:ecs_factor}
A^{J_u}_j(T) &= \frac{6 + \left(0.129 j B_0 l_c \right)^2 {\mu_{{\rm SiO-H}_2}}/T}{6 + \left(0.129 J_u B_0 l_c \right)^2 {\mu_{{\rm SiO-H}_2}}/T}, 
\end{align}
where $B_0 = 0.724$ is the rotation constant of SiO (${\varv} = 0$) in ${\rm cm}^{-1}$, $l_c \approx 3$ is the typical inelastic impact parameter in {\AA}, and ${\mu_{{\rm SiO-H}_2}} = 1.93$ is the reduced mass of SiO--H$_2$ in atomic mass unit. To obtain the full set of rate coefficients up to $J_u = 69$, we need to know the coefficients $k_{j0}$ for $j$ from 1 up to 137. The SiO--H$_2$ rate coefficients (scaled from SiO--He values) computed by \citet{dayou2006} are up to $j=26$ only. We extrapolate the coefficients $k_{j0}$ to higher rotational levels by fitting the \citeauthor{dayou2006} coefficients to the function
\begin{align}
\label{eq:high_jf_fit}
\ln k_{j0} = a + b j.
\end{align}

The fitting function in Eq. (\ref{eq:high_jf_fit}) does not adopt a second-order term ($cj^2$) as in \citet{lamda} in order to get consistent fitting for all values of temperatures. The values of $k_{j0}$ become vanishingly small for $j \gg J_u$, and hence their contributions to the summation of the coefficient $k_{J_uJ_l}$ are insignificant. Odd and even $j$ are fitted together, so the propensity rule is ignored for the extrapolated $j$-levels. 

However, the rate coefficients computed with Eq. (\ref{eq:ratecoeff}) and the extrapolated $k_{j0}$ values are consistently smaller than those theoretically derived by \citet{dayou2006}. In view of the discrepancy between the \citeauthor{dayou2006} coefficients and the extrapolated coefficients, we make direct extrapolation of the rate coefficients $k_{J_uJ_l}$ instead of summing up the extrapolated $k_{j0}$ in Eq. (\ref{eq:ratecoeff}). For each $J_l \le 23$, we extrapolated $k_{J_uJ_l}$ to higher $J_u$ with
\begin{align}
\label{eq:low_jf_fit}
\ln k_{J_uJ_l} = a + b J_u.
\end{align}

For a particular $J_l$, the variation of the collisional rate coefficients with $J_u$ will then be smoother than in the LAMDA SiO datafile.

\subsubsection{Extrapolation to high $T$}

With the $J$-extrapolated collisional rate coefficients in the temperature range between 10\,K and 300\,K, we develop the following empirical relation to extrapolate the rate coefficients to higher temperatures:
\begin{align}
\label{eq:temp_fit}
\ln \left( \frac{k_{J_uJ_l} (T)}{10^{-10}\,{\rm cm}^{3}\,{\rm s}^{-1}} \right) &= \ln \left( \frac{m_{\Delta J}}{T} \right)^{\frac{1}{4}} - \left( \frac{n_{\Delta J}}{T} \right)^{\frac{1}{4}}.
\end{align}
The low-temperature (10--300\,K) coefficients of all transitions with the same $\Delta J = J_u-J_l$ are (globally) fitted to Eq. \ref{eq:temp_fit} with the same set of fitting parameters $m_{\Delta J}$ and $n_{\Delta J}$. The fitting parameters are dependent on $\Delta J$ and their values in the unit of Kelvin are given in Table \ref{tab:fit_params}. For individual transitions $J_u \rightarrow J_l$, we scale the extrapolated coefficient with a factor that ensures continuity with respect to temperature at $T=300\,{\rm K}$,
\begin{align}
\frac{k^{\rm DB06}_{J_uJ_l} (300\,{\rm K})}{\displaystyle \left( \frac{m_{\Delta J}}{300\,{\rm K}} \right)^{1/4} \exp\left[ -\left( \frac{n_{\Delta J}}{300\,{\rm K}} \right)^{1/4} \right] \times 10^{-10}\,{\rm cm}^{3}\,{\rm s}^{-1}},
\end{align}
where $k^{\rm DB06}_{J_uJ_l} (300\,{\rm K})$ is the SiO--H$_2$ rate coefficients (scaled from the SiO--He values) derived by \citet{dayou2006} for the transition from $J_u$ to $J_l$ at $T=300\,{\rm K}$. Our fitting function in Eq. (\ref{eq:temp_fit}) is modified from and has the similar form to the extrapolation functions empirically derived by \citet{dejong1975}, \citet{albrecht1983}, and \citet{bieging1998} respectively. The extrapolated rate coefficients up to $2000\,{\rm K}$ are generally consistent with the values in the LAMDA datafile, which are extrapolated by Eq. (13) of \citet{lamda}.

\begin{table}[!htbp]
\caption{Fitting parameters.}
\label{tab:fit_params}
\centering
\begin{tabular}{cccccc}
\hline\hline
$\Delta J$ & $m_{\Delta J} ({\rm K})$ & $n_{\Delta J} ({\rm K})$ & $\Delta J$ & $m_{\Delta J} ({\rm K})$ & $n_{\Delta J} ({\rm K})$ \\
\hline
1 & 8.91E+04 & 1.19E+03 & 36 & 9.67E+06 & 1.76E+07 \\
2 & 1.93E+04 & 3.98E+03 & 37 & 1.13E+07 & 1.95E+07 \\
3 & 3.86E+05 & 8.03E+03 & 38 & 1.05E+07 & 2.12E+07 \\
4 & 1.10E+03 & 1.06E+04 & 39 & 1.22E+07 & 2.34E+07 \\
5 & 1.10E+06 & 2.77E+04 & 40 & 1.12E+07 & 2.54E+07 \\
6 & 2.78E+03 & 2.95E+04 & 41 & 1.30E+07 & 2.79E+07 \\
7 & 1.68E+06 & 7.13E+04 & 42 & 1.17E+07 & 3.01E+07 \\
8 & 5.06E+04 & 8.16E+04 & 43 & 1.35E+07 & 3.29E+07 \\
9 & 1.48E+06 & 1.55E+05 & 44 & 1.18E+07 & 3.53E+07 \\
10 & 8.77E+05 & 1.97E+05 & 45 & 1.36E+07 & 3.85E+07 \\
11 & 6.19E+05 & 2.82E+05 & 46 & 1.16E+07 & 4.11E+07 \\
12 & 8.22E+06 & 4.11E+05 & 47 & 1.33E+07 & 4.47E+07 \\
13 & 1.12E+05 & 4.31E+05 & 48 & 1.11E+07 & 4.75E+07 \\
14 & 4.13E+07 & 7.65E+05 & 49 & 1.27E+07 & 5.15E+07 \\
15 & 2.19E+04 & 6.09E+05 & 50 & 1.03E+07 & 5.45E+07 \\
16 & 1.02E+08 & 1.28E+06 & 51 & 1.18E+07 & 5.89E+07 \\
17 & 2.14E+04 & 9.24E+05 & 52 & 9.28E+06 & 6.21E+07 \\
18 & 1.20E+08 & 1.94E+06 & 53 & 1.05E+07 & 6.69E+07 \\
19 & 9.18E+04 & 1.50E+06 & 54 & 8.02E+06 & 7.02E+07 \\
20 & 6.00E+07 & 2.69E+06 & 55 & 9.04E+06 & 7.55E+07 \\
21 & 5.15E+05 & 2.37E+06 & 56 & 6.67E+06 & 7.90E+07 \\
22 & 1.48E+07 & 3.47E+06 & 57 & 7.48E+06 & 8.46E+07 \\
23 & 2.12E+06 & 3.56E+06 & 58 & 5.31E+06 & 8.83E+07 \\
24 & 3.12E+06 & 4.32E+06 & 59 & 5.92E+06 & 9.44E+07 \\
25 & 4.68E+06 & 4.99E+06 & 60 & 4.04E+06 & 9.81E+07 \\
26 & 1.28E+06 & 5.42E+06 & 61 & 4.48E+06 & 1.05E+08 \\
27 & 5.48E+06 & 6.40E+06 & 62 & 2.92E+06 & 1.08E+08 \\
28 & 5.57E+06 & 7.22E+06 & 63 & 3.22E+06 & 1.15E+08 \\
29 & 6.59E+06 & 8.24E+06 & 64 & 2.00E+06 & 1.19E+08 \\
30 & 6.59E+06 & 9.22E+06 & 65 & 2.19E+06 & 1.27E+08 \\
31 & 7.77E+06 & 1.04E+07 & 66 & 1.29E+06 & 1.30E+08 \\
32 & 7.64E+06 & 1.16E+07 & 67 & 1.41E+06 & 1.38E+08 \\
33 & 8.98E+06 & 1.30E+07 & 68 & 7.83E+05 & 1.41E+08 \\
34 & 8.69E+06 & 1.44E+07 & 69 & 8.44E+05 & 1.50E+08 \\
35 & 1.02E+07 & 1.60E+07 &    &          &          \\
\hline
\end{tabular}
\end{table}

\subsection{Transitions in the vibrational excited states of SiO}
\label{sec:appendix_vibrotrates}

The only available collisional rate coefficients of vibrationally excited SiO in the literature are those derived by \citet{bg1983a,bg1983b}, based on the old PES model for the SiO--He system and in the temperature range of $1000$--$3000$\,K. Thus, for all transitions involving the vibrationally excited states of SiO, we extrapolate from the coefficients of \citet{bg1983a,bg1983b} (hereafter BG1983 coefficients).

\subsubsection{Extrapolation to high $J$ and high $T$}

The extrapolation of BG1983 coefficients works on the parameters presented in Table 1 of \citet{bg1983a,bg1983b}. In their Table 1, the parameters $(2J_i+1) k^{{\varv}_i,J_i}_{{\varv}_f,0} (T_0)$ and $q^{{\varv}_i,J_i}_{{\varv}_f,0}$ from $J_i = 0$ to $39$ are presented, where $k^{{\varv}_i,J_i}_{{\varv}_f,0} (T_0)$ is the rate coefficient from the initial level $({\varv}_i,J_i)$ to the ground rotational level of the final vibrational state $({\varv}_f,0)$ at $T_0 = 2000\,{\rm K}$, and $q^{{\varv}_i,J_i}_{{\varv}_f,0}$ is the power-law index of the same transition, which describes the temperature dependence of the rate coefficients in the range $1000$--$3000\,{\rm K}$ such that
\begin{align}
(2J_i+1) k^{{\varv}_i,J_i}_{{\varv}_f,0} (T) &= (2J_i+1) k^{{\varv}_i,J_i}_{{\varv}_f,0} (T_0) \left( \frac{T}{T_0} \right)^{q^{{\varv}_i,J_i}_{{\varv}_f,0}}.
\end{align}

We extrapolate the rate coefficient parameters and power law indices to high-$J$ with approximations similar to Eq. (\ref{eq:low_jf_fit}), i.e.,
\begin{align}
\ln \left[ (2J_i+1) k^{{\varv}_i,J_i}_{{\varv}_f,0} (T_0) \right] &= a + b J_i, 
\end{align}
and
\begin{align}
\ln \left( q^{{\varv}_i,J_i}_{{\varv}_f,0} \right) &= c + d J_i,
\end{align}
except for the rate coefficient parameters of $({\varv}_i,{\varv}_f) = (1,0)$, where we think the parameters are better fitted by a second-order polynomial,
\begin{align}
\ln \left[ (2J_i+1) k^{{\varv}_i,J_i}_{{\varv}_f,0} (T_0) \right] &= a + b J_i + c {J_i}^2.
\end{align}

After the extrapolation, we then compute the full set of rate coefficients at $2000\,{\rm K}$, $k^{{\varv}_i,J_i}_{{\varv}_f,J_f} (T_0)$, with the IOS approximation similar to Eq. (\ref{eq:ratecoeff}), with the only differences being that $J_u$ and $J_l$ in Eq. (\ref{eq:ratecoeff}) are replaced by $J_i$ (initial level) and $J_f$ (final level), respectively. We only apply the adiabaticity factor (Eq. \ref{eq:ecs_factor}) to the terms within the summation for the pure rotational transition within the same vibrational state, i.e., ${\varv}_i = {\varv}_f = 1$ or $2$. For transitions involving a change in the vibrational state, $\Delta {\varv} = {\varv}_i - {\varv}_f \ne 0$, \citet{depristo1979a} have derived a factorisation for the summation terms to correct for the inaccuracy of IOS approximation when the kinetic energy is no longer much greater than the energy of the transition. However, because the rate coefficients without the correction are not accurate anyway (see our discussion in Appendix \ref{sec:appendix_purerotrates}) and the corrected coefficients would not change by many orders of magnitude, we do not think it worthwhile to carry out the factorisation which is quite computationally complicated. So no adiabaticity correction has been applied to the rate coefficients of transitions with $\Delta {\varv} \ne 0$, i.e., $A^{J_i}_j(T) \equiv 1$.

Although the power law indices, $q^{{\varv}_i,J_i}_{{\varv}_f,0}$, are supposed to be valid only for $T = 1000$--$3000\,{\rm K}$, we still adopt the same indices for rate coefficients at $T < 1000\,{\rm K}$ and $T > 3000\,{\rm K}$ because alternative estimate is not available. New calculations of the collisional rate coefficients for a broad range of kinetic temperature and for the vibrationally excited states with the updated PES model will be very useful.

\subsection{Collisions with rotationally excited H$_2$}
\label{sec:appendix_h2rotation}

AGB stars typically have a surface temperature of ${\sim}3000$\,K. In the winds and circumstellar envelopes of these stars, the gas temperature ranges from a few hundred to ${\lesssim}3000$\,K. Molecular hydrogen in these environments may be significantly populated in the upper rotational levels \citep[e.g.][]{flower1989,doel1990,field1998}. For examples, the wavelengths of the lowest para-H$_2$ transition, $S(0)$ ($E_{\rm up}/k = 510\,{\rm K}$), and the lowest ortho-H$_2$ transition, $S(1)$ ($E_{\rm up}/k = 1015\,{\rm K}$), are 28.2\,$\mu$m and 17.0\,$\mu$m, respectively \citep{dabrowski1984}. The approximation of the collisional rate coefficients by the scaled (by ${\sim}1.4$) coefficients with atomic helium as the collision partner is only valid for the collisions with para-H$_2$ ($J=0$), but not for other rotational states of H$_2$ \citep{field1998}. \citet{lique2008} and \citet{klos2008} have found that the rate coefficients for the collisions between the diatomic molecule SiS with para-H$_2$ ($J=0$) are significantly lower than those for the collisions with ortho-H$_2$ ($J=1$) or para-H$_2$ ($J=2$) by up to an order of magnitude. Similarly, \citet[][and references therein]{daniel2011} have also found that the collisions between H$_2$O and rotationally excited H$_2$ in the $J=1$ and $2$ states are the dominant processes in the collisional (de-)excitation of H$_2$O molecule.

For SiO molecule, because the currently available collisional rate coefficients do not consider rotationally excited H$_2$ ($J=1$ or $2$) as the collision partners, the collisional (de-)excitations of SiO in the radiative transfer modelling could well be inaccurate and it is not possible to estimate the errors. In our modelling, it appears that most of the absorption/emission is contributed by the regions which are close to local thermodynamic equilibrium, otherwise there would not be enough absorption along the line-of-sight towards the continuum or there would be strong maser emission in the SiO ${\varv} = 0$ and $2$ transitions, which show very weak or no maser in the data. So the gas temperature and expansion/infall velocity profiles are probably not affected by the incomplete collisional rate coefficients, but the magnitudes of the H$_2$ gas density and SiO abundance may not be accurate.


\section{Infall-only models}
\label{sec:appendix_model}

We present here two models, Models 1 and 2, that fail to reproduce the observed SiO and H$_2$O spectra around Mira in ALMA Band 6. Models 1 and 2 exhibit global infall motion only, while our preferred model (Model 3; Sect. \ref{sec:model_preferred}) contains a globally expanding layer between the radio photosphere and the globally infalling region.

\begin{figure*}[!htbp]
\centering
\includegraphics[width=\modelwidth]{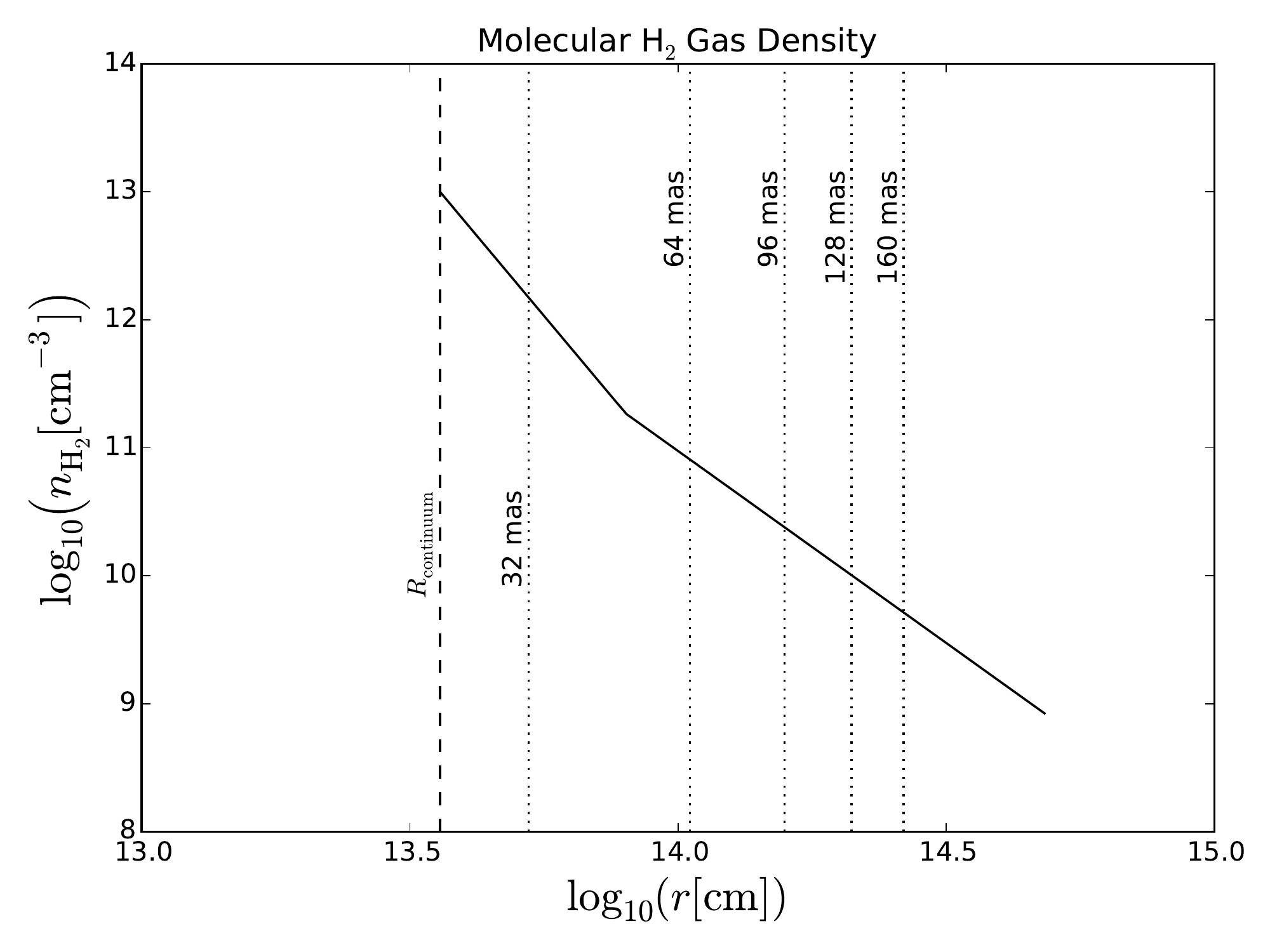}
\includegraphics[width=\modelwidth]{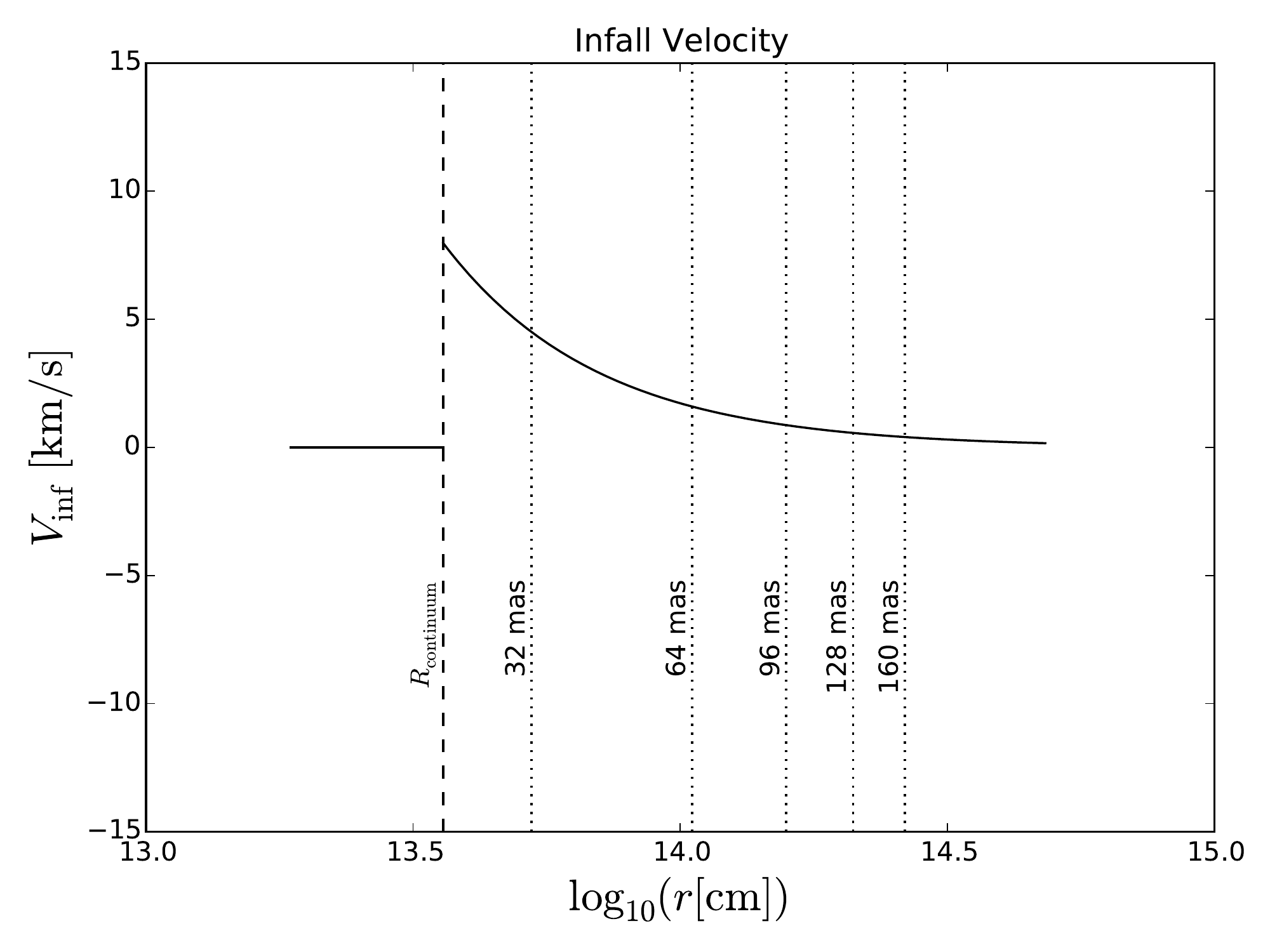}\\
\includegraphics[width=\modelwidth]{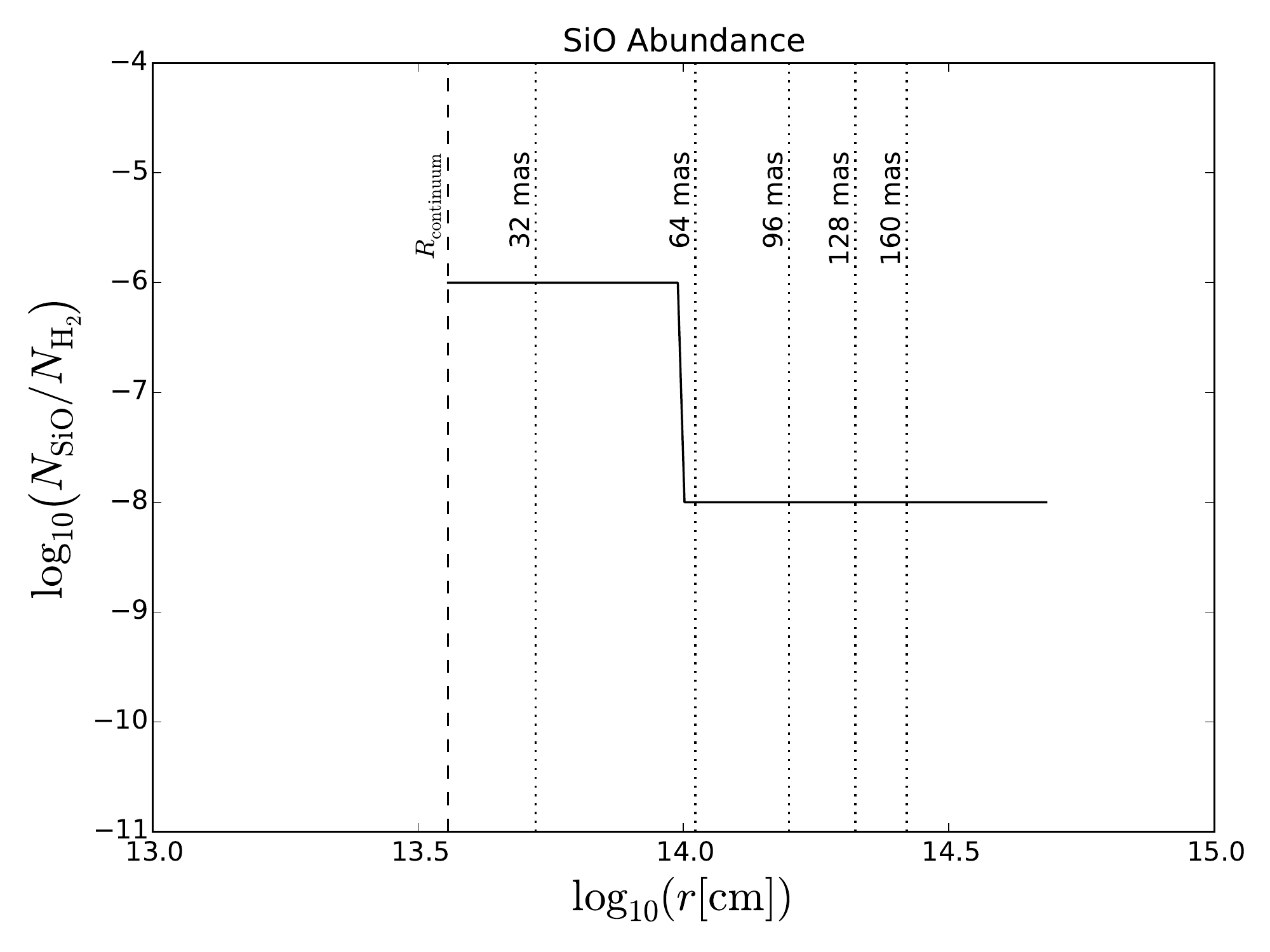}
\includegraphics[width=\modelwidth]{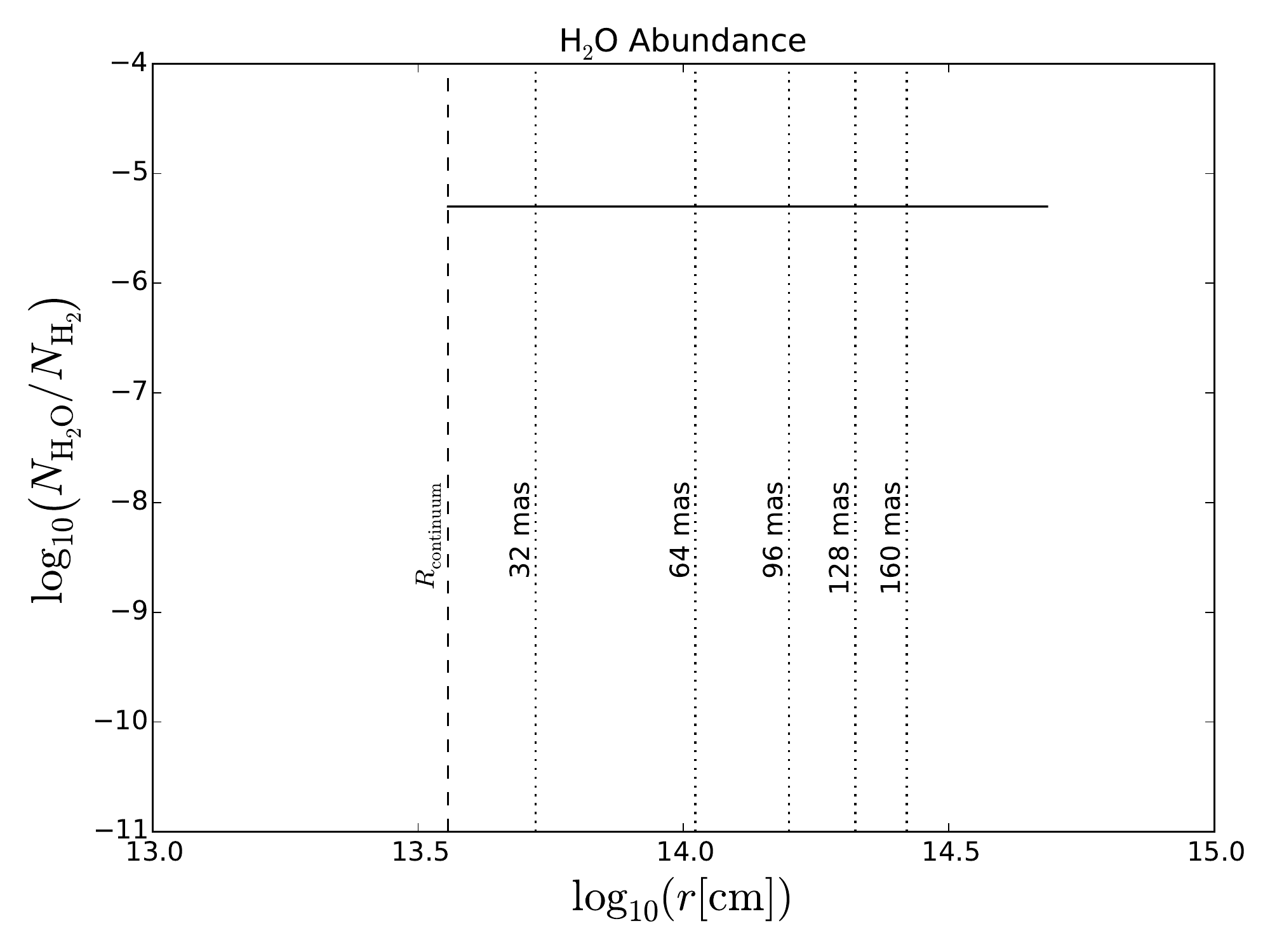}\\
\includegraphics[width=\modelwidth]{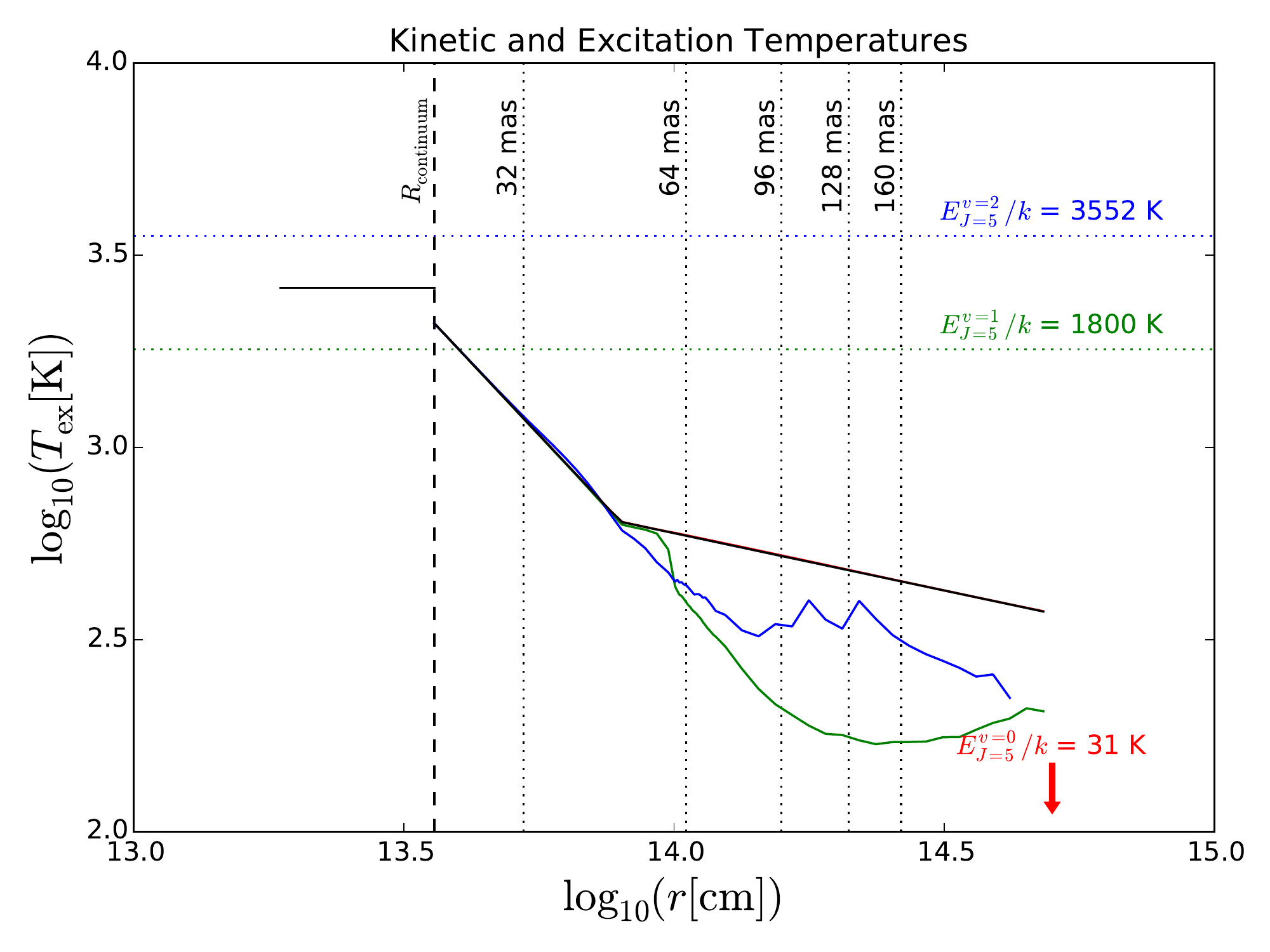}
\includegraphics[width=\modelwidth]{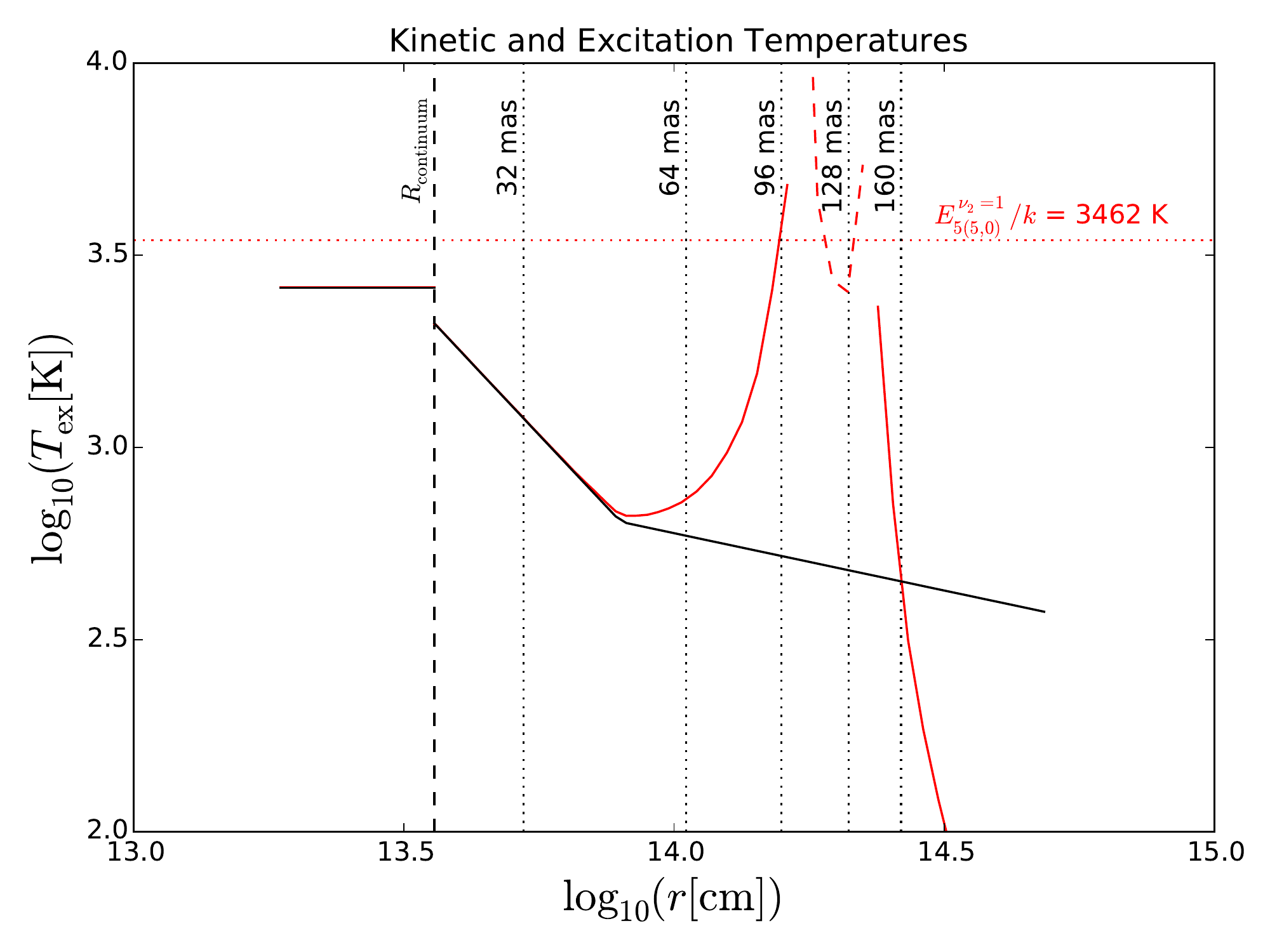}
\caption{Inputs for Model 1. Shown in the panels are the H$_2$ gas density (\textbf{top-left}) , infall velocity (negative represents expansion) (\textbf{top-right}), $^{28}$SiO abundance (\textbf{middle-left}), H$_2$O abundance (\textbf{middle-right}), and the kinetic temperature (in black) and excitation temperatures (in colours) of the three $^{28}$SiO transitions (\textbf{bottom-left}) and the H$_2$O transition (\textbf{bottom-right}). In the bottom-right panel, solid red line indicates positive excitation temperature (i.e., non-maser emission) of the H$_2$O transition, and the dashed red line indicates the absolute values of the negative excitation temperature (i.e., population inversion) between $1.7 \times 10^{14}$ and $2.2 \times 10^{14}\,{\rm cm}$. Small negative values for the excitation temperature would give strong maser emission. Vertical dotted lines mark the radii at which the spectra were extracted; coloured horizontal dotted lines in the bottom panels indicates the upper-state energy ($E_{\rm up}/k$) of the respective transitions. The innermost layer within $R_{\rm continuum}$ represents the grid cells for the \emph{pseudo}-continuum, in which the input values for H$_2$ gas density and molecular abundances are above the range of the plots.}
\label{fig:model1}
\clearpage
\end{figure*}

We first test the infall velocity profiles with monotonically increasing infall velocity (Model 1). Figure \ref{fig:model1} shows the profiles of the gas density (top-left), infall velocity (top-right), SiO and H$_2$O abundances (middle) and the kinetic temperature (bottom). The bottom row of Fig. \ref{fig:model1} also shows the excitation temperatures of the SiO and H$_2$O transitions (in colour).

\begin{figure*}[!htbp]
\centering
\includegraphics[trim=1.0cm 2.0cm 2.0cm 1.5cm, clip, width=\multispecwidth]{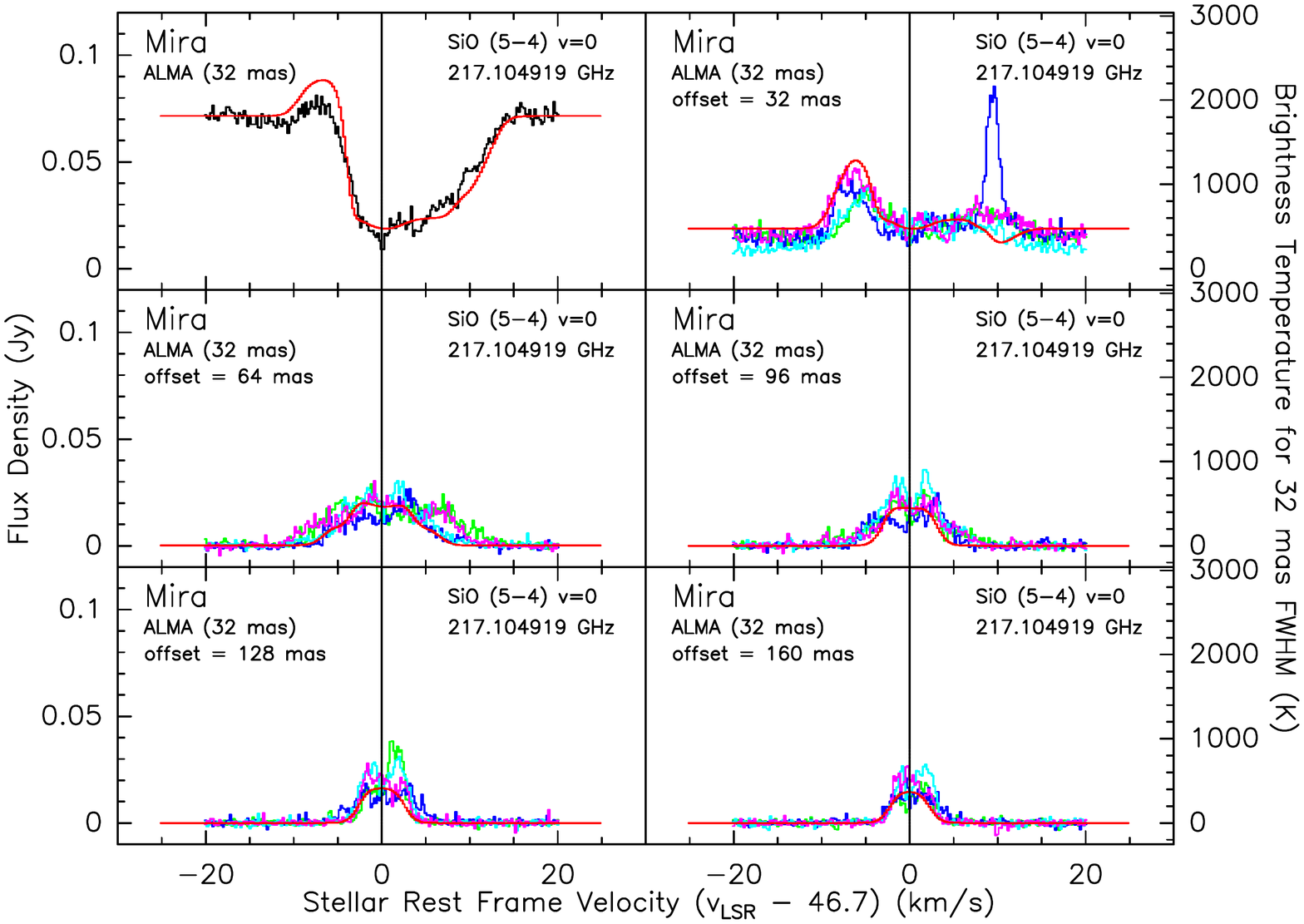}
\caption{Model 1: spectra of SiO ${\varv} = 0$ $J=5-4$ at various positions. The black histogram is the observed spectrum at the centre of continuum, green, blue, cyan and magenta histograms are the observed spectra along the eastern, southern, western and northern legs, respectively, at various offset radial distances as indicated in each panel. The red curves are the modelled spectra predicted by {\ratran}.}
\label{fig:m1siov0spec}
\end{figure*}

\begin{figure*}[!htbp]
\centering
\includegraphics[trim=1.0cm 7.3cm 2.0cm 1.5cm, clip, width=\multispecwidth]{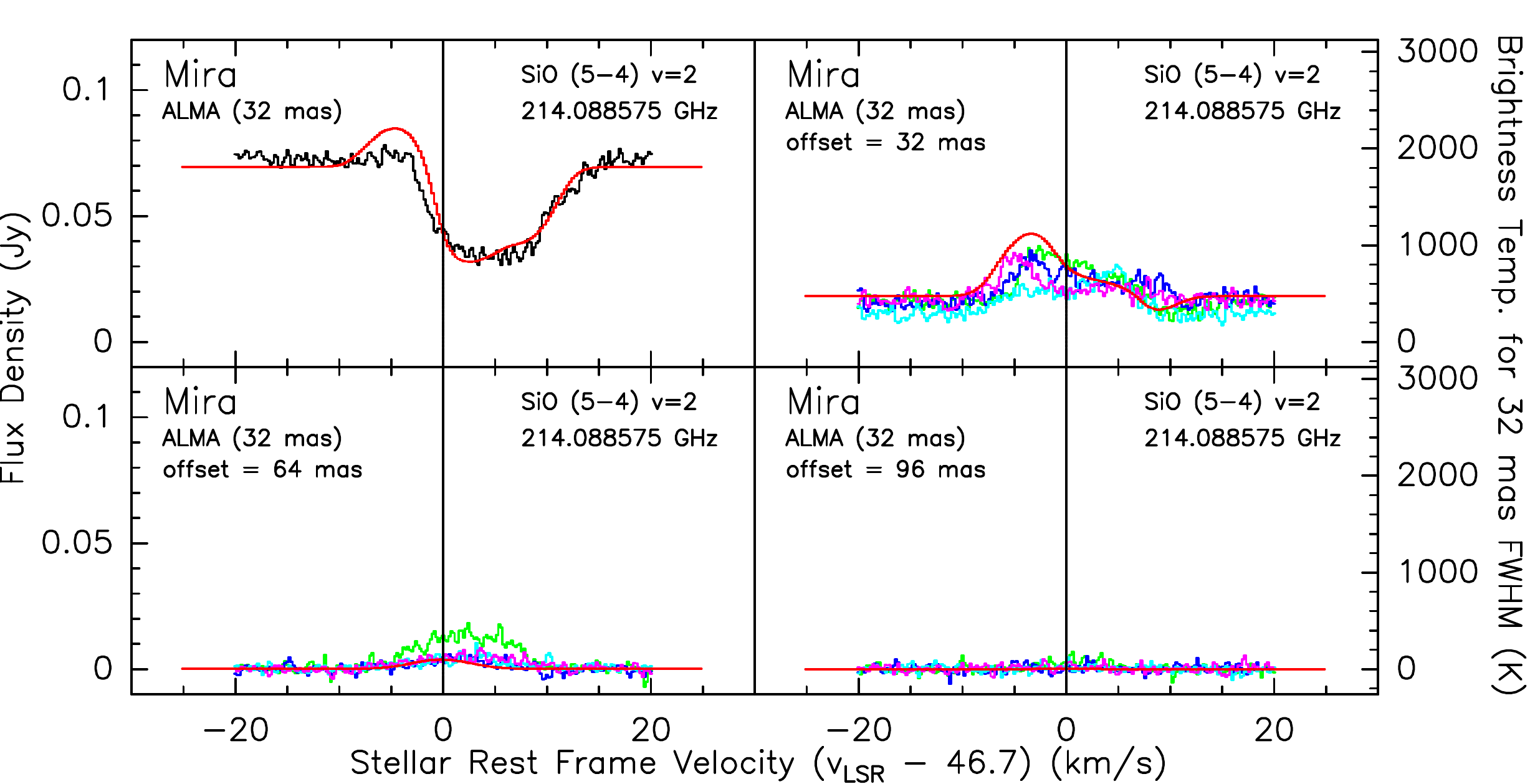}
\caption{Model 1: spectra of SiO ${\varv} = 2$ $J=5-4$ at various positions. The black histogram is the observed spectrum at the centre of continuum, green, blue, cyan and magenta histograms are the observed spectra along the eastern, southern, western and northern legs, respectively, at various offset radial distances as indicated in each panel. The red curves are the modelled spectra predicted by {\ratran}.}
\label{fig:m1siov2spec}
\end{figure*}

\begin{figure*}[!htbp]
\centering
\includegraphics[trim=1.0cm 7.3cm 2.0cm 1.5cm, clip, width=\multispecwidth]{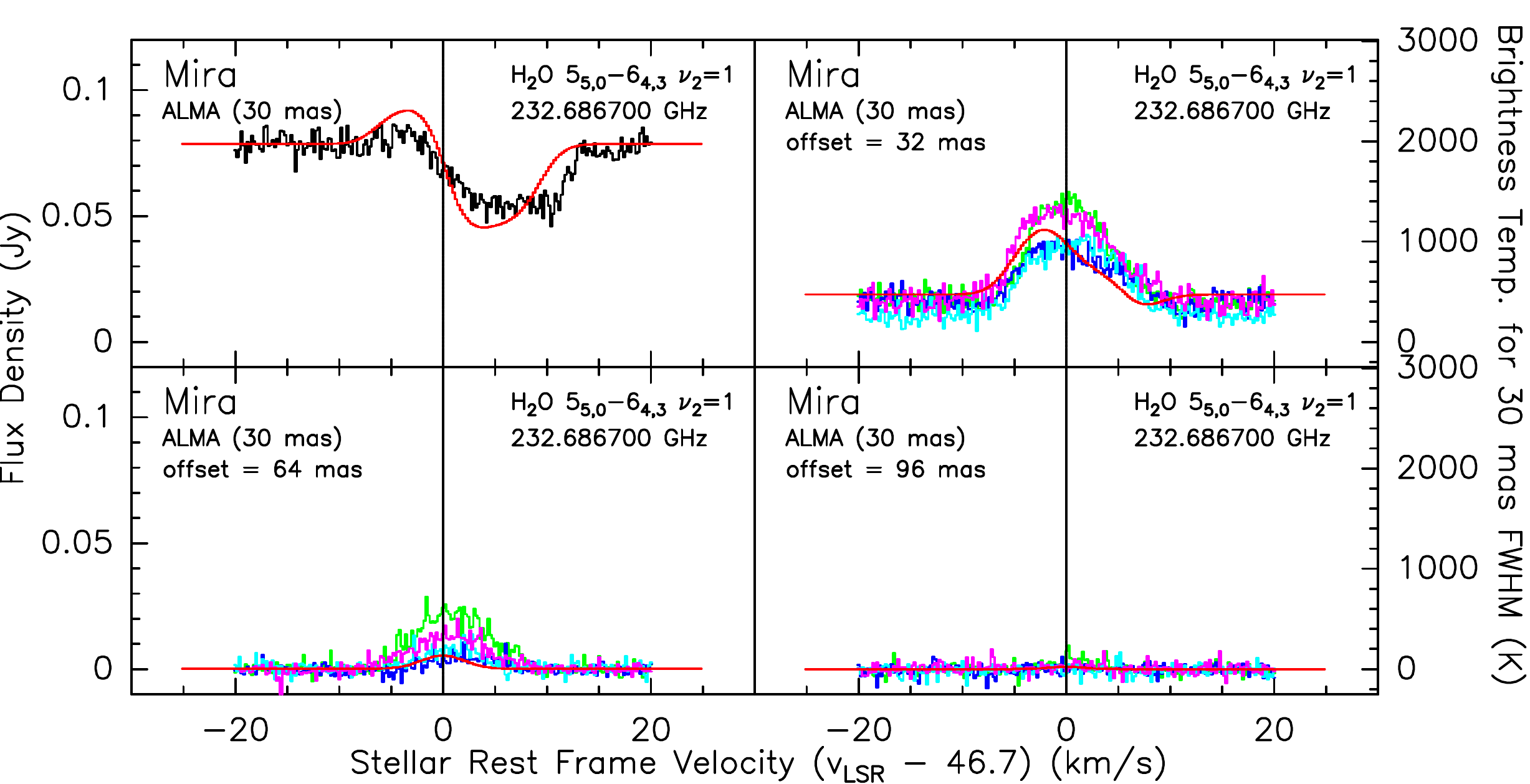}
\caption{Model 1: spectra of H$_2$O $v_2=1$ $J_{K_a,K_c}=5_{5,0}-6_{4,3}$ at various positions. The black histogram is the observed spectrum at the centre of continuum, green, blue, cyan and magenta histograms are the observed spectra along the eastern, southern, western and northern legs, respectively, at various offset radial distances as indicated in each panel. The red curves are the modelled spectra predicted by {\ratran}.}
\label{fig:m1h2ov1spec}
\end{figure*}

Figures \ref{fig:m1siov0spec}, \ref{fig:m1siov2spec}, and \ref{fig:m1h2ov1spec} show our modelled and observed spectra for SiO ${\varv} = 0$ $J=5-4$, SiO ${\varv} = 2$ $J=5-4$, and H$_2$O $v_2=1$ $J_{K_a,K_c}=5_{5,0}-6_{4,3}$, respectively. The top-left panel of these figures show the spectra extracted from the line-of-sight towards the continuum centre. Although Model 1 may fit well to the redshifted wings (about $+10$ to $+15\,\kms$) of the SiO spectra, there is significant excess emission from the blueshifted velocities between about $-10$ and $-4\,\kms$. The excess blueshifted emission is also seen in the spectra extracted from $32\,{\rm mas}$. Additionally, in this spectra extracted from $32\,{\rm mas}$, there is an absorption feature against the continuum near the redshifted velocity of about $+10\,\kms$, which is not present in the observed spectra.


As we have mentioned in Sect. \ref{sec:results}, the observed weak blueshifted emission feature around the velocities from $-10$ to $-3\,\kms$ originates from the emission from the far side of Mira's extended atmosphere. As the line-of-sight moves away from the continuum disk, the emission flux of the radio continuum, which is represented by a circular uniform disk in our input model and appears as a flat line in the spectra, falls off with radius rapidly; the flux of the far side molecular emission, on the other hand, decreases more gradually. Hence, the inverse P Cygni profile of the molecular transition emerges from the decreasing continuum level. The blueshifted emission feature in the resultant spectra after beam convolution represents the combined effect of the inner continuum-dominated emission and the outer line-dominated emission. The excess blueshifted emission in Model 1 therefore suggests two possibilities: (1) the gas temperature at the innermost radii in this model is significantly overestimated, and/or (2) this model does not reproduce sufficient absorption in the blueshifted velocities.

\begin{figure*}[!htbp]
\centering
\includegraphics[width=\modelwidth]{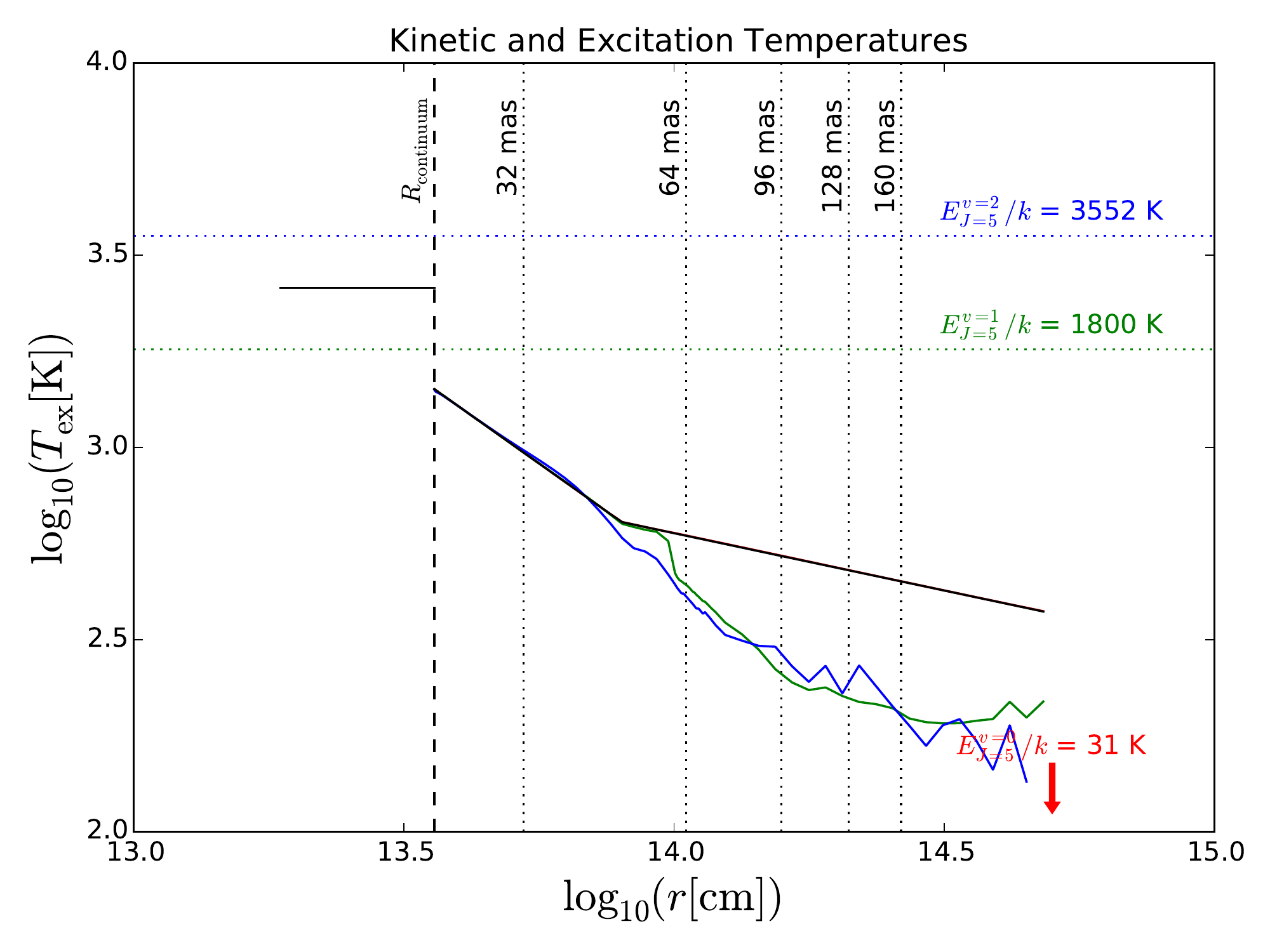}
\includegraphics[width=\modelwidth]{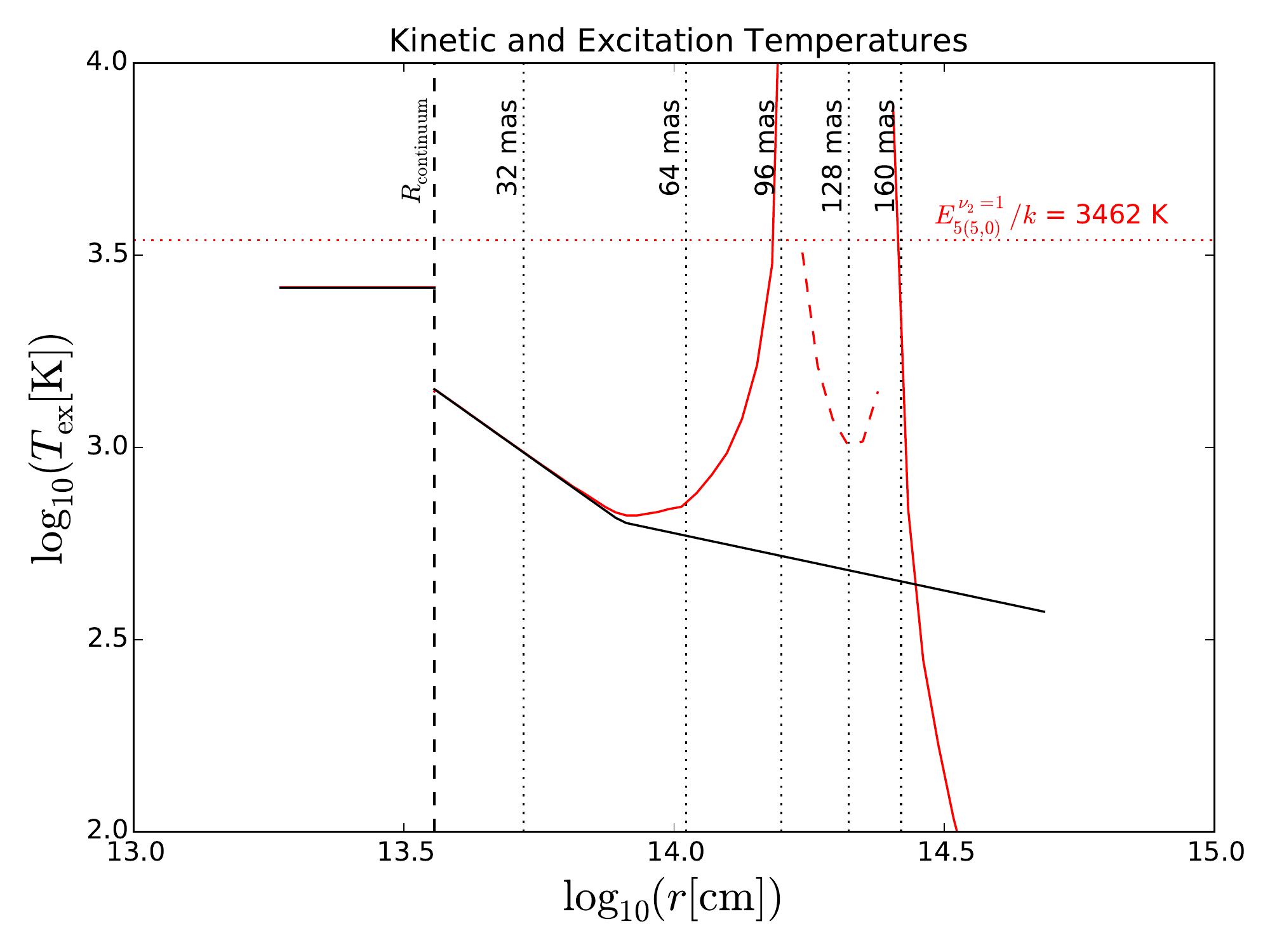}
\caption{Input temperature for Model 2. The panels show the kinetic temperature (in black) and excitation temperatures (in colours) of the three $^{28}$SiO transitions (\textbf{left}) and the H$_2$O transition (\textbf{right}). In the right panel, solid red line indicates positive excitation temperature (i.e., non-maser emission) of the H$_2$O transition, and the dashed red line indicates the absolute values of the negative excitation temperature (i.e., population inversion) between $1.7 \times 10^{14}$ and $2.4 \times 10^{14}\,{\rm cm}$. Small negative values for the excitation temperature would give strong maser emission. Vertical dotted lines mark the radii at which the spectra were extracted; coloured horizontal dotted lines in the bottom panels indicates the upper-state energy ($E_{\rm up}/k$) of the respective transitions. The innermost layer within $R_{\rm continuum}$ represents the grid cells for the \emph{pseudo}-continuum.}
\label{fig:model2}
\end{figure*}
\begin{figure*}[!htbp]
\centering
\includegraphics[trim=1.0cm 2.0cm 2.0cm 1.5cm, clip, width=\multispecwidth]{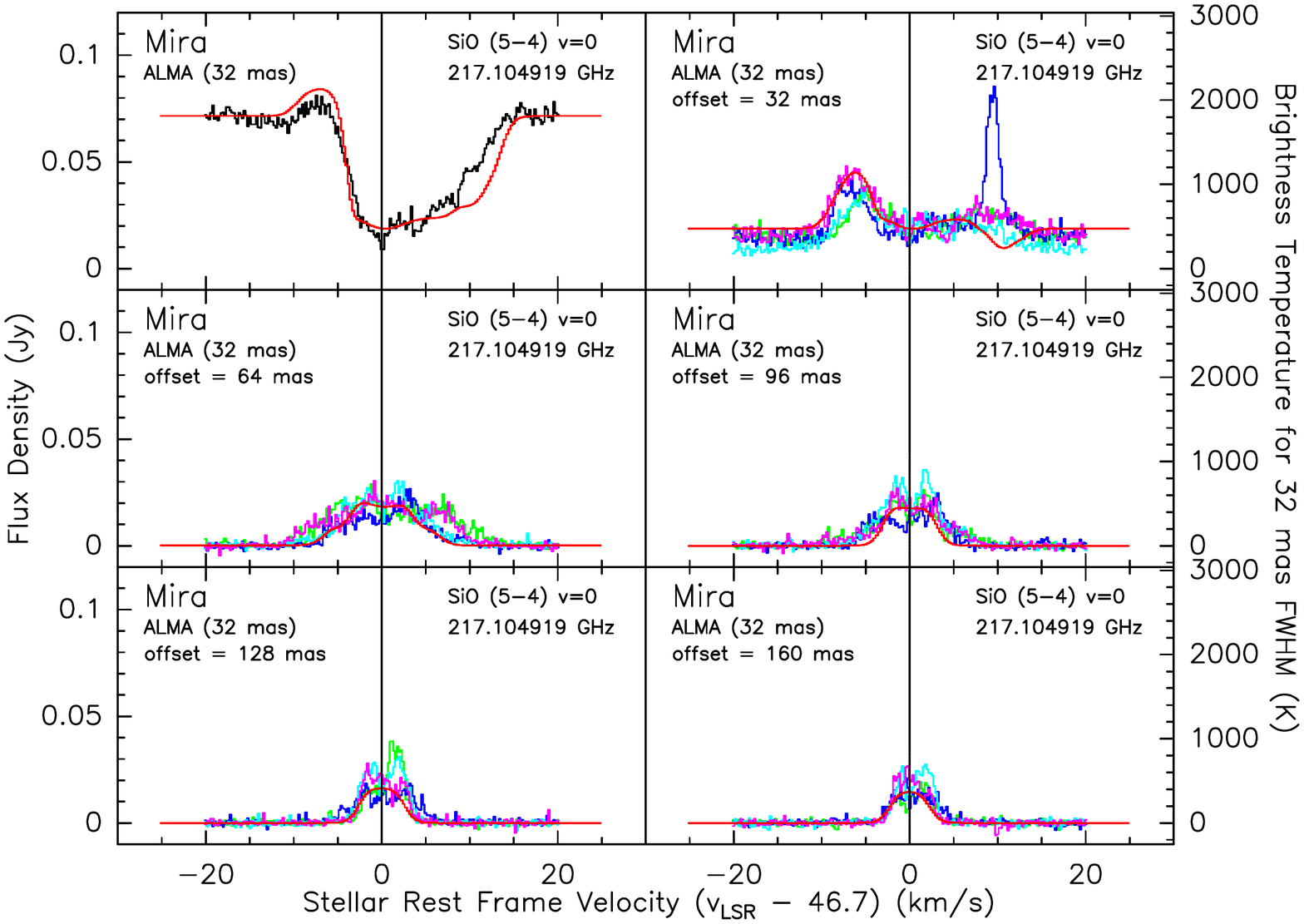}
\caption{Model 2: spectra of SiO ${\varv} = 0$ $J=5-4$ at various positions. The black histogram is the observed spectrum at the centre of continuum, green, blue, cyan and magenta histograms are the observed spectra along the eastern, southern, western and northern legs, respectively, at various offset radial distances as indicated in each panel. The red curves are the modelled spectra predicted by {\ratran}.}
\label{fig:m2siov0spec}
\end{figure*}
\begin{figure*}[!htbp]
\centering
\includegraphics[trim=1.0cm 7.3cm 2.0cm 1.5cm, clip, width=\multispecwidth]{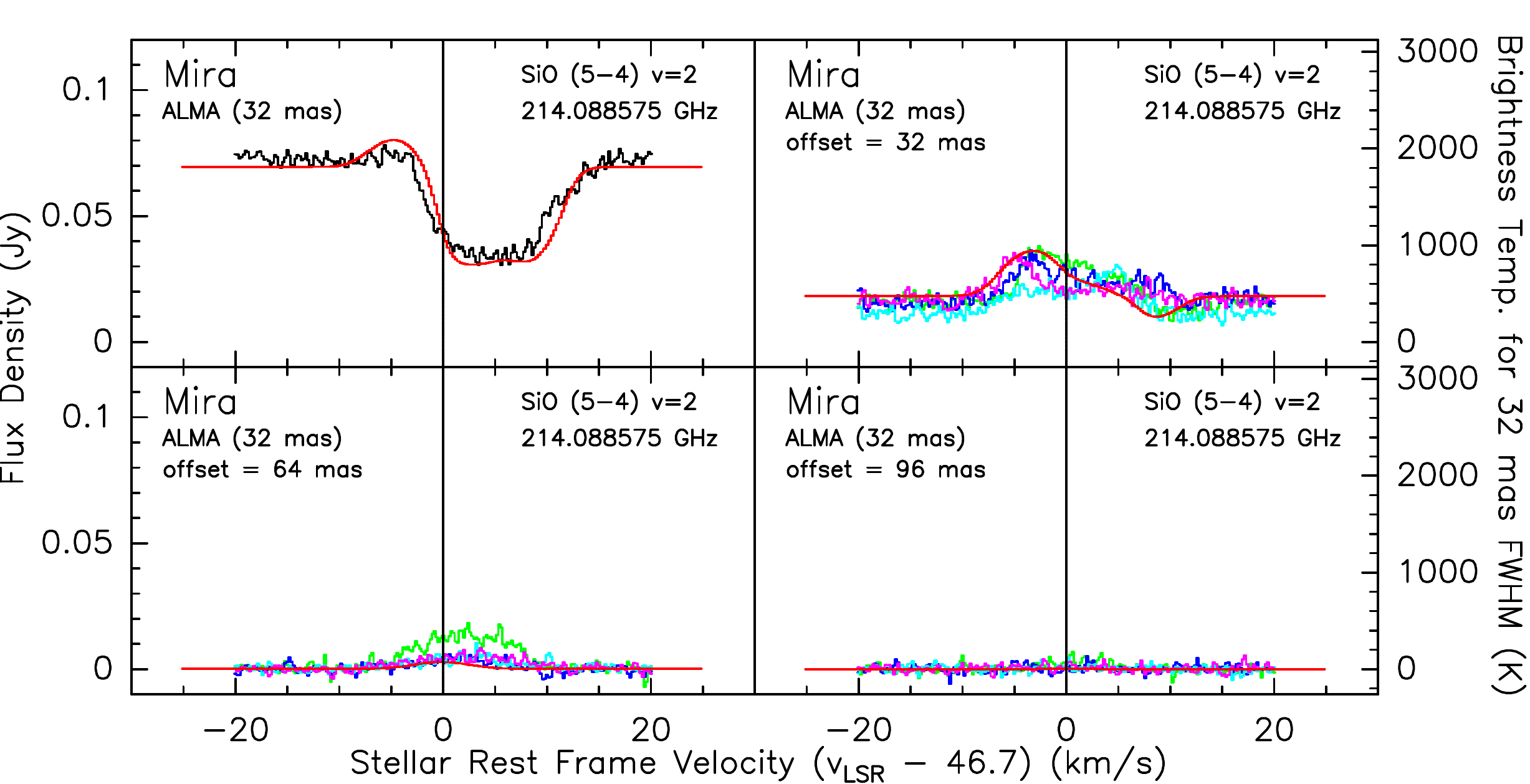}
\caption{Model 2: spectra of SiO ${\varv} = 2$ $J=5-4$ at various positions. The black histogram is the observed spectrum at the centre of continuum, green, blue, cyan and magenta histograms are the observed spectra along the eastern, southern, western and northern legs, respectively, at various offset radial distances as indicated in each panel. The red curves are the modelled spectra predicted by {\ratran}.}
\label{fig:m2siov2spec}
\end{figure*}
\begin{figure*}[!htbp]
\centering
\includegraphics[trim=1.0cm 7.3cm 2.0cm 1.5cm, clip, width=\multispecwidth]{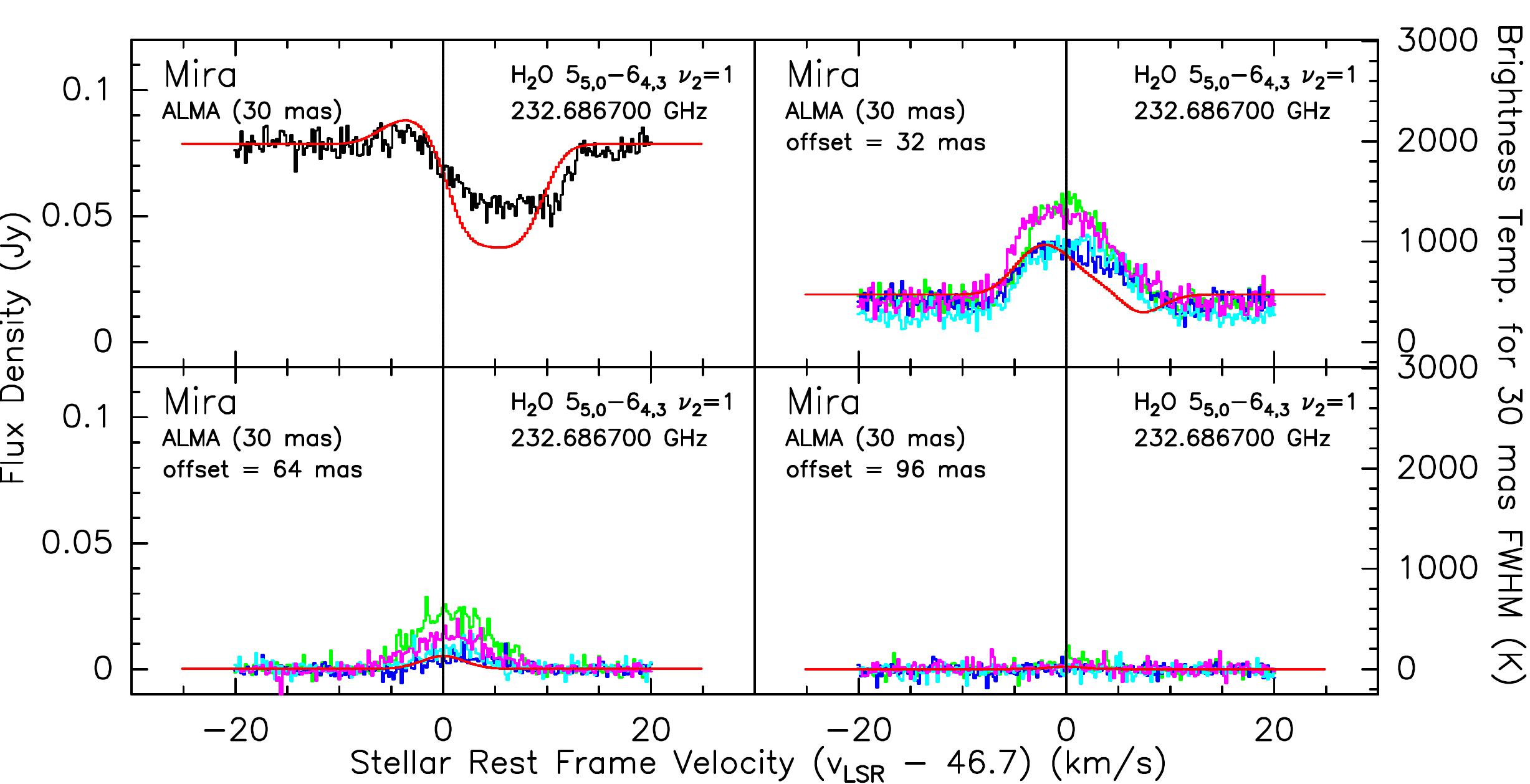}
\caption{Model 2: spectra of H$_2$O $v_2=1$ $J_{K_a,K_c}=5_{5,0}-6_{4,3}$ at various positions. The black histogram is the observed spectrum at the centre of continuum, green, blue, cyan and magenta histograms are the observed spectra along the eastern, southern, western and northern legs, respectively, at various offset radial distances as indicated in each panel. The red curves are the modelled spectra predicted by {\ratran}.}
\label{fig:m2h2ov1spec}
\end{figure*}

For the first postulation, we have constructed Model 2, which is identical to Model 1 with the only exception being that the gas temperature near the radio photosphere is significantly reduced from $2100\,{\rm K}$ in the original model to about $1400\,{\rm K}$. Figure \ref{fig:model2} shows the input gas temperature and the modelled excitation temperature profiles for the SiO and H$_2$O transitions. Figures \ref{fig:m2siov0spec}, \ref{fig:m2siov2spec} and \ref{fig:m2h2ov1spec} show the modelled spectra with reduced input gas temperature.

Our modelling results in Figs. \ref{fig:m2siov0spec}, \ref{fig:m2siov2spec} and \ref{fig:m2h2ov1spec} show that even if we reduce the kinetic temperature to $\lesssim 1400\,{\rm K}$ in the entire inner wind of Mira, there is still an excess in the blueshifted emission. In fact, we have found that the gas temperature near the radio photosphere has to be much lower than $1000\,{\rm K}$ in order to get rid of this excess emission. In addition, when the kinetic temperature in the proximity of the continuum is reduced, excess absorption relative to the observed spectra would appear in both the SiO and H$_2$O spectra. Although the excess absorption may be compensated by adopting different gas density and molecular abundance profiles, we do no consider the low-temperature atmosphere to be a likely solution.


We therefore speculate that, in the hot and dense part of the extended atmosphere of Mira (presumably at the innermost radius), there should be a relatively low-infall velocity, or even expanding, component contributing to the absorption in the blueshifted part of the spectra. In other words, the infall velocity the modelled region should not increase monotonically towards the star, but has to decelerate at some radii. In our preferred model (Model 3), we modify the velocity profile such that the motion of the bulk material switches from infall to expansion rapidly at $3.75 \times 10^{13}\,{\rm cm}$, which is just above our adopted radio photosphere ($R_{\rm continuum} = 3.60 \times 10^{13} {\rm cm}$). This expansion zone lies beneath the infall region and hence a large scale shock of shock velocity $\Delta V \lesssim 12\,\kms$ (constrained by the width of the line profile) could be created near this radius.


\end{appendix}


\end{document}